\documentclass[11pt,a4paper]{article} 
\usepackage{jheppub,multirow}
\usepackage{amsthm}
\usepackage{xcolor}
\usepackage{float}
\usepackage{url}
\usepackage{microtype}
\usepackage{subcaption}
\usepackage{empheq}
\usepackage{booktabs}  
\usepackage[makeroom]{cancel}
\usepackage{dsfont} 
\usepackage{ytableau} 
\usepackage{setspace}
\usepackage[skins]{tcolorbox}

\newcommand{\beq}{\begin{equation}}
\newcommand{\eeq}{\end{equation}}

\newcommand{\fg}{\mathfrak{g}}

\newcommand{\fm}{\mathfrak{m}}
\newcommand{\fq}{\mathfrak{q}}

\newcommand{\cC}{\mathcal{C}}
\newcommand{\cN}{\mathcal{N}}

\newcommand{\cW}{\mathcal{W}}

\newcommand{\cY}{\mathcal{Y}}

\def\ie{\begin{equation}\begin{aligned}}
\def\fe{\end{aligned}\end{equation}}

\newtheoremstyle{fullit}
{\topsep}      
{\topsep}      
{\normalfont}  
{0pt}          
{\itshape}     
{.\ }          
{0pt}          
{\thmname{#1} \thmnumber{#2}}             

\newtheorem{thm}{Theorem}[section]

\theoremstyle{definition}

\theoremstyle{fullit}
\newtheorem{example}[thm]{Example}

\DeclareCaptionSubType*[alph]{figure}
\title{Supersymmetric Wilson Loops, Instantons, and Deformed ${\cal W}$-Algebras}
\author{Nathan Haouzi$^a$,}
\author{Can Koz\c{c}az$^b$}
	\affiliation{$^a$Simons Center for Geometry and Physics, State University of New York,
		Stony Brook, NY 11794}
	\affiliation{$^b$Department of Physics, Bo\u{g}azi\c{c}i University, Istanbul, Turkey}
	\emailAdd{nhaouzi@scgp.stonybrook.edu,can.kozcaz@boun.edu.tr}

\abstract{Let $\fg$ be a simple Lie algebra. We study 1/2-BPS Wilson loops of supersymmetric 5d $\fg$-type quiver gauge theories on a circle, in a non-trivial instanton background. The Wilson loops are codimension 4 defects of the gauge theory, and their interaction with self-dual instantons is captured by a modified 1d ADHM quantum mechanics. We compute the partition function as its Witten index. This index is a  ``$qq$-character" of a finite-dimensional irreducible representation of the quantum affine algebra $U_q(\widehat{\fg})$. Using gauge/vortex duality, we can understand the 5d physics in 3d gauge theory terms. Namely, we reinterpret the 5d theory with vortex flux from the point of view of the vortices themselves. This vortex perspective has an advantage: it has yet another dual description in terms of deformed $\fg$-type Toda Theory on a cylinder, in free field formalism. We show that the gauge theory partition function is equal to a chiral correlator of the deformed Toda Theory, with stress tensor and higher spin operator insertions. 
We derive all the above results from type IIB string theory, compactified on a resolved $ADE$ singularity times a cylinder with punctures, with various branes wrapping the blown-up 2-cycles.}
	
\begin{document}
		\maketitle
		\setlength{\parindent}{0pt}
		\clearpage


\newpage

\section{Introduction}

\subsection{1/2-BPS Wilson Loops and Instantons}

An important class of observables in gauge theory is called the Wilson loop.
It is a non-local and gauge-invariant operator, which encodes essential aspects of the strongly-coupled regime. A Wilson loop is formulated as the trace of a holonomy matrix, where a quark is parallel transported along a closed curve in spacetime, and the trace is evaluated in a given representation of the gauge group. The vacuum expectation of such a loop then gives the phase shift of the quark wavefunction. \\

A natural question to ask is what happens when the quark moves in a non-trivial instanton background. Instantons are solutions of the self-dual Yang Mills equations on $\mathbb{R}^4$, and a powerful way to identify such solutions is the celebrated ADHM construction \cite{Atiyah:1978ri}. 
Adding Wilson lines, one expects this construction to be generalized, since when an instanton moves in the presence of a quark, it now also experiences a Lorentz force; this is because a quark is electrically charged under the gauge field, while an instanton is magnetically charged. The endeavor of understanding the dynamics of instantons in this modified background was initiated recently with supersymmetry \cite{Tong:2014yna,Tong:2014cha}.\\

Instantons and  Wilson loops both admit realizations in string theory. For definiteness, consider maximal supersymmetric Yang-Mills (SYM) in 4+1 dimensions, with gauge group $U(n)$. This theory appears as the low energy effective field theory on $n$ D4 branes in type IIA. There, $k$ instantons are realized as $k$ D0 branes nested inside the D4 branes \cite{Douglas:1995bn}. Meanwhile, supersymmetric Wilson loops were first analyzed in a stringy picture in the works \cite{Maldacena:1998im,Rey:1998ik}, in the context of holography; namely, a loop in the first fundamental representation of $U(n)$ is described as a fundamental string whose worldsheet ends at the loop, located at the boundary of $AdS$. Later, a description of the loops was given in terms of branes instead \cite{Drukker:2005kx,Gomis:2006sb,Yamaguchi:2006tq}, allowing for loops in more general representations. In particular, one defines a 1/2-BPS Wilson loop of 4+1 SYM using additional D$4'$ branes orthogonal to the original $n$ D4 branes. After integrating out the degrees of freedom associated to the D$4'$ branes, the path integral becomes the generating function of Wilson loop vevs, valued in irreducible representations of $U(n)$.

Naturally, the string theory setup that captures the dynamics of instantons in the presence of such Wilson loops is simply the superposition of the above two configurations: the resulting ``modified" ADHM prescription amounts to studying the one-dimensional quantum mechanics of the D0 branes nested inside D4 branes, in the presence of orthogonal D$4'$ branes. The instanton partition function is then the Witten index of this quantum mechanics.\\

In this paper, we aim to study 1/2-BPS Wilson loops in the instanton background of a large class of five-dimensional quiver gauge theories, which we denote as $T^{5d}$. These theories are labeled by a simple Lie algebra $\fg$, have $\cN=1$ supersymmetry, and are defined on the manifold $S^1\times \mathbb{C}^2$. The gauge group of $T^{5d}$ is a product of unitary groups, and the shape of the quiver is that of the Dynkin diagram of $\fg$. The Wilson loops will wrap the circle $S^1$, and sit at the origin of $\mathbb{C}^2$. Our first result is the following: 

\begin{tcolorbox}
The instanton partition function $\left[\chi^{\fg}\right]^{5d}$ of the quiver gauge theory $T^{5d}$ in the presence of a Wilson loop can be computed as the Witten index of a modified ADHM quantum mechanics. We provide an integral representation for the index, and explicitly evaluate it.
\end{tcolorbox}
	
This requires  specifying the contours, which we do using a generalized Jeffrey-Kirwan residue prescription. For previous work in this direction, the case $\fg=A_1$ was studied in \cite{Kim:2016qqs}, and some computations in the case $\fg=A_m$ appeared in \cite{Assel:2018rcw}. The case $\fg=A_1$ subjected to an orientifold projection was analyzed in \cite{Chang:2016iji}.\\

Our second result is to derive the 5d quivers and Wilson loops from string theory (in a setup $T$-dual to the one reviewed above). Namely, we study type IIB string theory on $X\times\cC\times\mathbb{C}^2$, where $X$ is a resolved $ADE$ singularity, and $\cC$ is a cylinder. The 5d quiver gauge theory description emerges after introducing D5 branes wrapping compact 2-cycles of $X$, following the seminal work \cite{Douglas:1996sw}. The Wilson loops are realized as D1 branes wrapping non-compact 2-cycles of $X$. Non simply-laced theories will come about from considering a non-trivial fibration of $X$ over $\cC\times\mathbb{C}^2$. 
We manually decouple gravity and retain only the degrees of freedom supported near the origin of $X$ by sending the string coupling to 0, $g_s\rightarrow 0$; this limit is also known as the $(2,0)$ little string theory. We can therefore phrase our analysis in a purely stringy perspective, as follows: 

\begin{tcolorbox}
$\left[\chi^{\fg}\right]^{5d}$ is the partition function of the IIB little string on $\cC\times\mathbb{C}^2$, in the presence of codimension 2 defects (D5 branes) and point-like defects (D1 branes). The $\fg$-type quiver gauge theory lives on D5 branes, while the Wilson loop is realized as D1 branes.
\end{tcolorbox}

The D1 branes are located at points on the cylinder $\cC$, and as we will see, the coordinates of these points are (complexified) masses for the loop quarks.\\

Two important remarks are in order. 
First, independently of the above results, Nekrasov introduced the ADHM of so-called crossed instantons \cite{Nekrasov:2015wsu}, in a setup intimately related to ours. There, the author computes the instanton partition function of 3+1 SYM in the presence of point defects, for a class of quiver gauge theories labeled by a simply-laced Lie algebra. In that work, such a partition function is nicknamed a $qq$-character of $ADE$ type, a generalization of the usual characters which appear in the study of the representation theory of Yangians.\\

Our results can be understood as a K-theoretic lift of that construction. In particular, we will see that when $\fg$ is an arbitrary simple Lie algebra, not necessarily simply-laced, 
\begin{tcolorbox} 
$\left[\chi^{\fg}\right]^{5d}$ is a $qq$-character of a finite-dimensional irreducible representation of $U_q(\widehat{\fg})$.
\end{tcolorbox} 
Taking the size of the circle $S^1$ to zero, and specializing to the case where $\fg$ is simply-laced,  we recover the $ADE$ $qq$-characters of \cite{Nekrasov:2015wsu}; see also the work \cite{Kimura:2015rgi}.\\

Recently, a generalization of the characters has been proposed for formal ``fractional" quiver gauge theories \cite{Kimura:2017hez}\footnote{Quivers labeled by a simple Lie algebra are a subset of fractional quivers, which can have an arbitrary high lacing number $r$. These are mathematically well-defined, but physics imposes some restrictions on which theories are allowed: we can have $r=1$ (simply-laced case), $r=2$ ($B_N, C_N, F_4$), or $r=3$ ($G_2$). Certain twisted affine algebras can arise as well in Physics, and they can perfectly be studied using the formalism we develop in this paper, though we leave the explicit analysis to future work.}. 
Our construction can be seen as the string theory realization of the purely mathematical arguments invoked there, in the case where the fractional quiver is labeled by a simple Lie algebra. The formulas we present are still applicable to compute formal partition functions for arbitrary fractional quivers, but because the purpose of this work is to provide a physical construction, we will limit ourselves to the case where $\fg$ labels a simple Lie algebra.

\subsection{A 3d Interpretation}

It turns out that we can reinterpret the 5d physics in 3d gauge theory terms. This comes about by studying the vortices of $T^{5d}$, and realizing that the theory has an effective description on the vortices themselves, which we call $G^{3d}$. This phenomenon is known as gauge/vortex duality. $G^{3d}$ is a 3d quiver gauge theory which enjoys $\cN=2$ supersymmetry, defined on $S^1\times \mathbb{C}$. The shape of the quiver is again the Dynkin diagram of $\fg$, just like the parent theory $T^{5d}$.\\

We want to elucidate the role played by the Wilson loops in the vortex theory $G^{3d}$. We show the following: 

\begin{tcolorbox} 
The partition function $\left[\chi^{\fg}\right]^{3d}$ of the vortex theory $G^{3d}$ in the presence of a Wilson loop is a ``3d $qq$-character" of some finite-dimensional irreducible representation of  $U_q(\widehat{\fg})$. We define such a character from the 3d gauge theory and compute it. 
\end{tcolorbox}

In particular, we show that the 3d $qq$-character can be understood as a truncation of the 5d partition function at values of the Coulomb moduli tuned to some hypermultiplet masses\footnote{In a related context, a \emph{two-dimensional} $qq$-character was defined in \cite{Nekrasov:2017rqy}, again related to our three-dimensional partition function (in the simply-laced case) by a circle reduction. However, we want to keep the circle size finite here, since the 3d perspective has a crucial feature: as we will see, it has a dual description in terms of observables in a so-called deformed ${\cW}$-algebra theory on the cylinder.}.\\

In string theory,  vortices are D3 branes which are points on the cylinder $\cC$, and wrap compact 2-cycles of the resolved singularity $X$. The low energy gauge theory on these branes is precisely $G^{3d}$. From a string theory standpoint, then, $\left[\chi^{\fg}\right]^{3d}$ is the partition function of the theory on D3 branes in the $(2,0)$ little string, in the presence of D5 and D1 branes.\\

We use this vortex perspective to make contact with certain chiral algebras defined on the cylinder $\cC$, called ${\cW}(\fg)$-algebras. These algebras, labeled by a simple Lie algebra $\fg$\footnote{${\cW}(\fg)$-algebras are also labeled by a choice of a nilpotent orbit, which in this paper will always be the maximal one.}, realize the symmetry of Toda theory. The particular case $\fg=A_1$ is also called Liouville theory, which enjoys Virasoro symmetry. When $\fg\neq A_1$, the Virasoro stress tensor is still present in the theory, but there are also higher spin currents. That such 2d theories should be related at all to 4d $\cN=2$  theories is a formidable conjecture first made precise by Alday-Gaiotto-Tachikawa \cite{Alday:2009aq}. More precisely, the partition function of a 4d ${\cN}=2$ theory whose origin is a 6d $(2,0)$ superconformal field theory (SCFT) compactified on a punctured Riemann surface $\cC$, is predicted to be equal to the conformal block of the 2d Toda CFT on this Riemann surface.
In the 90's, Frenkel and Reshetikhin introduced a two-parameter deformation of the ${\cW}$-algebras, denoted as ${\cW}_{q,t}(\fg)$ \cite{Frenkel:1998}; we will refer to them as deformed ${\cW}$-algebras. Crucially, while an ordinary ${\cW}$-algebra has conformal symmetry, its deformation does not. The deformed conformal blocks are defined in the free field formalism, as integrals over the positions of some deformed screening currents on  $\cC$.\\

As was first shown in \cite{Aganagic:2013tta}, these algebras happen to be naturally related to the quiver gauge theories $T^{5d}$ and $G^{3d}$ under study. Namely, in the absence of Wilson loops and at \textit{integer} values of the Coulomb moduli, the partition function of $T^{5d}$ becomes a conformal block of deformed $\fg$-type Toda, with insertion of certain vertex operators at points on the cylinder. Gauge/vortex duality therefore implies that the partition function of the 3d theory $G^{3d}$ is in fact equal to a deformed conformal block.  This 5d/3d duality is the $(2,0)$ little string version of the AGT correspondence \cite{Aganagic:2015cta}.
Because of the free field formalism, the conformal blocks are constrained to have some integral momenta. This is not so much a restriction as it is a key feature, since these integers play the physical role of the ranks of the gauge groups in the quiver $G^{3d}$.\\

What happens to the deformed Toda conformal blocks when we include Wilson loops in the gauge theory? The answer we find takes an elegant form: 
a Wilson loop is realized as the insertion of a generating current operator inside a deformed Toda correlator. These operators are nothing but the deformed stress tensor and higher spin currents of the ${\cW}_{q,t}(\fg)$ algebra, and they are constructed in free field formalism as the commutant of the screening currents. There are ${\text{rank}}({\fg})$ independent generators constructed in this way, with spin $2\leq s \leq {\text{rank}}({\fg})+1$.\\

\begin{tcolorbox} 
We compute the ${\cW}_{q,t}(\fg)$-algebra  correlators and show, after proper normalization, that they are equal to the vortex partition function $\left[\chi^{\fg}\right]^{3d}$. More precisely, the expectation value of the spin $s$ generating current is equal to the $(s-1)$-th fundamental 3d $qq$-character of $U_q(\widehat{\fg})$.  
\end{tcolorbox} 

In this way, we obtain a triality of relations between the Wilson loop physics of 5d gauge theories, 3d gauge theories, and ${\cW}_{q,t}(\fg)$ algebras.\\

The various actors appearing in deformed Toda correlators again have a natural interpretation in string theory, which follows from the chain of dualities: The D5 branes are vertex operators labeled by a collection of coweights of $\fg$, or equivalently, labeled by a collection of weights of the Langlands dual algebra $^L\fg$. The D3 branes are the screening charges, and the D1 branes are the stress tensor and higher spin currents of the algebra.

Note that it was anticipated in \cite{Kimura:2017hez} that the partition function of $T^{5d}$ with defects should equal some ${\cW}_{q,t}(\fg)$-algebra correlator. However, no 3d gauge theory interpretation was possible there, since the number of screening charges (which in our dictionary is the rank of the 3d gauge groups) was formally infinite in that work. Specializing our results to an infinite number of screenings, and further setting our vertex operators to be trivial, we lose the 3d gauge theory interpretation, and recover the result of \cite{Kimura:2017hez}.

\subsection{Outline}

The paper is organized as follows: in Section 2, we compute the instanton partition function of the 5d $\fg$-type quiver gauge theory with a 1/2-BPS Wilson loop insertion. We motivate the construction from type IIB string theory. In Section 3, we study gauge/vortex duality and reinterpret the partition function in the language of a 3d  $\cN=2$ quiver gauge theory. In Section 4, we make contact with the deformed algebra $\cW_{q,t}(\fg)$: we show that the 3d vortex partition function of section 3 is a free-field correlator of the $\cW_{q,t}(\fg)$-algebra, with insertion of generating current operators.  In Section 5, we present a variety of explicit examples to showcase our results.

\newpage

\section*{Notations}

In this section, we collect the notations, conventions and some definitions used later in the paper.\\

$m$ is the rank of the simple Lie algebra $\fg$.

$r$ is the lacing number of $\fg$: the maximum number of arrows linking two adjacent nodes in the Dynkin diagram of $\fg$.

$\alpha_a$ is the $a$-th positive simple root of $\fg$.

$\lambda_a$ is the $a$-th fundamental weight of $\fg$.

$\alpha^\vee_a$ is the $a$-th positive simple coroot of $\fg$. The simple coroots of $\fg$ are defined through the relation $\alpha_a^{\vee}=2\alpha_a/\langle\alpha_a,\alpha_a\rangle$. They are dual to the fundamental weights: $\langle \lambda_a, \alpha_b^{\vee}\rangle=\delta_{ab}$.

$\lambda^\vee_a$ is the $a$-th fundamental coweight of $\fg$. The fundamental coweights of $\fg$ are dual to the simple roots: $\langle \lambda_a^{\vee}, \alpha_b\rangle=\delta_{ab}$.

The Cartan matrix of $\fg$ is defined as $C_{ab}=\langle\alpha_a,\alpha_b^{\vee}\rangle$.

Square length of a long root: $\langle\alpha_a,\alpha_a\rangle=2$.

Square length of a short root: $\langle\alpha_a,\alpha_a\rangle=2/r$.

$r^{(a)}\equiv r\,\langle\alpha_a, \alpha^\vee_a\rangle/2$. Put differently, $r^{(a)}= 1$ if the node $a$ labels a short root, and $r^{(a)}= r$ if the node $a$ labels a long root.

$r^{(a b)}\equiv\text{gcd}(r^{(a)}, r^{(b)})$, the greatest common divisor of $r^{(a)}$ and $r^{(b)}$. For two adjacent nodes $a$ and $b$, $r^{(ab)}=1$ if either node $a$ or node $b$ denotes a short root; otherwise, $r^{(ab)}=r$. 

Definition of the $q$-Pochhammer symbol: $\left( x\, ; q\right)_{\infty}=\prod_{l=0}^{\infty}\left(1-q^l\, x\right)$.

Definition of the theta function: $\Theta\left(x\,;q\right)= \left(x \,;\, q\right)_\infty\,\left(q/x \,;\, q\right)_\infty$.

$v^{(a)}\equiv\sqrt{q^{r^{(a)}}/t}$

$v^{(ab)}\equiv\sqrt{q^{r_{ab}}/t}$

${\overline{r^{(ab)}}}\equiv r$ if node $a$ labels a short root \textit{and} node $b$ labels a long root, and ${\overline{r^{(ab)}}}\equiv 1$ otherwise.

$f(x)\equiv 2 \sinh\left(x/2\right)$

$\Delta_{a,b}$ is the upper-diagonal incidence matrix.

$X$ is a resolved $ADE$ singularity.

$n$ is the total number of D5 branes wrapping the compact 2-cycles of $X$.

$N_f$ is the total number of D5 branes wrapping the non-compact 2-cycles of $X$.

$D$ is the total number of  D3 branes wrapping the compact 2-cycles of $X$.

$L$ is the total number of D1 branes wrapping the non-compact 2-cycles of $X$.

\newpage

\section{The $\fg$-type quiver gauge theory $T^{5d}$ and Wilson Loops}
\label{sec:5dtheory}

\subsection{String Theory Construction}
\label{ssec:string5d}

We start by considering the ten-dimensional type IIB string theory compactified on a complex surface $X$, where $X$ is a resolution of a singularity of type ${\mathbb C}^2/\Gamma$; here, $\Gamma$ is one of the discrete subgroups of $SU(2)$, and the McKay correspondence guarantees that such a discrete subgroup is labeled by one of the simply-laced Lie algebras ${\fg=A, D, E}$, of rank $m=\text{rank}(\fg)$. It is well known that the resolved surface $X$ is a hyperk\"ahler manifold. Explicitly, the singularity is resolved by being blown up: one obtains 2-spheres $S_a$, $a=1,\ldots,m$, that organize themselves in the shape of the Dynkin diagram of $\fg$.
Furthermore, we decouple gravity and focus only on the degrees of freedom supported near the origin of $X$ by sending the string coupling to $g_s\rightarrow 0$. In this limit, type IIB string theory on $X$ is referred to as the six-dimensional $(2,0)$ little string theory of type $\fg$. The $(2,0)$ little string is labeled by an $\fg=ADE$ Lie algebra. It is not a local QFT; the little strings have finite tension $m_s^2$, the square of the string mass. Taking further the limit $m_s\rightarrow \infty$, we lose the one scale of the theory and end up with a $(2,0)$ SCFT, labeled by the same Lie algebra $\fg$;  see \cite{Seiberg:1997zk,Losev:1997hx} and \cite{Aharony:1999ks} for a review.  
Its moduli space is 
\beq
\left(\mathbb{R}^4\times S^1\right)^{m}/W(\fg)\; ,
\eeq
where $W(\fg)$ is the Weyl group of $\fg$. The moduli come from periods of various 2-forms along the 2-cycles $S_a$ of the resolved singularity $X$: the $S^1$ modulus is the scalar obtained the R-R 2-form $C^{(2)}$ of the ten-dimensional type IIB string theory, integrated on $S_a$. The $\mathbb{R}^4$ moduli come from the NS-NS B-field $B^{(2)}$, and a triplet of self-dual 2-forms $\omega_{I,J,K}$, which arise since $X$ is a hyperk\"ahler manifold. To get the correct NS-NS and R-R normalizations, we need to recall the low energy action of the type IIB superstring. In particular, note that the R-R field should not be accompanied by any power of $g_s$. Furthermore, the mass dimension of a scalar in a theory of 2-forms should be 2. We therefore obtain
\beq\label{periods}
\frac{m_s^4}{g_s}\int_{S_a}\omega_{I,J,K}\, ,\qquad \frac{m_s^2}{g_s}\int_{S_a}B^{(2)}\, , \qquad m_s^2\int_{S_a}C^{(2)}\, .
\eeq

When we take the limit $g_s\rightarrow 0$, we require that the above periods remain fixed.\\

To make contact with lower dimensional physics, we further compactify the type IIB string theory on a Riemann surface $\cC$. It is necessary that $\cC$ has a flat metric, or else the space $X\times\cC$ would not be a solution of type IIB string theory. In this paper, we fix the Riemann surface to be an infinite cylinder of radius $R$, namely $\cC= \mathbb{R} \times S^1(R)$.
As is, this background preserves 16 supercharges, which is too much supersymmetry  to produce interesting dynamics.
For instance, bosons and fermions are paired up in such a way that they cancel out in a supersymmetric index. To produce non-trivial dynamics, we need to break supersymmetry, at least locally. A natural way to achieve this in type IIB string theory is by adding various D-branes. In our setup, we can wrap branes around various 2-cycles of the resolved singularity in $X$. For our purpose, we will be interested in a certain configuration of D5 and D1 branes wrapping 2-cycles, which we now turn to.

\subsubsection{5d $ADE$ Quiver Theories}
\label{sssec:simplylaced}

To be more quantitative, we need to introduce some notations: According to the McKay correspondence, the second homology group $H_2(X, \mathbb{Z})$ of $X$ is identified with the root lattice $\Lambda$ of $\fg$. As we have briefly discussed, $H_2(X, \mathbb{Z})$ is spanned by $m$ 2-cycles $S_a$ that blow up the singularity, and correspondingly,  $\Lambda$ is spanned by positive simple roots $\alpha_a$. The intersection pairing in homology is further identified with the Cartan Killing metric of $\fg$, up to a sign:
\beq
\# (S_a \cap S_b)=-C_{ab}\; ,
\eeq
where $C_{ab}$ is the Cartan matrix of $\fg$.
We introduce a total of $n$ D5 branes wrapping the compact 2-cycles of $X$ and $\mathbb{C}^2$. This results in a net non-zero D5 brane charge, measured by a class $[S]\in H_2(X, \mathbb{Z})$. We expand $[S]$ in terms of simple roots as
\beq\label{compact}
[S] = \sum_{a=1}^m  \,n^{(a)}\,\alpha_a\;\;  \in  \,\Lambda \; ,
\eeq
with $n^{(a)}$  non-negative integers.\\

We also need to consider the second relative homology group $H_2(X, \partial X, \mathbb{Z})$. Its elements are 2-cycles of $X$ that can have a boundary at infinity $\partial X$. The group is spanned by non-compact 2-cycles $S_a^*$, with $a=1,\ldots, m$. Each $S_a^*$ is constructed as the fiber of the cotangent bundle $T^*S_a$ over a generic point on the $a$-th 2-sphere $S_a$. Then, we have
\beq
\# (S_a \cap S^*_b)=\delta_{ab}\; .
\eeq
The group $H_2(X, \partial X, \mathbb{Z})$ is identified with the weight lattice $ \Lambda_*$ of $\fg$; correspondingly, the 2-cycle $S^*_a$ can be understood as the $a$-th fundamental weight $\lambda_a$ of $\fg$.
Note that $H_2(X, \mathbb{Z})\subset H_2(X, \partial X, \mathbb{Z})$, since compact 2-cycles can be understood as elements of $H_2(X, \partial X, \mathbb{Z})$ with trivial boundary at infinity. This is just the homological version of the familiar statement  that the root lattice of $\fg$ is a sublattice of the weight lattice, $\Lambda\subset \Lambda_*$.\\

\begin{figure}[h!]
	\emph{}
	\centering
	\includegraphics[trim={0 0 0 3cm},clip,width=0.8\textwidth]{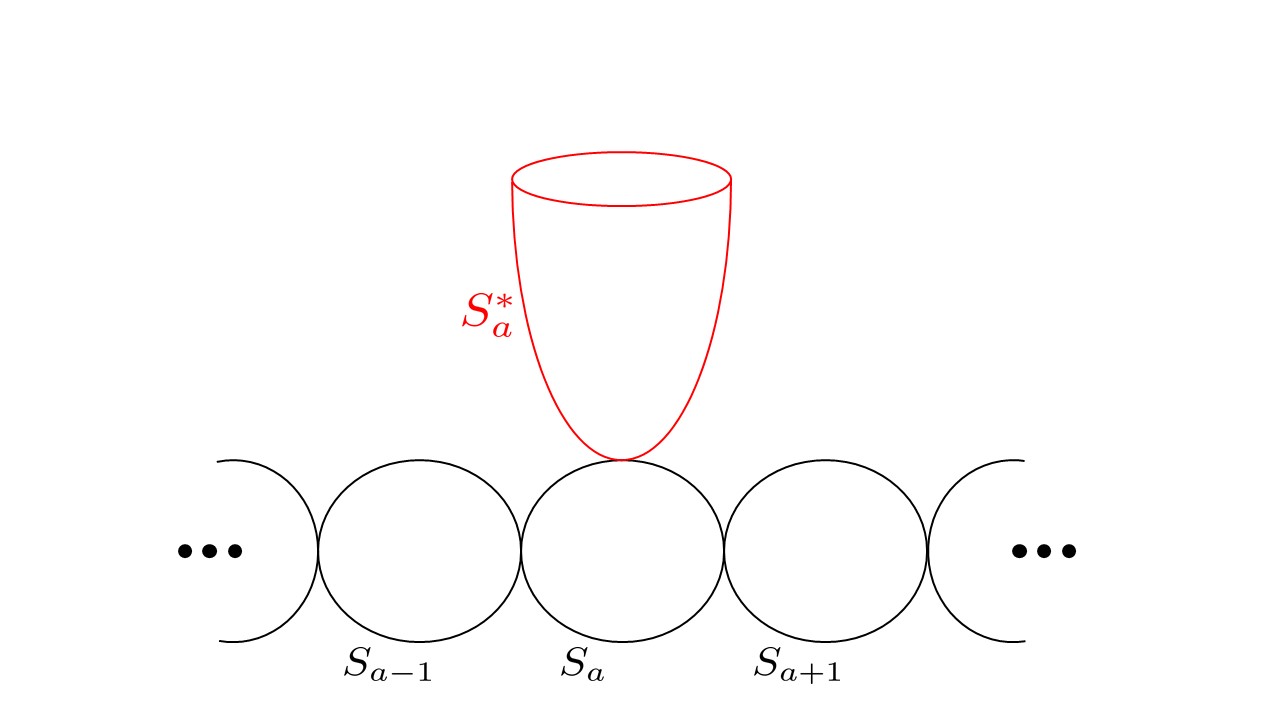}
	\vspace{-1pt}
	\caption{A vanishing 2-cycle of an $A_m$ singularity, labeled by $S_a$ (the black 2-sphere), and the dual non-compact 2-cycle $S_a^*$ (the red cigar).} 
	\label{fig:quivercycles}
\end{figure}

Next, we consider a total of $N_f$ D5 branes wrapping non-compact 2-cycles in $X$, along with $\mathbb{C}^2$. The charge for these branes is measured by a class $[S^*]\in H_2(X, \partial X, \mathbb{Z})$. We expand $[S^*]$ in terms of fundamental weights as
\beq\label{noncompact}
[S^*] = - \sum_{a=1}^{m} \, N^{(a)}_f \, \lambda_a \;\;  \in\, \Lambda_*\; ,
\eeq  
where $N^{(a)}_f$ are non-negative integers called the Dynkin labels. In this basis, the fundamental weight $\lambda_a$ is conveniently written as an $m$-sized vector, with the entry 1 in the $a$-th entry and 0 everywhere else: $\lambda_a=[0,\ldots , 0, 1, 0, \ldots, 0]$. We can therefore rewrite \eqref{noncompact} as the vector  $[S^*]=-[N^{(1)}_f, N^{(2)}_f, N^{(3)}_f, \ldots, N^{(m)}_f]$. Of course, a compact D5 brane class $[S]$ can also be written in this basis, with the caveat that some of its Dynkin labels may be negative integers.

The weight $w$ can be identified with a class $[S^*]$ in the second relative homology group $H_2(X, \partial X, \mathbb{Z})$, by the McKay correspondence. Our notation will not differentiate between a weight and a class in $H_2(X, \partial X, \mathbb{Z})$. For instance, we write $[S_a^*]=- \lambda_a$ for (minus) the $a$-th fundamental weight of $\fg$. 

The total D5 brane charge is then $[S+S^*]$, understood as a class in the weight lattice.\\

Lastly, we consider a total of $L$ D1 branes wrapping the non-compact 2-cycles in $X$, and sitting at the origin of $\mathbb{C}^2$. The charge for these D1 branes is  measured by a class $[S^*_{D1}]\in H_2(X, \partial X, \mathbb{Z})$, expanded in terms of fundamental weights as:
\begin{equation}
[S^*_{D1}] = - \sum_{a=1}^{m} \, L^{(a)} \, \lambda_a \;\;  \in\, \Lambda_*\; ,
\end{equation} 
where $L^{(a)}$ are non-negative integers.\\

\begin{figure}[h!]
	\emph{}
	\centering
	\includegraphics[width=1.0\textwidth]{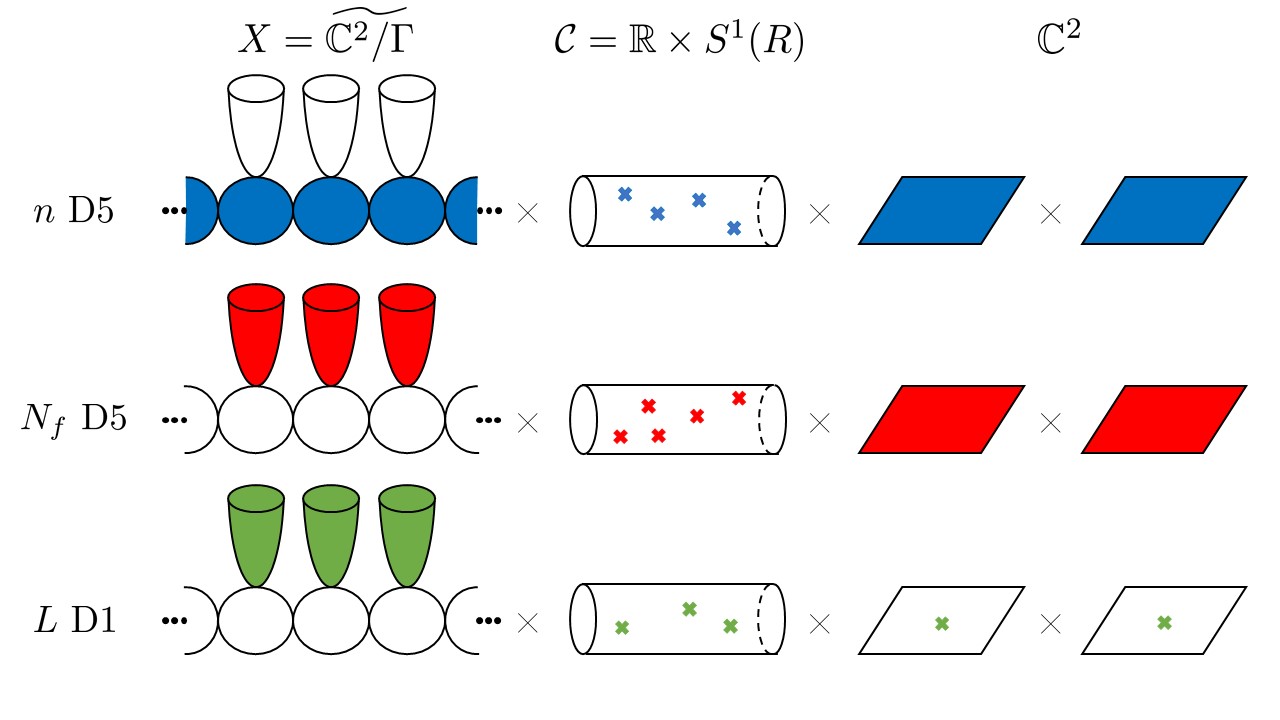}
	\vspace{-10pt}
	\caption{Brane configuration: there are $n$ D5 branes wrapping compact 2-cycles $S_{a}$'s (blue),  $N_{f}$ D5 branes wrapping non-compact 2-cycles $S_{a}^{*}$'s (red). All D5 branes are points on the cyclinder $\cC$ and extend in $\mathbb{C}^2$. There are also $L$ D1 branes wrapping the non-compact 2-cycles $S_{a}^{*}$'s (green), sitting at the origin of $\mathbb{C}^2$. All branes are points on the cylinder. Later, we will consider the quantum mechanics of $k$ D1$_{inst}$ branes (not pictured) wrapping the compact 2-cycles $S_{a}$'s.} 
	\label{fig:branes5d}
\end{figure}

Is supersymmetry preserved at all? To answer that, first recall we have defined in the last section \eqref{periods} the periods of a triplet $\vec{\omega}=(\omega_I, \omega_J, \omega_K)$ of self-dual 2-forms. The non-compact D1 and D5 branes can preserve supersymmetry only if the vectors $\int_{S^*_a} \vec{\omega}$ point in the same direction, for all $a=1, \ldots, m$. We can always choose  $\vec{\omega}$ to satisfy,  for all $a$,
\beq
\int_{S^*_a}\omega_{I}>0\, ,\qquad \int_{S^*_a}\omega_{J}=0\, ,\qquad \int_{S^*_a}\omega_{K}=0\,  .
\eeq
Likewise, the supersymmetry preserved by the D5 branes wrapping the compact 2-cycles is determined by the periods of the 2-forms through the 2-cycles $S_a$, corresponding to the choice of a metric on $X$. For all $a=1, \ldots, m$, we choose
\beq\label{couplings}
\tau^{(a)}\equiv \int_{S_a}\left(\frac{m^2_s}{g_s}\,\omega_{I}+i\, C^{(2)}\right)\, ,\qquad \int_{S_a}\omega_{J}=0\, ,\qquad \int_{S_a}\omega_{K}=0\;  ,
\eeq
along with
\beq\label{NSNS}
\int_{S_a}B^{(2)}=0\; .
\eeq
For our purposes, it will be important that the complex numbers $\tau^{(a)}$ have $\text{Re}(\tau^{(a)})>0$\footnote{This is because we will carry out an instanton expansion in the next section; this condition ensures the convergence of the series.}.
With our choice of periods, the D5 branes wrapping the compact and non-compact 2-cycles break the same half supersymmetry, resulting in 8 supercharges. Introducing the D1 branes further breaks half the supersymmetry, so only 4 supercharges are preserved in total.\\

We end this section by mentioning that our brane construction has appeared in various forms in the recent literature, related by string dualities. Most notably, performing two T-dualities, one finds the configuration of branes \ref{table:nekrasovtable}. 
That setup was first studied in our context in \cite{Nekrasov:2015wsu}. There, D(-1) branes bound to either stack of D3 branes are referred to as crossed instantons. It is argued that after integrating out the degrees of freedom due to the D$3'$ branes, one ends up with a stack of D3 branes with point-like defects on them. The low-energy theory on the D3 branes is then a 4d $\cN=2$ quiver gauge theory of $ADE$-shape, with $L$ point defects. The D7 branes encode the flavor symmetry of the gauge theory. The instanton partition function of the system was nicknamed an $ADE$ $qq$-character. We will come back to this terminology in the next section.
\begin{table}
\centering
	\begin{tabular}{|l|c|c|c|c|c|c|c|c|c|r|}
		\hline
		& 0 & 1 & 2 & 3 & 4 & 5 & 6 & 7 & 8 & 9 \\ \hline
		$ADE$ & & & & & & & $\times$ & $\times$ & $\times$ & $\times$ \\ \hline
	$n$	D3 & $\times$ & $\times$ & $\times$ & $\times$ & & & & & & \\ \hline
	$N_f$	D7 & $\times$ & $\times$ & $\times$ & $\times$ & & &  $\times$ & $\times$ & $\times$ & $\times$ \\ \hline
	$L$	D$3'$ & & & & & \phantom{$\times$} & \phantom{$\times$} & $\times$ & $\times$ & $\times$ & $\times$ \\ \hline
	\end{tabular}\caption{The brane configuration appearing in \cite{Nekrasov:2015wsu}, related to ours by two T-dualities. Crossed instantons are realized as D(-1) branes.}
\label{table:nekrasovtable}
\end{table}

The case $\fg=A_m$ with $m\geq 1$ was also studied in the literature using $(p,q)$-webs of 5-branes, again related by two T-dualities to our setup. Namely, the action of the type IIB S-duality action on the web, also called fiber-base duality, was initiated in the presence Wilson loops in \cite{Assel:2018rcw}. The case with adjoint matter was subsequently treated in \cite{Agarwal:2018tso}.

Naturally, the special case $\fg=A_1$  has received the most attention, and was first considered in \cite{Tong:2014cha}, in a type IIA T-dual picture. In that same setup, the instanton partition function of $U(n)$ SYM (living on a stack of D4 branes) with a fundamental Wilson loop insertion (a single orthogonal D$4'$ brane) was computed in \cite{Kim:2016qqs}, reproducing the type IIB result of \cite{Nekrasov:2015wsu}  when $\fg=A_1$. 
Finally, we mention that yet another T-dual brane configuration was proposed in \cite{Tong:2014yna}, with the aim of proposing a holographic dual to $AdS_3\times S^3 \times S^3 \times S^1$. A careful quantization of the various open strings in that setup requires turning on $B$-fields, and the analysis was carried out in \cite{Nekrasov:2016gud}.

\subsubsection{5d $BCFG$  Quiver Theories}
\label{sssec:nonsimplylaced}

The six-dimensional $(2,0)$ little string itself is labeled by an $ADE$ Lie algebra. However, once we introduce branes, the low energy quiver gauge theory on those branes can be labeled by more general algebras than the simply-laced ones. In particular, one can engineer five-dimensional quiver gauge theories labeled by a non simply-laced Lie algebra.\\ 

\begin{figure}[h!]
	\emph{}
	\centering
	\includegraphics[width=0.8\textwidth]{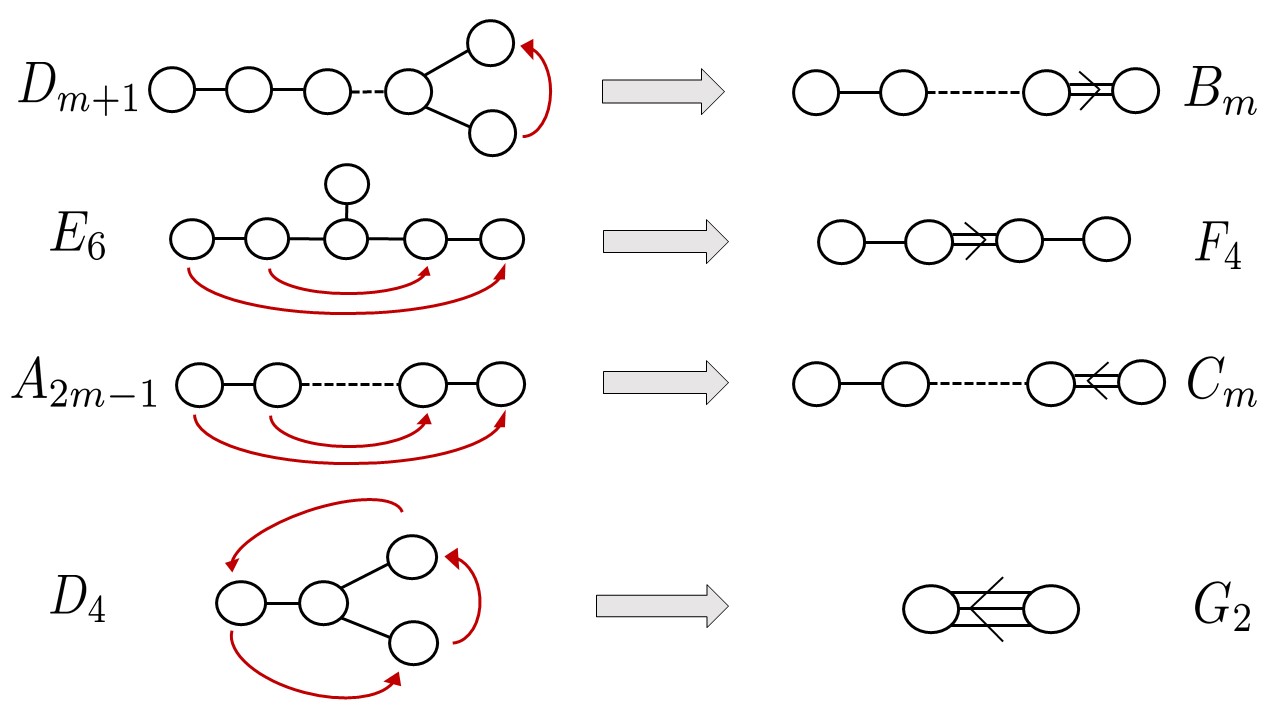}
	\vspace{-1pt}
	\caption{A non simply-laced Lie algebra $\fg$ is constructed as a subalgebra of a simply-laced Lie algebra $\fg_0$ that is invariant under the $A$-action on $\fg_0$. Note that $D_4$ is the only simply-laced Lie algebra that admits a $\mathbb{Z}_3$ outer automorphism action, resulting in the non simply-laced $G_2$.} 
	\label{fig:nonsimplylaced}
\end{figure}

We deem it useful to first remind the reader of some elementary facts about non simply-laced Lie algebras. In what follows, we let $\fg_0$ be a simply-laced Lie algebra, and we denote the Cartan-Killing form by $\langle\cdot, \cdot\rangle$. The length squared $\langle\alpha_a,\alpha_a\rangle$ of a simple root $\alpha_a$ in $\fg_0$ is assumed to be 2.  Let $A$ be a non-trivial outer automorphism group of $\fg_0$. The outer automorphisms of $\fg_0$ are isomorphic to the automorphisms of the Dynkin diagram of $\fg_0$.
Define $\fg$ to be a subalgebra of $\fg_0$ invariant under the $A$-action on $\fg_0$. This subalgebra  $\fg$ is called non simply-laced. The group $A$ is abelian, and one finds that $A=\mathbb{Z}_2$ or $A=\mathbb{Z}_3$, as shown in figure \ref{fig:nonsimplylaced}:\\

Let $a\in A$, and let $\Delta$ be the set of simple roots of $\fg_0$. Then, the simple roots of $\fg$ split into two sets: first, we have the long roots of $\fg$
\begin{align}
\Delta_l = \{\alpha \; | \;\alpha\in\Delta,\; \alpha=a(\alpha) \}\; .
\end{align}
These are the simple roots in $\fg_0$ left invariant under $a$-action. Throughout this paper, we stick to the convention that the long roots have length squared $2$. Second, we have the short roots of $\fg$, defined by:
\begin{align}
\Delta_s &= \{\frac{1}{r}\left(\alpha+a(\alpha)+\ldots+a^{r-1}(\alpha)\right)\; | \; \alpha\in\Delta,\; \alpha\neq a(\alpha) \}
\end{align}
In our conventions, the length squared of the short roots is fixed to be $2/r$, where $r$ the lacing number of $\fg$. The lacing number is the maximal number of links between two adjacent nodes in the Dynkin diagram of $\fg$. Namely, $r=2$ if $A=\mathbb{Z}_2$ and $r=3$ if $A=\mathbb{Z}_3$.\\

The simple coroots of $\fg$ are defined through the relation $\alpha_a^{\vee}=2\alpha_a/\langle\alpha_a,\alpha_a\rangle$. They are dual to the fundamental weights: $\langle \lambda_a, \alpha_b^{\vee}\rangle=\delta_{ab}$. The Cartan matrix of $\fg$ is defined as $C_{ab}=\langle\alpha_a,\alpha_b^{\vee}\rangle$. Finally, one defines the fundamental coweights of $\fg$, as dual to the simple roots: $\langle \lambda_a^{\vee}, \alpha_b\rangle=\delta_{ab}$.\\

We are now ready to engineer non simply-laced theories from type IIB on $X\times\mathbb{C}^2\times \cC$, following \cite{Aspinwall:1996nk,Bershadsky:1996nh}. Let us distinguish the two complex lines as  $\mathbb{C}^2=\mathbb{C}_q\times\mathbb{C}_t$. As we go around the origin of one of the lines, say $\mathbb{C}_q$\footnote{Choosing the other complex line $\mathbb{C}_t$ will in fact result in distinct physics. It would be interesting to investigate this further. We will encounter again the choice of a preferred line when discussing deformed $\cW$-algebra labeled by a non-simply Lie algebra.}, we let $X$ come back to itself up to the action of a generator $a\in A$ of the outer automorphism group. This is a non-trivial action on the root lattice of $\fg_0$, and therefore a non-trivial action on the homology group $H_2(X, \mathbb{Z})$, by the McKay correspondence. Put differently, the D5 (and D1) branes are permuted by the action of $a$. A necessary condition to engineer a non simply-laced theory is therefore to only allow D5 branes left invariant under the action of $a$. In terms of the low energy quiver gauge theory $T^{5d}$, this implies that the ranks of the flavor and gauge groups which belong in a given $A$-orbit must be equal.\\

We can now discuss the D5 and D1 brane charges:
A fundamental coweight  $\lambda_a^{\vee}$ of $\fg$ is a sum of fundamental weights of $\fg_0$ where all the weights are in a single $A$-orbit. We conclude that a set of D5 branes (or D1 branes) wrapping non-compact 2-cycles of the fibered geometry has a net charge measured by a coweight of $\fg$. Likewise, a simple coroot of $\fg$ is a sum of simple roots of $\fg_0$, where all the roots  are in a single $A$-orbit. Correspondingly, a set of D5 branes wrapping  compact 2-cycles of the geometry has a charge measured by a coroot of $\fg$\footnote{Alternatively, the net non-compact D5 brane charge is labeled by a weight in the Langlands dual algebra $^L\fg$, and the net compact D5 brane charge is labeled by a root in $^L\fg$. The Langlands dual algebra of $\fg$ is defined as the Lie algebra with the transpose Cartan matrix of $\fg$. See \cite{Haouzi:2017vec} for details when only D5 branes are present, and \cite{Aganagic:2017smx} for a discussion where a configuration of D3 branes is studied.}.\\

This means that just as in the simply-laced case, we can still denote the total D5 brane charge as $[S+S^*]$, but $[S]$ and $[S^*]$ are to be understood as classes in the coroot and coweight lattices of $\fg$, respectively.\\

\begin{example}
Consider an $A_{3}$ singularity, with $L$ D1 branes wrapping the non-compact 2-cycle $S^*_1$, and $L$ D1 branes wrapping the non-compact 2-cycle $S^*_3$. Suppose that as we go around the origin of $\mathbb{C}_q$, the singularity goes back to itself, up to a $\mathbb{Z}_2$ action on the $A_{3}$ quiver. Imposing that the D1 branes remain invariant under this $\mathbb{Z}_2$ action is possible, since there is an equal number of D1 branes on nodes 1 and 3. As a result, the total D1 brane charge can be written in the coweight lattice of the algebra $C_2$, as $[S^*_{D1}]=L \, \lambda^\vee_1$. Here, by definition, $\lambda^\vee_1$ is the fundamental coweight dual to the simple (positive) short root of $C_2$. Alternatively, the D1 brane charge can be expanded in the weight lattice of $^L C_2 = B_2$.
\end{example}

\subsection{Gauge Theory Description}
\label{ssec:5dgauge}

At energies $E$ well below the string scale, $E/m_s\ll 1$, when the $\tau^{(a)}$ of  \eqref{couplings} are non-zero, the theory on the D5 branes can be described by a five-dimensional gauge theory with $\cN=1$ supersymmetry (since 8 supercharges are preserved by them, as we have just seen). The $\tau^{(a)}$ are not moduli of the theory; this is because the 2-forms in the definition of $\tau^{(a)}$ live in all six dimensions of $\cC\times \mathbb{C}^2$, so they are not dynamical. Instead, the $\tau^{(a)}$ are parameters; they determine the inverse gauge couplings of the gauge theory. More precisely, in five dimensions, the Yang-Mills inverse gauge coupling $1/g^2_{YM}$ has mass dimension 1, so the dimensionless $\tau^{(a)}$'s scale as $1/m_s g^2_{YM}$. We will be interested in the regime where the gauge theory is weakly coupled, $E\, g^2_{YM}\ll 1$. In terms of $\tau^{(a)}$, this reads $E/m_s\ll  \tau^{(a)}$.\\

The characterization of the gauge theory on the D5 branes was determined by Douglas and Moore \cite{Douglas:1996sw}. When $\fg$ is simply-laced of rank $m$, it is a quiver gauge theory of shape the Dynkin diagram of $\fg$\footnote{The analysis in \cite{Douglas:1996sw} predicts a quiver with the shape of an affine Dynkin diagram. However, here we are considering the limit $g_s\rightarrow 0$, which effectively decouples the affine node.}. The gauge group is
\beq\label{gaugegroup}
G=\prod_{a=1}^m U(n^{(a)})\; ,
\eeq
where the ranks $n^{(a)}$ were defined in  \eqref{compact} as the number of D5 branes wrapping the compact 2-cycle $S_a$
\[
[S] = \sum_{a=1}^m  \,n^{(a)}\,\alpha_a\;\;  \in  \,\Lambda \; .
\]
To be precise, because of the Green-Schwarz mechanism, each $U(n^{(a)})$ gauge group contains a massive $U(1)$, so the gauge groups really are  $SU(n^{(a)})$. This means one of the Coulomb moduli is actually frozen in each gauge group. Nevertheless, we will write the gauge groups as $U(n^{(a)})$ in this paper, since we will be mainly interested in the partition function of the theory, which in our background includes the extra $U(1)$'s.

The flavor symmetry is
\beq\label{flavorgroup}
G_F=\prod_{a=1}^m U(N^{(a)}_f)\; ,
\eeq
where the ranks $N^{(a)}_f$ were defined in  \eqref{noncompact} as the number of D5 branes wrapping the non-compact 2-cycle $S^*_a$
\[
[S^*] = - \sum_{a=1}^{m} \, N^{(a)}_f \, \lambda_a \;\;  \in\, \Lambda_*\; .
\] 
This gives $ N^{(a)}_f$ hypermultiplets on node $a$, in the fundamental representation of the gauge group $U(n^{(a)})$. This comes about from quantizing the strings coming from the intersection of D5 branes wrapping the compact 2-cycle $S_a$ and the non-compact 2-cycle $S^*_a$.

Finally, we have hypermultiplets coming from the intersection of 2-cycles $S_a$ and $S_b$ at a point. Open strings with one end on the $a$-th D5 brane and the other end on the $b$-th D5 brane results in a hypermultiplet in the bifundamental representation $(n^{(a)}, \overline{n^{(b)}})$.\\

When $\fg$ is non simply-laced, the theory on the D5 branes can still be interpreted as a quiver gauge theory of shape the Dynkin diagram of $\fg$ \cite{Haouzi:2017vec,Kimura:2017hez}. In particular, the gauge group is \eqref{gaugegroup}, where the ranks $n^{(a)}$ are again the number of D5 branes wrapping compact 2-cycles
\[
[S^\vee] = \sum_{a=1}^m  \,n^{(a)}\,\alpha^\vee_a\;\;  \in  \,\Lambda^\vee \; .
\]	
This equation is now understood as valued in the coroot lattice of $\fg$. Similarly, the flavor symmetry  \eqref{flavorgroup} is determined from the number of non-compact D5 branes, whose charge is now measured in the coweight lattice $\fg$. 

The implications for the various fields of the quiver gauge theory are the following: let $\psi_{long}$ (respectively, $\psi_{short}$) be a field involving long roots (respectively, short roots), and let $(z_1, z_2)\in\mathbb{C}_q\times\mathbb{C}_t$. Then
\begin{align}
\psi_{long}(e^{2\pi i}z_1, z_2) =\psi_{long}(z_1, z_2)\; ,\qquad
\psi_{short}(e^{2\pi i}z_1, z_2) =a\cdot \psi_{short}(z_1, z_2)\; ,
\end{align} 
where the right-hand side is the image of the field $\psi$ under the action of $a$. Namely, the $a$-action is trivial for fields depending on long roots. For short roots, a single-valued field is constructed on the $r$-fold cover of $\mathbb{C}_q$ as the sum $\psi_{short}+a\cdot \psi_{short}+\ldots+a^{r-1}\cdot \psi_{short}$; this has integer mode expansion in the variable $z_1'=z_1^{1/r}$, which is a good coordinate on the cover. All in all, the various fields arising from strings between the various branes are only defined on the $r$-fold cover of $\mathbb{C}_q$.\\

What happens when we include the D1 branes in the picture? Recall that those branes only wrap the non-compact 2-cycles of the geometry. As such, they are not dynamical. They represent point defects on the $\mathbb{C}^2$ where the D5 brane gauge theory lives. The number of such D1 branes is $L=\sum_{a=1}^m L^{(a)}$. Correspondingly, the defects have a flavor symmetry of their own,
\beq\label{defectflavorgroup}
G^{D1}_F=\prod_{a=1}^m U(L^{(a)})\;.
\eeq
Before we go any further, let us comment on the dimension of the gauge theory and the defects. At first, our brane setup may suggest that the quiver gauge theory we obtain is only four dimensional, with  $\cN=2$ supersymmetry. However, it is not so: the D5 branes are points on the cylinder $\cC=\mathbb{R}\times S^1(R)$, and by T-duality, they are D6 branes wrapping the circle of the T-dual cylinder $\cC'=\mathbb{R}\times S^1(\widehat{R})$, with $\widehat{R}=1/m^2_s R$.
Since T-duality is a perturbative duality, the low energy theory on the branes coming from those two descriptions is guaranteed to be the same. The second description makes it manifest that the theory under study is really five dimensional, defined on $S^1(\widehat{R})\times\mathbb{C}^2$. We call this gauge theory $T^{5d}$ in what follows.

\begin{figure}[h!]
	\emph{}
	\centering
	\includegraphics[trim={0 0 0 2cm},clip,width=0.8\textwidth]{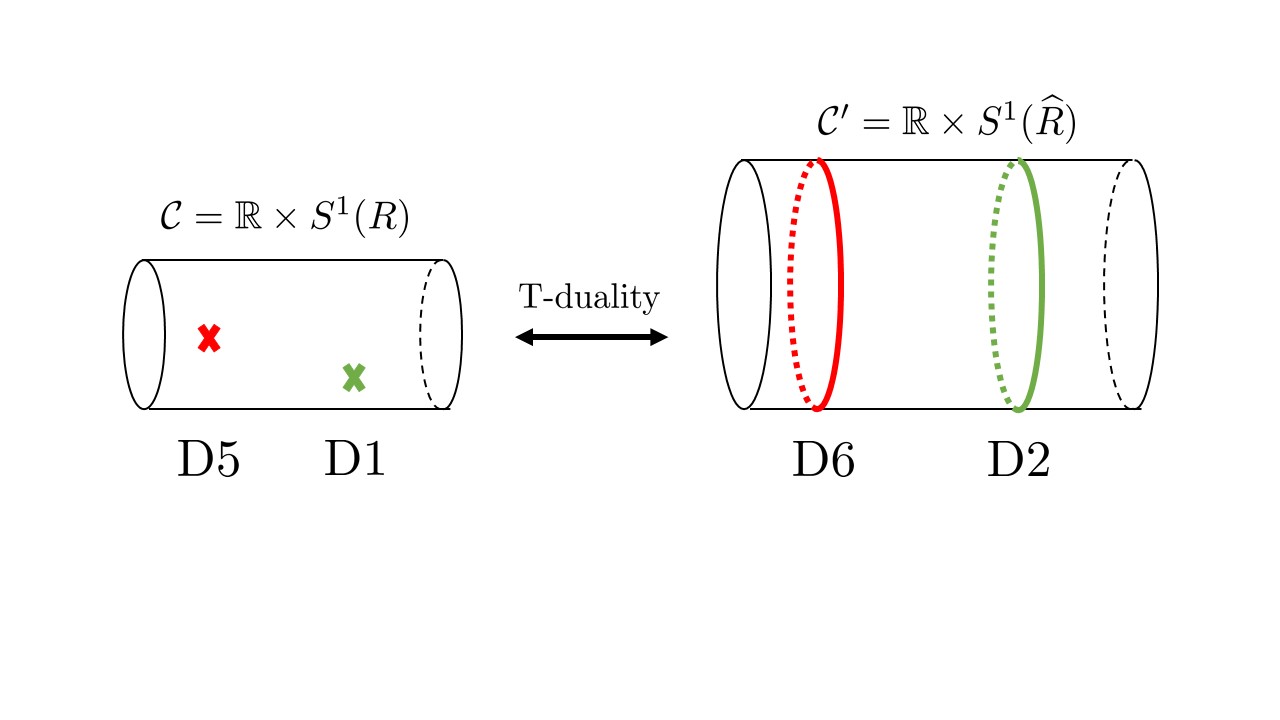}
	\vspace{-40pt}
	\caption{T-duality tells us that the D5 and D1 branes at points on the cylinder $\cC$ in type IIB are the same as D6 and D2 branes wrapping the T-dual cylinder $\cC$ in type IIA.  } 
	\label{fig:Tduality}
\end{figure}

In the same way, the D1 branes are points on $\cC$, or equivalently, D2 branes wrapping the circle $S^1(\widehat{R})$ of $\cC'$. Thus, we see that from of the point of view of the gauge theory $T^{5d}$, the D1 branes are 1/2-BPS defects wrapping the $S^1(\widehat{R})$, and at the origin of $\mathbb{C}^2$; they make up a  Wilson loop of $T^{5d}$.\\

For a given gauge group $U(n^{(a)})$, introducing such a 1/2-BPS Wilson loop can be done with the use of one-dimensional fermion field $\chi^{(a)}$ \cite{Gomis:2006sb}, transforming in the fundamental representation of $U(n^{(a)})$ and in the fundamental representation of $U(L^{(a)})$, coupled to the 5d gauge field in the bulk as 
\beq
\label{1dfermion}
S^{1d/5d}=\int dt\; {\chi_{i,\rho}^{(a)}}^\dagger\, \left( \delta_{ij}\, \partial_t - i\, A^{(a)}_{t, ij}   + \Phi^{(a)}_{ij}   - \delta_{ij}\, M_{\rho}^{(a)}  \right)\, \chi_{j,\rho}^{(a)} \; .
\eeq
Above, $A^{(a)}_t$ and $\Phi^{(a)}$ are the pullback of the 5d gauge field and the adjoint scalar of the vector multiplet, respectively. $i$ and $j$ are indices for the fundamental representation of $U(n^{(a)})$, while $\rho$ is an index for the fundamental representation of $U(L^{(a)})$. The variable $t$ is periodic, with period $2\pi \,\widehat{R}$. The parameters $M_\rho^{(a)}$ are (large) masses for the fermions, and can be thought of as a background gauge field for the $U(L^{(a)})$ symmetry acting on them. Those parameters set the energy scale for the excitation of the fermions.\\

The full dictionary from geometry to gauge theory is as follows:
As we already mentioned, the periods $\tau^{(a)}$  \eqref{couplings} are the gauge couplings of $T^{5d}$. The periods of $\omega_J$, $\omega_K$, and $B^{(2)}$ through the 2-cycles $S^2_a$ are the Fayet-Iliopoulos (F.I.) parameters, set to zero for now. As far as the $n$ D5 branes wrapping the compact 2-cycles are concerned, their position on the cylinder $\cC$ are the Coulomb moduli of $T^{5d}$. The position on $\cC$ of the $N_f$ non-compact D5 branes are the mass parameters for the fundamental hypermultiplets of $T^{5d}$. Finally, the position on $\cC$ of the $L$ D1 branes are the fermion masses \eqref{1dfermion} used to define the Wilson loop coupling to 5d. All of the above moduli and parameters are complexified, due to the presence of the circle $S^1(\widehat{R})$.\\

Our main object of study will be the evaluation of the path integral of the coupled 5d/1d system
\beq\label{5d1dpathintegral}
\left[\chi^{\fg}\right]^{5d} =\int D\psi D\chi \, e^{i(S^{5d}[\psi]+ S^{1d/5d}[\psi, \chi])} \; .
\eeq
Above, $\psi$ denotes collectively all the fields of the bulk 5d theory, written as $S^{5d}$, while $S^{1d/5d}$ denotes the coupling term \eqref{1dfermion}.

\subsection{The Partition Function of $T^{5d}$ with Wilson Loops}
\label{ssec:5dpartition}

As advertised, we now compute the partition function of $T^{5d}$ in the presence of a Wilson Loop wrapping $S^1(\widehat{R})$.\\

For the $a$-th node, we define an integer $r^{(a)}= r\langle \alpha_a , \alpha_a \rangle/2$, with $r$ the lacing number of $\fg$, with the convention that $\langle \alpha_a , \alpha_a \rangle=2$ if $\alpha_a$ is a long root. Then, in this convention, the integer $r^{(a)}$ is simply equal to 1 if $a$ denotes a short root, while it is equal to $r$ if $a$ denotes a long root.
In particular, if $\fg$ is simply-laced, all $r^{(a)}$'s are equal to 1.\\

We place the theory on the $\Omega$-background $S^1(\widehat{R})\times\mathbb{C}_q\times\mathbb{C}_t$, where as we go around the circle, the two complex lines are rotated respectively as:
\begin{align}\label{longorshort}
z_{1}\mapsto q^{r^{(a)}}\,z_{1}, \qquad z_{2}\mapsto  t^{-1}\,z_{2}\; .
\end{align}

Let us introduce $k=\sum_{a=1}^m k^{(a)}$ D1 branes wrapping the compact 2-cycles of $X$; we denote them as D1$_{inst}$ branes. The partition function can then be expressed as the Witten index of the  $\cN=(0,4)$ ADHM gauged quantum mechanics\footnote{In what follows, when we talk about a $\cN=(0,2)$ or $\cN=(0,4)$ multiplet in the context of the quantum mechanics, what we really mean is the reduction of a 2d $\cN=(0,2)$ or $\cN=(0,4)$ multiplet to 1d.}  living on those $k$ branes. This is best expressed in the T-dual IIA picture, where the $k$ branes now wrap $S^1(\widehat{R})$:
\beq
\left[\chi^{\fg}\right]^{5d}  = {\rm Tr}\, (-1)^F \; \fm \;\; .
\eeq
We define $\fm=q^{r^{(a)}(S_1 - S_R)} t^{-S_2 + S_R}\, \fm_{G,\; G_F,\; G^{D1}_F}$; here, $S_1$ and $S_2$ are the generators of the rotations of the  complex lines ${\mathbb C}_{q}\times {\mathbb C}_{t}$ as we go around the circle $S^1(\widehat{R})$. The fugacity $q$ keeps track of which roots are long and which roots are short, according to \eqref{longorshort}. $F$ is the fermion number. $S_R$ is the generator of a $U(1)_R$ charge which is a subset of the $R$-symmetry, required here in order to preserve supersymmetry. The factor $\fm_{G,\; G_F,\; G^{D1}_F}$ implicit denotes the product of fugacities associated with the Cartan generators of the various groups $G$, $G_F$ and $G^{D1}_F$, all understood as flavor symmetries from the point of view of the quantum mechanics on the $k$ compact D1 branes.

The index is the grand canonical ensemble of all instanton BPS states. It takes the form of a product of a perturbative factor involving the classical and the 1-loop contributions, and of an instanton factor. The perturbative part will play no role in our story, so we will safely ignore it. 
The computation of the index reduces to a zero mode integral of various 1-loop determinants: 
\begin{align}
\label{5dintegral}
&\left[\chi^{\fg}\right]_{(L^{(1)},\ldots, L^{(m)})}^{5d}  =\sum_{k^{(1)},\ldots, k^{(m)}=0}^{\infty}\;\prod_{a=1}^{m}\frac{e^{\widehat{R}\,\tau^{(a)}\cdot k^{(a)}}}{k^{(a)}!} \; \\ 
&\;\;\qquad\qquad\qquad\qquad\times\oint  \left[\frac{d\phi^{{(a)}}_I}{2\pi i}\right]Z^{(a)}_{vec}\cdot Z^{(a)}_{fund}\cdot Z^{(a)}_{CS}\cdot \prod_{b>a}^{m} Z^{(a,b)}_{bif}\cdot \prod_{\rho=1}^{L^{(a)}}  Z^{(a)}_{D1}  \; .\nonumber
\end{align}
The parameter $e^{\widehat{R}\,\tau^{(a)}}$ is the gauge coupling for the $a$-th gauge node. In 5d, it is also the instanton counting parameter.

The factor
\begin{align}
&Z^{(a)}_{vec} =\prod_{I, J=1}^{k^{(a)}} \frac{{\widehat{f}}\left(\phi^{(a)}_{I}-\phi^{(a)}_{J}\right)f\left(\phi^{(a)}_{I}-\phi^{(a)}_{J}+ 2\,\epsilon_+ + (r^{(a)}-1)\,\epsilon_1\right)}{f\left(\phi^{(a)}_{I}-\phi^{(a)}_{J}+\epsilon_1 + (r^{(a)}-1)\,\epsilon_1\right)f\left(\phi^{(a)}_{I}-\phi^{(a)}_{J}+\epsilon_2\right)}\nonumber\\
&\qquad\times\prod_{I=1}^{k^{(a)}} \prod_{i=1}^{n^{(a)}} \frac{1}{f\left(\phi^{(a)}_I-a^{(a)}_i+\epsilon_+ + (r^{(a)}-1)\,\epsilon_1/2\right)f\left(\phi^{(a)}_I-a^{(a)}_i-\epsilon_+ - (r^{(a)}-1)\,\epsilon_1/2\right)}
\end{align}
is the contribution of a vector multiplet\footnote{In this paper, whenever the quiver is non simply-laced, what we mean by a supersymmetric ``multiplet" is the folding of proper multiplets of a simply-laced theory, following the discussion in section \ref{sssec:nonsimplylaced}.} corresponding to the gauge group $U(n^{(a)})$. The parameters $a^{(a)}_i$ are the $U(n^{(a)})$-equivariant parameters, or Coulomb parameters of this gauge group, valued in the Cartan subalgebra. This contribution $Z^{(a)}_{vec}$ arises from quantizing D1$_{inst}$/D1$_{inst}$ strings and D1$_{inst}$/D5 strings, where all the branes are wrapping one and the same compact 2-cycle $S_a$\footnote{The quantization of the above strings actually gives more than what we have described, as it is the ADHM quantum mechanics of the so-called 5d $\cN=1^*$ SYM, with an extra hypermultiplet in the adjoint representation. In this paper, we decouple the adjoint hypermultiplet by sending its mass to infinity.}. 
The expressions we write down become quickly involved, so we introduced the notation $f(x)\equiv 2 \sinh\left(x/2\right)$. Furthermore, we defined $\widehat{f}(x)$ to be equal to $f(x)$, unless $x=0$, in which case $\widehat{f}(0)=1$.

The factor 
\begin{align}
Z^{(a)}_{fund} =\prod_{I=1}^{k^{(a)}} \prod_{d=1}^{N^{(a)}_f} f\left(\phi^{{(a)}}_I-m^{{(a)}}_d\right) \equiv\prod_{I=1}^{k^{(a)}} Q^{(a)}(\phi^{{(a)}}_I)\label{fundQ}
\end{align}
is the contribution of a hypermultiplet, associated to the flavor symmetry $U(N^{(a)}_f)$. The parameters $m^{{(a)}}_d$ are the corresponding masses, valued in the Cartan subalgebra. This comes about from the quantization of D1$_{inst}$/D5 strings, where the D1$_{inst}$ branes are wrapping the compact 2-cycle $S_a$, while the D5 branes are wrapping the non-compact 2-cycle $S^*_a$. This results in extra Fermi multiplets; the associated fermion fields transform in the representation $(k^{(a)}, \overline{N_f^{(a)}})$ of $U(k^{(a)})\times U(N_f^{(a)})$, while being singlets of  $U(n^{(a)})$.

The factor
\begin{align}
Z^{(a)}_{CS}= \prod_{I=1}^{k^{(a)}} e^{k^{(a)}_{CS}\, \phi^{(a)}_I}
\end{align} 
is the contribution of the effective Chern-Simons term on the $a$-th node. In this paper, we set the bare Chern-Simons level to be zero on all nodes, which  uniquely fixes the parameters $k^{(a)}_{CS}$ in terms of the ranks $n^{(a)}$ and $N^{(a)}_f$ of the gauge and flavor groups, respectively\footnote{For instance, in the case $\fg=A_1$, $T^{5d}$ has a UV fixed point if $N_f + 2 \,|k_{CS, bare}| \leq 2\, n$. We take $k_{CS, bare}=0$ here, and allow for all theories that still satisfy the inequality. When we later make contact with 3d physics, we will choose to saturate this inequality to simplify the analysis. Such inequalities exist for all $\fg$-type quivers.}.

The factor 
\begin{align}
&Z^{(a, b)}_{bif} =\left[\prod_{J=1}^{k^{(b)}}\prod_{i=1}^{n^{(a)}} \prod_{p=0}^{\overline{r^{(a b)}}-1} f\left(\phi^{(b)}_J- m^{(a)}_{bif} - a^{(a)}_i - \epsilon_+ - (r^{(b)}-1 - 2\, p)\,\epsilon_1/2 \right)\right.\nonumber\\
&\left.\qquad \times\prod_{I=1}^{k^{(a)}}\prod_{j=1}^{n^{(b)}} \prod_{s=0}^{\overline{r^{(b a)}}-1} f\left(\phi^{(a)}_I+ m^{(a)}_{bif} - a^{(b)}_j + \epsilon_+ + (r^{(a)}-1 -2 \, s)\,\epsilon_1/2 \right)\right.\nonumber\\
&\qquad \times \left.\prod_{I=1}^{k^{(a)}}\prod_{J=1}^{k^{(b)}} \frac{f\left(\phi^{(a)}_{I}-\phi^{(b)}_{J}+ m^{(a)}_{bif}+\epsilon_1 + (r^{(a)}-1)\,\epsilon_1 - (\overline{r^{(a b)}}+\overline{r^{(b a)}}-2)\epsilon_1/2\right)}{f\left(\phi^{(a)}_{I}-\phi^{(b)}_{J} +m^{(a)}_{bif}  - (\overline{r^{(a b)}}+\overline{r^{(b a)}}-2)\epsilon_1/2\right)}\right.\nonumber\\
&\qquad \times \left.\prod_{I=1}^{k^{(a)}}\prod_{J=1}^{k^{(b)}} \frac{f\left(\phi^{(a)}_{I}-\phi^{(b)}_{J}+ m^{(a)}_{bif}+\epsilon_2  - (\overline{r^{(a b)}}+\overline{r^{(b a)}}-2)\epsilon_1/2 \right)}{f\left(\phi^{(a)}_{I}-\phi^{(b)}_{J}+ m^{(a)}_{bif}+2\, \epsilon_+ + (r^{(a)}-1)\,\epsilon_1  - (\overline{r^{(a b)}}+\overline{r^{(b a)}}-2)\epsilon_1/2\right)}\right]^{\Delta_{a,b}}
\end{align}
is the contribution of a bifundamental hypermultiplet in representation $(n^{(a)}, \overline{n^{(b)}})$ of $U(n^{(a)})\times U(n^{(b)})$.  In the definition of $Z^{(a, b)}_{bif}$, we have introduced an upper diagonal incidence matrix $\Delta_{a,b}$, with entries equal to 1 if there is a link connecting nodes $a$ and $b$, and 0 otherwise. Correspondingly, we have introduced $\text{rank}(\fg)-1$ bifundamental masses $m^{(a)}_{bif}$, one for each link in the quiver. The contribution $Z^{(a, b)}_{bif}$ comes again from D1$_{inst}$/D1$_{inst}$  and D1$_{inst}$/D5 strings, but this time the left D1$_{inst}$ branes wrap the compact 2-cycle $S_a$, while the right D1$_{inst}$ and D5 branes wrap an adjacent compact 2-cycle $S_b$.
We have also defined a new integer for the non simply-laced bifundamentals: ${\overline{r^{(ab)}}}\equiv r$ if node $a$ labels a short root \textit{and} node $b$ labels a long root, while  ${\overline{r^{(ab)}}}\equiv 1$ otherwise.

The factor 
\begin{align}
&Z^{(a)}_{D1} =\resizebox{.95\hsize}{!}{$\left[{\displaystyle\prod_{i=1}^{n^{(a)}}}  f\left(a^{{(a)}}_i-M^{{(a)}}_\rho\right) {\displaystyle\prod_{I=1}^{k^{(a)}}} \frac{f\left(\phi^{{(a)}}_I-M^{(a)}_\rho+ \epsilon_- + (r^{(a)}-1)\,\epsilon_1/2\right)f\left(\phi^{{(a)}}_I-M^{(a)}_\rho- \epsilon_- - (r^{(a)}-1)\,\epsilon_1/2\right)}{f\left(\phi^{{(a)}}_I-M^{(a)}_\rho+ \epsilon_+ + (r^{(a)}-1)\,\epsilon_1/2\right)f\left(\phi^{{(a)}}_I-M^{(a)}_\rho- \epsilon_+ - (r^{(a)}-1)\,\epsilon_1/2\right)}\right]$}
\end{align}
is the contribution of the Wilson loop defect, associated to the symmetry group $U(L^{(a)})$. The parameters $M^{(a)}_\rho$ are the corresponding fermion masses, valued in the Cartan subalgebra. In the second product, the numerator is the contribution of a $\cN=(0,4)$ Fermi hypermultiplet, while the denominator is the contribution of a $\cN=(0,4)$ twisted hypermultiplet, both arising from the $D1/D1_{inst}$ strings.

At first sight, it may seem like all the relevant strings have been accounted for, but there are extra fermions to consider, from $D1/D5$ strings; these result in chiral fermions which sit in a $\cN=(0,2)$ Fermi multiplet, and can be made compatible with $\cN=(0,4)$ supersymmetry \cite{Tong:2014yna}. These fermions are neutral under $U(k^{(a)})$, and transform in the representation $(n^{(a)}, \overline{L^{(a)}})$ of $U(n^{(a)})\times U(L^{(a)})$. They are precisely the fields we called $\chi$ when we introduced the coupled 1d/5d action \eqref{1dfermion}. In the partition function integrand, the corresponding factor is 
\beq
\label{wilsonclassical}
\prod_{i=1}^{n^{(a)}}  f\left(a^{{(a)}}_i-M^{{(a)}}_\rho\right).
\eeq

Notice that in the $k=0$ sector, meaning in the absence of instanton corrections, the partition function becomes equal to this product only, as was first computed in \cite{Gomis:2006sb}.\\

Some remarks are in order:

The above integral may look puzzling, but all factors other than the 5d bifundamentals can be obtained by ``folding" the various contributions that appear in simply-laced quiver quantum mechanics. Namely, starting from a simply-laced theory, one performs identifications that are natural under the outer automorphism action. 
In particular, the form of the bifundamental hypermultiplets is new: we conjecture that a bifundamental between a short node (respectively long node)  and a long node (respectively short node) results in $r-1$ extra Fermi multiplets in the integrand. After performing the integration, with the contours defined in the next section, it is not hard to show one recovers the non simply-laced index formula presented in \cite{Haouzi:2017vec,Kimura:2017hez}, in a 5d picture. Put differently, the integral representation introduced here can be interpreted as the quantum mechanical version of the equivariant index written there. It would be nice to have a microscopic derivation of our formalism without resorting to a folding of simply-laced theory argument, and we leave this important question to future work. 

When $\fg$ is simply-laced, meaning the lacing number $r$ is equal to 1, the integrand reduces to the familiar simply-laced quiver quantum mechanics (see for instance the appendix of \cite{Assel:2018rcw} for a recent review).\\

Because the theory is valued on a circle of radius $\widehat{R}$, it is useful in what follows to introduce K-theoretic fugacities for each of the equivariant parameters:
\begin{align}\label{fugacities5d}
\widetilde{\fq}^{(a)}&=e^{\widehat{R}\,\tau^{(a)}}\, \qquad q=e^{\widehat{R}\,\epsilon_1},\qquad t=e^{-\widehat{R}\,\epsilon_2},\qquad e^{(a)}_i=e^{-\widehat{R}\,a^{(a)}_i},\\
f^{(a)}_d&=e^{-\widehat{R}\,m^{{(a)}}_d},\qquad z^{(a)}_\rho=e^{-\widehat{R}\,M^{(a)}_\rho},\qquad \mu^{(a)}_{bif}=e^{-\widehat{R}\,m^{(a)}_{bif}}\; .
\end{align}

\subsubsection{Integration Contours}
\label{sssec:5dcontours}

One still needs to specify the contours for the  $k=\sum_{a=1}^m k^{(a)}$ integration variables $\phi^{{(a)}}_I$. This can be done, for instance, with the Jeffrey-Kirwan (JK) residue prescription \cite{Jeffrey:1993}. Let us first briefly review the argument in the case where $\fg$ is simply-laced, and no Wilson loop is present, following \cite{Hwang:2014uwa,Cordova:2014oxa,Hori:2014tda} and \cite{Benini:2013xpa}. The strategy is the following:
the computation of the partition function $\left[\chi^{\fg}\right]^{5d}$ will a priori depend on the sign of the F.I. parameters in the quiver quantum mechanics, and we should expect wall crossing between different chambers of the F.I.-parameter space.  For our purposes, we take all the F.I. parameters to be strictly positive, and work in the associated chamber. Correspondingly, we define a reference vector of size $k$, called $\eta$;  here, we choose $\eta=\left(1, 1, \ldots, 1, 1\right)$ . According to the JK prescription, the contours should then pick up poles $k$ poles coming from 1d chiral multiplets whose charge is measured by a vector $O_I$, $I=1, \ldots, k$, but only those obeying the constraint:
\beq
\left(1, 1, \ldots, 1, 1\right)=\sum_{I=1}^{k} d_I O_I\; ,
\eeq
with $d_I$  strictly positive integers. In the 1d terminology, such chiral multiplets can originate from $\cN=(0,4)$ hypermultiplets, in which case the charge vector $O_I$ satisfies
\beq\label{hyperpole}
O_I(\phi_I)+\epsilon_+ + \ldots=0 \; ,
\eeq
or the chiral multiplets can originate from $\cN=(0,4)$ twisted hypermultiplets, in which case the charge vector $O_I$ satisfies
\beq\label{twistedhyperpole}
O_I(\phi_I)-\epsilon_+ + \ldots=0 \; .
\eeq
Summing over all allowed poles $\phi_*$, the partition function takes the form
\beq\label{JK}
\left[\chi^{\fg}\right]^{5d}=\sum_{k^{(1)},\ldots, k^{(m)}=0}^{\infty}\widetilde{\fq}^{(1)}{}^{k^{(1)}}\ldots\widetilde{\fq}^{(m)}{}^{k^{(m)}}\, \frac{1}{k^{(1)}!\ldots k^{(m)}!}\sum_{\phi_*}\text{JK-res}_{\phi_*}(Q_*, \eta)\,Z_{vec}\cdot Z_{fund} \cdot Z_{bif}  \; ,
\eeq
with residue equal to:
\beq\label{JKrule}
\text{JK-res}_{\phi_*}(Q_*, \eta)\frac{d^k \phi}{Q_{I_1}(\phi)\ldots Q_{I_k}(\phi)}=\begin{cases}
	\frac{1}{\left|\text{det}\left(Q_{I_1}\ldots Q_{I_k}\right)\right|} \;\; \text{if}\;\; \eta\in\text{cone}\left(Q_{I_1}\ldots Q_{I_k}\right)\\
	0 \qquad\qquad\qquad\, \text{otherwise}
\end{cases}
\eeq
The condition $\eta\in\text{cone}\left(Q_{I_1}\ldots Q_{I_k}\right)$ means that the vector $\eta$ should lie in the cone spanned by the $k$ vectors $Q_{I}$.
The poles end up being classified by $n^{(a)}$-tuples of Young diagrams $\overrightarrow{\boldsymbol{\mu}^{(a)}}=\{\boldsymbol{\mu}^{(a)}_1, \boldsymbol{\mu}^{(a)}_2, \ldots, \boldsymbol{\mu}^{(a)}_{n^{(a)}}\}$. Therefore, the JK residue rule reproduces the results of \cite{Nekrasov:2002qd,Nekrasov:2003rj}, where the partition function was first computed. Namely, the poles are located at 
\beq
\label{simplylacedyoungtuples}
\phi^{(a)}_I=a^{(a)}_i +\epsilon_+ - s_1\, \epsilon_1 - s_2\, \epsilon_2,\qquad \text{with}\; (s_1, s_2)\in \boldsymbol{\mu}^{(a)}_i \; .
\eeq
Note this is consistent with the prescription  \eqref{hyperpole}, since all the above poles occur in some hypermultiplet coming from $Z_{vec}$ in the integrand  \eqref{5dintegral}. There are additional poles to include due to the bifundamental hypermultiplets, but they turn out to have zero residue, so the above list of poles is exhaustive.

The JK prescription needs to be generalized in two ways for our purposes: first, we want to be able to discuss the case where $\fg$ is a non simply-laced Lie algebra. Second, we have to take into account the effect of Wilson loops defect in the integrand, namely the factor $Z^{(a)}_{D1}$. We discuss these two points in turn.\\

We first consider the case when $\fg$ is non simply-laced. Does the JK rule \eqref{JKrule} continue to make sense?  The answer is affirmative, but the pole prescription \eqref{hyperpole}, \eqref{twistedhyperpole}, needs to be slightly modified, to account for the rescaling of $\epsilon_1$. The contours should now enclose poles of the form
\beq\label{NEWhyperpole}
O_I(\phi_I)+\epsilon_+ +(r^{(a)}-1)\epsilon_1/2 + \ldots=0,
\eeq
if those arise from the folding of $r^{(a)}$  $\cN=(0,4)$ hypermultiplets in the associated simply-laced theory, and the contours should enclose poles of the form 
\beq\label{NEWtwistedhyperpole}
O_I(\phi_I)-\epsilon_+ -(r^{(a)}-1)\epsilon_1/2 + \ldots=0,
\eeq
if those arise from the folding of $r^{(a)}$  $\cN=(0,4)$ twisted hypermultiplets in the associated simply-laced theory.
All in all, the poles are still labeled by Young diagrams, but with $\epsilon_1$ weighted differently from $\epsilon_2$ \cite{Haouzi:2017vec,Kimura:2017hez}:
\beq
\label{generalyoungtuples}
\phi^{(a)}_I = a^{(a)}_i +\epsilon_+ -(r^{(a)}-1)\epsilon_1/2 - r^{(a)}\, s_1\, \epsilon_1 - s_2\, \epsilon_2 , \text{with}\; (s_1, s_2)\in \boldsymbol{\mu}^{(a)}_i \; .
\eeq

We now address the second point and include Wilson loops. 
After including the contribution of the defect D1 branes in the integrand, the Young diagram rule  \eqref{generalyoungtuples} is no longer valid, since there are additional poles depending on the fermion masses $M^{(a)}_\rho$ to be enclosed by the contours. To be precise, in the $\cN=(0,2)$ language, there are now new chiral multiplets making up the denominator of $Z^{(a)}_{D1}$, which are responsible for additional poles depending on the fermion masses, following \eqref{NEWtwistedhyperpole}. Moreover,  the bifundamental factors $Z^{(a, b)}_{bif}$ now have poles with nonzero residue, and such poles will necessarily depend on some bifundamental mass $m^{(a)}_{bif}$. All in all, following the modified JK prescription, the new potential poles are of the form:
\begin{align}
&\phi^{(a)}_I=M^{(a)}_\rho+\epsilon_+ +(r^{(a)}-1)\epsilon_1/2, \label{newpole1}\\
&\phi^{(a+1)}_J=\phi^{(a)}_I + m^{(a)}_{bif} +2\, \epsilon_+ + (r^{(a)}-1)\,\epsilon_1 - (\overline{r^{(a b)}}+\overline{r^{(b a)}}-2)\epsilon_1/2, \label{newpole2}\\
&\phi^{(a-1)}_J=\phi^{(a)}_I - m^{(a-1)}_{bif} + (\overline{r^{(a b)}}+\overline{r^{(b a)}}-2)\epsilon_1/2.\label{newpole3}
\end{align}
In the above, the poles  \eqref{newpole2}, \eqref{newpole3}, will depend on the bifundamental masses, but also on one of the fermion masses $M^{(a)}_\rho$, picked up by a previous contour of type \eqref{newpole1}. All the remaining poles are of the form  \eqref{generalyoungtuples}.\\

Evaluating the contour integral by residues then becomes a straightforward exercise, though computationally involved in practice. We can be more explicit: denoting the defect fermion masses as $z^{(a)}_\rho$ and the 5d Coulomb parameters as $e^{(a)}_{i}$ (see our notations  \eqref{fugacities5d}), the partition function is an expansion in the Wilson loop vevs of the theory:
\beq\label{partitionfunctionloop}
\left[\chi^{\fg}\right]_{(L^{(1)},\ldots, L^{(m)})}^{5d}(\{z^{(a)}_\rho\})=\sum_{(\textbf{R}, \textbf{Q})} d_{\textbf{R}, \textbf{Q}}\cdot \langle W_{\textbf{R}}(e^{(a)}_{i})\rangle \cdot X_{\textbf{Q}}(\{z^{(a)}_\rho\}).
\eeq
In the above, $d_{\textbf{R}, \textbf{Q}}$ are integer coefficients, and
\beq\label{WilsonLoop}
\langle W_{\textbf{R}}\rangle = \text{Tr}_{\textbf{R}}\, {\cal{P}}\, \text{exp}\left(i\int dt (A^{(a)}_t + i\,\Phi^{(a)})\right),
\eeq
is a Wilson loop valued in the representation ${\textbf{R}}=\left({\textbf{R}}^{(1)}, \ldots , {\textbf{R}}^{(m)}\right)$ of $\prod_{a=1}^m SU(n^{(a)})$. For each node $a$, the representation $\textbf{R}^{(a)}$ is a tensor product of fundamental representations of $SU(n^{(a)})$\footnote{Which tensor product representation exactly is dictated by a 1d Chern-Simons level $K^{(a)}$ we are allowed to turn on for the fermion field action on each node, and which acts as a Lagrange multiplier term in the path integral \eqref{5d1dpathintegral}. This corresponds to a background of $K^{(a)}$ units of electric charge localized on the defect D1 branes, meaning there are $K^{(a)}$ fundamental strings stretching between the D5 and D1 branes. One still needs to distribute the $K^{(a)}$ units of string charge among the $L^{(a)}$ D1 branes; in other words, one needs to choose a partition $(K^{(a)}_1, \ldots, K^{(a)}_{L^{(a)}})$ of $L^{(a)}$. This in turn specifies a representation ${\textbf{R}}^{(a)}={\textbf{K}}^{(a)}_1\otimes \ldots \otimes {\textbf{K}}^{(a)}_{L^{(a)}}$ of $SU(n^{(a)})$, where each ${\textbf{K}}^{(a)}_\rho$ is a fundamental representation of $SU(n^{(a)})$.
In this paper, we do not include such a 1d Chern Simons term; the partition function then becomes a sum over all possible tensor products of  $SU(n^{(a)})$ fundamental representations.}. The function $X_{\textbf{Q}}(\{z^{(a)}_\rho\})$ does not depend on the 5d gauge theory fugacities, but depends instead on the defect fugacities, and on the representation $\textbf{Q}=\left(\textbf{Q}^{(1)}, \ldots , \textbf{Q}^{(m)}\right)$ of $\prod_{a=1}^m SU(L^{(a)})$. For each node $a$, the representation $\textbf{Q}^{(a)}$ is a tensor product of fundamental representations of $SU(L^{(a)})$\footnote{As written, there are instanton corrections contained in $X_{\textbf{Q}}(\{z^{(a)}_\rho\})$. By sending such corrections to 0, $X_{\textbf{Q}}(\{z^{(a)}_\rho\})$ literally becomes the character of the representation $\textbf{Q}$. These  instanton corrections have important physics of their own: they are related after two $T$-dualities to the monopole bubbling of 't Hooft loops in 4d $U(L^{(a)})$ SYM \cite{Brennan:2018yuj,Assel:2019iae}.}. This presentation of the partition function is simply the instanton-corrected version of the ``classical" contribution \eqref{wilsonclassical}.\\

There exists a more illuminating presentation of the partition function, first exhibited by Nekrasov \cite{Nekrasov:2015wsu} in the simply-laced case $\fg=A, D, E$, using different methods. Here, our presentation will apply to an arbitrary simple Lie algebra $\fg$. One introduces a defect operator expectation value, one for each node $a=1, \ldots, m$, in the $\fg$-shaped quiver of rank $m$. The expectation value of the $a$-th defect operator, with corresponding fermion mass $z^{(a)}_\rho\equiv z$, is defined as
\begin{align}
\label{Yoperator}
&\left\langle \left[Y^{(a)}_{5d}\right]^{\pm 1} (z)\right\rangle =
\sum_{k^{(1)},\ldots, k^{(m)}=0}^{\infty}\;\prod_{b=1}^{m}  \frac{\left(\widetilde{\fq}^{(b)}\right)^{k^{(b)}}}{k^{(b)}!} \; \cdot\\  &\;\;\qquad\qquad\qquad\qquad\cdot\oint_{\{\overrightarrow{\boldsymbol{\mu}}\}}  \left[\frac{d\phi^{{(b)}}_I}{2\pi i}\right]Z^{(b)}_{vec}\cdot Z^{(b)}_{CS}\cdot Z^{(b)}_{fund}\cdot \prod_{c>b}^{m} Z^{(b, c)}_{bif}  \cdot \left[Z^{(a)}_{D1}(z)\right]^{\pm 1}\, . \nonumber
\end{align}
Importantly,  the above contour integral is defined to \emph{only} enclose poles labeled by Young diagrams, of the form  \eqref{generalyoungtuples}. This is in contrast to the actual physical partition function \eqref{5dintegral}, where the extra poles  \eqref{newpole1},  \eqref{newpole2}, and \eqref{newpole3}, are required by the JK prescription, contribute with nonzero residue. We have denoted the $Y$-operator integral by the symbol $\oint_{\{\overrightarrow{\boldsymbol{\mu}}\}}$ to emphasize the different contour.\\

Put differently, the product of operator expectation values $\prod_{a=1}^m \prod_{\rho=1}^{L^{(a)}} \left\langle {Y^{(a)}_{5d}(z^{(a)}_\rho)} \right\rangle$, and the partition function $\left[\chi^{\fg}\right]_{(L^{(1)},\ldots, L^{(m)})}^{5d}$ have identical integrands, but the contours in the second expression enclose more poles, some of which will necessarily depend on the fermion masses $\{z^{(a)}_\rho\}$. 

Remarkably, the  partition function has a simple expression in terms of such operator expectation values $\left\langle \left[Y^{(a)}_{5d}\right]^{\pm 1}\right\rangle$, and has a beautiful connection to the representation theory of quantum affine algebras. Namely, as a function of the fermion masses $\{z^{(a)}_\rho\}$, we compute
\begin{align}\label{character5d}
\boxed{\left[\chi^{\fg}\right]_{(L^{(1)},\ldots, L^{(m)})}^{5d}(\{z^{(a)}_\rho\})=\sum_{\omega\in V(\lambda)}\prod_{b=1}^m \left({\widetilde{\fq}^{(b)}}\right)^{d_b^\omega}\; c_{d_b^\omega}(q, t)\; \left(Q^{(b)}(\{z_{* \rho}^{(a)}\})\right)^{d_b^\omega}\,  \left[{Y}_{5d}(\{z^{(a)}_\rho\})\right]_{\omega} \, .}
\end{align}
In other words, the partition function is a twisted $qq$-character of a finite dimensional irreducible representation $V(\lambda)$ of $U_q(\widehat{\fg})$\footnote{The literature on the representation theory of quantum affine algebras is quite rich, and remains an active subject of research to this day. For details on finite dimensional representations of quantum affine algebras, there are two main presentations, one due to Jimbo \cite{jimbo}, and the other due to Drinfeld \cite{Drinfeld:1987sy}. In our context, it is the latter presentation that is relevant. See also the works   \cite{Chari:1994pf, Chari:1994pd}. Characters of representations of quantum affine algebras appeared in the literature under the name $q$-characters, in the work of Frenkel and Reshetikhin \cite{Frenkel:qch}. $(q,t)$-characters are a generalization of those characters, defined by the same authors \cite{Frenkel:1998} as Ward identities satisfied by deformed $\cW_{q,t}$-algebras. In the context of the BPS/CFT correspondence, those objects have recently appeared in a higher dimensional gauge theory context as $qq$-characters \cite{Nekrasov:2015wsu,Kimura:2015rgi}, where ``$qq$" stands for the two equivariant parameters $q$ and $t$ of the $\Omega$-background. In particular, one recovers the usual $q$-characters in the so-called Nekrasov-Shatashvili limit  $\epsilon_1\rightarrow 0$ \cite{Nekrasov:2013xda,Bullimore:2014awa}.  For related work on $t$-analogues of $q$-characters, see \cite{Nakajima:tanalog,Frenkel:2010wd}.}$^{,}$\footnote{In the context of integrable systems, it is well known that XXZ spin chains have quantum affine symmetry. The fact that such algebras appear in the study of five dimensional theories on  $\mathbb{C}^2\times S^1(\widehat{R})$ is expected from the gauge/Bethe correspondence \cite{Nekrasov:2009rc}. This will be true again in three dimensions, by construction, as we will see explicitly in the next section.}, with highest weight $\lambda=\sum_{a=1}^m L^{(a)}\, \lambda_a$.\\

We will explain in the next section how to obtain the above result. For now, let us unpack the notation: 
$\{z^{(a)}_\rho\}$ denotes collectively the $L$ fermion masses $z^{(a)}_\rho\equiv e^{-\widehat{R} M^{(a)}_\rho}$.
$\omega$ runs over all the weights of the representation $V(\lambda)$.  The label $d_b^\omega$ is a positive integer that is determined by solving
\beq\label{sl2string}
\omega=\lambda -\sum_{b=1}^m d_b^\omega\, \alpha_b\; .
\eeq
Namely, any weight $\omega$ is reached by lowering  a finite number of times the highest weight $\lambda$ of the representation, using the positive simple roots $\alpha$. This procedure is sometimes referred to as building the weight $\omega$ out of  $sl_2$ strings.

The factor $\widetilde{\fq}^{(b)}$ is the 5d gauge coupling for the $b$-th gauge group. The factors $c_{d_b^\omega}(q, t)$ are coefficients depending only on $q$ and $t$. The function $Q^{(b)}(\{z^{(a)}_\rho\})$ was defined as the contribution of a fundamental hypermultiplet in  \eqref{fundQ}. Let us rewrite it here in terms of the redefined fugacities:
\beq
Q^{(b)}(z_{* \rho}^{(a)})=\prod_{d=1}^{N^{(b)}_f} \left(1-f^{(b)}_d/z_{* \rho}^{(a)}\right)\; .
\eeq

The variables $\{z_{* \rho}^{(a)}\}$ are the fermion masses $\{z^{(a)}_\rho\}$, shifted by various powers of $q$ and $t$. More precisely, they are the residues at the poles \eqref{newpole1},  \eqref{newpole2} and  \eqref{newpole3} in the integrand of the partition function.

Finally, the operator $\left[{Y}_{5d}(\{z^{(a)}_\rho\})\right]_{\omega}$, for a given weight $\omega$, is in general the expectation value of products and ratios of various defect $Y$-operators  $\langle \prod_a \left[Y^{(a)}_{5d}\right]^{\pm 1} \rangle $, where each operator $\left[Y^{(a)}_{5d}\right]^{\pm 1}$ is a function of a fermion mass $z^{(a)}_\rho$. The arguments of each factor is shifted by various powers of $q$ and $t$, uniquely determined by  \eqref{sl2string}. The $qq$-character then organizes itself as a finite Laurent polynomial in the $Y$-operator vevs.

Occasionally, the operator $\left[{Y}_{5d}(\{z^{(a)}_\rho\})\right]_{\omega}$ may also consist of derivatives (with respect to the fermion masses) of such products, though we will not encounter them in the examples we study here\footnote{An example where such a derivative term can appear is the $D_4$ partition function $\left[\chi^{\fg}\right]_{(0,1,0,0)}^{5d}(z^{(2)}_1)$, meaning there is only one D1 brane wrapping the non-compact 2-cycle $S^*_2$ in a resolved $D_4$ singularity. The partition function is then a sum of 29 terms, one of which involves derivatives of $Y^{(a)}_{5d}$ operators. The attentive reader may wonder why there are 29 terms in the first place, since the second fundamental representation of $D_4$ is 28-dimensional. However, finite dimensional irreducible representations of quantum affine algebras are in general bigger than their non affine counterpart. Indeed, the second fundamental representation $V(\lambda_2)$ of $U_q(\widehat{D_4})$ decomposes into irreducible representations of $U_q(D_4)$ as $V(\lambda_2) = \textbf{28}\oplus \textbf{1}$: one necessarily needs to add the trivial representation \textbf{1} (an extra null weight) to the \textbf{28} to obtain an irreducible representation of $U_q(\widehat{D_4})$.}.

We call the resulting $qq$-character \textit{twisted} because of the presence of the 5d gauge couplings $\widetilde{\fq}^{(b)}$ and the matter factors $Q^{(b)}$.

\begin{example}
The $qq$-character for the fundamental representation of $U_q(\widehat{A_1})$  is written as
\beq
\left[\chi^{A_1}\right]^{5d}_{(1)}(z)=\langle Y(z) \rangle + \widetilde\fq \, Q(z \,v^{-1}) \left\langle\frac{1}{Y(z \, v^{-2})}\right\rangle \; .
\eeq
Here, we wrote $v=\sqrt{q/t}$. We will derive this formula from the JK prescription in the Examples section.
\end{example}

\begin{example}
	The $qq$-character for the first fundamental representation of $U_q(\widehat{A_2})$  is
\begin{align}
&\left[\chi^{A_m}\right]_{(1,0)}^{5d}(z) =\left\langle Y^{(1)}_{5d}(z) \right\rangle\nonumber \\
&\qquad\qquad + \widetilde{\fq}^{(1)} \; Q^{(1)}(z\, v^{-1}) \left\langle\frac{Y^{(2)}_{5d}(z\, v^{-2}\, \mu^{(1)}_{bif})}{Y^{(1)}_{5d}(z\, v^{-2})}\right\rangle\nonumber\\
&\qquad\qquad + \widetilde{\fq}^{(1)}\widetilde{\fq}^{(2)} \; Q^{(1)}(z\, v^{-1})\, Q^{(2)}(z\, v^{-3}\, \mu^{(1)}_{bif}) \left\langle\frac{1}{Y^{(2)}_{5d}(z\, v^{-4}\,  \mu^{(1)}_{bif})}\right\rangle\; .
\end{align}
It corresponds to having a single D1 brane wrapping the non-compact 2-cycle $S^*_1$ in the resolved $A_2$ geometry. This partition function will  be derived explicitly in the Examples section.
\end{example}

\subsubsection{Evaluation of the Integrals}

The expression  \eqref{character5d} suggests that the evaluation of the partition function rests on the computation of various $Y$-operators integrals  \eqref{Yoperator}. The procedure is recursive, and goes as follows: As before, let $a=1, \ldots, m$, with $m=\text{rank}(\fg)$. In the case where $\fg$ is simply-laced, the 5d quiver gauge theory $T^{5d}$ is defined by fixing, for each $a$, the gauge couplings $\widetilde{\fq}^{(a)}$ and F.I. parameters (i.e. the periods  \eqref{periods} in the geometry), the ranks $n^{(a)}$ of the gauge groups (i.e. the number of compact D5 branes), and the  ranks $N_f^{(a)}$ of the flavor groups (i.e. the number of non-compact D5 branes). We further fix the bare Chern Simons level to be trivial. Once this is done, we specify a Wilson loop representation by fixing the integers $L^{(a)}$ (i.e. the number of non-compact D1 branes).\\

As we explained, the non simply-laced case can be defined by further setting the integers to be equal on nodes that lie in a single orbit of the outer automorphism group action. Then, the partition function has the universal form:
\beq\label{firstterm}
\left[\chi^{\fg}\right]_{(L^{(1)},\ldots, L^{(m)})}^{5d}(\{z^{(a)}_\rho\})= \prod_{a=1}^m \prod_{\rho=1}^{L^{(a)}} \left\langle {Y_{5d}^{(a)}(z^{(a)}_\rho)} \right\rangle +\ldots\; ,
\eeq
where the dots stand for more terms that we will describe below. That is, the first term is always a product of $Y$-operators.  Performing the integrals over the poles  \eqref{generalyoungtuples}, we find
\beq\label{5ddefectexpression2}
 \left\langle {Y_{5d}^{(a)}(z^{(a)}_\rho)} \right\rangle = \sum_{\{\overrightarrow{\boldsymbol{\mu}}\}}\left[ Z^{5d}_{bulk}\; \cdot Y^{(a)}_{5d}(z^{(a)}_\rho)\right]\, .
\eeq
The sum is over a collection of 2d partitions, one for each $U(1)$ Coulomb modulus
\beq
\{\overrightarrow{\boldsymbol{\mu}}\}=\{\boldsymbol{\mu}^{(a)}_i\}_{a=1, \ldots, m\, ; \,\; i=1,\ldots,n^{(a)}}\; .
\eeq
Every such partition describes a configuration of instantons at the fixed point of $\mathbb{C}_q\times \mathbb{C}_t$. In the above, we have defined variables
\beq\label{ydef}
y^{(a)}_{i,k} = e^{(a)}_i\, q^{r^{(a)}{\boldsymbol{\mu}}^{(a)}_{i, k}} \, t^{-k}\;\;\;,\; k=1, \ldots, \infty\;,
\eeq 
where  ${\boldsymbol{\mu}}^{(a)}_{i, k}$ is the length of the $k$-th row of the partition ${\boldsymbol{\mu}}^{(a)}_{i}$. The factor  $Z^{5d}_{bulk}$ encodes all the 5d bulk physics. It is simply given by the instanton partition function in the absence of Wilson loops, written as
\beq\label{bulk5d}
Z^{5d}_{bulk}=  \prod_{a=1}^m \widetilde{\fq}^{(a)}{}^{\sum_{i=1}^{n^{(a)}}{\left|\boldsymbol{\mu}^{(a)}_i\right|}}\, Z^{(a)}_{bulk,vec} \cdot Z^{(a)}_{bulk,fund}\cdot Z^{(a)}_{bulk,CS} \cdot \prod^n_{b>a}
Z^{(a, b)}_{bulk,bif}\; .
\eeq

Each factor above is naturally expressed in terms of the following function
\beq\label{nekrasovN}
N_{\boldsymbol{\mu}^{(a)}_i\boldsymbol{\mu}^{(b)}_j}(Q\, ;q^{r^{(ab)}}) \equiv \prod\limits_{k,s = 1}^{\infty} 
\frac{\big( Q \, q^{r^{(a)}\boldsymbol{\mu}^{(a)}_{i,k}-r^{(b)}\boldsymbol{\mu}^{(b)}_{j,s}} \,t^{s - k + 1}\,;q^{r^{(ab)}} \big)_{\infty}}{\big( Q\,  q^{r^{(a)}\boldsymbol{\mu}^{(a)}_{i,k}-r^{(b)}\boldsymbol{\mu}^{(b)}_{j,s}}\, t^{s - k}\, ;q^{r^{(ab)}}\big)_{\infty}} \,
\frac{\big( Q\,  t^{s - k}\, ;q^{r^{(ab)}} \big)_{\infty}}{\big( Q\,  t^{s - k + 1}\, ;q^{r^{(ab)}}\big)_{\infty}}\, .
\eeq
We have introduced a standard notation for the $q$-Pochhammer symbol:
\beq
\left( x\, ; q\right)_{\infty}=\prod_{l=0}^{\infty}\left(1-q^l\, x\right)\; ,
\eeq
and as before, $r^{(a)}= r$ if the node $a$ labels a long root, and $r^{(a)}= 1$ if the node $a$ labels a short root. We further defined $r^{(a b)}=\text{gcd}(r^{(a)}, r^{(b)})$, the greatest common divisor of $r^{(a)}$ and $r^{(b)}$.
Going back to the bulk partition function expression \eqref{bulk5d}, the various factors are as follows: the gauge couplings keep track of the total instanton charge, via the factor
\begin{align}\label{5dbulkgauge}
\widetilde{\fq}^{(a)}{}^{\sum_{i=1}^{n^{(a)}}{\left|\boldsymbol{\mu}^{(a)}_i\right|}}.
\end{align}

For each node $a$, the vector multiplets contribute the factor
\begin{align}\label{5dbulkvec}
Z^{(a)}_{bulk,vec} = \prod_{i,j=1}^{n^{(a)}}\left[N_{\boldsymbol{\mu}^{(a)}_i\boldsymbol{\mu}^{(a)}_j}\left(e^{(a)}_{i}/e^{(a)}_{j};q^{r^{(a)}}\right)\right]^{-1}\; .
\end{align}

At each node $a$, we also couple $N^{(a)}_f$ hypermultiplets charged in the fundamental representation of the $U(N^{(a)}_f)$ group, with associated masses $f^{(a)}_d$'s. They contribute to the factor
\begin{align}\label{5dbulkmatter}
Z^{(a)}_{bulk,fund} = \prod_{d=1}^{N^{(a)}_f} \prod_{i=1}^{n^{(a)}} N_{\boldsymbol{\emptyset}\, \boldsymbol{\mu}^{(a)}_i}\left( v^{(a)}\, f^{(a)}_{d}/e^{(a)}_{i};\, q^{r^{(a)}}\right)\; .
\end{align}
Above, we have introduced the non simply-laced notation $v^{(a)}\equiv\sqrt{q^{r^{(a)}}/t}$. When the $a$-th node denotes a short root, meaning $r^{(a)}=1$, we will use the more common notation $v\equiv\sqrt{q/t}$. 

The bifundamental hypermultiplets (with corresponding masses $\mu^{(a)}_{bif}$, one for each link in the quiver) contribute the factor
\begin{align}\label{5dbulkbif}
Z^{(a, b)}_{bulk,bif}= \prod_{i=1}^{n^{(a)}}\prod_{j=1}^{n^{(b)}}\left[N_{\boldsymbol{\mu}^{(a)}_i \boldsymbol{\mu}^{(b)}_j}\left(\mu^{(a)}_{bif}\,e^{(a)}_{i}/e^{(b)}_{j};\, q^{r^{(ab)}}\right)\right]^{\Delta_{a,b}}\; .
\end{align}

Finally, the contribution of $k^{(a)}_{CS}$ units of the Chern-Simons term on node $a$ gives a contribution
\begin{align}\label{5dbulkCS}
Z^{(a)}_{bulk,CS} = \prod\limits_{i=1}^{n^{(a)}} \left(T_{\boldsymbol{\mu}^{(a)}_i}\right)^{k^{(a)}_{CS}}\; .
\end{align}
Here, $T_{\boldsymbol{\mu}^{(a)}}$ is defined as  $ T_{\boldsymbol{\mu}^{(a)}} =(-1)^{|\boldsymbol{\mu}^{(a)}|} q^{r^{(a)}\Arrowvert \boldsymbol{\mu}^{(a)}\Arrowvert^{2}/2}t^{-\Arrowvert \boldsymbol{\mu}^{(a)\, t}\Arrowvert^{2}/2}$. As we mentioned before, because we fix the bare Chern-Simons levels to zero, the parameters  $k^{(a)}_{CS}$ are fully determined by the ranks of the gauge and flavor groups.\\

Now that we have explained each factor in the 5d bulk $Z^{5d}_{bulk}$, we move on to the remaining factors in \eqref{5ddefectexpression2}, which encode the physics of the Wilson loop. We compute
\beq\label{Wilsonfactor}
Y^{(a)}_{5d}(z)\equiv\prod_{i=1}^{n^{(a)}}\prod_{k=1}^{\infty}\frac{1-t\, y^{(a)}_{i,k}/z}{1- y^{(a)}_{i,k}/z}\; .
\eeq
As a reminder, we defined  variables $y^{(a)}_{i,k}=e^{(a)}_i\, q^{r^{(a)} \boldsymbol{\mu}^{(a)}_{i, k}} \, t^{-k}$. For completeness, we note that these operators have different equivalent expressions, though in practice we will only really need the first one:
\begin{align}
Y^{(a)}_{5d}(z)&=\prod_{i=1}^{n^{(a)}}\prod_{k=1}^{\infty}\frac{1-t\, y^{(a)}_{i,k}/z}{1- y^{(a)}_{i,k}/z}\label{Wilsonfactor2}\\
&=\prod_{i=1}^{n^{(a)}}\left[\left(1-e^{(a)}_i/z\right)\prod_{(k,s)\in\boldsymbol{\mu}^{(a)}_i}\frac{(1-q^{s-1}t^{-k}\; e^{(a)}_i/z)(1-q^{s}t^{1-k}\;e^{(a)}_i/z)}{(1-q^{s}t^{-k}\; e^{(a)}_i/z)(1-q^{s-1}t^{1-k}\;e^{(a)}_i/z)}\right]\label{Wilsonfactor3}\\
&=\prod_{i=1}^{n^{(a)}}\left[\prod_{(k,s)\in\partial_+\boldsymbol{\mu}^{(a)}_i}\left(1-t^{1-k}q^{s-1}\; e^{(a)}_i/z\right)\prod_{(k,s)\in\partial_-\boldsymbol{\mu}^{(a)}_i}\left(1-t^{-k}q^{s}\; e^{(a)}_i/z\right)\right]\, .\label{Wilsonfactor4}
\end{align}
In the last expression,  $\partial_+\boldsymbol{\mu}^{(a)}_i$ is the outer boundary of the Young tableau $\boldsymbol{\mu}^{(a)}_i$, while $\partial_-\boldsymbol{\mu}^{(a)}_i$ is its inner boundary.\\

Now, we would like to fill in the dots in \eqref{firstterm}. The key argument is to note that we have missed some poles, as dictated by the JK prescription. As we advertised in the last section, the missing poles all belong to the following list: 
\begin{align}
&\phi^{(a)}_I=M^{(a)}_\rho+\epsilon_+ +(r^{(a)}-1)\epsilon_1/2\; ,\\
&\phi^{(a+1)}_J=\phi^{(a)}_I+m^{(a)}_{bif} +2\, \epsilon_+ + (r^{(a)}-1)\,\epsilon_1\; ,\\
&\phi^{(a-1)}_J=\phi^{(a)}_I-m^{(a-1)}_{bif}\; ,
\end{align}
for some fermion mass $M^{(a)}_\rho$. The first set of poles occurs in the Wilson loop factor  $Z^{(a)}_{D1}$ in the index integral, while the other two are solely due to the bifundamentals  $Z^{(a, b)}_{bif}$. Fixing the algebra $\fg$ and the number of fermion masses $\sum_{a=1}^m\sum_{\rho=1}^{L^{(a)}}M^{(a)}_\rho$ fully fixes the above set of missing poles. Importantly, this set is always of \textit{finite} size $k'$. Having identified the $k'$ new poles above, the remaining $k-k'$ poles should be taken among Young diagrams \eqref{generalyoungtuples} as usual.

Now, a remarkable fact comes into play, which can be proved by direct computation: each missing residue due to one of the $k'$ new poles can be traded for an integration contour which encloses $k-j'$ poles \textit{only}, where $j'=1,\ldots,k'$. In particular, these contours are chosen in a way that they do not enclose any new pole,  at the expense of introducing in the integrand extra $Y$-operators and $k'$ extra fundamental matter factors  $Z_{fund}$.

\begin{figure}[h!]
	\emph{}
	\centering
	\includegraphics[trim={0 0 0 3cm},clip,width=0.8\textwidth]{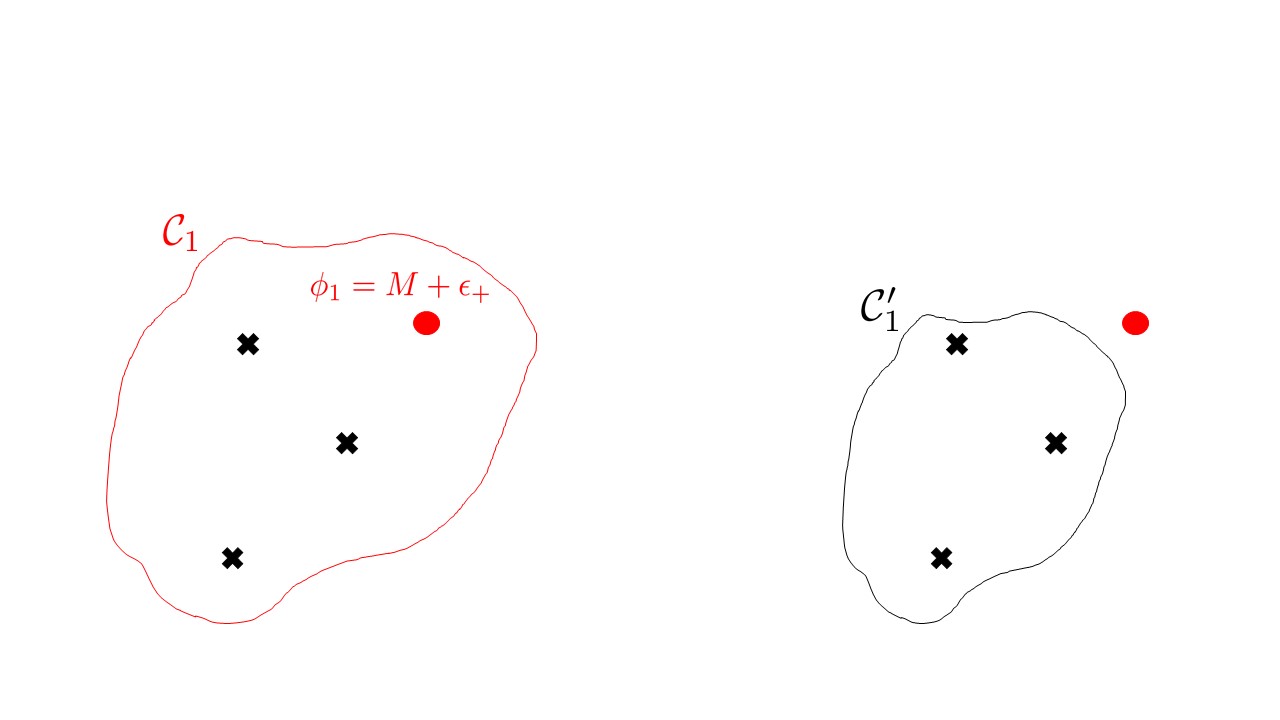}
	\vspace{-12pt}
	\caption{The black crosses denote poles labeled by Young diagrams, while the red dot denotes a new pole due to a D1 brane insertion, resulting in the factor $Z^{(a)}_{D1}$ in the integrand. On the left, we show a possible contour for the computation of the 5d partition function, say for a $SU(3)$ gauge theory at instanton number $k$. By the JK residue prescription, we must in particular enclose the new pole in red. It turns out it is equivalent to trade this contour for the one on the right, at instanton number $k-1$, which only encloses the usual poles labeled by Young diagrams; this comes at the expense of inserting in the integrand new $Y$-operators and fundamental matter, with an instanton shift of one unit to account for the missing pole.} 
	\label{fig:contours}
\end{figure}

Since $k'$ is finite, we are guaranteed that the number of terms filling the dots in \eqref{firstterm} is finite. Carrying out the computation, the $k'$ terms make up a finite dimensional irreducible representation of the quantum affine algebra $U_q(\widehat{\fg})$. In this language, the first term in \eqref{firstterm} denotes the highest weight $\lambda=\sum_{a=1}^m L^{(a)}\, \lambda_a$ of the representation. We derived here from the JK prescription the $qq$-character formula found in \cite{Nekrasov:2015wsu} for the $ADE$ case.

In section \ref{sec:examples}, we will be very explicit in carrying out this procedure to compute \eqref{character5d} in many examples, including for non simply-laced quiver gauge theories.

As a final remark, we want to stress the two-fold role of the integers $L^{(a)}$. First,  they can be understood as Dynkin labels in the fundamental \textit{coweight} basis of $\fg$, where they label the net D1 brane charge in the geometry. Second, they can by understood as Dynkin labels in the fundamental \textit{weight} basis of $\fg$, where they label (the highest weight of) a representation in the quantum affine algebra $U_q(\widehat{\fg})$.\\

Having analyzed the quiver gauge theory $T^{5d}$ in the presence of a Wilson loop, we now show that our results can be reinterpreted in the language of a 3d quiver gauge theory with $\cN=2$ supersymmetry, with the same 1/2-BPS Wilson loop.

\newpage

\section{The $\fg$-type quiver gauge theory $G^{3d}$ and Wilson Loops}

\subsection{String Theory Construction}
\label{ssec:string3d}

Just like in the previous section, our starting point to derive three-dimensional physics will be type IIB string theory. We first construct 3d quiver gauge theories labeled by a simple Lie algebra, before moving on to the non simply-laced case.

\subsubsection{3d $ADE$ Quiver Theories}
\label{Sssec:ADEstring3d}

Consider the same setup as before, namely type IIB on $X\times\mathbb{C}_q\times\mathbb{C}_t\times \cC$, with a collection of $n$ D5 branes wrapping compact 2-cycles of $X$, $N_f$ D5 branes and $L$ D1 branes wrapping the non-compact 2-cycles of $X$. From now on and in the rest of this paper, we make a specialization that will make the analysis easier to follow. Namely, we impose a constraint on the total D5 brane charge, and require that it vanishes:
\beq\label{confzero}
[S+S^*] =0 \, .
\eeq
This implies the following vanishing condition in homology: $\# (S_a \cap (S+S_*))=0$, for all nodes $a=1, \, \ldots \, , m$. This last condition is itself equivalent to
\beq\label{betazero}
\sum_{b=1}^m C_{ab} \;n^{(a)} = N^{(a)}_f\; ,
\eeq
with $C_{ab}$  the Cartan matrix of $\fg$. Note that this constraint may look more familiar in four dimensions where it is a conformality condition, i.e. the vanishing of the beta function.
Relaxing the constraint \eqref{confzero}, as we did in the previous section, results in the running of the gauge coupling on the D5 branes. This will not produce new physics as far as the Wilson loop is concerned. Namely, \eqref{confzero} will simply ensure that 3d Chern-Simons levels vanish and simplify the computation of the partition function. Therefore, we safely impose the constraint for the rest of this paper\footnote{A classification of D5 brane defects subject to \eqref{confzero} was carried out in \cite{Haouzi:2016ohr,Haouzi:2016yyg}. It is shown there that the constraint has an important connection to nilpotent orbits. Namely, the Coulomb branch of the low energy theory on D5 branes, in the CFT limit $m_s\rightarrow \infty$, is a nilpotent orbit, with Bala-Carter label directly readable from the Dynkin labels of $[S+S^*]$.}.

\begin{figure}[h!]
	\emph{}
	\centering
	\includegraphics[trim={0 0 0 3cm},clip,width=0.8\textwidth]{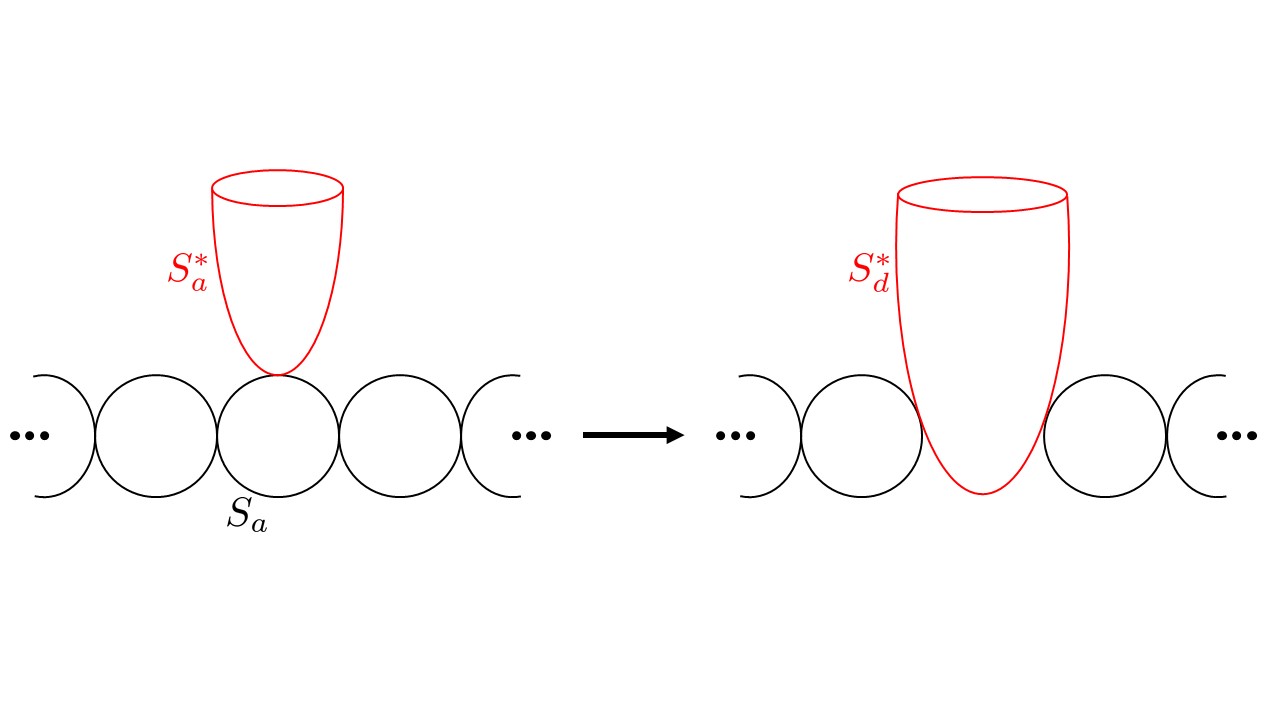}
	\vspace{-40pt}
	\caption{Higgsing the theory $T^{5d}$ in the geometry: a D5 brane wrapping a compact 2-cycle $S_a$ and a D5 brane wrapping a non-compact 2-cycle $S^*_a$ recombine into a single D5 brane, which wraps the non-compact 2-cycle $S^*_d$.} 
	\label{fig:higgsing}
\end{figure}

The three-dimensional physics emerges on the Higgs branch of the theory on the D5 branes, where the $U(n^{(a)})$ gauge groups get broken to $U(1)$'s, one for each $a=1, \ldots, m$.

Now, at this Higgs branch locus, the $n$ D5 branes wrapping compact 2-cycles and the $N_f$ D5 branes wrapping the non-compact 2-cycles rearrange themselves in a configuration of $N_f$ D5 branes wrapping non-compact 2-cycles only, which we denote by  $S_d^*$.
This is possible because, as we already mentioned, the group $H_2(X, \mathbb{Z})$ is contained in $H_2(X, \partial X, \mathbb{Z})$. 

The homology classes of the non-compact 2-cycles $S_d^*$ will be henceforth denoted by
\beq\label{weightsnew}
[S_d^*] = \omega_d  \qquad \in\, \Lambda_*\; ,
\eeq
for $d=1, \ldots, N_f$. Each $\omega_d$ is a weight belonging in some fundamental representation of $\fg$. Written in terms of those weights, the constraint of vanishing D5 brane charge \eqref{confzero}  becomes
\beq\label{weightsadduptozero}
\sum_{d=1}^{N_f} \omega_d =0\; .
\eeq
The weights $\omega_d$ are determined from the previous data as follows: since we now require the various D5 branes to bind, the positions of the $n$ compact and $N_f$ non-compact branes must coincide on the cylinder $\cC$. A given weight $\omega_d$ can then always be written as
\beq\label{weightexpression}
\omega_d=-\lambda_a + \sum_{b=1}^m h^{(b)}_{d}\, \alpha_b\; ,
\eeq
where the notations were introduced in section \ref{sec:5dtheory}; namely, $-\lambda_a$ is the negative of the $a$-th fundamental weight, $h^{(b)}_{d}$ are non-negative integers, and $\alpha_b$ is a positive simple root. The expression \eqref{weightexpression} reflects the fact that a total of $\sum_{b=1}^m h^{(b)}_{d}$ compact D5 branes are now bound to the $a$-th non-compact brane, labeled by $-\lambda_a$. In this way, all the compact branes will bind to at least one of the non-compact D5 branes.

\begin{figure}[h!]
	\emph{}
	\centering
	\includegraphics[trim={0 0 0 1cm},clip,width=0.9\textwidth]{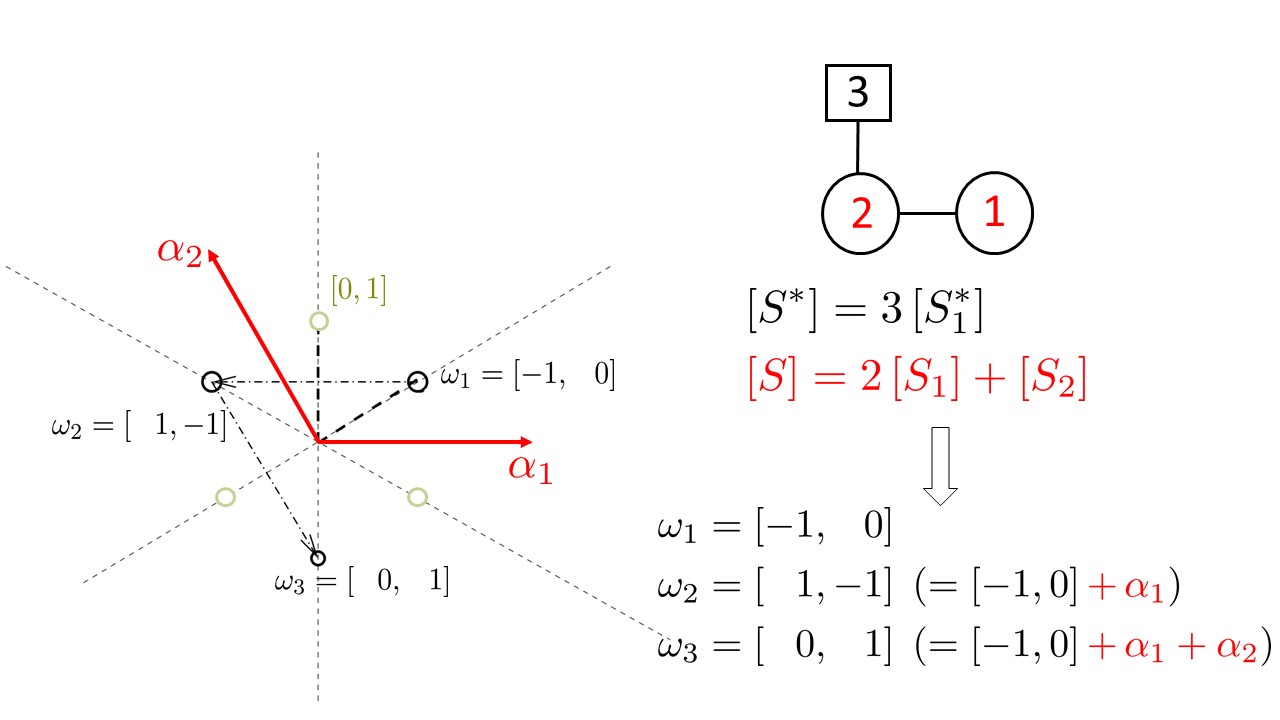}
	\vspace{-10pt}
	\caption{The $A_2$ 5d quiver gauge theory is defined by D5 brane charges, measured in fundamental weights and simple roots (or equivalently by classes in homology). On the Higgs branch, we bind the branes in such a way that the theory is now described by a set $\{\omega_d\}$ of weights that add up to zero, as shown at the bottom. On the left, we display the weight lattice of $A_2$, where the relevant weights are pictured. The set in question characterizes a codimension 2 defect of the $(2,0)$ little string theory; in this example, the defect in question is called the ``full puncture" on the cylinder.} 
	\label{fig:A2example}
\end{figure}

The above geometric picture is perfectly consistent with the gauge theory description. Indeed, recall that the position of a compact brane on $\cC$ is a Coulomb parameter of $T^{5d}$, while the position of a non-compact brane on $\cC$ is a mass parameter of $T^{5d}$. When \eqref{weightexpression} is satisfied, $\sum_{b=1}^m h^{(b)}_{d}$ Coulomb parameters are frozen to the value of one of the masses. The associated fundamental hypermultiplets become massless, and can therefore acquire a nonzero vacuum expectation value, which is precisely how the Higgs branch arises.

If no proper subset of weights adds up to zero within the set $\{\omega_d\}_{d=1}^{N_f}$, we say that $\{\omega_d\}_{d=1}^{N_f}$ characterizes a D5 brane defect on the cylinder $\cC$\footnote{The ``position" of the defect on $\cC$ is then the center of mass of the $N_f$ D5 branes. Since we are setting $g_s\rightarrow 0$, this is in fact a codimension 2 defect of the $(2,0)$ little string on $\cC$.}.

Recall that the defect D1 branes used to wrap the non-compact 2-cycles $S^*_a$ of $X$; this is still the case after going to the Higgs branch. We will analyze their physics in detail in the next section.\\

Having described the Higgs branch of $T^{5d}$, we can now make contact with three-dimensional physics  through the introduction of (codimension 2) 1/2-BPS vortices in the theory. In type IIB string theory, such vortices are $D$ D3 branes at points on the cylinder $\cC$, wrapping compact 2-cycles of $X$ and one of the two  complex lines, say $\mathbb{C}_q$. The D3 branes end on the D5 branes; this turns on magnetic flux on the D5 branes, in a direction transverse to the D3 branes. Crucially, we need to ensure that supersymmetry is not completely broken by the introduction of such D3 branes on the Higgs branch of the D5 brane theory. Vortices are mutually supersymmetric as long as the 5d F.I. parameters are aligned. Correspondingly, we consider a background where 
\beq\label{periods3d}
\frac{m_s^4}{g_s}\int_{S_a}\omega_{J}>0\, ,\qquad \frac{m_s^4}{g_s}\int_{S_a}\omega_{K}=0\, ,\qquad \frac{m_s^2}{g_s}\int_{S_a}B^{(2)}=0\, .
\eeq
The vortices are then guaranteed to be supersymmetric vacua of the D3 brane theory.

We denote the charge of such D3 branes by the following homology class in $H_2(X, \mathbb{Z})$:
\beq\label{d3compact}
[D] = \sum_{a=1}^m  \,D^{(a)}\,\alpha_a\;\;  \in  \,\Lambda \; ,
\eeq
where $D^{(a)}$ are strictly positive integers, and $D=\sum_{a=1}^m D^{(a)}$.

\subsubsection{3d $BCFG$ Quiver Theories}
\label{Sssec:BCFGstring3d}

\begin{figure}[h!]
	\emph{}
	\centering
	\includegraphics[trim={0 0 0 1cm},clip,width=0.9\textwidth]{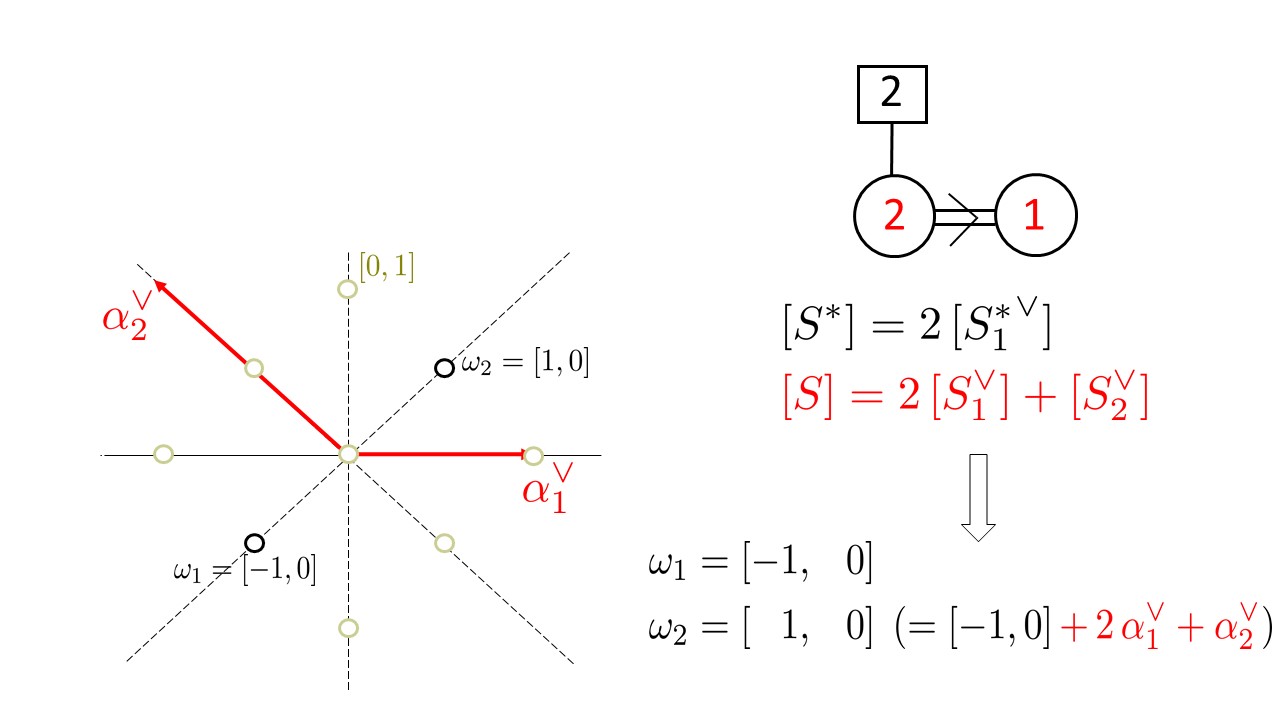}
	\vspace{-10pt}
	\caption{The $B_2$ 5d quiver gauge theory is defined by D5 brane charges, measured in fundamental coweights and simple coroots. We bind the branes in such a way that the theory is now described by a set $\{\omega_d\}$ of coweights that add up to zero, as shown at the bottom. On the left, we display the coweight lattice of $B_2$, where the relevant coweights are pictured.} 
	\label{fig:B2example}
\end{figure}

The previous discussion easily carries through to the non simply-laced case $\fg=BCFG$. Namely, for $\fg_0=ADE$, we restrict ourselves to configurations of D5 branes left invariant under the outer automorphism group action of $\fg_0$. The construction was reviewed in section \ref{sssec:nonsimplylaced}. We can still make sense of the vanishing brane charge constraint $[S+S^*]=0$ as:
\beq\label{betazeroBCFG}
\sum_{b=1}^m C_{ab} \;n^{(a)} = N^{(a)}_f\; ,
\eeq
where $C_{ab}$ is understood as the Cartan matrix of $\fg$. In particular, $C_{ab}$ is no longer symmetric. This constraint can again be understood in terms of coweights $\omega_d$ of $\fg$ (i.e. weights belonging in fundamental representations of $^L\fg$), which are decomposed as:
\beq\label{weightsBCFG}
\omega_d=-\lambda_a^\vee+ \sum_{b=1}^m h^{(b)}_{d}\, \alpha_b^\vee\; .
\eeq
Above, $-\lambda_a^\vee$ is the negative of the $a$-th fundamental coweight, $h^{(b)}_{d}$ are non-negative integers, and $\alpha_b^\vee$ is a positive simple coroot. The condition \eqref{betazeroBCFG} then translates to the following vanishing condition on coweights:
\beq\label{weightsadduptozeroBCFG}
\sum_{d=1}^{N_f} \omega_d =0\; .
\eeq
The net defect D1 brane charge is likewise measured in the coweight lattice. Introducing 1/2-BPS vortices is done as in the simply-laced case, restricting ourselves to configurations of D3 branes left invariant under the outer automorphism group action of $\fg_0$.  The net D3 brane charge is now measured in the coroot lattice of $\fg$, as
\beq\label{d3compactnsl}
[D^\vee] = \sum_{a=1}^m  \,D^{(a)}\,\alpha^\vee_a\;\;  \in  \,\Lambda^\vee \; .
\eeq	
A crucial subtlety is that unlike the Coulomb branch, there is no notion of Higgs branch for a 5d $BCFG$-type quiver gauge theory, so it is a priori unclear what it means to study its vortices. Nevertheless, the formal procedure of freezing the Coulomb moduli to some masses is algebraically sound, and we will see a fortiori it is the correct picture to make contact with $\cW$-algebras of non simply-laced type. We therefore \emph{define} a non simply-laced vortex to be a vortex of a simply-laced theory left invariant under outer automorphism group action.

\subsection{Gauge Theory Description}
\label{ssec:3dgauge}

Following the same argument we used in section \ref{ssec:5dgauge}, the quiver gauge theory on the D3 branes is not two-dimensional: the D3 branes are points on the cylinder $\cC=\mathbb{R}\times S^1(R)$, so by T-duality, they are equivalently D4 branes wrapping the T-dual cylinder  $\cC'=\mathbb{R}\times S^1(\widehat{R})$, with $\widehat{R}=1/m^2_s R$.
This makes it clear that the theory on the D3 branes is actually three-dimensional, defined on on a circle $S^1(\widehat{R})$. At low energies compared to the string scale $m_s$, an effective description of the theory on the D3 branes was described in \cite{Douglas:1996sw}. In the limit $g_s\rightarrow 0$, we obtain a 3d quiver gauge theory, of shape the Dynkin diagram of $\fg$, just like the parent theory $T^{5d}$ \footnote{Once again, note that the quiver is really the one corresponding to $\fg$, and not to $\widehat{\fg}$, since we are taking the $g_s\rightarrow 0$ limit, which decouples the affine node.}.  We denote this gauge theory by $G^{3d}$ in the rest of this paper.

\begin{figure}[h!]
	\emph{}
	\centering
	\includegraphics[width=1.0\textwidth]{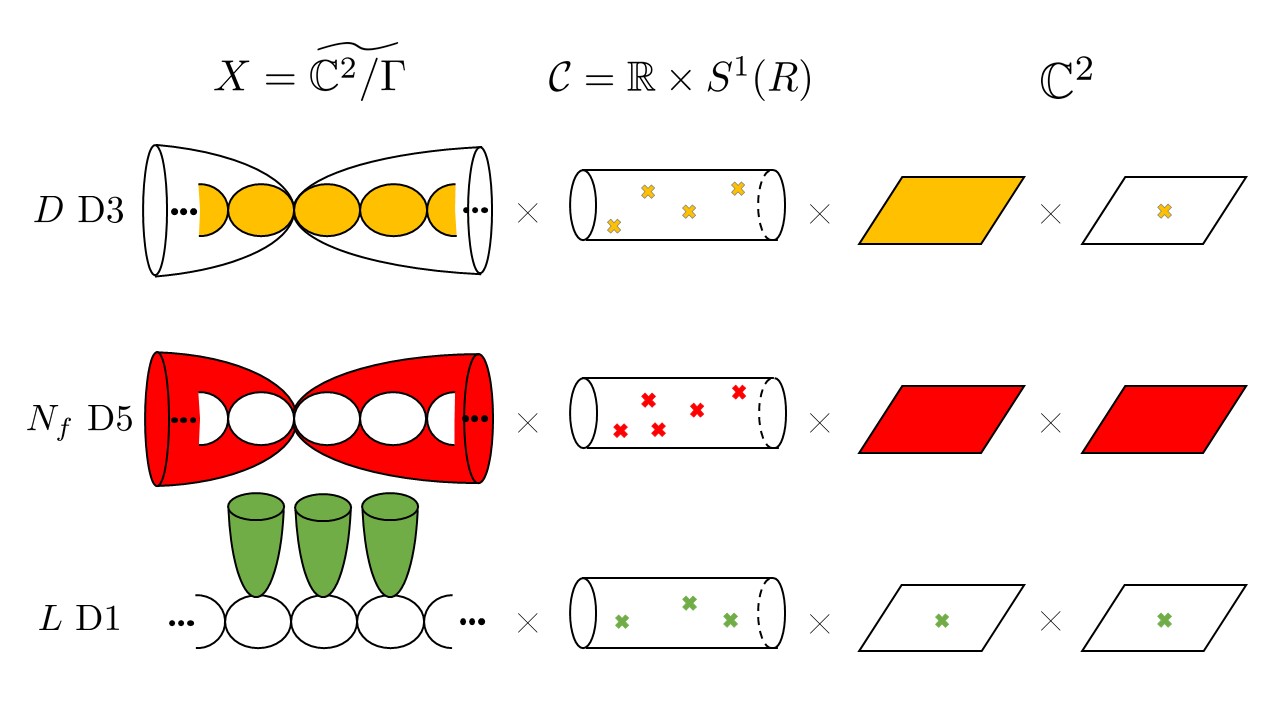}
	\vspace{-10pt}
	\caption{Brane configuration for the vortex theory. The $D$ D3 branes wrap compact 2-cycles $S_a$, and preserve 8 supercharges (yellow). The $N_f$ D5 branes wrapping the non-compact 2-cycles $S^*_d$ further break supersymmetry by half (red). The $L$ D1 branes wrap the non-compact 2-cycles $S^*_a$ (green); only 2 supercharges are preserved by this setup. A configuration of $k$ vortices corresponds to $k$ D1 branes (not pictured here) wrapping the compact 2-cycles $S_{a}$'s.} 
	\label{fig:branes3d}
\end{figure}

Let us first assume there are no D1 branes present. The theory $G^{3d}$ then has $\cN=2$ supersymmetry. This is consistent with the fact that the D3 branes are 1/2-BPS vortices of the 5d theory, which had 8 supercharges.
The gauge group is
\beq\label{3dgaugegroup}
G'=\prod_{a=1}^m U(D^{(a)})\; ,
\eeq
where the ranks $D^{(a)}$ were defined in \eqref{d3compactnsl}.
This implies that on the Coulomb branch of $G^{3d}$, where the theory is abelianized, one counts a total of $\sum_{a=1}^m\sum_{i=1}^{D^{(a)}}$ moduli, which we denote by $y^{(a)}_i$.

There is also matter coming from the intersection of 2-cycles $S_a$ and $S_b$ that touch each other at a point. Open strings with one end on the $a$-th D3 brane and the other end on the $b$-th D3 brane result in a hypermultiplet in the bifundamental representation $(D^{(a)}, \overline{D^{(b)}})$. We write the corresponding bifundamental mass parameters as $v^{(ab)}=\sqrt{q^{r^{(ab)}}/t}$. So far, the vector and matter multiplets are those of 3d $\cN=4$ supersymmetry. Adding the D5 branes results in additional matter multiplets, which breaks supersymmetry to $\cN=2$.\\

\begin{figure}[h!]
	\emph{}
	\centering
	\includegraphics[trim={0 0 0 1cm},clip,width=0.8\textwidth]{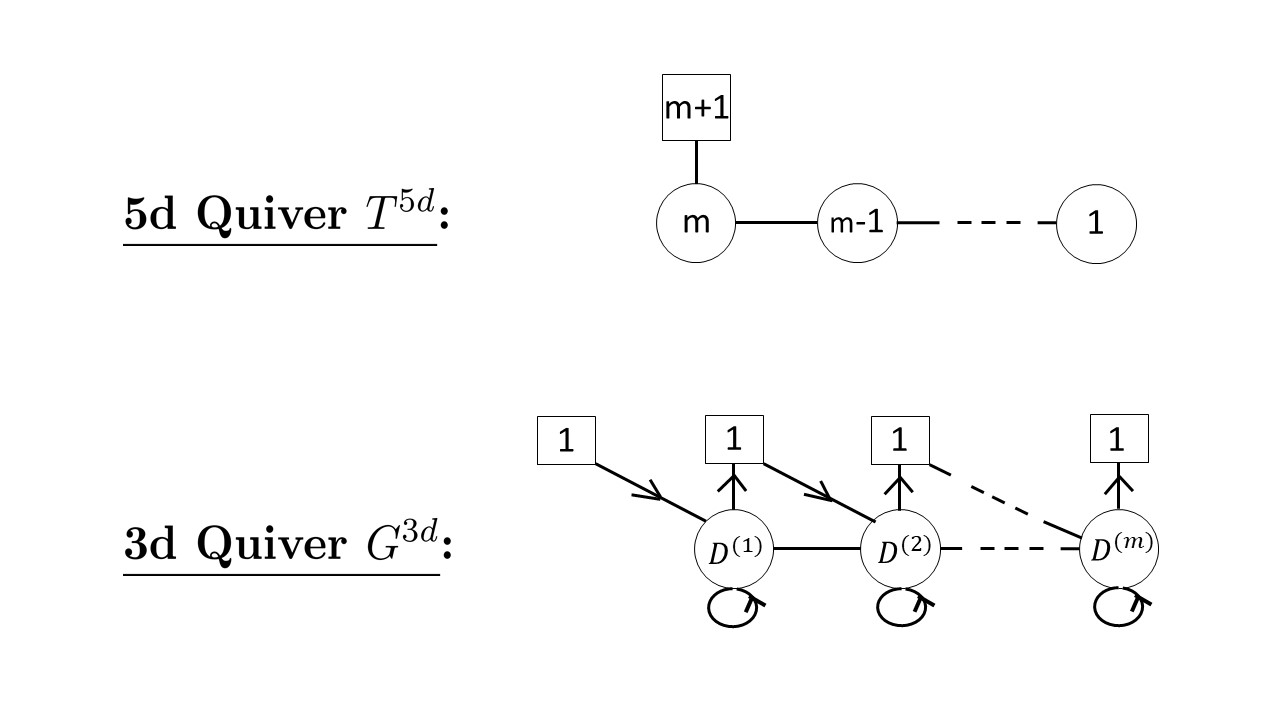}
	\vspace{-10pt}
	\caption{Example of a 5d theory $T^{5d}$ and the 3d theory on its vortices, $G^{3d}$. The spikes in the 3d theory are responsible for breaking the supersymmetry to $\cN=2$, and represent chiral/antichiral multiplets. Each loop represents a chiral multiplet in the adjoint representation of the gauge group $U(D^{(a)})$.}
	\label{fig:5dto3d}
\end{figure}

Let us first analyze  the simply-laced case $\fg=ADE$: recall that a single D5 brane is now labeled by a weight $\omega_d$ taken from the set $\{\omega_d\}_{d=1}^{N_f}$, and accordingly wraps a non-compact 2-cycle  $S_d^*$. We write the corresponding mass parameter as $x_d$. The strings stretching between the $D^{(a)}$ D3 branes and the D5 brane wrapping $S_d^*$ produce chiral and antichiral hypermultiplets, conveniently encoded in the Dynkin weights of $\omega_d$. Indeed, we have the following intersection pairing in homology
\beq
\#\left( S_a \cap S_d^* \right) = \langle \alpha_a , \omega_d \rangle\; ,
\eeq
which is explicitly computed from \eqref{weightexpression}. The weight $\omega_d$ is in a fundamental representation of $\fg$, with lowest weight $-\lambda_a$, say. If $\omega_d$ also happens to be in the Weyl group orbit of $-\lambda_a$, then it turns out we can read off the number of chiral and antichiral multiplets directly from the Dynkin labels of $\omega_d$. Namely, in the fundamental weight basis, we write $\omega_d=[\omega^{(1)}_d, \ldots, \omega^{(m)}_d]$, where each $\omega^{(a)}_d$ is an integer Dynkin weight; the strings stretching between the D5 brane labeled by $\omega_d$ and the $D^{(a)}$ D3 branes result in $\left|\omega^{(a)}_i\right|$ chiral multiplets (resp. antichiral multiplets) on node $a$ when $\omega^{(a)}_i$ is strictly positive (resp. strictly negative). In particular, a zero Dynkin weight $\omega^{(a)}_i=0$ for the $a$-th entry of $\omega_d$ means that the D5 brane will contribute  no matter multiplets on node $a$.
If the weight $\omega_d$  \textit{does not} belong in the Weyl group orbit of $-\lambda_a$, then one has to do more work to figure out the matter content of $G^{3d}$; see \cite{Haouzi:2016ohr, Haouzi:2016yyg} for details.\\

It now remains to understand the effect of the D1 branes, which break supersymmetry further down to 2 supercharges. The D1 branes are point defects on the line $\mathbb{C}_q$ where the D3 brane gauge theory lives. The number of such defects is $L=\prod_{a=1}^m L^{(a)}$, with associated flavor symmetry
\beq\label{3ddefectflavorgroup}
G^{D1}_F=\prod_{a=1}^m U(L^{(a)})\; .
\eeq
From the point of view of $G^{3d}$, which is defined on $S^1(\widehat{R})\times\mathbb{C}_q$, the D1 branes make up a 1/2-BPS Wilson loop wrapping the circle $S^1(\widehat{R})$ and sitting at the origin of $\mathbb{C}_q$. 

The analysis is fully analogous to the one we made for the Wilson loops of $T^{5d}$, so we will be brief. Indeed, in the case of a single gauge group $U(D^{(a)})$, introducing a 1/2-BPS Wilson loop can be done with the use of a one-dimensional fermion field $\chi^{(a)}$, transforming in the fundamental representation of $U(D^{(a)})$ and in the fundamental representation of $U(L^{(a)})$, coupled to the 3d gauge field in the bulk as \cite{Assel:2015oxa}:
\beq
\label{3d1dfermion}
S^{1d/3d}=\int dt\; {\chi_{i,\rho}^{(a)}}^\dagger\, \left( \delta_{ij}\, \partial_t - i\, A^{(a)}_{t, ij}   + \sigma^{(a)}_{ij}   - \delta_{ij}\, M_{\rho}^{(a)}  \right)\, \chi_{j,\rho}^{(a)} \; .
\eeq
Above, $A^{(a)}_t$ and $\sigma^{(a)}$ are the pullback of the 3d gauge field and the adjoint scalar of the vector multiplet, respectively. $i$ and $j$ are indices for the fundamental representation of $U(D^{(a)})$, while $\rho$ is an index for the fundamental representation of $U(L^{(a)})$. The variable $t$ is periodic, with period $2\pi \,\widehat{R}$. Just like in 5d, the parameters $M_\rho^{(a)}$ denote masses for the fermions, and can be thought of as a background gauge field for the $U(L^{(a)})$ symmetry acting on them.

The path integral of the coupled 3d/1d system is therefore
\beq\label{3d1dpathintegral}
\left[\chi^{\fg}\right]^{3d}= \int D\psi D\chi \, e^{i(S^{3d}[\psi]+ S^{1d/3d}[\psi, \chi])} \; ,
\eeq
where $\psi$ denotes collectively all the fields of the bulk 3d theory, written as $S^{3d}$, while $S^{1d/3d}$ denotes the action term \eqref{3d1dfermion}.

The full dictionary from geometry to gauge theory is as follows:
The periods $\tau^{(a)}$ in \eqref{couplings}, which used to be the gauge couplings of $T^{5d}$, now become 3d real F.I. terms of $G^{3d}$, complexified because of the presence of the circle $S^1(\widehat{R})$. The periods of the NS-NS  B-field $B^{(2)}$ and of the 2-form $\omega_K$ are the complex F.I. parameters, set to zero. The gauge couplings on the D3 branes are also set to zero. 
The positions on the cylinder $\cC$ of the $D$ D3 branes are the Coulomb moduli of $G^{3d}$. The positions on $\cC$ of the $N_f$ non-compact D5 branes are the mass parameters for the chiral and antichiral multiplets of  $G^{3d}$. Finally, the positions on $\cC$ of the $L$ D1 branes are the fermion masses introduced in \eqref{3d1dfermion} to define the Wilson loop coupling to 3d.\\

The discussion generalizes at once to 3d $BCFG$-type quiver theories, after restricting the analysis to configurations of branes invariant under the action of the $ADE$ outer automorphism. Just like in 5d, the various 3d fields $\psi$ of $G^{3d}$ which depend on nodes labeling short roots are single-valued only when written as $\psi_{short}+a\cdot \psi_{short}+\ldots+a^{r-1}\cdot \psi_{short}$, where $r$ is the lacing number of $\fg$ and $a$ an element of the outer automorphism group.

\subsection{Gauge/Vortex Duality}
\label{ssec:duality}

Gauge/vortex duality is a correspondence between the gauge theory $T^{5d}$ on the D5 branes, and the gauge theory $G^{3d}$ on D3 branes\footnote{Originally, the duality was phrased between a 4d $\cN=2$ gauge theory placed in 2d $\Omega$-background, and a two-dimensional theory with ${\cN}=(2,2)$ supersymmetry, living on the vortices of the former. 
The first evidence of this relation was given in \cite{Dorey:1998yh}, and further developed in \cite{Dorey:1999zk}, there it was recognized that the BPS spectra of the vortices living in the parent 4d theory matched the BPS spectrum of the 2d theory. 
This relation was also verified at the level partition functions. After specializing the 4d Coulomb parameters, it was observed that the superpotentials of both theories were the same \cite{Dorey:2011pa,Chen:2011sj}.}. On the one hand, one can describe $G^{3d}$ from the point of view of vortices in $T^{5d}$. On the other hand, one can describe $G^{3d}$ from the point of view of the vortices themselves. The dimensions of $T^{5d}$ and $G^{3d}$ do not agree, but nonetheless, turning on the $\Omega$-background transverse to the vortex has drastic effects. For instance,  this ensures that the two theories now preserve same supersymmetries, and are both effectively described by a 3d theory on $S^1(\widehat{R})\times \mathbb{C}_q$, making the duality possible. 

Remarkably, the vortex theory captures some features of the parent theory at special values of the 5d Coulomb moduli. To see this, we note that in the presence of transverse $\Omega$-background, the effect of turning on $D^{(a)}_i$ units of vortex flux is not to introduce additional defects in the theory, as one might have expected, but rather to simply shift the  5d Coulomb moduli by $D^{(a)}_i$ units of $t$, meaning $e^{(a)}_i \rightarrow e^{(a)}_i\, t^{(D^{(a)}_i)}$ \cite{Nekrasov:2010ka}. This, in turn, pushes the 5d theory  back onto the Coulomb branch, on an integer-valued $t$-lattice.
The reason why the equivariant parameter $t$ appears here is because it rotates the line $\mathbb{C}_t$ where we turned on the $\Omega$-background, which is precisely transverse to where $G^{3d}$ lives.

In what follows, the observation that $T^{5d}$ has a dual interpretation at specialized values of its the Coulomb moduli will be crucial to prove the equality of 5d and 3d partition functions, this time in full $\Omega$-background.

\subsection{The Partition Function of $G^{3d}$ with Wilson Loops}
\label{ssec:polar}

We now focus on computing the partition function of the vortex theory $G^{3d}$. We will do it two ways.

\subsubsection{Witten Index of the Vortex Quantum Mechanics}
\label{sssec:vortexwitten} 

We saw that the instanton partition function of the five-dimensional theory $T^{5d}$ was computable as the Witten index of a gauged quantum mechanics. The same is true of the vortex partition function for the three-dimensional theory $G^{3d}$.
First, since $G^{3d}$ is defined on $\mathbb{C}\times S^1{(\widehat{R})}$, with coordinate $z_1$ on $\mathbb{C}$, we can introduce an $\Omega$-background. In practice, this means that as we go around the circle, the coordinate $z_1$ is mapped to
\begin{align}
z_{1}\mapsto q^{r^{(a)}}\,z_{1} \; .
\end{align}

The partition function is then given by
\beq\label{3dindex}
\left[\chi^{\fg}\right]^{3d}  = {\rm Tr}\, (-1)^F \; \fm \;\; ,
\eeq
where $\fm=q^{r^{(a)}(S_1 - S_R)} t^{-S_2 + S_R}\, \fm_{G',\; G_F,\; G^{D1}_F}$; here, $S_1$ and $S_2$ are the generators of the rotations of the  lines ${\mathbb C}_{q}\times {\mathbb C}_{t}$ as we go around the circle $S^1(\widehat{R})$. The fugacity $q$ keeps track of which roots are long and which roots are short. $F$ is the fermion number. $S_R$ is the generator of a $U(1)_R$ R-symmetry charge which is a subset of the R-symmetry. The factor $\fm_{G',\; G_F,\; G^{D1}_F}$  denotes the product of fugacities associated with the Cartan generators of the various global symmetry groups $G'$, $G_F$ and $G^{D1}_F$. In practice, the contribution of $k$ vortices to the index is the  partition function of the theory on $k$ D1$_{vortex}$ branes wrapping the compact 2-cycles of $X$. The various multiplets arise from strings stretching between these D1$_{vortex}$ branes and all the other branes pictured in figure \ref{fig:branes3d}.\\

We preemptively set the 5d bifundamental masses $m^{(a)}_{bif}$ to be equal to
 \beq\label{bifspecialized}
 m^{(a)}_{bif}= - \epsilon_+ - (r^{(ab)}-1)\epsilon_1/2\; ,
 \eeq
 where $m^{(a)}_{bif}$ is a mass for the multiplet between nodes $a$ and $b$. Equivalently, in terms of K-theoretic variables $\mu{(a)}_{bif}=e^{-R\,m^{(a)}_{bif}}$, $q=e^{R\,\epsilon_1}$, and $t=e^{-R\,\epsilon_2}$, this reads
 \beq\label{bifspecialized2}
 \mu^{(a)}_{bif}=v^{(ab)}\; .
 \eeq
This specialization of bifundamental masses is by no means necessary, and is performed here out of convenience to match our results more easily to the existing literature. In particular, it will make the next part of our discussion easier to follow when we make contact with $\cW$-algebras.\\
 
Now, a convenient shortcut to compute the index \eqref{3dindex}  is to make use of the gauge/vortex duality we reviewed in the previous section. Indeed, it is enough to start with the quantum mechanics index for the 5d partition function \eqref{5dintegral}, and specialize the 5d $G$-equivariant parameters as 
\beq\label{specialize5d}
e^{(a)}_{i}=x_{d_i}\,t^{D^{(a)}_i} v^{-\#^{(a)}_{i,d}}q^{-\widetilde{\#}^{(a)}_{i,d}/2} \; .
\eeq
Above, $D^{(a)}_i$ is a strictly positive integer denoting units of vortex flux, and $x_{d_i}$ is one of the 5d masses $f^{(b)}_i$, written in 3d notation using \eqref{weightexpression}. $\#^{(a)}_{i,d}$ is a strictly positive integer, and $\widetilde{\#}^{(a)}_{i,d}$ is a positive integer which is nonzero only when $\fg$ is a non simply-laced Lie algebra.
The integers $\#^{(a)}_{i,d}$ and $\widetilde{\#}^{(a)}_{i,d}$, are all determined from the R-symmetry charges of the various $G^{3d}$ multiplets.
In practice, after writing the masses as $x_{d}= e^{-\widehat{R}\,m_d}$, this amounts to substituting 
\beq\label{specialize4d}
a^{(a)}_i = m_{d_i}+D^{(a)}_i \epsilon_2+\#^{(a)}_{i,d}\,\epsilon_+ + \widetilde{\#}^{(a)}_{i,d}\, \epsilon_1/2\,,
\eeq
in the integrand  \eqref{5dintegral}. After these replacements, one recognizes the partition function as the quantum mechanics index of a 3d $\cN=2$ quiver gauge theory (see for instance \cite{Hwang:2017kmk}), with a 1d Wilson loop defect inserted: 
\beq\label{3dintegral}
\left[\chi^{\fg}\right]_{(L^{(1)},\ldots, L^{(m)})}^{3d}  =\sum_{k^{(1)},\ldots, k^{(m)}=0}^{\infty}\;\prod_{a=1}^{m}\frac{\left[\widetilde{\fq}^{(a)}\right]{}^{k^{(a)}}}{k^{(a)}!} \, \oint  \left[\frac{d\phi^{{(a)}}_I}{2\pi i}\right]Z^{(a), 3d}_{vec}\cdot Z^{(a), 3d}_{matter}\cdot \prod_{b>a}^{m} Z^{(a,b), 3d}_{bif}\cdot \prod_{\rho=1}^{L^{(a)}}  Z^{(a), 3d}_{D1}\,.
\eeq
Above, the parameter $\widetilde{\fq}^{(a)}$, which used to be the 5d gauge coupling for the $a$-th gauge group, or instanton counting parameter, is now understood as the $a$-th 3d F.I. parameter, or vortex counting parameter. The vector and bifundamental multiplets contributions are those of a 3d $\cN=4$ theory: 
\begin{align}
&Z^{(a), 3d}_{vec} =\prod_{I, J=1}^{k^{(a)}} \frac{{\widehat{f}}\left(\phi^{(a)}_{I}-\phi^{(a)}_{J}\right)f\left(\phi^{(a)}_{I}-\phi^{(a)}_{J}+ 2\,\epsilon_+ + (r^{(a)}-1)\,\epsilon_1\right)}{f\left(\phi^{(a)}_{I}-\phi^{(a)}_{J}+\epsilon_1 + (r^{(a)}-1)\,\epsilon_1\right)f\left(\phi^{(a)}_{I}-\phi^{(a)}_{J}+\epsilon_2\right)}\label{3dintegral2}\\
&Z^{(a, b), 3d}_{bif} = \left[\prod_{I=1}^{k^{(a)}}\prod_{J=1}^{k^{(b)}} \frac{f\left(\phi^{(a)}_{I}-\phi^{(b)}_{J}- \epsilon_+ +\epsilon_1  - (\overline{r^{(a b)}}+\overline{r^{(b a)}}+r^{(ab)}-2\, r^{(a)}-1)\epsilon_1/2\right)}{f\left(\phi^{(a)}_{I}-\phi^{(b)}_{J} - \epsilon_+ - (\overline{r^{(a b)}}+\overline{r^{(b a)}}+r^{(ab)}-3)\epsilon_1/2\right)}\right.\nonumber\\
&\qquad \times \left.\prod_{I=1}^{k^{(a)}}\prod_{J=1}^{k^{(b)}} \frac{f\left(\phi^{(a)}_{I}-\phi^{(b)}_{J}- \epsilon_+ +\epsilon_2  - (\overline{r^{(a b)}}+\overline{r^{(b a)}}+r^{(ab)}-3)\epsilon_1/2 \right)}{f\left(\phi^{(a)}_{I}-\phi^{(b)}_{J}+ \epsilon_+  - (\overline{r^{(a b)}}+\overline{r^{(b a)}}+r^{(ab)}-2\, r^{(a)}-1)\epsilon_1/2\right)}\right]^{\Delta_{a,b}}\; .
\end{align}
All the notations used here were introduced in section \ref{ssec:5dpartition}, and as we mentioned, we have explicitly written  the bifundamental masses as $m^{(a)}_{bif}= - \epsilon_+ - (r^{(ab)}-1)\epsilon_1/2$.
A quick glance at the 5d integrand reveals that the vector and bifundamental multiplet contributions have more factors than their 3d counterpart. Indeed, the 5d vector contains extra $(0,2)$ chiral multiplets, while the 5d bifundamental contains extra  $(0,2)$ Fermi multiplets. These extra multiplets combine together with the Fermi multiplets coming from the 5d fundamental matter factor $Z^{(a)}_{fund}$ to make up the $\cN=2$ matter content $Z^{(a), 3d}_{matter}$ of the 3d quiver gauge theory.
Finally,  after imposing \eqref{specialize4d}, the Wilson loop contribution which we previously called $Z^{(a)}_{D1}$ in 5d now becomes a Wilson loop contribution in 3d, denoted as $Z^{(a), 3d}_{D1}$.\\

\subsubsection{Integration Contours}
\label{sssec:3dcountours}

To fully determine the partition function \eqref{3dintegral}, one still needs to specify the  $k=\sum_{a=1}^m k^{(a)}$ contours of integration. This can be done once again using the JK residue prescription. Just as before, the computation of the partition function will depend on the sign of the F.I. parameters in the quiver quantum mechanics. We take the sign of these F.I. parameters to be strictly positive, as we did before in 5d. Then, the contour prescription is exactly as it was in section \ref{sssec:5dcontours}.\\

Let us first consider the poles in the absence of the Wilson loop. The poles are no longer classified by  $n^{(a)}$-tuples of arbitrary Young diagrams as was the case in 5d. Instead, one should label the poles using  \textit{truncated} Young diagrams, meaning that a given Young digram $\boldsymbol{\mu}^{(a)}_i$ cannot have more than $D^{(a)}_i$ rows, where this integer was defined in \eqref{specialize5d} as  vortex flux.
We can justify this directly from the vortex quantum mechanics integral, as follows: the integrand necessarily contains the factor
\beq\label{prooftruncated1}
\prod_{I=1}^{k^{(a)}} \frac{f\left(\phi^{(a)}_I-m_d\right)}{f\left(\phi^{(a)}_I-m_d-D^{(a)}_i \epsilon_2 \right)}\; ,
\eeq
for some node $a$ and mass $m_d$. Such a factor appears because of the 3d matter $Z^{(a), 3d}_{matter}$, and manifests itself here in the quantum mechanics as a  $(0,2)$ Fermi multiplet in the numerator and a $(0,2)$ chiral multiplet in the denominator. Another way to understand \eqref{prooftruncated1} is directly from 5d: the numerator is part of what used to be the 5d fundamental matter $Z^{(a)}_{fund}$ contribution, while the denominator is one of two $(0,2)$ chiral multiplets that used to sit inside a $(0,4)$ hypermultiplet, as part of the 5d vector multiplet $Z^{(a)}_{vec}$. This chiral multiplet contribution used to depend on some 5d Coulomb parameter $a^{(a)}_i$, but in 3d, it depends instead on the mass $m_d$, thanks to \eqref{specialize4d}. Explicitly, there exists at least one 5d Coulomb parameter of the form
\beq\label{prooftruncated2}
a^{(a)}_i= m_{d_i} + D^{(a)}_i \epsilon_2 + \epsilon_+ + (r^{(a)}-1)\,\epsilon_1/2\; ,
\eeq
resulting in the denominator \eqref{prooftruncated1}. Now, by the JK prescription, one must enclose the pole in this denominator, so, for definiteness, let
\beq\label{prooftruncated3}
\phi^{(a)}_1 = m_d + D^{(a)}_i \epsilon_2 \; .
\eeq

Meanwhile, the 3d $\cN=4$ vector multiplet on node $a$, which we wrote explicitly in \eqref{3dintegral2}, contributes the following $(0,2)$ chiral multiplets to the integral:
\beq\label{prooftruncated4}
\prod_{I, J=1}^{k^{(a)}} \frac{1}{f\left(\phi^{(a)}_{I J}+\epsilon_1 + (r^{(a)}-1)\,\epsilon_1\right)f\left(\phi^{(a)}_{I J}+\epsilon_2\right)}\; .
\eeq
The JK prescription dictates to enclose poles from both of these chiral multiplets. The first factor, which depends on $\epsilon_1$, results in a sequence of poles labeled by 1d partitions. At first glance, one may think that a similar sequence of poles will occur due to the second factor, depending on $\epsilon_2$. This was the case for the 5d integral,  where the poles ended up in labeled by 2d partitions. However, in this 3d setting, the numerator in \eqref{prooftruncated1} prevents the poles from ``growing too much" in $\epsilon_2$. Indeed, starting from $\phi^{(a)}_1$ and considering the successive poles shifted by $-\epsilon_2$, we have
\begin{align}
\phi^{(a)}_1 &= m_d + D^{(a)}_i \epsilon_2 \nonumber\\
\phi^{(a)}_2 &= m_d + D^{(a)}_i \epsilon_2 -\epsilon_2\nonumber\\
\vdots\;\;\; &=\;\;\; \vdots\nonumber\\
\phi^{(a)}_{D^{(a)}_i} &= m_d + D^{(a)}_i \epsilon_2 - (D^{(a)}_i-1)\,\epsilon_2
\end{align}
The sequence of poles terminates here and we cannot have another pole shifted further by $-\epsilon_2$: this would make the whole partition function vanish, because of the numerator. All in all, the contour in question encloses poles that are labeled by a Young diagram with an arbitrary number of columns (the $\epsilon_1$ label), but no more than $D^{(a)}_i$ rows (the $\epsilon_2$ label).

Though the above proof was given fo a single contour only, the argument generalizes to all allowed contours, for all possible sequences of poles. In particular, starting from the $a$-th node as we did, one should now consider the effects of the bifundamentals, which will truncate Young diagrams on all the other nodes.\\

We now consider the integrand in the presence of the Wilson loop factor $Z^{(a), 3d}_{D1}$. Just like in 5d, this factor introduces new poles in the integrand, which modifies the previous truncated Young diagram rule. The additional poles will therefore depend on the fermion masses $M^{(a)}_\rho$. In fact, the exhaustive list of new poles is the same as the one we had identified in 5d, with the redefinition \eqref{bifspecialized} of the bifundamental masses:
\begin{align}
&\phi^{(a)}_I=M^{(a)}_\rho+\epsilon_+ +(r^{(a)}-1)\epsilon_1/2 \label{3dnewpole1}\\
&\phi^{(a+1)}_J=\phi^{(a)}_I + \epsilon_+  - (\overline{r^{(a b)}}+\overline{r^{(b a)}}+r^{(ab)}-2\, r^{(a)}-1)\epsilon_1/2 \label{3dnewpole2}\\
&\phi^{(a-1)}_J=\phi^{(a)}_I  + \epsilon_+ + (\overline{r^{(a b)}}+\overline{r^{(b a)}}+r^{(ab)}-3)\epsilon_1/2\label{3dnewpole3}
\end{align}
Having identified the poles to be enclosed by the contours, the vortex partition function can be evaluated in a straightforward way.
In the spirit of the computation we carried out in 5d, we will now argue that this vortex partition function can also be expressed as the character of some finite-dimensional irreducible representation of the quantum affine algebra $U_q(\widehat{\fg})$. To prove this, we  use the same strategy as in 5d, and introducing a defect operator expectation value, one for each node $a=1, \ldots, m$ of $G^{3d}$. The expectation value of the $a$-th defect operator, with corresponding fermion mass $z^{(a)}_\rho\equiv z$, is defined as
\begin{align}
	\label{3dYoperator}
	\left\langle \left[Y^{(a)}_{3d}\right]^{\pm 1} (z)\right\rangle =
	\sum_{k^{(1)},\ldots, k^{(m)}=0}^{\infty}\;\prod_{b=1}^{m}\frac{\left(\widetilde{\fq}^{(b)}\right)^{k^{(b)}}}{k^{(b)}!} \, \oint_{\{\overrightarrow{\boldsymbol{\mu}}\}}  \left[\frac{d\phi^{{(b)}}_I}{2\pi i}\right]Z^{(b), 3d}_{vec}\cdot Z^{(b), 3d}_{fund}\cdot \prod_{c>b}^{m} Z^{(b, c), 3d}_{bif}  \cdot \left[Z^{(a), 3d}_{D1}\right]^{\pm 1}\, . 
\end{align}
Importantly,  the above contour integral is defined to only enclose poles labeled by (truncated) Young diagrams. This is in contrast to the actual physical partition function \eqref{3dintegral}, where the extra poles \eqref{3dnewpole1}, \eqref{3dnewpole2}, and \eqref{3dnewpole3}, required by the JK prescription, contribute with nonzero residue. We have denoted the $Y$-operator integral by the symbol $\oint_{\{\overrightarrow{\boldsymbol{\mu}}\}}$ to emphasize the different contour. Then, as a function of the fermion masses $z^{(a)}_\rho$, we find
\begin{align}\label{character3d}
	\boxed{\left[\chi^{\fg}\right]_{(L^{(1)},\ldots, L^{(m)})}^{3d}(z^{(a)}_\rho)=\sum_{\omega\in V(\lambda)}\prod_{b=1}^m \left({\widetilde{\fq}^{(b)}}\right)^{d_b^\omega}\; c_{d_b^\omega}(q, t)\; \left(Q^{(b)}(z_{* \rho}^{(a)})\right)^{d_b^\omega}\,  \left[{Y}_{3d}(z^{(a)}_\rho)\right]_{\omega} \, .}
\end{align}
The vortex partition function is a new twisted $qq$-character of a finite dimensional irreducible representation $V(\lambda)$ of $U_q(\widehat{\fg})$, with highest weight $\lambda=\sum_{a=1}^m L^{(a)}\, \lambda_a$. This time, however, the twist is to be interpreted in the language of a 3d gauge theory. 
As before, $\omega$ runs over all the weights of the representation $V(\lambda)$. The label $d_b^\omega$ is a positive integer that is determined by solving
\beq\label{3dsl2string}
\omega=\lambda -\sum_{b=1}^m d_b^\omega\, \alpha_b\; .
\eeq
Namely, any weight $\omega$ is reached by lowering  a finite number of times the highest weight $\lambda$ of the representation, using the positive simple roots $\alpha_b$. The factor $\widetilde{\fq}^{(b)}$ is the 3d F.I. parameter for the $b$-th gauge group. The factors $c_{d_b^\omega}(q, t)$ are coefficients depending only on $q$ and $t$. The function $Q^{(b)}(z^{(a)}_\rho)$ is understood as the contribution of  (anti)chiral multiplets in fundamental representation:
\beq
Q^{(b)}(z_{* \rho}^{(a)})=\prod_{d=1}^{N^{(b)}_f} \left(1-x_d/z_{* \rho}^{(a)}\right)\; .
\eeq

The variables $z_{* \rho}^{(a)}$ are the fermion masses $z^{(a)}_\rho$, shifted by various powers of $q$ and $t$. They are precisely the residues at the poles \eqref{3dnewpole1}, \eqref{3dnewpole2}, \eqref{3dnewpole3} in the integrand of the partition function \eqref{3dintegral}.

Finally, the operator $\left[{Y}_{3d}(z^{(a)}_\rho)\right]_{\omega}$, for a given weight $\omega$, is in general the expectation value of a product of various defect $Y$-operators $\langle \prod_a \left[{Y^{(a)}_{3d}}\right]^{\pm 1} \rangle $, where each operator $\left[{Y^{(a)}_{3d}}\right]^{\pm 1}$ is a function of a fermion mass $z^{(a)}_\rho$. The arguments of each factor is shifted by various powers of $q$ and $t$, uniquely determined by \eqref{3dsl2string}. We once again call the $qq$-character twisted because of the presence of the 3d F.I. parameters $\widetilde{\fq}^{(b)}$ and the matter factors $Q^{(b)}$.\\

At this point, we could proceed and evaluate the integrals in the partition function, just like we did in five dimensions. Because of the contour prescription in \eqref{3dYoperator}, each pole is labeled by a Young diagram, with at most $D^{(a)}_i$ rows. As a result, the integral evaluates to a sum over (truncated) partitions. We would find that the summand is made of a bulk 3d $\cN=2$ contribution, written in terms of various $q$-Pochhammer functions, and some $Y$-operator insertions. The partition function is once again a generating function of Wilson loops, where each loop is valued in the representation ${\textbf{R}}=\left({\textbf{R}}^{(1)}, \ldots , {\textbf{R}}^{(m)}\right)$ of the gauge groups $\prod_{a=1}^m SU(D^{(a)})$. For a given node $a$, the representation $\textbf{R}^{(a)}$ is a tensor product of fundamental representations of $SU(D^{(a)})$.

In conclusion, starting from the quantum mechanics of $T^{5d}$ and imposing the specialization of the Coulomb parameters \eqref{specialize4d} in the index, we were able to derive the partition function of $G^{3d}$. We will now give an alternate derivation of the 3d partition function, perhaps more illuminating, without the need to compute the Witten index of a vortex quantum mechanics. Indeed, in three dimensions, there exists a different integral representation of the partition function\footnote{This is sometimes called 3d Coulomb branch localization.}. The advantage of this alternate presentation is that it is directly related to certain correlators in deformed Toda theory, as we will soon see.

\subsubsection{3d $\cN=2$ Index Presentation}
\label{sssec:3dindex}

In the absence of D1 branes,  one can define the following twisted index  on  $S^1(\widehat{R})\times\mathbb{C}_q$\footnote{It is also referred to as a holomorphic block \cite{Beem:2012mb}. Sometimes, this index is also defined on the manifold  $S^1(\widehat{R})\times D$, where $D$ is the disk. Because this new manifold has a boundary, it becomes important to specify the boundary conditions of the various fields at the edge of the disk. In particular, what we call antichiral multiplets are now understood as 3d chiral multiplets with Dirichlet boundary conditions, while our chiral multiplets are really chiral multiplets with Neumann boundary conditions. The gauge fields have Neumann boundary conditions, and the appearance of theta functions in the 3d vector multiplet is understood as the contribution of the 2d elliptic genus on the  boundary torus. For details, see \cite{Gadde:2013wq,Yoshida:2014ssa,Dimofte:2017tpi}, and the discussion on boundary conditions in our context in \cite{Aganagic:2016jmx,Aganagic:2017smx}.}, with $\mathbb{C}_q$ placed in the $\Omega$-background:
\beq\label{new3dindex}
\left[\widetilde{\chi}^{\fg}\right]^{3d}  = {\rm Tr}\, (-1)^F \; \fm \;\; ,
\eeq
where  $\fm=q^{r^{(a)}(S_1 - S_R)} t^{-S_2 + S_R}\, \fm_{G_F}$; here, $S_1$ and $S_2$ are the generators of the rotations of the  lines ${\mathbb C}_{q}\times {\mathbb C}_{t}$ as we go around the circle $S^1({\widehat{R}})$. The fugacity $q$ keeps track of which roots are long and which roots are short. $F$ is the fermion number. $S_R$ is once again the generator of a $U(1)_R$ R-symmetry charge which is a subset of the R-symmetry. The factor $\fm_{G_F}$ stands for the global symmetries, which here includes an adjoint chiral multiplet. This index happens to have a natural representation as an integral over the 3d Coulomb moduli, which we will review next.

We would like to introduce the Wilson loop, which is a codimension 2 defect from the point of view of $G^{3d}$. We conjecture that the index in the presence of a Wilson loop can be computed by inserting a 1d defect factor inside the 3d Coulomb integral. The 3d/1d coupling is achieved by partially gauging the 1d defect and integrating over it. To achieve this, we start by defining a renormalized $Y$-operator expectation value, written as an integral over the Coulomb moduli of the 3d theory 
\begin{align}\label{3ddefectexpression2}
\left\langle\left[{\widetilde{Y}^{(a)}_{3d}}(z)\right]^{\pm 1} \right\rangle \equiv \left[{\widetilde{Y}^{(a)}_{D1/D5}}(\{x_d\}, z)\right]\; \oint_{\{\overrightarrow{\boldsymbol{\mu}}\}} d{y}\,\left[I^{3d}_{bulk}(y)\, \left[{\widetilde{Y}^{(a)}_{D1/D3}}(y, z)\right]^{\pm 1}\right]  \; .
\end{align}
Above, the integration variables denoted collectively by $y$ stand for the 3d Coulomb parameters of the gauge groups, meaning they parameterize the Cartan subalgebra of $G=\prod_{a=1}^{m}U(D^{(a)})$): 
\beq\label{yvariables}
``\int dy\; "\equiv\frac{1}{|W_{G^{3d}}|}\prod_{a=1}^{m}\int d^{D^{(a)}}y^{(a)} \; ,
\eeq
where $W_{G^{3d}}$ is the Weyl group of the gauge group $G$. We ask that the contours enclose poles labeled by Young diagrams $\{\overrightarrow{\boldsymbol{\mu}}\}$ with no more than $D^{(a)}_i$ rows. This choice of contours is in one-to-one correspondence with a choice of a vacuum for $G^{3d}$, and results in the breaking of the gauge symmetry
\beq\label{breaking}
G=\prod_{a=1}^{m}U(D^{(a)})\qquad\longrightarrow \qquad \prod_{a=1}^{m}\prod_{i=1}^{n^{(a)}}U(D^{(a)}_i)\; .
\eeq
This implies $D^{(a)}=\sum_{i=1}^{n^{(a)}}D^{(a)}_i$. Correspondingly, the integration variables $y$ break up as 
\beq\label{ybreaking}
\prod_{a=1}^{m}\prod_{l=1}^{D^{(a)}} y^{(a)}_l \longrightarrow \prod_{a=1}^{m}\prod_{i=1}^{n^{(a)}}\prod_{k=1}^{D^{(a)}_i}  y^{(a)}_{i,k}\; .
\eeq 

$I^{3d}_{bulk}(y)$ stands for the contribution of all the bulk multiplets to the partition function, in the absence of Wilson loops.  These can be read off from the 3d $\cN=2$ quiver description of the theory.  Generically, this bulk contribution has the following form: 
\begin{align}\label{bulk3d}
I^{3d}_{bulk}(y)=\prod_{a=1}^{m}\prod_{l=1}^{D^{(a)}}{y^{(a)}_l}^{\left(\tau^{(a)}-1\right)}\;I^{(a)}_{bulk, vec}\cdot I^{(a)}_{bulk, matter}\cdot\prod_{b>a}I^{(a,b)}_{bulk, bif}\; .
\end{align}

The factor 
\begin{align}\label{FI3d}
\prod_{a=1}^{m}\prod_{l=1}^{D^{(a)}}{y^{(a)}_l}^{\left(\tau^{(a)}\right)}
\end{align}
is the contribution of the 3d F.I. parameters. In type IIB, it comes about from turning on the periods (\ref{couplings}) and (\ref{NSNS}). The factor
\begin{align}\label{vec3d}
I^{(a)}_{bulk, vec}(y^{(a)})=\prod_{1\leq i\neq j\leq D^{(a)}}\frac{\left(y^{(a)}_{i}/y^{(a)}_{j};q^{r^{(a)}}\right)_{\infty}}{\left(t\, y^{(a)}_{i}/y^{(a)}_{j};q^{r^{(a)}}\right)_{\infty}}\;\prod_{1\leq i<j\leq D^{(a)}} \frac{\Theta\left(t\,y^{(a)}_{i}/y^{(a)}_{j};q^{r^{(a)}}\right)}{\Theta\left(y^{(a)}_{i}/y^{(a)}_{j};q^{r^{(a)}}\right)}
\end{align}
stands for the contribution of a ${\cN}=4$ vector multiplet for the gauge group $U(D^{(a)})$. We used the following definition of the theta function: $\Theta\left(x\,;q^{r^{(a)}}\right)= \left(x \,;\, q^{r^{(a)}}\right)_\infty\,\left(q^{r^{(a)}}/x \,;\, q^{r^{(a)}}\right)_\infty$.  In type IIB, this vector multiplet contribution comes about from quantizing D3/D3 strings, with the D3 branes wrapping the $a$-th compact 2-cycle. 

The factor
\begin{align}
\label{bif3d}
I^{(a,b)}_{bulk, bif}(y^{(a)}, y^{(b)})=\prod_{1\leq i \leq D^{(a)}}\prod_{1\leq j \leq D^{(b)}}\left [ \frac{(v^{(ab)} t\, y^{(a)}_{i}/y^{(b)}_{j};q^{r^{(ab)}})_{\infty}}{(v^{(ab)} \, y^{(a)}_{i}/y^{(b)}_{j};q^{r^{(ab)}})_{\infty}}\right]^{\Delta_{a, b}}
\end{align}
is the contribution of $\cN=4$ bifundamental hypermultiplets. Once again, $\Delta_{a, b}$ is an upper-diagonal incidence matrix,   $r^{(ab)}\equiv\text{gcd}(r^{(a)},r^{(b)})$, and $v^{(ab)}\equiv\sqrt{q^{r^{(ab)}}/t}$.  In type IIB, it comes about from quantizing D3/D3 strings, with one set of D3 branes wrapping the $a$-th compact 2-cycle, and the other set of D3 branes wrapping the $b$-th compact 2-cycle.

The factor 
\begin{align}\label{matter3d}
I^{(a)}_{bulk, matter}(y^{(a)}, \{x_d\})
\end{align}
stands for the contribution of $\cN=2$ chiral multiplets in the fundamental representation of the $a$-th gauge group.  A chiral multiplet with $S_{R}$ R-charge $-r/2$ (not to confuse with the lacing number of $\fg$) contributes $\prod_{1\leq l\leq D^{(a)}}({v^{(a)}}^r\,x_d/y^{(a)}_l ;q^{r^{(a)}})^{-1}_{\infty}$ to the partition function, while an anti-chiral multiplet contributes $\prod_{1\leq l\leq D^{(a)}}({v^{(a)}}^r\,x_d/y^{(a)}_l ;q^{r^{(a)}})_{\infty}$; the associated mass is denoted as $x_d$. The R-charge symmetry of $G^{3d}$ is responsible for all the ${v^{(a)}}^r$ factors. These are fixed by requiring superconformal symmetry of the 3d theory in the IR.
In type IIB, this matter contribution comes from D3/D5 strings, where the D3 branes are wrapping the $a$-th 2-cycle.

Now that the bulk contributions $I^{3d}_{bulk}$ have been explained, we move on to the remaining factors in \eqref{3ddefectexpression2}, which will encode the physics of the Wilson loop. First, we conjecture the following factor inside the integrand:
\beq\label{3dWilsonfactor}
{\widetilde{Y}^{(a)}_{D1/D3}}(y^{(a)},z)=\prod_{l=1}^{D^{(a)}}\frac{1-t\, y^{(a)}_{l}/z}{1- y^{(a)}_{l}/z}\; .
\eeq
In the type IIB picture, this is consistent with a contribution due to D1/D3 strings.

Second, we conjecture an additional factor outside the integral:
 \beq\label{3dWilsonfactor2}
 {\widetilde{Y}^{(a)}_{D1/D5}}(\{x_{d}\}, z)=\prod_{i=1}^{n^{(a)}}\left(1- x_{d_i}v^{-\#^{(a)}_{i,d}}q^{-\widetilde{\#}^{(a)}_{i,d}/2}/z\right)\; .
 \eeq
The integer $n^{(a)}$ in the product is determined from the D5 brane charge constraint \eqref{betazero}, which we rewrite here for convenience:
 \beq
 \sum_{b=1}^m C_{ab} \;n^{(a)} = N^{(a)}_f\; .
 \eeq
 $C_{ab}$ is the Cartan matrix of $\fg$. All the other integers were defined below equation \eqref{specialize5d}. In the type IIB picture,  this is consistent with a contribution due to D1/D5 strings\footnote{Since we are interested in computing the partition function on the D3 branes, one may wonder why we should care about D1/D5 strings in the first place. The answer is that they quantize the fermions we called $\chi$ in the action term \eqref{3d1dfermion}. Those strings also played a role in 5d, as we discussed in section \ref{ssec:5dpartition}.}.\\

To complete the definition of the $Y$-operator \eqref{3ddefectexpression2}, we need to provide a contour prescription. If one follows the JK prescription, there is a priori a pole in the fermion fugacity $z$ coming from the Wilson loop factor $Y^{(a)}_{D1/D3}$. However, we will explicitly ignore this pole here, as part of the contour definition. We claim that the resulting poles are then all labeled by truncated Young diagrams. Let us pause and see how truncated Young diagrams come about, as we did when we studied the quantum mechanics integral \eqref{3dintegral}.
 
Here, the index is three-dimensional, and the various bulk factors are all infinite $q$-Pochhammer symbols, so the classification of poles comes about differently.
There exists at least one chiral multiplet contribution from the bulk matter
\beq\label{newproof1}
\prod_{1\leq l\leq D^{(a)}}\frac{1}{\left(x_d/y^{(a)}_l ;q^{r^{(a)}}\right)_{\infty}}\; , 
\eeq   
for some node $a$. Recall the definition $\left(x ;q\right)_{\infty}\equiv \prod_{n=0}^\infty\left(1-q^n\, x\right)$. Let us focus for now on the $n=0$ factor. Then, there is a pole, say, at
\beq\label{newproof2}
y^{(a)}_1= x_d\; .
\eeq 
Now, recalling the form of the bulk vector multiplet\footnote{Throughout this discussion, we are assuming $|q|, |t| < 1$ and $y^{(a)}_{i}/y^{(a)}_{j}\leq 1$ when $i\leq j$.}
\begin{align}\label{newproof3}
\prod_{1\leq i\neq j\leq D^{(a)}}\frac{\left(y^{(a)}_{i}/y^{(a)}_{j};q^{r^{(a)}}\right)_{\infty}}{\left(t\, y^{(a)}_{i}/y^{(a)}_{j};q^{r^{(a)}}\right)_{\infty}},
\end{align}
we have $t$-shifted poles at
\beq\label{newproof4}
y^{(a)}_{i}/y^{(a)}_{i+1}= t\; .
\eeq  
Putting \eqref{newproof2} and \eqref{newproof4} together, we find that there are poles at:
\beq\label{newproof5}
y^{(a)}_{i,k}=  x_d\, t^{D^{(a)}_i-k}\; ,\qquad\;\; k=1, \ldots, D^{(a)}_i\; .
\eeq  
Note that our notation reflects the breaking of the gauge symmetry, as in \eqref{ybreaking}. 
Now, the infinite product ($n>1$) in the $q$-Pochhammer symbols generates new poles, all $q$-shifted with respect to the ones above. Namely, the poles are located at
\beq\label{newproof6}
y^{(a)}_{i,k}=  x_d\, q^{r^{(a)} \boldsymbol{\mu}^{(a)}_{i,k}}\, t^{D^{(a)}_i-k}\; ,\qquad\;\; k=1, \ldots, D^{(a)}_i\; ,
\eeq 
where $\boldsymbol{\mu}^{(a)}_{i}$  is a Young diagram with at most $D^{(a)}_i$ rows, and $\boldsymbol{\mu}^{(a)}_{i,k}$ is the length of the $k$-th row. Similarly, after taking the bifundamentals into account, the discussion generalizes to all nodes and all other poles. Note that these are exactly the pole we had found from the quantum mechanics.\\

It follows that the conjectured index \eqref{new3dindex} takes a familiar form
\begin{align}\label{NEWcharacter3d}
\boxed{\left[\widetilde{\chi}^{\fg}\right]_{(L^{(1)},\ldots, L^{(m)})}^{3d}(\{z^{(a)}_\rho\})=\sum_{\omega\in V(\lambda)}\prod_{b=1}^m \left({\widetilde{\fq}^{(b)}}\right)^{d_b^\omega}\; c_{d_b^\omega}(q, t)\; \left(Q^{(b)}(\{z_{* \rho}^{(a)}\})\right)^{d_b^\omega}\,  \left[\widetilde{Y}_{3d}(\{z^{(a)}_\rho\})\right]_{\omega} \, .}
\end{align}
In the above, the factor $\left[\widetilde{Y}_{3d}(\{z^{(a)}_\rho\})\right]_{\omega}$ is defined as the vev of a product of our newly defined $Y$-operators \eqref{3ddefectexpression2}.
Namely, the index is once again a twisted $qq$-character of a finite dimensional irreducible representation $V(\lambda)$ of $U_q(\widehat{\fg})$, with highest weight $\lambda=\sum_{a=1}^m L^{(a)}\, \lambda_a$. A priori, the index  $\widetilde{\chi}^{3d}$ differs from the previously computed quantum mechanics Witten index ${\chi}^{3d}$ by the normalization.  We address this point next, and in the process, we derive the conjectured expression \eqref{NEWcharacter3d} directly from five dimensions. Later, we will provide more evidence for our 3d index being well-defined, by further deriving it from Toda theory.

\begin{example}
	The 3d $qq$-character for the fundamental representation of $U_q(\widehat{A_1})$ is
	\begin{align}\label{3dpartitionfunctionexample}
	\left[\widetilde{\chi}^{A_1}\right]_{(1)}^{3d}(z)= \left(1-v^{-1}x_2/z\right)\left\langle \widetilde{Y}_{D1/D3}(z) \right\rangle + \widetilde{\fq} \; \left(1-v\,x_1/z\right) \left\langle\frac{1}{\widetilde{Y}_{D1/D3}(z\, v^{-2})}\right\rangle\; .
	\end{align}
We will prove this expression in the Examples section.
\end{example}

\vspace{8mm}

\subsection{Derivation of $\widetilde{\chi}^{3d}$ from 5d}

In order to derive the 3d index, we first evaluate the various integrals in the 3d index by residue:
\begin{align}
\left\langle\left[{\widetilde{Y}^{(a)}_{3d}(z)}\right]^{\pm 1} \right\rangle &= \left[{\widetilde{Y}^{(a)}_{D1/D5}}(\{x_d\}, z)\right] \oint_{\{\overrightarrow{\boldsymbol{\mu}}\}} d{y}\,\left[I^{3d}_{bulk}(y)\, \left[{\widetilde{Y}^{(a)}_{D1/D3}}(y, z)\right]^{\pm 1}\right]\\
&=\left[{\widetilde{Y}^{(a)}_{D1/D5}}(\{x_d\}, z)\right]\,\sum_{\{ \overrightarrow{\boldsymbol{\mu}} \}}\text{res}_{\overrightarrow{\boldsymbol{\mu}}}\left[I^{3d}_{bulk}(y)\, \left[{\widetilde{Y}^{(a)}_{D1/D3}}(y, z)\right]^{\pm 1}\right].\label{residuesum}
\end{align}

As we have just pointed out, we need to be careful about the normalization of these $Y$-operators before we can make contact with the 5d physics. Indeed, evaluating the integral above by residue yields an infinite result. To obtain a finite answer, it proves useful to introduce a constant $c_{3d}$ which stands for the trivial residue of the 3d bulk contribution\footnote{For more details on the meaning of this constant, see \cite{Aganagic:2013tta}.}; so we let
\beq\label{c3d}
c_{3d}\equiv \text{res}_{\boldsymbol{\emptyset}}I^{3d}_{bulk}(y)\; , 
\eeq
the residue at an empty partition. We can then write
\begin{align}
\left\langle\left[{\widetilde{Y}^{(a)}_{3d}(z)}\right]^{\pm 1} \right\rangle &=\left[{\widetilde{Y}^{(a)}_{D1/D5}}(\{x_d\}, z)\right]\,\sum_{\{ \overrightarrow{\boldsymbol{\mu}} \}}\text{res}_{\overrightarrow{\boldsymbol{\mu}}}\left[I^{3d}_{bulk}(y)\, \left[{Y^{(a)}_{D1/D3}}(y, z)\right]^{\pm 1}\right]\label{Yresidue}\\
&=\left[{\widetilde{Y}^{(a)}_{D1/D5}}(\{x_d\}, z)\right]\,c_{3d}\sum_{\{ \overrightarrow{\boldsymbol{\mu}}  \}}\left[\frac{I^{3d}_{bulk}(y_{\{\overrightarrow{\boldsymbol{\mu}}\}})}{I^{3d}_{bulk}(y_{\{\boldsymbol{\boldsymbol{\emptyset}}\}})}\, \left[{Y^{(a)}_{D1/D3}}(y_{\{\overrightarrow{\boldsymbol{\mu}}\}}, z)\right]^{\pm 1}\right].\label{residuesum2}
\end{align}

In the last line, we made use of the fact that the ratio of the residues is equal to the (finite) ratio of the integrands, evaluated at the points $\{\overrightarrow{\boldsymbol{\mu}}\}$ and $\{\boldsymbol{\boldsymbol{\emptyset}}\}$. Explicitly, the variables $y_{\{\overrightarrow{\boldsymbol{\mu}}\}}$ denote collectively the residues at:
\beq\label{3dto5d}
y^{(a)}_{i,k}=  e^{(a)}_{i}\, q^{r^{(a)} \boldsymbol{\mu}^{(a)}_{i, k}}\, t^{-k} \; ,\qquad\;\; k=1, \ldots, D^{(a)}_i\; ,
\eeq  
$\boldsymbol{\mu}^{(a)}_i$ is a Young diagram with at most $D^{(a)}_i$ rows, and $\boldsymbol{\mu}^{(a)}_{i, k}$ is the length of the $k$-th row.

To each Young diagram $\boldsymbol{\mu}^{(a)}_i$, we associate a Coulomb modulus $e^{(a)}_{i}$ of the 5d theory $T^{5d}$, specialized as follows:
\beq\label{coulomb5dspecialized}
e^{(a)}_{i}=x_{d_i}\,t^{D^{(a)}_i} v^{-\#^{(a)}_{i,d}}q^{-\widetilde{\#}^{(a)}_{i,d}/2} \; .
\eeq
Namely, as we already advertised in \eqref{specialize5d}, the 5d Coulomb moduli are set equal to some mass $x_d$, and then shifted by $D^{(a)}_i$ integer units of vortex flux. From the perspective of $T^{5d}$, this gives a description of the 5d Coulomb branch on an integer lattice. The remaining integers $\#^{(a)}_{i,d}$,  $\widetilde{\#}^{(a)}_{i,d}$ are fully determined from the requirement that the 5d partition function should truncate. This is because the 5d bulk function \eqref{nekrasovN} has the following property: if $N$ is a non-negative integer, then 
\beq\label{truncateNekrasov}
N_{\boldsymbol{\mu}^{(a)}_i\boldsymbol{\mu}^{(b)}_j}\left(\left(v^{(b)}\right)^2\,t^{-N} ;q^{r^{(ab)}}\right)=0\;\; \text{unless}\;\; l\left(\boldsymbol{\mu}^{(b)}_j\right)\leq l\left(\boldsymbol{\mu}^{(a)}_i\right)+ N
\eeq
To make contact with the 3d index, we will need two more facts about the function $N_{\boldsymbol{\mu}^{(a)}_i\boldsymbol{\mu}^{(b)}_j}$.

First, suppose that a Young diagram  $\boldsymbol{\mu}^{(a)}_i$ has no more than $D^{(a)}_i$ rows, $l\left(\boldsymbol{\mu}^{(a)}_i\right)\leq D^{(a)}_i$, and suppose further that a Young diagram  $\boldsymbol{\mu}^{(b)}_j$ has no more than $D^{(b)}_j$ rows, $l\left(\boldsymbol{\mu}^{(b)}_j\right)\leq D^{(b)}_j$. Then, the 5d bulk functions can be rewritten as
\begin{align}\nonumber
N_{\boldsymbol{\mu}^{(a)}_i\boldsymbol{\mu}^{(b)}_j}(Q\, ;q^{r^{(ab)}}) &= \prod_{k=1}^{D^{(a)}_i}\prod_{s=1}^{D^{(b)}_j} 
\frac{\big( Q \, q^{r^{(a)}\boldsymbol{\mu}^{(a)}_{i,k}-r^{(b)}\boldsymbol{\mu}^{(b)}_{j,s}} \,t^{s - k + 1}\,;q^{r^{(ab)}} \big)_{\infty}}{\big( Q\,  q^{r^{(a)}\boldsymbol{\mu}^{(a)}_{i,k}-r^{(b)}\boldsymbol{\mu}^{(b)}_{j,s}}\, t^{s - k}\, ;q^{r^{(ab)}}\big)_{\infty}} \,
\frac{\big( Q\,  t^{s - k}\, ;q^{r^{(ab)}} \big)_{\infty}}{\big( Q\,  t^{s - k + 1}\, ;q^{r^{(ab)}}\big)_{\infty}}\\
&\;\;\times N_{\boldsymbol{\mu}^{(a)}_i\, \boldsymbol{\emptyset}}(Q\, t^{D^{(b)}_j};q^{r^{(ab)}})\,N_{\boldsymbol{\emptyset}\,\boldsymbol{\mu}^{(b)}_j}(Q\,t^{-{D^{(a)}_i}};q^{r^{(ab)}})\; .\label{NEWnekrasovN}
\end{align}

Second, the bulk 5d functions have the property
\beq\label{Nekrasovpropoerties}
N_{\boldsymbol{\mu}^{(a)}_i\boldsymbol{\mu}^{(b)}_j}(Q\, ;q)=\prod_{k=0}^{r^{(ab)}-1}N_{\boldsymbol{\mu}^{(a)}_i\boldsymbol{\mu}^{(b)}_j}(q^k\, Q\, ;q^{r^{(ab)}})\; ,
\eeq
due to the fact they are products and ratios of infinite $q$-Pochhammer symbols.\\

Armed with these facts, we are ready to derive the 3d partition function $\widetilde{\chi}^{3d}$.The first step is to define the 5d theory $T^{5d}$. This is done by specifying a simple Lie algebra $\fg$ and a set $\{\omega_d\}_{d=1}^{N_f}$ of coweights of $\fg$; this is the non-compact D5 brane charge. We do not further need to specify the gauge content of the theory, that is to say the compact D5 brane charge. Indeed, recall that we impose an extra constraint \eqref{betazero}, which determines the rank of all the gauge groups, given the flavor content of the theory. Likewise, the effective Chern-Simons contribution is entirely fixed by the constraint. Put differently, the set of coweights $\{\omega_d\}_{d=1}^{N_f}$ uniquely determines the theory $T^{5d}$.

Its partition function $\chi^{5d}$  is expressed as a Laurent polynomial in $Y$-operator expectation values, so we should analyze those first. The expectation value of the $Y$-operator in five dimensions was defined in \eqref{5ddefectexpression2}, which we rewrite here for convenience:
\beq\label{5ddefectexpression2AGAIN}
\left\langle {Y_{5d}^{(a)}(z)} \right\rangle = \sum_{\{\overrightarrow{\boldsymbol{\mu}}\}}\left[ Z^{5d}_{bulk}\cdot Y^{(a)}_{5d}(z)\right]\;,\; \text{where}\;\;Y^{(a)}_{5d}(z)=\prod_{i=1}^{n^{(a)}}\prod_{k=1}^{\infty}\frac{1-t\, y^{(a)}_{i,k}/z}{1- y^{(a)}_{i,k}/z}\; .
\eeq
We specialize all the Coulomb parameters as \eqref{coulomb5dspecialized}. This ensures that all the partitions $\{\overrightarrow{\boldsymbol{\mu}}\}$ truncate. Indeed, the fundamental matter $Z^{(a)}_{bulk,fund}$ inside the bulk factor $Z^{5d}_{bulk}$ will now have factors of the form
\beq\label{fundtruncation}
N_{\boldsymbol{\emptyset}\,\boldsymbol{\mu}^{(b)}_j}\left(\left(v^{(b)}\right)^2\,t^{-D^{(b)}_j} ;q^{r^{(b)}}\right)\; ,
\eeq
for some nonnegative integer $D^{(b)}_j$. 
The first fact guarantees that this expression vanishes unless the partition $\boldsymbol{\mu}^{(b)}_j$ has no more than $D^{(b)}_j$ rows. All the remaining partitions are truncated in this fashion, either directly from fundamental matter  $Z^{(a)}_{bulk,fund}$ as above, or from bifundametal hypermultiplets $Z^{(a, b)}_{bulk,bif}$. Therefore, the sums \eqref{5ddefectexpression2AGAIN}  and \eqref{residuesum2} are over the same set of truncated partitions. We now show that the summands are term-by-term the same, up to the proportionality constant $c_{3d}$.\\

First, we look at the factors in $Z^{5d}_{bulk}$, which are all the contributions independent of the Wilson loop factor. 
The 5d gauge couplings \eqref{5dbulkgauge} are the 3d F.I. parameters \eqref{FI3d}. Note that the 3d variables $y^{(a)}_l$ are periodic, valued on the circle $S^1(\widehat{R})$. 

Next, the 5d vector multiplet factors \eqref{5dbulkvec} become
\begin{align}\label{5dbulkvecto3d}
Z^{(a)}_{bulk,vec} = \frac{I^{(a)}_{bulk, vec}(y_{\{\boldsymbol{\mu^{(a)}}\}})}{I^{(a)}_{bulk, vec}(y_{\{\boldsymbol{\emptyset}\}})}\cdot V_{vec} \; ,
\end{align}
where the right-hand side is precisely the contribution of the 3d vector multiplets \eqref{vec3d}, and some leftover factors we denote as $V_{vec}$.

The 5d bifundamental hypermultiplet factors \eqref{5dbulkbif} become
\begin{align}\label{5dbulkbifto3d}
Z^{(a, b)}_{bulk,bif} = \frac{I^{(a,b)}_{bulk, bif}(y_{\{\boldsymbol{\mu^{(a)}}\}}, y_{\{\boldsymbol{\mu^{(b)}}\}})}{I^{(a,b)}_{bulk, bif}(y_{\{\boldsymbol{\emptyset}\}}, y_{\{\boldsymbol{\emptyset}\}})}\cdot V_{bif} \; ,
\end{align}
where the right-hand side is the contribution of the 3d bifundamental hypermuliplets \eqref{bif3d}, and some leftover factors we denote as $V_{bif}$.

We collect all the 5d leftover factors, along with the fundamental hypermultiplets and the Chern-Simons contributions, and find after some algebra
\begin{align}\label{5dbulkmatterto3d}
Z^{(a)}_{bulk,fund}\cdot Z^{(a)}_{bulk,CS} \cdot V_{vec} \cdot V_{bif} = \frac{I^{(a)}_{bulk, matter}(y_{\{\boldsymbol{\mu^{(a)}}\}}, \{x_d\})}{I^{(a)}_{bulk, matter}(y_{\{\boldsymbol{\emptyset}\}}, \{x_d\})}\; .
\end{align}
That is, the remaining 5d factors combine to produce the matter content of $G^{3d}$, \eqref{matter3d}. The constraint \eqref{betazero} ensures that this matter content has an equal number of chiral and antichiral multiplets.

Let us now look at the contribution of the Wilson loop in 5d, subject to the identification \eqref{3dto5d}. In particular, since we have shown that a given partition $\boldsymbol{\mu}^{(a)}_{i}$ cannot have more than $D^{(a)}_i$ rows, we have
\begin{align}\label{truncatevariables}
y^{(a)}_{i,k} &=  e^{(a)}_{i}\, q^{r^{(a)} \boldsymbol{\mu}^{(a)}_{i, k}}\, t^{-k} \; ,\qquad\;\; k=1, \ldots, D^{(a)}_i\\
y^{(a)}_{i,k} &=  e^{(a)}_{i}\,t^{-k} \; ,\qquad\qquad\;\;\;\;\;\;\;\,  k > D^{(a)}_i\; .
\end{align}
After some algebra, the 5d $Y$-operator becomes:
\begin{align}\label{fromY5dtoY3d}
Y^{(a)}_{5d}(z) &=\prod_{i=1}^{n^{(a)}}\prod_{k=1}^{\infty}\frac{1-t\, y^{(a)}_{i,k}/z}{1- y^{(a)}_{i,k}/z}\nonumber\\
&= \prod_{i=1}^{n^{(a)}}\prod_{k=1}^{D^{(a)}_i}\frac{1-t\, y^{(a)}_{i,k}/z}{1- y^{(a)}_{i,k}/z}\cdot  \prod_{i=1}^{n^{(a)}}\left(1- e^{(a)}_{i}\, t^{-D^{(a)}_i}/z\right)\nonumber\\
&= \prod_{i=1}^{n^{(a)}}\prod_{k=1}^{D^{(a)}_i}\frac{1-t\, y^{(a)}_{i,k}/z}{1- y^{(a)}_{i,k}/z}\cdot  \prod_{i=1}^{n^{(a)}}\left(1- x_{d_i}\, v^{-\#^{(a)}_{i,d}}q^{-\widetilde{\#}^{(a)}_{i,d}/2} /z\right)\nonumber\\
&= \left[{\widetilde{Y}^{(a)}_{D1/D3}}(y_{\{\overrightarrow{\boldsymbol{\mu}}\}}, z)\right]\cdot \left[{\widetilde{Y}^{(a)}_{D1/D5}}(\{x_d\}, z)\right]
\end{align}
In the second line, we have performed an infinite number of telescopic cancellations using \eqref{truncatevariables}. In the third line, we have specialized the 5d Coulomb moduli to be \eqref{coulomb5dspecialized}.
Remarkably, the Wilson loop operator in five dimensions factorizes into the Wilson loop operator in three dimensions, made up of the two factors \eqref{3dWilsonfactor} and \eqref{3dWilsonfactor2}.\\

Putting it all together, we have shown that after specialization of the 5d Coulomb parameters, the following expectation values are equal:
\beq\label{Yequality}
\left\langle{\widetilde{Y}^{(a)}_{3d}(z)} \right\rangle = c_{3d} \, \left\langle {Y_{5d}^{(a)}(z)} \right\rangle_{e^{(a)}_{i}\propto\; x_{d_i} t^{D^{(a)}_i}}
\eeq

Since the building blocks of the 5d partition function \eqref{character5d} are precisely $Y$-operator expectation values, we derive the 3d partition function \eqref{NEWcharacter3d} at once:
\begin{align}\label{CHIequality}
\boxed{\left[\widetilde{\chi}^{\fg}\right]_{(L^{(1)},\ldots, L^{(m)})}^{3d}(\{z^{(a)}_\rho\})=c_{3d}\;  \left[\chi^{\fg}\right]_{(L^{(1)},\ldots, L^{(m)})}^{5d}(\{z^{(a)}_\rho\})_{e^{(a)}_{i}\propto\; x_{d_i}t^{D^{(a)}_i}}\, .}
\end{align}
The proportionality constant $c_{3d}$ was defined above in \eqref{c3d} as the trivial residue contribution.\\

In the next section, we show that this partition function $\left[\chi^{\fg}\right]^{3d}$ is in fact equal to a  (chiral) correlator of a certain deformation of Toda CFT, defined on the cylinder.

\newpage

\section{Wilson Loops from deformed $W$-algebra correlators}

Let us first review some facts about the more common Toda CFT and its free field formulation. We will then briefly review the construction of the deformed theory, sometimes referred to as $q$-Toda. Throughout this section, $\fg$ denotes a simple Lie algebra.

\subsection{Review: Free Field Toda CFT} 
\label{ssec:Todafree}

$\fg$-type Toda field theory is a two-dimensional theory  of $m=\text{rank}(\fg)$ free bosons. The action reads:
\beq\label{Todaaction}
S_{Toda} =  \int dz d{\bar z} \;\sqrt g \; g^{z{\bar z}}\left[\langle\partial_z \phi,  \partial_{\bar z} \phi\rangle+  \langle Q, \phi\rangle\, R + \sum_{a=1}^m e^{\langle\alpha_a^{\vee},\phi\rangle/b} \right].
\eeq
The bosons are denoted by the field $\phi$, which is a vector in the $m$-dimensional coweight space. The bracket $\langle\cdot,\cdot\rangle$ is the Cartan-Killing form on the Cartan subalgebra of $\fg$, as in the previous sections. The $\alpha_a^{\vee}$ label the simple positive coroots of $\fg$. $R$ is the scalar curvature of a sphere, with metric $g$. The quantity $Q=\rho\, b+ \rho^\vee/b$ is the background charge, which is coupled to the bosons through the exponential potential term in the action. $\rho$ is the Weyl vector of $\fg$, while $\rho^\vee$ is the Weyl vector of $^L\fg$. The central charge of the theory is
\beq\label{centralcharge}
c=m+12 \left\langle Q , Q \right\rangle \; .
\eeq

In what follows, we will consider the free field formalism of the Toda CFT. This is sometimes called the Coulomb gas formalism, or the Dotsenko-Fateev formalism, named after the authors of \cite{Dotsenko:1984nm}. For a modern treatment of the topic, see \cite{Dijkgraaf:2009pc,Itoyama:2009sc,Mironov:2010zs,Morozov:2010cq,Maruyoshi:2014eja}. Let us focus our attention on the holomorphic sector of the theory, meaning the various fields and operators have no $\bar z$ dependence. The primary operators are (normal-ordered) vertex operators, labeled by a $m$-dimensional vector of momentum $\eta\, \omega$, with $\omega$ a coweight of $\fg$.
\beq\label{primary}
V_{\eta}(u) = \; :e^{\eta\,\langle\omega, \varphi(u)\rangle}: \; .
\eeq
The free field formalism requires the consideration of $m$ additional nonlocal operators, called screening charges:
\beq\label{screeningcharge}
Q^{(a)} \equiv \int dy \,S^{(a)}(y) \, ,
\eeq

where the charges are integrals of the $m$ screening current operators $S^{(a)}(y)$:
\beq\label{screeningcurrent}
S^{(a)}(y) = \;:e^{\langle\alpha^{\vee}_{a}, \phi(y)\rangle/b}: \; .
\eeq

There is a momentum conservation constraint, which in the presence of the background charge $Q$, takes the form:
\beq\label{constraint}
\eta_0 + \eta_\infty + \eta + \sum_{a=1}^m D^{(a)} \alpha^{\vee}_{a}/b = 2\,Q \; .
\eeq 
The above constraint fixes the momentum of the vertex operators, in terms of the momenta $\eta_0$ and $\eta_\infty$ at 0 and $\infty$, as well the number $D^{(a)}$ of screening charges.

The Toda CFT enjoys a ${\cW}({\fg})$-algebra symmetry\footnote{An extensive review can be found in \cite{Bouwknegt:1992wg}.}. The generators $W^{(s)}(z)$ of this algebra, which we label according to their spin, are constructed as the commutant of the screening charges of the theory:
\beq\label{generators}     
[W^{(s)}(z),Q^{(a)}]=0\; , \qquad \text{for all}\;\; a=1, \ldots, m, \;\;\text{and}\;\; s=2, \ldots,m+1\; .
\eeq
Equivalently, the commutator of the generators with the screening currents $S^{(a)}(y)$ is a total derivative with respect to the variable $y$. 
The special case $\fg= A_1$ is called the Liouville CFT, and ${\cW}({A_1})$ is more commonly called the Virasoro symmetry, generated by the spin 2 stress energy tensor $W^{(2)}(z)$. When $\fg$ is a higher rank algebra, the stress tensor $W^{(2)}(z)$ is still present, but there are also more currents $W^{(s)}(z)$ of higher spin $s>2$.\\

We will be interested in evaluating the following correlators: 
\beq\label{correlator}
\left\langle \psi'\left|\prod_{d=1}^{N_f}V_{\eta_d}(x_d)\; \prod_{a=1}^{m} (Q^{(a)})^{D^{(a)}}\; \prod_{s=2}^{m+1}\prod_{\rho=1}^{L^{(s-1)}}W^{(s)}(z^{(a)}_\rho) \right| \psi \right\rangle \, .
\eeq
Above, we wrote the incoming and outgoing states as $|\psi\rangle$ and  $|\psi'\rangle$ respectively, instead of the trivial vacuum $|0\rangle$. In the absence of the currents $W^{(s)}(z)$, these correlators are nothing but the conformal blocks of the Toda CFT \cite{Fateev:2007ab}. Because we are using the free-field formalism here, a correlator can be evaluated using straightforward Wick contractions. Namely, \eqref{correlator}  can be written as an integral over the positions $y$ of the $D^{(a)}$ screening currents:
\beq\label{conf1}
\int d_{Haar}y \;I_{Toda}(y) \; ,
\eeq
where the integrand $I_{Toda}(y)$ is the product of all possible two-point functions, and
\beq\label{Haar}
d_{Haar}y=\prod_{a=1}^{m}\prod_{i=1}^{D^{(a)}}\frac{dy^{(a)}_i}{y^{(a)}_i}\; .
\eeq
We now review the deformation of the Toda CFT, where such two-point functions will be reinterpreted as building blocks of the 3d gauge theory index $\left[\widetilde{\chi}^{\fg}\right]^{3d}$.

\subsection{The Deformed ${\cW}_{q,t}({\fg})$-Algebra} 
\label{ssec:qToda}

In the work \cite{Frenkel:1998}, a deformation of the above ${\cW}({\fg})$ algebra was proposed, based on a certain canonical deformation of the screening currents; see also \cite{Shiraishi:1995rp} for the special case $\fg=A_1$, and \cite{Feigin:1995sf,Awata:1995zk} for the case $\fg=A_m$.)

The starting point is to define a $(q,t)$-deformed Cartan matrix\footnote{The way we write the deformed Cartan matrix follows the conventions of \cite{Frenkel:1998}. This corresponds to setting the bifundamental masses to be  $\mu^{(a)}_{bif}=v^{(ab)}$ in the corresponding $\fg$-shaped 3d quiver gauge theory, see section \ref{sssec:vortexwitten}. As we pointed out there, if one wishes, one can leave the bifundamental masses arbitrary instead; the price to pay is to slightly modify the definition of the deformed Cartan matrix in this section, which will now contain explicit bifundamental masses in its  off-diagonal entries. For details, we refer the reader to \cite{Kimura:2015rgi}, \cite{Kimura:2017hez}.}:
\begin{align}\label{CartanToda}
C_{ab}(q,t)= \left(q^{r^{(a)}}t^{-1} +q^{-r^{(a)}}t\right) \, \delta_{ab}- [I_{ab}]_q\; .
\end{align} 
Let us explain the notation: a number in square brackets is called a quantum number, defined as
\begin{align}\label{quantumnumber}
[n]_q = \frac{q^{n}-q^{-n}}{q-q^{-1}} \; ,
\end{align}
and the incidence matrix is $I_{ab}= 2 \, \delta_{ab} - C_{ab}$. We use here the same notation as in the gauge theory sections: $r^{(a)}\equiv r\,\langle\alpha_a, \alpha^\vee_a\rangle/2$, where $r$ the lacing number of $\fg$.

When $\fg$ is non simply-laced, its Cartan matrix $C_{ab}$ is not symmetric. One can symmetrize it 
by defining a matrix $B_{ab}$ as 
\beq\label{Bmatrix}
B_{ab}=r^{(a)} \, C_{ab} \; .
\eeq
In the deformed theory, one introduces a symmetrization of $C_{ab}(q,t)$, which we call $B_{ab}(q,t)$:
\beq\label{Bmatrixdef}
B(q,t)= D(q,t)\, C(q,t) \; .
\eeq
In the above, we have defined the diagonal matrix $D=\text{diag}\left([r^{(1)}]_q, \ldots, [r^{(m)}]_q\right)$, which is the identity if $\fg$ is simply-laced.
Later, we  will also need  the inverse of the Cartan matrix:
\beq\label{Mmatrixdef}
M(q,t)\equiv D(q,t)C(q,t)^{-1}= D(q,t)B(q,t)^{-1} D(q,t) \; .
\eeq

We can now construct a deformed Heisenberg algebra, generated by $m$ positive simple coroots, satisfying
\begin{align}\label{commutatorgenerators}
[\alpha_a[k], \alpha_b[n]] = {1\over k} (q^{k\over 2} - q^{-{k\over 2}})(t^{{k\over 2} }-t^{-{k\over 2} })B_{ab}(q^{k\over 2} , t^{k\over 2} ) \delta_{k, -n} \; .
\end{align}
In the above, it is understood that the zero-th generator commutes with all others: $[\alpha_a[k], \alpha_b[0]]=0$, for $k$ an arbitrary integer.

The Fock space representation of this algebra is given by acting on the vacuum state $|\psi\rangle$:
\begin{align}
\alpha_a[0] |\psi\rangle &= \langle\psi, \alpha_a\rangle |\psi\rangle\label{eigenvalue}\nonumber\\
\alpha_a[k] |\psi\rangle &= 0\, , \qquad\qquad\;\; \mbox{for} \; k>0\; .
\end{align}

Then, we can define deformed screening operators as\footnote{The screening operators we write down are called ``magnetic" in \cite{Frenkel:1998}, and are defined with respect to the parameter $q$. Another ``electric" set of screening and vertex operators can be constructed using the parameter $t$ instead, which rotates the complementary line $\mathbb{C}_t\subset\mathbb{C}_q\times\mathbb{C}_t$, but these will not enter our discussion.}
\beq\label{screeningdef}
S^{(a)}(y) = y^{-\alpha_a[0]/r^{(a)}}\,: \exp\left(\sum_{k\neq 0}{ \alpha_a[k] \over q^{k\, r^{(a)}\over 2} - q^{-\,{k \, r^{(a)} \over 2}}} \, y^k\right): \; .
\eeq
Note all operators in this section are written up to a center of mass zero mode, whose effect is simply to shift the momentum of the vacuum. Up to redefinition of the vacuum $|\psi\rangle$, we safely ignore such factors.

The ${\cW}_{q,t}({\fg})$-algebra is defined as the associative algebra whose generators are the Fourier modes of operators which commute with the screening charges 
\beq\label{screeningchargedef}
Q^{(a)} =\int dy\, S^{(a)}(y) \; .
\eeq 
We borrow the notation from the undeformed theory and name the generating currents $W^{(s)}(z)$, labeled by their ``spin" $s$. We therefore have
\beq\label{generatorsdef}     
[W^{(s)}(z),Q^{(a)}]=0\; , \qquad \text{for all}\;\; a=1, \ldots, m, \;\;\text{and}\;\; s=2, \ldots,m+1\; .
\eeq
In this way, one finds that each such generating current can be written as a Laurent polynomial in certain vertex operators, which we call $\cY$-operators for reasons that will soon be clear:
\beq\label{YoperatorToda}
{\cY^{(a)}}(z)= e^{\lambda_a[0]}\,: \exp\left(-\sum_{k\neq 0} w_a[k]\, t^{-k/2}  \, z^k\right): \; .
\eeq

Degenerate vertex operators are constructed out of $m$ fundamental coweight generators,
\begin{align}\label{commutator2}
[\alpha_a[k], w_b[n]] ={1\over k} (q^{k \, r^{(a)}\over 2}  - q^{-{k \, r^{(a)}\over 2} })(t^{{k \over 2}}-t^{-{k \over 2} })\,\delta_{ab}\,\delta_{k, -n} \, .
\end{align}
These are dual to the operators $\alpha_a[k]$. Put differently,
\beq\label{etow}
\alpha_a[k] = \sum_{b=1}^n C_{ab}(q^{k\over 2},t^{k \over 2}) w_b[k]\; .
\eeq
For completeness, we also write the commutator of two coweight generators,
\begin{align}\label{commutator3}
[w_a[k], w_b[n]] ={1\over k} (q^{k\over 2}  - q^{-{k \over 2} })(t^{{k \over 2}}-t^{-{k \over 2} })\,M_{ab}(q^{k\over 2} , t^{k\over 2})\,\delta_{k, -n} \, .
\end{align}

There is no definition of what a ${\cW}_{q,t}({\fg})$ algebra  primary vertex operator should be in the Mathematics literature, since there is no formal definition of what a deformed chiral algebra is in general. In this paper, the vertex operators we consider were first defined in \cite{Aganagic:2015cta,Haouzi:2017vec}.
Namely, one defines a set of $N_f$ coweights $\{\omega_d\}$ in the coweight lattice of $\fg$, satisfying the following three conditions:
\begin{itemize}
	\item For all $1\leq d \leq N_f$, the coweight $\omega_d$ belongs in a fundamental representation of $^L\fg$.
	\item $\sum_{d=1}^{N_f} \omega_d=0$.
	\item No proper subset of coweights in $\{\omega_d\}$ adds up to $0$. 
\end{itemize}
The first condition ensures that each coweight is represented by a D5 brane wrapping a non-compact 2-cycle in the $ADE$ geometry. The second condition ensures that the constraint \eqref{betazero} is satisfied\footnote{In gauge theory terms, this was a ``conformality" condition for the quiver $T^{5d}$, and the vanishing of the  Chern Simons levels in $G^{3d}$.}. The third condition ensures that each set of coweights describes a single defect on the Riemann surface $\cC$, as opposed to composite defects. These conditions can all be relaxed, at the expense of introducing more vertex operators, but we will not do so here to keep the presentation clear.

Once these conditions are satisfied, we construct the ``deformed primary" vertex operator
\beq\label{genericvertex}
V_\eta (u)= \, :\prod_{d=1}^{N_f} V_{\omega_d}(x_d): \; .
\eeq
The parameters $\eta$ and $u$ are to be understood as center of mass momentum and position respectively. They  are defined through the relations
\begin{align}\label{commomentum}
&\eta \equiv\sum_{d=1}^{N_f} \eta_d\, \omega_d\; ,\nonumber\\
&x_d \equiv u \, q^{-\eta_d}\; ,\qquad\;\; d=1, \ldots, N_f \; .
\end{align}
In the above, each $V_{\omega_d}(x_d)$ operator is a normal ordered product of ``fundamental coweight" operators  
\beq\label{coweightvertexdef}
\Lambda^{(a)}(x) \equiv\; : \exp\left(\, \sum_{k\neq 0}{w_a[k] \over (q^{k\, r^{(a)}\over 2} - q^{-\,{k \, r^{(a)} \over 2}})(t^{k\over 2} - t^{-\,{k \over 2}})} \,t^{-\,{k \over 2}}\, x^k\right): \; .
\eeq

Such an operator $\Lambda^{(a)}(x)$ can be thought of as a quantization of the $a$-th fundamental coweight in the Lie algebra $\fg$. The argument of these operators will contain $q$ and $t$ shifts, which we will later fix uniquely using inputs from the gauge theory.

\begin{example}
	Let $\fg=A_2$. Let us consider the  set of coweights
	\beq
	\{\omega_1=[-1, 0] \; , \;\;\; \omega_2=[1, -1]  \; , \;\;\; \omega_2=[0, 1] \}\; .
	\eeq
This set clearly satisfies the three conditions we enunciated above. Correspondingly, we construct the vertex operators
\begin{align}\label{3vertex}
V_{\omega_1}(x_1) &=\; :\left[\Lambda^{(1)}(x_1)\right]^{-1}: \nonumber\\
V_{\omega_2}(x_2) &=\; :\Lambda^{(1)}(v^{-2}\,x_2)\left[\Lambda^{(2)}(v^{-1}\,x_2)\right]^{-1}:\nonumber\\
V_{\omega_3}(x_3) &=\; :\Lambda^{(2)}(v^{-3}\,x_3): 
\end{align}
The vertex operator that we will consider inside a correlator is then
\beq\label{genericvertexexmaple}
V_\eta (u)= \, :\prod_{d=1}^{3} V_{\omega_d}(x_d): \; .
\eeq
\end{example}
\vspace{9mm}

Equivalently, each operator $V_{\omega_d}(x_d)$ can be written as a normal ordered product of a \textit{single} fundamental coweight operator $[\Lambda^{(a)}(x)]^{-1}$ and a number of ``simple coroot" operators 
\beq\label{rootvertexdef}
E^{(a)}(x) \equiv\; : \exp\left(\sum_{k\neq 0}{ \alpha_a[k] \over (q^{k\, r^{(a)}\over 2} - q^{-\,{k \, r^{(a)} \over 2}})(t^{k\over 2} - t^{-\,{k \over 2}})} \, x^k\right): \; .
\eeq
This operator product can be thought of as a quantization of the decomposition \eqref{weightexpression},
\beq\label{rememberweights}
\omega_d=-\lambda_a + \sum_{b=1}^m h^{(b)}_{d}\, \alpha_b\; .
\eeq 
The content of this equation is that a coweight in the set $\{\omega_d\}$ can always be written as (minus) a fundamental coweight, plus some (positive) simple coroots.

\begin{example}
Let $\fg=A_2$. We consider again the same set of three coweights
\beq
\{\omega_1=[-1, 0] \; , \;\;\; \omega_2=[1, -1]  \; , \;\;\; \omega_2=[0, 1] \}\; .
\eeq
The vertex operators we previously wrote down can equivalently be written as:
\begin{align}\label{3vertexnew}
V_{\omega_1}(x_1) &=\; :\left[\Lambda^{(1)}(x_1)\right]^{-1}: \nonumber\\
V_{\omega_2}(x_2) &=\; :\left[\Lambda^{(1)}(x_2)\right]^{-1}E^{(1)}(v^{-1}\,x_2):\nonumber\\
V_{\omega_3}(x_3) &=\; :\left[\Lambda^{(1)}(x_3)\right]^{-1}E^{(1)}(v^{-1}\,x_3)\, E^{(2)}(v^{-2}\,x_3): 
\end{align}
This reflects the fact that the $A_2$ coweights can be written as
\begin{align}\label{3vertexnew2}
\omega_1 &= -\lambda_1 \nonumber\\
\omega_2 &= -\lambda_1+\alpha_1\nonumber\\
\omega_3 &= -\lambda_1+\alpha_1+\alpha_2
\end{align}
\end{example}

\vspace{8mm}

We now comment on the conformal limit from the deformed ${\cW}_{q,t}({\fg})$-algebra to the usual ${\cW}({\fg})$-algebra, with central charge $c=m+12 \left\langle Q,Q\right\rangle$ and background charge  $Q=\rho\, b+ \rho^\vee/b$. Namely, the undeformed theory is recovered by setting 
$q= \exp(R \epsilon_1),\, t=\exp(-R \epsilon_2)$, taking the limit $R\rightarrow 0$, and further setting $b\equiv -\epsilon_2/\epsilon_1$.  For the Heisenberg algebra to keep making sense, the coroot and coweight generators should also be rescaled by $R$ and $\epsilon_1$ in this limit.

The individual vertex operators $V_{\omega_d}(x_d)$ in \eqref{genericvertex} do not have a good conformal limit, but their product $V_{\eta}(u)$ does, and becomes the vertex operator \eqref{primary}. In particular, the center of mass momentum and position \eqref{commomentum} of the deformed vertex operator become \textit{the} momentum and position of the primary in the conformal theory. 

Furthermore, in the limit, the deformed screenings \eqref{screeningdef} become the screening currents \eqref{screeningcurrent}, and the deformed generators $W^{(s)}(z)$ become the stress tensor and higher spin currents of the ${\cW}({\fg})$ algebra.

\begin{example}
The deformed stress tensor of ${\cW}_{q,t}({A_1})$ is the following Laurent polynomial in the $\cY$ operators defined in \eqref{YoperatorToda}: 
\beq\label{examplestress}
W^{(2)}(z)= \cY(z) + \left[\cY(v^{-2}z)\right]^{-1} \; .
\eeq
This can be checked explicitly by computing the commutator $[W^{(s)}(z),Q^{(a)}]$, and noticing that it vanishes. In the limit $q= \exp(R \epsilon_1),\, t=\exp(-R \epsilon_2)$, and further rescaling of the Heisenberg algebra generators, we find
\beq\label{examplestress2}
W^{(2)}(z) \qquad \longrightarrow \qquad-\frac{1}{2}:\left(\partial_z\phi(z)\right)^2:+ Q\cdot :\partial_z^2\phi(z):\; .
\eeq
This is nothing but the Liouville stress energy tensor.
\end{example}

\vspace{8mm}

\subsection{$\left[\chi^{\fg}\right]^{3d}$ is a Deformed ${\cW}_{q,t}({\fg})$-algebra correlator}

We now focus our attention on the correlator
\beq\label{correlatordef}
\left\langle \psi'\left|\prod_{d=1}^{N_f}V_{\omega_d}(x_d)\; \prod_{a=1}^{m} (Q^{(a)})^{D^{(a)}}\; \prod_{s=2}^{m+1}\prod_{\rho=1}^{L^{(s-1)}}W^{(s)}(z^{(s-1)}_\rho) \right| \psi \right\rangle \, ,
\eeq
and will show that it reproduces the 3d index $\left[\widetilde{\chi}^{\fg}\right]^{3d}$ of the quiver gauge theory $G^{3d}$.

In what follows, we use the shorthand notation  $\langle \mathellipsis\rangle$ for a vacuum expectation value, and we introduce the theta function $\theta_{q^{r^{(a)}}}(x)= (x \,;\, q^{r^{(a)}})_\infty\,(q^{r^{(a)}}/x \,;\, q^{r^{(a)}})_\infty$. 
We will also make heavy use of the non simply-laced notation from the previous sections: $v^{(a)}\equiv\sqrt{q^{r^{(a)}}/t}$ and $v^{(ab)}\equiv\sqrt{q^{r^{(ab)}}/t}$.

After taking into account the normal ordering of the various operators, the correlator \eqref{correlatordef} becomes the integral
\beq\label{conf1def}
\int d_{Haar}y \;I_{Toda}(y) \; ,
\eeq
where the Haar measure is given by
\beq\label{Haardef}
d_{Haar}y=\prod_{a=1}^{m}\prod_{i=1}^{D^{(a)}}\frac{dy^{(a)}_i}{y^{(a)}_i} \; .
\eeq
The integrand $I_{Toda}(y)$ is made up of various factors. First, we have 
\beq\label{TodaFI}
\prod_{a=1}^{m}\prod_{i=1}^{D^{(a)}} \left(y^{(a)}_i\right)^{\langle\psi, \alpha_a\rangle}\; ,
\eeq
where $\langle\psi, \alpha_a\rangle$ is the eigenvalue of the state $|\psi\rangle$, as we have defined it in \eqref{eigenvalue}. In 3d gauge theory language, this is nothing but the F.I. term \eqref{FI3d} contribution to the index $\left[\widetilde{\chi}^{\fg}\right]^{3d}$.

\begin{figure}[h!]
	\emph{}
	\centering
	\includegraphics[trim={0 0 0 2cm},clip,width=0.9\textwidth]{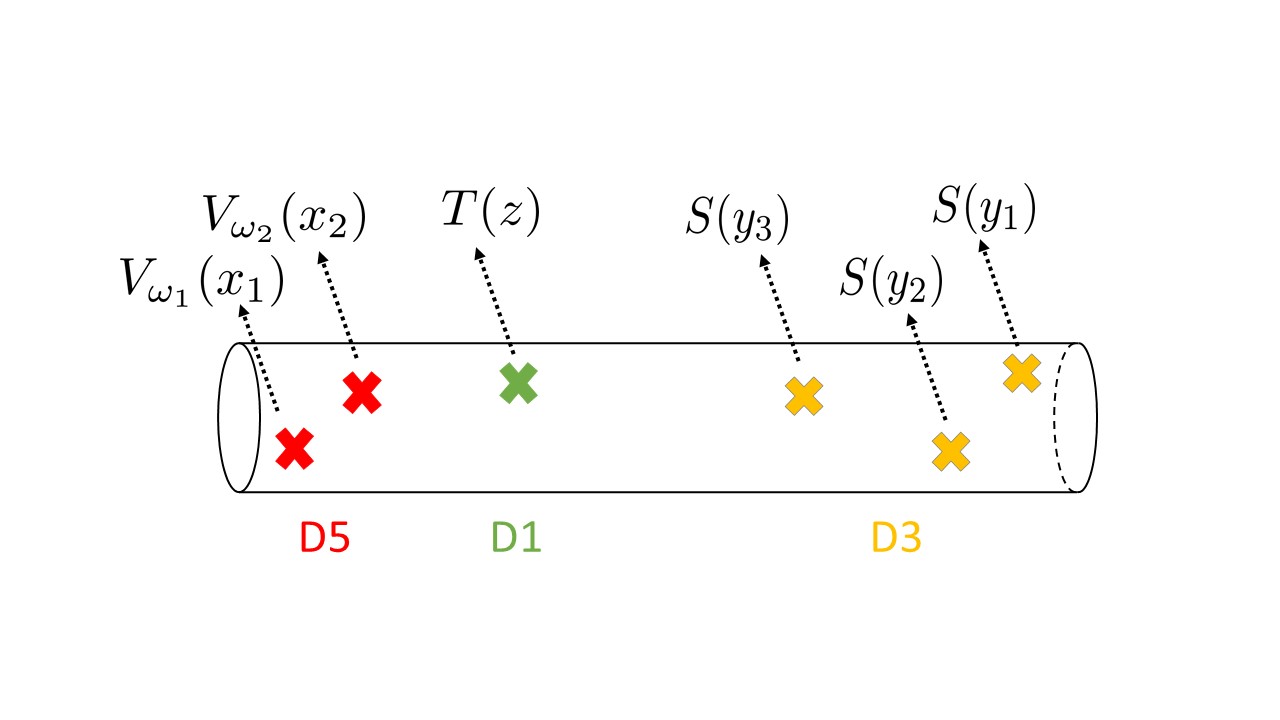}
	\vspace{-40pt}
	\caption{Example of a correlator in $q$-Liouville, along with the corresponding D branes at points on the cylinder. The specific correlator pictured here is $\langle \psi' |  V_{\eta}(u)\,  T(z) \, (Q)^{3}  | \psi \rangle$; for the D5 brane vertex operator, the center of mass momentum is $\eta=\eta_1\omega_1+\eta_2\omega_2$, where the center of mass position $u$ is defined through $x_1=u\, q^{-\eta_1}$ and  $x_2=u\, q^{-\eta_2}$.}
	\label{fig:cylinderLiouville}
\end{figure}

There are also various two-point functions: for a given $a=1,\ldots, m$, we find by direct computation
\beq\label{screena}
 \prod_{1\leq i< j\leq D^{(a)}}\left\langle S^{(a)}(y^{(a)}_i)\, S^{(a)}(y^{(a)}_j) \right\rangle = \prod_{1\leq i\neq j\leq D^{(a)}}\frac{\left(y^{(a)}_{i}/y^{(a)}_{j};q^{r^{(a)}}\right)_{\infty}}{\left(t\, y^{(a)}_{i}/y^{(a)}_{j};q^{r^{(a)}}\right)_{\infty}}\;\prod_{1\leq i<j\leq D^{(a)}} \frac{\Theta\left(t\,y^{(a)}_{j}/y^{(a)}_{i};q^{r^{(a)}}\right)}{\Theta\left(y^{(a)}_{j}/y^{(a)}_{i};q^{r^{(a)}}\right)}\; .
\eeq
We recognize the vector multiplet contribution \eqref{vec3d} to the index $\left[\widetilde{\chi}^{\fg}\right]^{3d}$, due to D3/D3 strings on the $a$-th 2-cycle.

For $a$ and $b$ two distinct nodes, we compute
\beq\label{screenab}
\prod_{1\leq i \leq D^{(a)}}\prod_{1\leq j \leq D^{(b)}}\left\langle S^{(a)}(y^{(a)}_i)\; S^{(b)}(y^{(b)}_j) \right\rangle = \prod_{1\leq i \leq D^{(a)}}\prod_{1\leq j \leq D^{(b)}}\left [ \frac{(v^{(ab)} t\, y^{(a)}_{i}/y^{(b)}_{j};q^{r^{(ab)}})_{\infty}}{(v^{(ab)} \, y^{(a)}_{i}/y^{(b)}_{j};q^{r^{(ab)}})_{\infty}}\right]^{\Delta_{a, b}}\; .
\eeq
We recognize the bifundamental contribution \eqref{bif3d} to the index $\left[\widetilde{\chi}^{\fg}\right]^{3d}$, due to D3/D3 strings on neighboring 2-cycles.

The two-point of a fundamental coweight vertex operator with a screening current equals
\beq\label{vertexscreening}
\left\langle \left[\Lambda^{(a)}(x_d)\right]^{\pm 1}\, S^{(b)}(y^{(b)}_i)\right\rangle = \left[(v^{(a)} x_d/y^{(a)}_{i};q^{r^{(a)}})_{\infty}\right]^{\mp \delta_{ab}}\; .
\eeq
We recognize chiral/antichiral matter contributions \eqref{matter3d} to the index $\left[\widetilde{\chi}^{\fg}\right]^{3d}$, due to D3/D5 strings. A chiral multiplet contribution is reproduced  by the two-point $\left\langle \left[\Lambda^{(a)}(x_d)\right]^{-1}\, S^{(b)}(y^{(b)}_i)\right\rangle$, while an antichiral contribution is reproduced by the two-point $\left\langle \Lambda^{(a)}(x_d)\, S^{(b)}(y^{(b)}_i)\right\rangle$.

We come to the two-point function of a screening with a $\cY$-operator, which is equal to\footnote{When the vacuum is $|\psi\rangle$ instead of $|0\rangle$, as it is in the correlator, this two-point is also responsible for a relative shift of one unit of 3d F.I. parameter between the various terms making up the generating current $W^{(s)}(z)$. This happens because the zero mode in the  $\cY$-operator \eqref{YoperatorToda} acts nontrivially on the vacuum $|\psi>$.}
\beq\label{Yopscreening}
\prod_{i=1}^{D^{(a)}}\left\langle S^{(a)}(y^{(a)}_i)\, \cY^{(b)}(z) \right\rangle = \prod_{i=1}^{D^{(a)}} \left[\frac{1-t\, y^{(a)}_{i}/z}{1- y^{(a)}_{i}/z}\right]^{\delta_{ab}}\; .
\eeq
We recognize part of the Wilson loop contribution \eqref{3dWilsonfactor} to the index $\left[\widetilde{\chi}^{\fg}\right]^{3d}$, due to D3/D1 strings. 

Now, each operator $V_{\omega_d}(x_d)$ is made up of various fundamental coweight operators \eqref{coweightvertexdef}. The two-point function of a single such fundamental coweight operator  with a $\cY$-operator is not as pretty:
\beq\label{Yopvertex}
\left\langle \Lambda^{(a)}(x_d)\, \cY^{(b)}(z) \right\rangle = \exp\left(-\sum_{k>0}\, \frac{1}{k} \,M_{ab}(q^{k\over 2} , t^{k\over 2})\, x_d^k/z^k\right)\; .
\eeq
Nonetheless, when we consider a (normal-ordered) product of such coweight vertex operators, making up the operator $V_\eta (u)$, we can make sense of this contribution to the correlator. Indeed, we have
\beq\label{Vopvertex}
\left\langle :\prod_{d=1}^{N_f}V_{\omega_d}(x_d):\, \cY^{(b)}(z) \right\rangle = B(\{x_d\}, z)\; \prod_{i=1}^{n^{(b)}}\left(1- x_{d_i}v^{-\#^{(b)}_{i,d}}q^{-\widetilde{\#}^{(b)}_{i,d}/2}/z\right).
\eeq
Up to the prefactor $B(\{x_d\},z)$, we recognize above the other part of the Wilson loop contribution \eqref{3dWilsonfactor2} to the index $\left[\widetilde{\chi}^{\fg}\right]^{3d}$, due to D5/D1 strings. The function $B(\{x_d\},z)$ has the following canonical form: it is a product of $N_f$ exponentials, each looking like \eqref{Yopvertex}, divided by an overall product $\prod_{i=1}^{n^{(b)}}\left(1-x_{d_i}v^{-\#^{(b)}_{i,d}}q^{-\widetilde{\#}^{(b)}_{i,d}/2}/z\right)$, which can be thought of as an overall normalization. It is a highly nontrivial fact that when we consider the generating current $W^{(s)}(z)$, which is a Laurent polynomial in $\cY$-operators, each term in this polynomial is of the form \eqref{Vopvertex}, with the \textit{same} function $B(\{x_d\},z)$ for each term. This implies that the prefactor $B(\{x_d\},z)$ can be factorized out of the correlator integral altogether.\\

To fully specify the correlator integral, we also need to make a choice of contour (i.e. a choice of a vacuum in the language of the 3d gauge theory $G^{3d}$). Here, we take the contours to simply be those described in section \ref{sssec:3dcountours}. In particular, all poles should be labeled by truncated Young diagrams, and none of the poles depending on the fermion masses $z^{(a)}$ variables should be enclosed by the contours.\\  

We can now claim
\begin{empheq}[box=\fbox]{align}
	&\left\langle \psi'\left|\prod_{d=1}^{N_f}V_{\omega_d}(x_d)\; \prod_{a=1}^{m} (Q^{(a)})^{D^{(a)}}\; \prod_{s=2}^{m+1}\prod_{\rho=1}^{L^{(s-1)}}W^{(s)}(z^{(s-1)}_\rho) \right| \psi \right\rangle \nonumber \\
	&\qquad\qquad\qquad=A\left(\{x_d\}\right)\, B\left(\{x_d\},\{z^{(s-1)}_\rho\}\right)\, \left[\widetilde{\chi}^{\fg}\right]_{(L^{(1)},\ldots, L^{(m)})}^{3d}(\{z^{(s-1)}_\rho\})\; .\label{correlatoris3dindex}
\end{empheq}

Namely, the index of the gauge theory $G^{3d}$ is the $q$-Toda correlator \eqref{correlatordef}, up to proportionality constants $A(\{x_d\})$ and $B(\{x_d\},\{z^{(s)}_\rho\})$. The constant $A(\{x_d\})$ stands for all the two-points of vertex operators $\Lambda^{(a)}(x_d)$ with themselves, which we have omitted here since they are not related to vortex physics in $G^{3d}$. The constant $B(\{x_d\},\{z^{(s)}_\rho\})$ is purely due to the two-point \eqref{Vopvertex}. Recall that this prefactor contains an exponential factor; we mention in passing that this exponential has a natural interpretation if $\fg=A_m$, in the more general framework of Ding-Iohara-Miki (DIM) algebras \cite{Ding:1996,Miki:2007}; see for instance \cite{Mironov:2016yue} for a presentation of the case $\fg=A_1$ in our context. Indeed, in the DIM formalism, the vertex operators $V_{\omega_d}(x_d)$ are built out of certain intertwiners, given by the product of a ${\cW}_{q,t}({A_m})$ vertex operator, as we have already written, but also a new Heisenberg algebra operator, with its own Fock space, and whose contribution is precisely the prefactor $B(\{x_d\},\{z^{(s)}_\rho\})$. Our results suggest that this structure continues to exist in the case where $\fg$ is an arbitrary simple Lie algebra. It would be important to understand this point in more details, specifically in the context of the growing DIM literature.

Finally, note that the constants $A(\{x_d\})$ and $B(\{x_d\},\{z^{(s)}_\rho\})$ stand outside the correlator integral, since they do not depend on the $y$ integration variables.\\

We can give the $q$-Toda correlator a 5d gauge theory interpretation, making use of the gauge/vortex duality between $T^{5d}$ and $G^{3d}$. Putting \eqref{CHIequality} and \eqref{correlatoris3dindex} together, we deduce at once that the correlator is proportional to the partition function of the 5d gauge theory $T^{5d}$, with specialized values of Coulomb parameters:
\begin{empheq}[box=\fbox]{align}
&\left\langle \psi'\left|\prod_{d=1}^{N_f}V_{\omega_d}(x_d)\; \prod_{a=1}^{m} (Q^{(a)})^{D^{(a)}}\; \prod_{s=2}^{m+1}\prod_{\rho=1}^{L^{(s-1)}}W^{(s)}(z^{(s-1)}_\rho) \right| \psi \right\rangle \nonumber\\ 
&\;\;\;\;\;\;\;\;\;\;\;\;\;=A\left(\{x_d\}\right)\, B\left(\{x_d\},\{z^{(s-1)}_\rho\}\right)\,c_{3d}\left[\chi^{\fg}\right]_{(L^{(1)},\ldots, L^{(m)})}^{5d}(\{z^{(s-1)}_\rho\})_{e^{(a)}_{i}\propto\; x_{d_i} t^{D^{(a)}_i}}\; .\label{correlatoris5d}
\end{empheq}

The proportionality constant $c_{3d}$ was defined in \eqref{c3d}. This equality can be proved directly by evaluating the left-hand side integrals using the pole prescription in section \ref{sssec:3dcountours}; the right-hand side is nothing but the residue sum.\\

At this point, one could ask what happens if we increase the number of D5 branes, meaning we would increase the number of punctures on the cylinder to be greater than one. The answer is not too interesting,  and does not substantially modify our previous discussion, since defects simply ``factorize"; in 5d gauge theory language, adding D5 branes simply means one has additional fundamental hypermultiplet contributions to the partition function; this slightly modifies the definition of the $Y$-operator vevs in an obvious way. In 3d gauge theory language, the extra D5 branes manifest themselves as additional chiral and antichiral contributions to the partition function. In the language of deformed $\cW$ algebras, this amounts to considering correlators with more than a single vertex operator insertion. At any rate, the formalism we have developed in this paper enables one to compute exactly the partition function in all three pictures with an arbitrary number of D5 brane defects, if one wishes. Such computations are omitted here since it is sufficient to look at a single defect to extract the Wilson loop physics. In particular, the partition function with more than one D5 brane puncture remains a twisted $qq$-character of the same finite dimensional irreducible representation of $U_q(\widehat{\fg})$ as in the single puncture case; the only change is the twist itself.\\

As a final remark, let us comment on a special case where we take all the matter vertex operators $V_{\omega_d}(x_d)$ to be trivial, and further set the number of screening charges to be infinite. The equality $\eqref{correlatoris3dindex}$ does not make sense anymore, since the positive integers $D^{(a)}$ on the left-hand side, now taken to be infinite, used to be the ranks of the 3d gauge groups on the right-hand side. However, we can still make sense of \eqref{correlatoris5d}, if we now leave the 5d Coulomb parameters to be arbitrary:
\beq\label{correlatoris5dKimura}
\left\langle \psi'\left| \prod_{a=1}^{m} \lim_{D^{(a)}\rightarrow\infty} (Q^{(a)})^{D^{(a)}}\; \prod_{s=2}^{m+1}\prod_{\rho=1}^{L^{(s-1)}}W^{(s)}(z^{(s-1)}_\rho) \right| \psi \right\rangle=c_{3d}\,\left[\chi^{\fg}\right]_{(L^{(1)},\ldots, L^{(m)})}^{5d}(\{z^{(s-1)}_\rho\}) \; .
\eeq
In this way, we recover an identity first proved recently in \cite{Kimura:2017hez}.\\

We will now present detailed examples to illustrate the various statements made in this paper.

\newpage

\section{Examples}
\label{sec:examples}

In the examples below, we use the following Cartan matrices when $\fg$ is a non simply-laced algebra:

\begin{align*}
& \qquad\qquad C_{ab}^{G_2} = \left( \begin{array}{cc}
2 & -1\\ 
-3 & 2
\end{array} \right) \qquad\qquad\qquad\qquad\;\;\;  C_{ab}^{F_4} = \left( \begin{array}{cccc}
2 & -1 & 0 & 0\\ 
-1 & 2 & -2 & 0\\
0 & -1 & 2 & -1\\
0 & 0 & -1 & 2
\end{array} \right)
\end{align*}

In all examples, the Riemann surface $\cC=\mathbb{R}\times S^1(R)$ is the infinite cylinder of radius $R$.  We write $\widehat{R}=1/m_s^2\, R$ for the T-dual radius, where $m_s$ is the string mass.\\

\vspace{10mm}

\subsection{Wilson Loop for $A_1$ Theories}
\label{ssec:exampleA1}

\vspace{8mm}

------- 5d Gauge Theory -------\\

We start by describing the case $\fg=A_1$ in detail. Namely, let $X$ be a resolved $A_1$ singularity, and consider type IIB string theory on $X\times\cC\times\mathbb{C}^2$. We introduce $n$ D5 branes wrapping the compact 2-cycle $S$ of $X$ and $\mathbb{C}^2$. We further introduce $N_f$ D5 branes and  wrapping the dual non-compact 2-cycle $S^*$ and $\mathbb{C}^2$. Finally, we add to this background $L$ D1 branes wrapping the same non-compact 2-cycle. We set $g_s\rightarrow 0$, which amounts to studying the $(2,0)$ $A_1$ little string on $\cC\times\mathbb{C}^2$ in the presence of codimension 2 defects (the D5 branes) and point-like defects (the D1 branes).

At energies below the string scale, the dynamics in this background is fully captured by the theory on the D5 branes, with D1 brane defects. The theory on the D5 branes is a 5d $U(n)$ gauge theory we call $T^{5d}$, defined on $S^1(\widehat{R})\times\mathbb{C}^2$, with $N_f$ flavors. The D1 branes make up a 1/2-BPS  Wilson loop wrapping the circle and sitting at the origin of $\mathbb{C}^2$. After  putting the theory on $\Omega$-background, the partition function of $T^{5d}$ with Wilson loop is the following Witten index:

\begin{align}
\label{5dintegralA1}
&\left[\chi^{A_1}\right]_{(L)}^{5d}  =\sum_{k=0}^{\infty}\;\frac{\widetilde{\fq}^{k}}{k!} \, \oint  \left[\frac{d\phi_I}{2\pi i}\right]Z_{vec}\cdot Z_{fund}\cdot Z_{CS}\cdot \prod_{\rho=1}^{L}  Z_{D1}  \; , \\
&Z_{vec} =\prod_{I, J=1}^{k} \frac{{\widehat{f}}\left(\phi_{I}-\phi_{J}\right)f\left(\phi_{I}-\phi_{J}+ 2\,\epsilon_+ \right)}{f\left(\phi_{I}-\phi_{J}+\epsilon_1 \right)f\left(\phi_{I}-\phi_{J}+\epsilon_2\right)}\nonumber\\
&\qquad\times\prod_{I=1}^{k} \prod_{i=1}^{n} \frac{1}{f\left(\phi_I-a_i+\epsilon_+ \right)f\left(\phi_I-a_i-\epsilon_+ \right)}\; ,\\
&Z_{fund} =\prod_{I=1}^{k} \prod_{d=1}^{N_f} f\left(\phi_I-m_d\right) \equiv\prod_{I=1}^{k} Q(\phi_I)\; ,\label{fundQA1}\\
&Z_{CS}= \prod_{I=1}^{k} e^{k_{CS}\, \phi_I}\\
&Z_{D1} =\left[\prod_{i=1}^{n}  f\left(a_i-M_\rho\right) \prod_{I=1}^{k} \frac{f\left(\phi_I-M_\rho+ \epsilon_- \right)f\left(\phi_I-M_\rho- \epsilon_- \right)}{f\left(\phi_I-M_\rho+ \epsilon_+ \right)f\left(\phi_I-M_\rho- \epsilon_+ \right)}\right]\, .
\end{align}

Note we have defined a function $Q$ in $Z_{fund}$, encoding the fundamental matter content
\begin{equation}
\label{A1matter}
Q(\phi_I)\equiv \prod_{d=1}^{N_f} f\left(\phi_I-m_d\right)\; .
\end{equation}

The integral \eqref{5dintegralA1} was performed for the case $L=1$ (one D1 brane) using the Jeffrey-Kirwan (JK) residue prescription in \cite{Kim:2016qqs}. In that case, by the JK pole prescription, the contour integral is required to pick up exactly one pole from the defect factor $Z_k^{D1}$ of the form
\begin{equation}
\phi_I-M-\epsilon_+=0 \; .
\end{equation}
The other $k-1$ poles are specified by $N$-colored Young diagrams $\overrightarrow{\boldsymbol{\mu}}=\{\boldsymbol{\mu}_1, \boldsymbol{\mu}_2, \ldots, \boldsymbol{\mu}_n\}$, such that $\left|\overrightarrow{\boldsymbol{\mu}}\right|=k-1$. For instance, we can choose $\phi_1=M+\epsilon_+$, and the other $k-1$ integration variables $\phi_2, \ldots, \phi_k$ pick up poles at
\beq
\label{A1youngtuples}
\phi_I=a_i +\epsilon_+ - s_1\, \epsilon_1 - s_2\, \epsilon_2 , \text{with}\; (s_1, s_2)\in \boldsymbol{\mu}_i \; .
\eeq
One can easily check that $\chi^{5d}$ is pole-free in the fermion mass fugacity $z\equiv e^{-\widehat{R}M}$, so the partition function is a polynomial in $z$. Furthermore, from the integrand expression, the asymptotics of $\left[\chi^{\fg}\right]_{(1)}^{5d}$ at $z\rightarrow 0$ and $z\rightarrow +\infty$ tell us that the degree of this polynomial is $n$.
The coefficients of the polynomial are $U(n)$ Wilson loop expectation values, evaluated in the fundamental (i.e. antisymmetric) representations of $U(n)$. In other words, the partition function is the generating function of $U(n)$ Wilson loops:
\beq\label{A1generatingfunction}
\left[\chi^{\fg}\right]_{(1)}^{5d}=z^{-n/2}\sum_{j=0}^n (-z)^j\, \left\langle W_{\Lambda^j}\right\rangle
\eeq
Let us now make contact with the representation theory of quantum affine algebras.
We introduce a defect operator expectation value:
\begin{align}
\label{YoperatorA1}
\left\langle \left[Y_{5d}(z)\right]^{\pm 1}\right\rangle =
\sum_{k=\left|\overrightarrow{\boldsymbol{\mu}}\right|=0}^{\infty}\frac{\widetilde{\fq}^{k}}{k!} \, \oint_{\{\overrightarrow{\boldsymbol{\mu}}\}}  \left[\frac{d\phi_I}{2\pi i}\right]Z_{vec}\cdot Z_{fund}\cdot Z_{CS} \cdot \left[Z_{D1}(z)\right]^{\pm 1}\, . 
\end{align}
As emphasized in the main text, the contour prescription used in defining this operator differs from the one used in writing down the partition function  \eqref{5dintegralA1} . Namely,  the partition function $\chi^{5d}$ is evaluated using the JK pole prescription, with a pole at $\phi_I=M+\epsilon_+$, whereas the expectation value $\langle Y_{5d}(M) \rangle$ ignores this new pole originating from  $Z_{D1}$, and only considers instead the poles labeled by Young diagrams, at \eqref{A1youngtuples}.

Then, the partition function can be expressed in terms of these $Y$-operators, as a sum of exactly two terms, to make up a twisted $qq$-character of the fundamental representation of $U_q(\widehat{A_1})$:
\begin{align}
\label{A1result2}
\left[\chi^{A_1}\right]_{(1)}^{5d}(z) = \left\langle Y_{5d}(z) \right\rangle + \widetilde{\fq} \; Q(z\, v^{-1}) \left\langle\frac{1}{Y_{5d}(z\, v^{-2})}\right\rangle\; .
\end{align}
The meaning of this expression is as follows: The first term on the right-hand side encloses almost all the ``correct" poles, but we are missing exactly one: this is the extra pole at $\phi_I-M-\epsilon_+=0$. The second term on the right-hand side makes up for this missing pole, and relies on the following key observation: one can trade a contour enclosing this extra pole for a contour which does not enclose it, at the expense of inserting the operator $Y_{5d}(z\, v^2)^{-1}$ inside the vev. The parameter $\widetilde{\fq}$ is there to make up for the deficit of this one pole we ignored. The function $Q$ that appears is the fundamental matter function \eqref{A1matter}, now written in K-theoretic notation:
\beq\label{A1matterK}
Q(z)=\prod_{d=1}^{N_f} \left(1-f_d/z\right)\; .
\eeq

Performing the integrals over the poles \eqref{A1youngtuples}, we find:
\beq\label{5ddefectexpression2A1}
\left\langle \left[Y_{5d}(z)\right]^{\pm 1} \right\rangle = \sum_{\{\overrightarrow{\boldsymbol{\mu}}\}}\left[ Z^{5d}_{bulk} \cdot Y_{5d}(z)^{\pm 1}\right]\, .
\eeq
The sum is over a collection of 2d partitions, one for each $U(1)$ Coulomb parameter:
\beq
\{\overrightarrow{\boldsymbol{\mu}}\}=\{\boldsymbol{\mu}_i\}_{i=1,\ldots,n}\; .
\eeq
The factor  $Z^{5d}_{bulk}$ encodes all the 5d bulk physics. It is given by:
\beq\label{bulk5dA1}
Z^{5d}_{bulk}=   \widetilde{\fq}^{\sum_{i=1}^{n}{\left|\boldsymbol{\mu}_i\right|}}\, Z_{bulk,vec} \cdot Z_{bulk,fund}\cdot Z_{bulk,CS} \; ,
\eeq

Each factor is written in terms of the function
\beq\label{nekrasovNA1}
N_{\boldsymbol{\mu}_i\boldsymbol{\mu}_j}(Q\, ;q) = \prod\limits_{k,s = 1}^{\infty} 
\frac{\big( Q \, q^{\boldsymbol{\mu}_{i,k}-\boldsymbol{\mu}_{j,s}} \,t^{s - k + 1}\,;q \big)_{\infty}}{\big( Q\,  q^{\boldsymbol{\mu}_{i,k}-\boldsymbol{\mu}_{j,s}}\, t^{s - k}\, ;q\big)_{\infty}} \,
\frac{\big( Q\,  t^{s - k}\, ;q \big)_{\infty}}{\big( Q\,  t^{s - k + 1}\, ;q\big)_{\infty}} \; ,
\eeq
as follows:
\begin{align}
&Z_{bulk,vec} = \prod_{i,j=1}^{n}\left[N_{\boldsymbol{\mu_i\boldsymbol{\mu}_j}}\left(e_{i}/e_{j};q\right)\right]^{-1}\label{5dbulkvecA1}\\
&Z_{bulk,fund} = \prod_{d=1}^{N_f} \prod_{i=1}^{n} N_{\boldsymbol{\emptyset}\, \boldsymbol{\mu}_i}\left( v\, f_{d}/e_{i};\, q\right) \label{5dbulkmatterA1}\\
&Z_{bulk,CS} = \prod\limits_{i=1}^{n} \left(T_{\boldsymbol{\mu}_i}\right)^{k_{CS}} \label{5dbulkCSA1}
\end{align}
Above, $T_{\boldsymbol{\mu}}$ is defined as $T_{\boldsymbol{\mu}} =(-1)^{|\boldsymbol{\mu}|} q^{\Arrowvert \boldsymbol{\mu}\Arrowvert^{2}/2}t^{-\Arrowvert \boldsymbol{\mu}^{t}\Arrowvert^{2}/2}$. 

Meanwhile, the Wilson loop factor has  the form
\beq\label{WilsonfactorA1}
Y_{5d}(z)=\prod_{i=1}^{n}\prod_{k=1}^{\infty}\frac{1-t\, y_{i,k}/z}{1- y_{i,k}/z}\; .
\eeq
Above, we have defined the variables
\beq\label{ydefA1}
y_{i,k} = e_i\, q^{\boldsymbol{\mu}_{i, k}} \, t^{-k}\;\;\;,\; k=1, \ldots, \infty\;,
\eeq 
where  $\boldsymbol{\mu}_{i, k}$ is the length of the $k$-th row of the partition $\boldsymbol{\mu}_{i}$. Then, having performed all the integrals, the 5d partition function with Wilson loop evaluates to \eqref{A1result2}, where each vev is now understood to be written in terms of the variables \eqref{ydefA1}.\\

It is straightforward to generalize this discussion to the case of an arbitrary number $L>1$ of D1 branes. This corresponds to considering a Wilson loop valued in the representation $\textbf{2}\otimes\ldots\otimes\textbf{2}$ of the gauge group, where the fundamental representation $\textbf{2}$ is tensored $L$ times with itself. In this case, the JK residue prescription dictates that for each value of $\rho$ in the set$\{1, 2, \ldots, L\}$, we should choose at most one pole
\begin{equation}
\phi_I-M_\rho-\epsilon_+=0 \; .
\end{equation}
Once again, the partition function can be expressed as a $qq$-character of $U_q(\widehat{A_1})$, with highest weight $[L]$ (the spin $L/2$ representation). An explicit expression in terms of the expectation values $\langle Y^{\pm 1}(z_\rho) \rangle$ can be found for instance in \cite{Nekrasov:2015wsu}, where it was computed using different methods.
Note that unlike the $L=1$ case, this time around the partition function is not a polynomial in any of the fermion masses $z_\rho$.\\

\vspace{8mm}

------- 3d Gauge Theory -------\\

We now make contact with the theory $G^{3d}$ on the vortices of $T^{5d}$. The D5 branes produce $N_f/2$ punctures on the cylinder. Since defects factorize, it is sufficient to focus our attention on a single defect, which means we specialize the number of D5 branes to $N_f=2$. Imposing the vanishing of D5 brane flux at infinity $[S+S^*]=0$ translates to the condition
\beq
2 \;n = N_f\; ,
\eeq
which then imposes $n=1$. The constraint also specifies the Chern-Simons term \eqref{5dbulkCSA1} to be $k_{CS}=n=1$. 

\begin{figure}[h!]
	\emph{}
	\centering
	\includegraphics[trim={0 0 0 1cm},clip,width=0.9\textwidth]{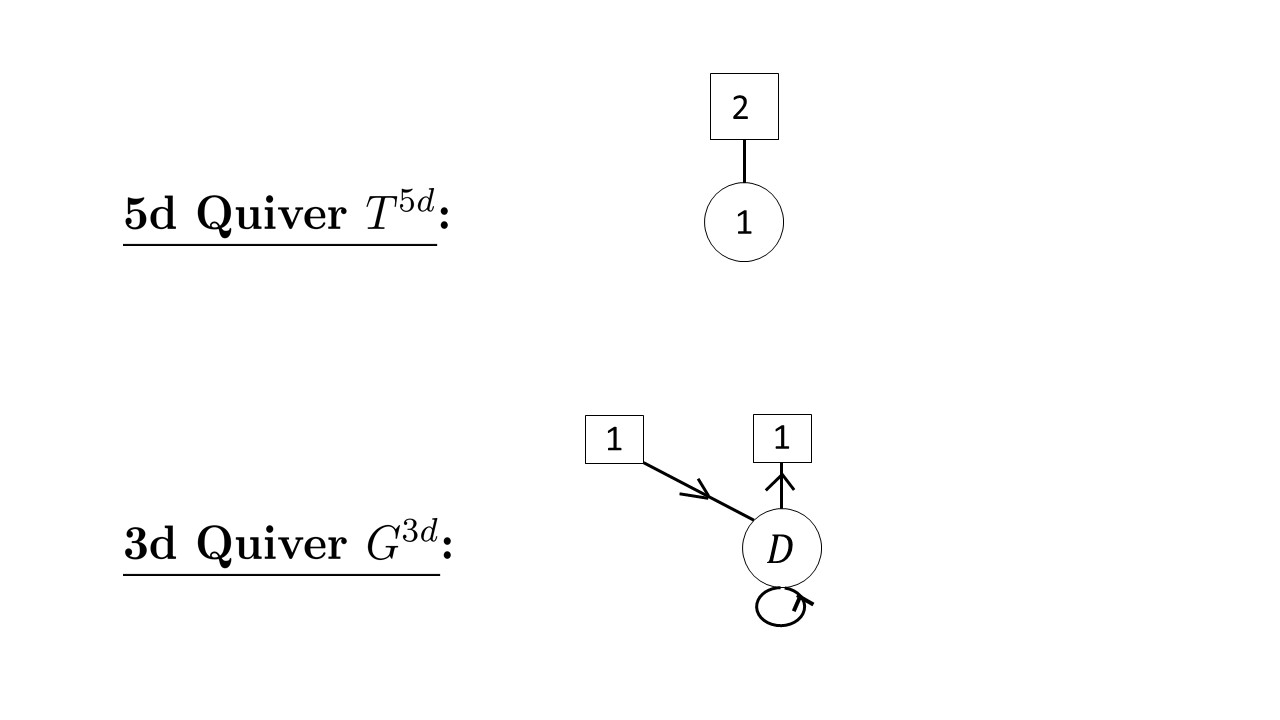}
	\vspace{-10pt}
	\caption{The $A_1$ theory $T^{5d}$ under study, along with the theory $G^{3d}$ on its vortices.} 
	\label{fig:A15d}
\end{figure}

Because $n=1$, only one D5 brane wrapping $S$, which we bind to one of the two D5 branes wrapping $S^*$. We can recast these statements in terms of the representation theory of $\fg$; namely, each of these two D5 brane is labeled by a weight in the fundamental representation of $A_1$, making up a set $\{\omega_d\}_{d=1}^{N_f=2}$:
\begin{align}
&\omega_1 = [-1] = -\lambda\nonumber\\
&\omega_2 = [\phantom{-}1] = -\lambda+\alpha\label{A1weight}
\end{align}
Here, we made the choice of having the first D5 brane wrap the non-compact 2-cycle $S^*$, while the second D5 brane wraps the non-compact 2-cycle $S^*+S$. Correspondingly, we now label the two masses by $x_1$ and $x_2$, respectively. The specific 5d gauge theory $T^{5d}$ engineered by these two weights is shown in figure \ref{fig:A15d}.
We further introduce $D$ D3 branes wrapping the compact 2-cycle $S$ and one of the two complex lines, say $\mathbb{C}_q$. 
In 5d gauge theory terms, we are at a special point on the moduli space where the Coulomb and Higgs branches  of $T^{5d}$ meet, and we further turn on $D$ units of vortex flux. Because of the Omega background on $\mathbb{C}_t$, this effectively shifts the Coulomb parameter $e_{1}$ by $D$ units of $t$-flux, since $\mathbb{C}_t$ is the line transverse to the vortex. As a result, we get off the root of the Higgs branch and probe the Coulomb branch of $T^{5d}$ at integer points on a $t$-lattice. For definiteness, let us set the Coulomb modulus to be:
\begin{align}
\label{specializeA1}
e_{1}=x_{2}\,t^{D}v^{-1}\; .
\end{align}
By definition of the function \eqref{nekrasovNA1}, we have 
\beq\label{truncationA1}
N_{\boldsymbol{\emptyset}\, \boldsymbol{\mu}}\left( v\, x_{2}/e_{1};\, q\right)=0 \qquad \text{unless $ \boldsymbol{\mu}$ has at most $D$ rows.} 
\eeq

Put differently, the 5d partition function is now a sum over truncated partitions, of length at most $D$. At this point, we introduce 3d Coulomb parameter variables $y_{i}$
\beq\label{3dto5dA1}
 y_{i}=  e_{1}\, q^{\boldsymbol{\mu}_{i}}\, t^{-i} \; ,\qquad\;\; i=1, \ldots, D\; .
 \eeq 
 
The 3d $\cN=2$ theory $G^{3d}$ is defined on $S^1(\widehat{R})\times\mathbb{C}$, and lives on the vortices of $T^{5d}$: it is a $U(D)$ gauge theory with 1 chiral multiplet and 1 antichiral multiplet, as pictured in figure \ref{fig:A15d}.\\

In order to compute the 3d partition function, we first define a 3d  Wilson loop operator expectation value, as an integral over the Coulomb moduli of $G^{3d}$, where it is understood that none of the contours below enclose poles depending on $z$:
 \begin{align}\label{3ddefectexpression2A1}
 \left\langle\left[{\widetilde{Y}_{3d}(z)}\right]^{\pm 1} \right\rangle \equiv \left[\widetilde{Y}_{D1/D5}(\{x_d\}, z)\right]\; \oint_{\{\overrightarrow{\boldsymbol{\mu}}\}} d{y}\,\left[I^{3d}_{bulk}(y)\, \left[{\widetilde{Y}_{D1/D3}}(y, z)\right]^{\pm 1}\right]  \; .
 \end{align}
The bulk contribution
\begin{align}\label{bulk3dA1}
I^{3d}_{bulk}(\{y\})=\prod_{k=1}^{D}{y_k}^{\tau-1}\;I_{bulk, vec}\cdot I_{bulk, matter}
\end{align}
is independent of  the Wilson loop physics. It contains the 3d F.I. term
\begin{align}\label{FI3dA1}
\prod_{l=1}^{D}{y_l}^{\tau} \; ,
\end{align}
the 3d $\cN=4$ vector multiplet contribution 
\begin{align}\label{vec3dA1}
I_{bulk, vec}(y) = \prod_{1\leq i\neq j\leq D} \frac{\left(y_{i}/y_{j};q\right)_{\infty}}{\left(t\, y_{i}/y_{j};q\right)_{\infty}}\;\prod_{1\leq i<j\leq D} \frac{\Theta\left(t\,y_{j}/y_{i};q\right)}{\Theta\left(y_{j}/y_{i};q\right)}\; ,
\end{align}
and the $\cN=2$ matter contribution
\begin{align}\label{matter3dA1}
I_{bulk, matter}(y, \{x_d\})=\prod_{i=1}^{D} \frac{\left(v\, x_{1}/y_{i};q\right)_{\infty}}{\left(v^{-1}\, x_{2}/y_{i};q\right)_{\infty}} \; .
\end{align}
Note the R-charge assignment in the matter factor is reflected in the various powers of $v$: specifically, a $U(1)_R$ charge  $S_R=-r/2$ for a chiral or antichiral multiplet corresponds to a shift of $v^r$ inside the corresponding $q$-Pochhammer symbol.

The Wilson loop contributions are given by
\beq\label{3dWilsonfactorA1}
{\widetilde{Y}_{D1/D3}}(z)=\prod_{k=1}^{D}\frac{1-t\, y_{k}/z}{1- y_{k}/z}
\eeq
and by
\beq\label{3dWilsonfactor2A1}
\widetilde{Y}_{D1/D5}(\{x_d\}, z)=\left(1-v^{-1}x_2/z\right) \; .
\eeq

All in all, the 3d partition function can be written as a twisted character of the fundamental representation of $U_q(\widehat{A_1})$:
\begin{align}\label{3dpartitionfunction}
\left[\widetilde{\chi}^{A_1}\right]_{(1)}^{3d}(z)= \left(1-v^{-1}x_2/z\right)\left\langle \widetilde{Y}_{D1/D3}(z) \right\rangle + \widetilde{\fq} \; \left(1-v\,x_1/z\right) \left\langle\frac{1}{\widetilde{Y}_{D1/D3}(z\, v^{-2})}\right\rangle\; .
\end{align}
As described in the main text, this can be proved in a straightforward way by Higgsing the 5d theory. Namely, using the specialization of the Coulomb parameter \eqref{specializeA1}, the first step is to show that the 5d $Y$-operator vev $\left\langle {Y_{5d}(z)} \right\rangle$ is the residue sum of the 3d $Y$-operator vev integral $\left\langle{\widetilde{Y}_{3d}(z)} \right\rangle$:
\beq\label{YequalityA1}
\left\langle{\widetilde{Y}_{3d}(z)} \right\rangle = c_{3d}\, \left\langle {Y_{5d}(z)} \right\rangle_{e_{1}=\; x_{2}\,t^{D}\,v^{-1}} 
\eeq
The proportionality constant $c_{3d}$ is the bulk residue at the empty partition, see \eqref{c3d}. The general proof of this statement was given in section in the main text, so we will not repeat it here.\\

The second step is to show that each of the two terms in the 5d character \eqref{A1result2} corresponds to a term in the 3d character \eqref{3dpartitionfunction}. Let us look in detail at the Wilson loop factors; in the first term, the Higgsing \eqref{specializeA1} triggers an infinite number of telescopic cancellations:
\begin{align}
Y_{5d}(z)&=\prod_{k=1}^{\infty}\frac{1-t\,y_k/z}{1- y_k/z}\nonumber \\
&=\prod_{k=1}^{D}\frac{1-t\,y_k/z}{1- y_k/z}\,\cdot\left(1-v^{-1} \,x_2/z\right)\nonumber\\
&=\widetilde{Y}_{D1/D3}(z) \, \left(1-v^{-1}x_2/z\right)
\end{align}
After taking a vacuum expectation value on both sides, it follows that the first term of the 5d partition function is in fact the first term of the 3d partition function.
Let us now look at the second term:
\begin{align}
\frac{Q(z\, v^{-1})}{Y_{5d}(z\, v^{-2})} &= \prod_{k=1}^{\infty}\frac{1-v^2 y_k/z}{1-t\,v^2\, y_k/z}\,\cdot\left(1-v \,x_1/z\right)\left(1-v \,x_2/z\right)\nonumber \\
&= \prod_{k=1}^{D}\frac{1-v^2 y_k/z}{1-t\,v^2\, y_k/z}\,\cdot\frac{\left(1-v \,x_1/z\right)\cancel{\left(1-v \,x_2/z\right)}}{\cancel{\left(1-v \,x_2/z\right)}}\nonumber\\
&=\frac{1}{\widetilde{Y}_{D1/D3}(z\, v^{-2})}\, \left(1-v\,x_1/z\right)
\end{align}

Taking a vev on both sides, we see that the second term in the 5d character once again agrees with the second term in the 3d character.\\
 
Therefore, after one further identifies the 5d gauge coupling with the 3d F.I. term, the equality of the partition functions follows:
\begin{align}\label{CHIequalityA1}
\left[\widetilde{\chi}^{A_1}\right]_{(1)}^{3d}(z)=c_{3d}\;  \left[\chi^{A_1}\right]_{(1)}^{5d}(z)_{e_{1}=\; x_{2}\,t^{D}\,v^{-1}} \, .
\end{align}

\vspace{8mm}

-----${\cW}_{q,t}(A_1)$ correlator -----\\

$q$-Liouville theory on the cylinder $\cC$ enjoys a ${\cW}_{q,t}(A_1)$ algebra symmetry, which is generated by the deformed stress tensor $W^{(2)}(z)$. This operator insertion at a point $z$ on $\cC$ corresponds to the location of a D1 brane, and it is constructed as the commutant of the screening charge. We find
\beq\label{A1stress}
W^{(2)}(z) = :\cY(z): + :\left[\cY(v^{-2}z)\right]^{-1}: \; ,
\eeq
where $\cY$ is the operator defined in \eqref{YoperatorToda}.\\

We now  build the vertex operator corresponding to the D5 brane content. Such an operator is fully determined by the set \eqref{A1weight} of weights $\{\omega_d\}$. In particular, the various $v$-shifts inside the argument of the operators are fixed by the R-symmetry of the 3d theory $G^{3d}$. Namely, we first define
\begin{align}\label{A1vertexweight}
V_{\omega_1}(x_1) &=\; :\left[\Lambda(x_1)\right]^{-1}: \nonumber \\
V_{\omega_2}(x_2) &=\; :\Lambda(v^{-2}\,x_2): \; ,
\end{align}
where $\Lambda(x)$ is the fundamental weight operator \eqref{coweightvertexdef}. Alternatively, we can define the D5 brane operator using the simple root operator \eqref{rootvertexdef}. The $v$-shift in the argument of this root operator is uniquely determined from the truncation locus \eqref{specializeA1}:
\begin{align}\label{A1vertexroot}
V_{\omega_1}(x_1) &=\; :\left[\Lambda(x_1)\right]^{-1}: \nonumber \\
V_{\omega_2}(x_2) &=\; :\left[\Lambda(x_2)\right]^{-1}E(v^{-1}\,x_2):
\end{align}
Either way, we are now able to construct the D5 brane operator as:
\beq\label{A1genericvertexexmaple}
V_\eta (u)= \, :\prod_{d=1}^{N_f=2} V_{\omega_d}(x_d): \; .
\eeq
The parameters $u$ and $\eta$ are defined through
\begin{align}\label{A1commomentum}
&\eta = \eta_1\, \omega_1 + \eta_2\, \omega_2\; ,\\
&x_d = u \, q^{-\eta_d}\; ,\qquad\;\; d=1, 2\; .
\end{align}

We consider the correlator 
\beq\label{A1correlatordef}
\left\langle \psi'\left|V_\eta (u)\; Q^{D}\; W^{(2)}(z) \right| \psi \right\rangle = \left\langle 0\left|\prod_{d=1}^{N_f=2} V_{\omega_d}(x_d)\; \prod_{i=1}^{D} \oint y_i^{\langle\psi, \alpha\rangle-1} S(y_i)\, dy_i\; W^{(2)}(z) \right| 0 \right\rangle\, .
\eeq
We impose that the contours should not enclose any pole in the $z$ variable.
The state $|\psi\rangle$ is defined such that:
\begin{align}
\alpha[0] |\psi\rangle &= \langle\psi, \alpha\rangle |\psi\rangle\label{eigenvalueA1}\\
\alpha[k] |\psi\rangle &= 0\, , \qquad\qquad\;\; \mbox{for} \; k>0,\nonumber
\end{align}
where the $\alpha[k]$ generate a deformed Heisenberg algebra:
\beq\label{A1commutator}
[\alpha[k], \alpha[n]] = {1\over k} (q^{k\over 2} - q^{-{k\over 2}})(t^{{k\over 2} }-t^{-{k\over 2} })(v^{k}+ v^{-k}) \delta_{k, -n} \; .
\eeq
We compute the various two-points making up the correlator; first, the bulk contributions
\begin{align}
&\prod_{1\leq i< j\leq D}\left\langle S(y_i)\, S(y_j) \right\rangle = \prod_{1\leq i\neq j\leq D} \frac{\left(y_{i}/y_{j};q\right)_{\infty}}{\left(t\, y_{i}/y_{j};q\right)_{\infty}}\;\prod_{1\leq i<j\leq D} \frac{\Theta\left(t\,y_{j}/y_{i};q\right)}{\Theta\left(y_{j}/y_{i};q\right)}\\
&\prod_{i=1}^{D}\left\langle :\prod_{d=1}^{N_f=2} V_{\omega_d}(x_d):\, S(y_j) \right\rangle = \prod_{i=1}^{D} \frac{\left(v\, x_{1}/y_{i};q\right)_{\infty}}{\left(v^{-1}\, x_{2}/y_{i};q\right)_{\infty}}\; .
\end{align}
We write the two-point of vertex operators with themselves as the factor $A(x_1, x_2)$:
\beq\label{vertexvertexA1}
\left\langle V_{\omega_1}(x_1)\,  V_{\omega_2}(x_2) \right\rangle = A(x_1, x_2) \; .
\eeq
We now come to the contributions involving the Wilson loop. First, the two-point of the stress tensor with the screening currents\footnote{In the actual correlator, the vacuum is labeled by $|\psi\rangle$ instead of $|0\rangle$, resulting in a relative shift of $\widetilde{\fq}$ between the two terms.}
\begin{align}
&\prod_{i=1}^{D}\left\langle S(y_i)\, W^{(2)}(z) \right\rangle = \prod_{i=1}^{D} \frac{1-t\, y_{i}/z}{1- y_{i}/z}+\prod_{i=1}^{D} \frac{1-v^2\, y_{i}/z}{1- t\, v^2\, y_{i}/z}
\end{align}
The two-point of the stress tensor with the D5 brane matter vertex operator is more subtle:
\begin{align}
&\left\langle :\prod_{d=1}^{N_f=2} V_{\omega_d}(x_d):\, W^{(2)}(z) \right\rangle\nonumber\\ 
&\;\;\;= \exp\left(\sum_{k>0}\, \frac{1}{k} \,\frac{1}{v^k+v^{-k}}\left[+\left(\frac{x_1}{z}\right)^k-\left(\frac{v^{-2}\,x_2}{z}\right)^k\right]\right)\nonumber\\
&\qquad\qquad+ \exp\left(\sum_{k>0}\, \frac{1}{k} \,\frac{1}{v^k+v^{-k}}\left[-\left(\frac{v^2\,x_1}{z}\right)^k+\left(\frac{x_2}{z}\right)^k\right]\right)\nonumber\\
&\;\;\;= \exp\left(\sum_{k>0}\, \frac{1}{k} \,\frac{1}{v^k+v^{-k}}\left[+\left(\frac{x_1}{z}\right)^k-\left(\frac{v^{-2}\,x_2}{z}\right)^k\right]\right)\cdot\left(1+\frac{1-v\, x_1/z}{1-v^{-1}\,x_2/z}\right)\nonumber\\
&\;\;\;= B\left(x_1, x_2, z\right)\cdot\left[\left(1-v^{-1}\,x_2/z\right)+\left(1-v\, x_1/z\right)\right]
\end{align}
In the second line, we used the commutator
\begin{align}
[w[k], w[n]] = {1\over k} (q^{k\over 2} - q^{-{k\over 2}})(t^{{k\over 2} }-t^{-{k\over 2} })\frac{1}{v^{k}+ v^{-k}}\delta_{k, -n} \; ,
\end{align}
which is dual to the relation \eqref{A1commutator}. In the third line, we used the identity $\exp(-\sum_{k>0}\frac{x^k}{k})=(1-x)$. In the last line, we defined the prefactor
\beq\label{BdefinedA1}
B\left(x_1, x_2, z\right)\equiv\frac{1}{1-v^{-1}\,x_2/z}\exp\left(\sum_{k>0}\, \frac{1}{k} \,\frac{1}{v^k+v^{-k}}\left[+\left(\frac{x_1}{z}\right)^k-\left(\frac{v^{-2}\,x_2}{z}\right)^k\right]\right) \; .
\eeq
All in all, we find that the ${\cW}_{q,t}(A_1)$ correlator is the 3d gauge theory partition function
\beq\label{A1correlatoris3dindex}
\left\langle \psi'\left|\prod_{d=1}^{N_f=2} V_{\omega_d}(x_d)\; Q^{D}\; W^{(2)}(z) \right| \psi \right\rangle=A(x_1, x_2)\, B(x_1, x_2, z)\, \left[\widetilde{\chi}^{A_1}\right]_{(1)}^{3d}(z)\; .
\eeq
The equality of the 5d and 3d partition functions \eqref{CHIequalityA1} further implies
\begin{align}
\left\langle \psi'\left|\prod_{d=1}^{N_f=2} V_{\omega_d}(x_d)\; Q^{D}\; W^{(2)}(z) \right| \psi \right\rangle=A(x_1, x_2)\, B(x_1, x_2, z)\, c_{3d}\left[\chi^{A_1}\right]_{(1)}^{5d}(z)_{e_{1}\,=\; x_{2}\,v^{-1} \,t^{D}}\; .\label{A1correlatoris5d}
\end{align}

As we already mentioned, a natural generalization is to increase the number of D1 branes present to $L>1$. The effect is that the 5d and 3d partition functions now become characters of the higher spin irreducible representations of $U_q(\widehat{A_1})$. In the deformed $\cW$ algebra picture, one simply considers a correlator with $L$ insertions of the deformed stress tensor:
\beq\label{A1correlatoris3dindexMORE}
\left\langle \psi'\left|\prod_{d=1}^{N_f=2} V_{\omega_d}(x_d)\; Q^{D}\; \prod_{\rho=1}^{L} W^{(2)}(z_{\rho}) \right| \psi \right\rangle=A(x_1, x_2)\, B(x_1, x_2, \{z_\rho\})\, \left[\widetilde{\chi}^{A_1}\right]_{(L)}^{3d}(\{z_\rho\})\; .
\eeq

\vspace{10mm}

\subsection{Wilson Loop for $A_m$ Theories ($m\geq 1$)}
\label{ssec:exampleAm}

\vspace{8mm}

------- 5d Gauge Theory -------\\

Let $X$ be a resolved $A_m$ singularity, and consider type IIB string theory on $X\times\cC\times\mathbb{C}^2$. 
For each $a=1, \ldots, m$, we introduce $n^{(a)}$ D5 branes wrapping the compact 2-cycle $S_a^2$ of $X$ and $\mathbb{C}^2$, for a total of $n=\sum_{a=1}^{m}n^{(a)}$ branes. We further introduce ${N^{(a)}_f}$ D5 branes wrapping the dual non-compact 2-cycle $S^*_a$ and $\mathbb{C}^2$, which make up a total of $N_f=\sum_{a=1}^{m}{N^{(a)}_f}$ non-compact branes. Finally, we add $L=\sum_{a=1}^{m}L^{(a)}$ D1 branes wrapping the same non-compact 2-cycles. We set $g_s\rightarrow 0$, which amounts to studying the $(2,0)$ $A_m$ little string on $\cC\times\mathbb{C}^2$ in the presence of codimension 2 defects (the D5 branes) and point-like defects (the D1 branes). 

At energies below the string scale, the little string dynamics in this background is fully captured by the theory on the D5 branes, with D1 brane defects. The theory on the D5 branes is a 5d quiver gauge theory of shape the Dynkin diagram of $A_m$. We call it $T^{5d}$, and it is defined on $S^1(\widehat{R})\times\mathbb{C}^2$, with gauge content  $\prod_{a=1}^{m}U(n^{(a)})$ and flavor matter content $\prod_{a=1}^{m}U({N^{(a)}_f})$. The D1 branes make up a 1/2-BPS a Wilson loop wrapping the circle and at the origin of $\mathbb{C}^2$.\\

After putting the theory on $\Omega$-background, the partition function of $T^{5d}$ with Wilson loop is the following Witten index:
{\allowdisplaybreaks
\begin{align}
\label{5dintegralAm}
&\left[\chi^{A_m}\right]_{(L^{(1)},\ldots, L^{(m)})}^{5d}  =\sum_{k^{(1)},\ldots, k^{(m)}=0}^{\infty}\;\prod_{a=1}^{m}\frac{\left(\widetilde{\fq}^{(a)}{}\right)^{k^{(a)}}}{k^{(a)}!} \;\;\nonumber\\
&\qquad\qquad\qquad\qquad\;\;\times \oint  \left[\frac{d\phi^{{(a)}}_I}{2\pi i}\right]Z^{(a)}_{vec}\cdot Z^{(a)}_{fund}\cdot Z^{(a)}_{CS}\cdot \prod_{b>a}^{m} Z^{(a,b)}_{bif}\cdot \prod_{\rho=1}^{L^{(a)}}  Z^{(a)}_{D1}  \; , \\
&Z^{(a)}_{vec} =\prod_{I, J=1}^{k^{(a)}} \frac{{\widehat{f}}\left(\phi^{(a)}_{I}-\phi^{(a)}_{J}\right)f\left(\phi^{(a)}_{I}-\phi^{(a)}_{J}+ 2\,\epsilon_+ \right)}{f\left(\phi^{(a)}_{I}-\phi^{(a)}_{J}+\epsilon_1\right)f\left(\phi^{(a)}_{I}-\phi^{(a)}_{J}+\epsilon_2\right)}\nonumber\\
&\qquad\times\prod_{I=1}^{k^{(a)}} \prod_{i=1}^{n^{(a)}} \frac{1}{f\left(\phi^{(a)}_I-a^{(a)}_i+\epsilon_+ \right)f\left(\phi^{(a)}_I-a^{(a)}_i-\epsilon_+ \right)}\; ,\\
&Z^{(a)}_{fund} =\prod_{I=1}^{k^{(a)}} \prod_{d=1}^{N^{(a)}_f} f\left(\phi^{{(a)}}_I-m^{{(a)}}_d\right) \equiv\prod_{I=1}^{k^{(a)}} Q^{(a)}(\phi^{{(a)}}_I)\; ,\label{fundQAm}\\
&Z^{(a)}_{CS}= \prod_{I=1}^{k^{(a)}} e^{k^{(a)}_{CS}\, \phi^{(a)}_I}\\
&Z^{(a, b)}_{bif} =\left[\prod_{J=1}^{k^{(b)}}\prod_{i=1}^{n^{(a)}}  f\left(\phi^{(b)}_J- m^{(a)}_{bif} - a^{(a)}_i - \epsilon_+\right) \prod_{I=1}^{k^{(a)}}\prod_{j=1}^{n^{(b)}} f\left(\phi^{(a)}_I+ m^{(a)}_{bif} - a^{(b)}_j + \epsilon_+  \right)\right.\nonumber\\
&\qquad \times \left.\prod_{I=1}^{k^{(a)}}\prod_{J=1}^{k^{(b)}} \frac{f\left(\phi^{(a)}_{I}-\phi^{(b)}_{J}+ m^{(a)}_{bif}+\epsilon_1 \right)}{f\left(\phi^{(a)}_{I}-\phi^{(b)}_{J} +m^{(a)}_{bif} \right)}\prod_{I=1}^{k^{(a)}}\prod_{J=1}^{k^{(b)}} \frac{f\left(\phi^{(a)}_{I}-\phi^{(b)}_{J}+ m^{(a)}_{bif}+\epsilon_2  \right)}{f\left(\phi^{(a)}_{I}-\phi^{(b)}_{J}+ m^{(a)}_{bif}+2\, \epsilon_+ \right)}\right]^{\Delta_{a,b}}\, ,\\
&Z^{(a)}_{D1} =\left[\prod_{i=1}^{n^{(a)}}  f\left(a^{{(a)}}_i-M^{{(a)}}_\rho\right) \prod_{I=1}^{k^{(a)}} \frac{f\left(\phi^{{(a)}}_I-M^{(a)}_\rho+ \epsilon_- \right)f\left(\phi^{{(a)}}_I-M^{(a)}_\rho- \epsilon_-\right)}{f\left(\phi^{{(a)}}_I-M^{(a)}_\rho+ \epsilon_+\right)f\left(\phi^{{(a)}}_I-M^{(a)}_\rho- \epsilon_+ \right)}\right]\, .
\end{align}}

The incidence matrix $\Delta_{a,b}$ in the bifundamental factor is the upper-diagonal matrix $\delta_{a+1, b}$. We have also defined a function $Q^{(a)}$ in $Z^{(a)}_{fund}$, which encodes the fundamental matter content on each node:
\begin{equation}
\label{Ammatter}
Q^{(a)}(\phi^{(a)}_I)\equiv \prod_{d=1}^{N^{(a)}_f} f\left(\phi^{(a)}_I-m^{(a)}_d\right)\; .
\end{equation}

We need to specify the contour for the integration variables $\phi^{{(a)}}_I$; in the absence of the Wilson loop factor $Z^{(a)}_{D1}$, the poles are classified by $n^{(a)}$-tuples of Young diagrams $\overrightarrow{\boldsymbol{\mu}^{(a)}}=\{\boldsymbol{\mu}^{(a)}_1, \boldsymbol{\mu}^{(a)}_2, \ldots, \boldsymbol{\mu}^{(a)}_{n^{(a)}}\}$. Namely, 
\beq
\label{Amyoungtuples}
\phi^{(a)}_I=a^{(a)}_i +\epsilon_+ - s_1\, \epsilon_1 - s_2\, \epsilon_2 , \text{with}\; (s_1, s_2)\in \boldsymbol{\mu}^{(a)}_i \; .
\eeq
In particular, note that in this case, the potential poles coming from the bifundamental factors have zero residue, so a fortiori, none of the poles depend on the bifundamental masses $m^{(a)}_{bif}$. 
After including $Z^{(a)}_{D1}$ in the integrand, there are additional poles to consider in the contours, dictated by the JK residue prescription:
\begin{align}
&\phi^{(a)}_I=M^{(a)}_\rho+\epsilon_+\\
&\phi^{(a+1)}_J=\phi^{(a)}_I+m^{(a)}_{bif}+2\epsilon_+\\
&\phi^{(a-1)}_J=\phi^{(a)}_I-m^{(a-1)}_{bif}
\end{align}
As a consequence, some of the residues will now explicitly depend on the bifundamental masses, and also necessarily on one of the fermion masses $M^{(a)}_\rho$. To be more quantitative, let us focus on a single D1 brane, $L=1$. Let us note in passing that in this case, the partition function is in fact a polynomial in the fermion mass fugacity $z=e^{-\widehat{R}M}$. This can be shown directly from the integral expression \eqref{5dintegralAm}. Furthermore, the asymptotics of $\chi^{5d}(z)$ at $z\rightarrow 0$ and $z\rightarrow +\infty$  tell us that the degree of this polynomial is $n^{(a)}$.\\

Now, we have $m$ choices for which non-compact 2-cycle $S^*_a$ this D1 brane should wrap. Let us choose the first 2-cycle $S^*_1$, as this will make it easy to write a closed form formula for the partition function. This means $L^{(1)}=1$, and $L^{(a)}=0$ for $a=2, \ldots, m$. 
Correspondingly, we denote the partition function as $\left[\chi^{A_m}\right]_{(L^{(1)},\ldots, L^{(m)})}^{5d}=\left[\chi^{A_m}\right]_{(1,0,\ldots,0)}^{5d}$. We write the corresponding fermion mass $M^{(1)}_1\equiv M$ to simplify the notations.

Then, by the JK residue prescription, an extra pole we need to consider comes from the Wilson loop factor $Z^{(a)}_{D1}$, say for instance
\beq 
\label{Amphi1}
\phi^{(1)}_1=M+\epsilon_+\; .
\eeq 
The remaining $k^{(1)}-1$ poles for the first gauge group should then be chosen among \eqref{Amyoungtuples},  along with the poles for the remaining gauge groups. 
Notice that the bifundamental factor $Z^{(1,2)}_{bif}$ has a denominator factor
\[
\prod_{J=1}^{k_{2}} \frac{1}{f\left(\phi^{(1)}_{1}-\phi^{(2)}_{J} +m^{(1)}_{bif}\right)f\left(\phi^{(1)}_{1}-\phi^{(2)}_{J}+ m^{(1)}_{bif}+2\, \epsilon_+\right)}\; ,
\]
so with the choice we made above for $\phi^{(1)}_1$, there is now an extra pole to consider from the first of these two factors, again by the JK residue prescription. For instance, 
\beq 
\label{Amphi2}
\phi^{(2)}_1=\phi^{(1)}_1+m^{(1)}_{bif}+2\epsilon_+=M+3\epsilon_+ + m^{(1)}_{bif}\; .
\eeq 
Then the remaining $k^{(1)} -1$ poles for the first gauge group and  $k^{(2)} -1$ poles for the second gauge group can be chosen  among \eqref{Amyoungtuples}, along with the poles for the remaining gauge groups. 
In fact, each bifundamental factor $Z^{(a,a+1)}_{bif}$, for $a=1, \ldots, m-1$, introduces a new pole. One easily checks that there are no extra poles beyond these. So, up to a permutation in $I$ of the $\phi^{(a)}_I$ variables, all the new poles are given by
\begin{align}
&\phi^{(1)}_1=M+\epsilon_+ \label{pole1Am}\\ 
&\phi^{(a+1)}_1=\phi^{(a)}_1+m^{(a)}_{bif}+2\epsilon_+\; , \qquad a=1, \ldots, m-1 \; .\label{pole2Am}
\end{align}
This is a total of $m$ new poles, which must be enclosed by the contours. We define the contours as follows: we choose $\cC^{(1)}_1$ to enclose the pole at \eqref{pole1Am} and the poles at \eqref{Amyoungtuples}, while the contours $\cC^{(1)}_I$ for $I>1$ will only enclose the poles in \eqref{Amyoungtuples}. For the other nodes $2\leq a \leq m$, we choose $\cC^{(a)}_1$ to enclose the pole at \eqref{pole2Am} as well as the poles at \eqref{Amyoungtuples}, while once again $\cC^{(a)}_I$ for $I>1$ will only enclose the poles at \eqref{Amyoungtuples}. This contour prescription fully specifies the partition function $\left[\chi^{A_m}\right]_{(1,0,\ldots,0)}^{5d}$, which can then be evaluated.\\

Let us now make contact with the representation theory of quantum affine algebras. We introduce a defect operator expectation value:
\begin{align}\label{YopAm}
&\left\langle \left[Y^{(a)}(z)\right]^{\pm 1} \right\rangle  \equiv\sum_{k^{(1)},\ldots, k^{(m)}=0}^{\infty}\frac{\left(\widetilde{\fq}^{(1)}{}\right)^{k^{(1)}}\ldots\,\left(\widetilde{\fq}^{(m)}{}\right)^{k^{(m)}}}{k_1!\ldots k_m!} \nonumber\\
&\qquad\qquad\qquad\times \oint_{\{\overrightarrow{\boldsymbol{\mu}}\}} \,\prod_{b=1}^{m} \left[\frac{d\phi^{{(b)}}_I}{2\pi i}\right]{Z^{(b)}_{vec}}\cdot Z^{(b)}_{fund} \cdot Z^{(b)}_{CS}\cdot \left[Z^{(a)}_{D1}(z)\right]^{\pm 1} \cdot \prod_{c=1}^{m-1} Z^{(c,c+1)}_{bif} \, .
\end{align}
It is understood in the above definition  that the contours enclose poles coming from Young diagrams only, at \eqref{Amyoungtuples}. As we have just seen, this is not the actual pole prescription of the partition function $\left[\chi^{A_m}\right]_{(1,0,\ldots,0)}^{5d}$, which is dictated by the JK prescription and has the extra poles \eqref{pole1Am}, \eqref{pole2Am}.  Nonetheless, we can write the partition function  as a Laurent polynomial in these $Y$-operator vevs, as a sum of exactly $m+1$ terms, to make up a $qq$-character of the first fundamental representation of $U_q(\widehat{A_m})$:
\begin{align}
\label{Amresult2}
&\left[\chi^{A_m}\right]_{(1,0,\ldots,0)}^{5d}(z) =\left\langle Y^{(1)}_{5d}(z) \right\rangle\nonumber\\
&\qquad\qquad + \widetilde{\fq}^{(1)} \; Q^{(1)}(z\, v^{-1}) \left\langle\frac{Y^{(2)}_{5d}(z\, v^{-2}\, \mu^{(1)}_{bif})}{Y^{(1)}_{5d}(z\, v^{-2})}\right\rangle\nonumber\\
&\qquad\qquad + \widetilde{\fq}^{(1)}\widetilde{\fq}^{(2)} \; Q^{(1)}(z\, v^{-1})\, Q^{(2)}(z\, v^{-3}\, \mu^{(1)}_{bif}) \left\langle\frac{Y^{(3)}_{5d}(z\, v^{-4}\, \mu^{(1)}_{bif}\,  \mu^{(2)}_{bif})}{Y^{(2)}_{5d}(z\, v^{-4}\,  \mu^{(1)}_{bif})}\right\rangle\nonumber\\
&\qquad\qquad + \widetilde{\fq}^{(1)}\widetilde{\fq}^{(2)}\widetilde{\fq}^{(3)} \; Q^{(1)}(z\, v^{-1})\, Q^{(2)}(z\, v^{-3}\, \mu^{(1)}_{bif}) \, Q^{(3)}(z\, v^{-5}\, \mu^{(1)}_{bif}\,\mu^{(2)}_{bif}) \left\langle\frac{Y^{(4)}_{5d}(z\, v^{-6}\, \mu^{(1)}_{bif}\,  \mu^{(2)}_{bif}\, \mu^{(3)}_{bif})}{Y^{(3)}_{5d}(z\, v^{-6}\,  \mu^{(1)}_{bif}\, \mu^{(2)}_{bif})}\right\rangle\nonumber\\
&\qquad\qquad+\ldots\nonumber\\
&\qquad\qquad+\prod_{a=1}^m \widetilde{\fq}^{(a)}\,Q^{(a)}\left(z\, v^{1-2a}\, \prod_{b=1}^{a-1} \mu^{(b)}_{bif}\right) \left\langle\frac{1}{Y^{(m)}_{5d}\left(z\, v^{-2m}\,  \prod_{b=1}^{m-1} \mu^{(b)}_{bif}\right)}\right\rangle \; .
\end{align}

Performing the integrals over the poles \eqref{Amyoungtuples}, we find
\beq\label{5ddefectexpression2Am}
\left\langle \left[Y^{(a)}_{5d}(z)\right]^{\pm 1} \right\rangle = \sum_{\{\overrightarrow{\boldsymbol{\mu}}\}}\left[ Z^{5d}_{bulk} \cdot Y^{(a)}_{5d}(z)^{\pm 1}\right]\, .
\eeq
The sum is over a collection of 2d partitions, one for each $U(1)$ Coulomb parameter:
\beq
\{\overrightarrow{\boldsymbol{\mu}}\}=\{\boldsymbol{\mu}^{(a)}_i\}_{a=1, \ldots, m\, ; \,\; i=1,\ldots,n^{(a)}}\; .
\eeq
The factor  $Z^{5d}_{bulk}$ encodes all the 5d bulk physics. It is given by
\beq\label{bulk5dAm}
Z^{5d}_{bulk}=  \prod_{a=1}^m \widetilde{\fq}^{(a)}{}^{\sum_{i=1}^{n^{(a)}}{\left|\boldsymbol{\mu}^{(a)}_i\right|}}\, Z^{(a)}_{bulk,vec} \cdot Z^{(a)}_{bulk,fund}\cdot Z^{(a)}_{bulk,CS} \cdot \prod^n_{b>a}
Z^{(a, b)}_{bulk,bif}\; ,
\eeq
Each factor is written in terms of the function \eqref{nekrasovN} as 
\begin{align}
&Z^{(a)}_{bulk,vec} = \prod_{i,j=1}^{n^{(a)}}\left[N_{\boldsymbol{\mu}^{(a)}_i\boldsymbol{\mu}^{(a)}_j}\left(e^{(a)}_{i}/e^{(a)}_{j};q\right)\right]^{-1}\label{5dbulkvecAm}\\
&Z^{(a)}_{bulk,fund} = \prod_{d=1}^{N^{(a)}_f} \prod_{i=1}^{n^{(a)}} N_{\boldsymbol{\emptyset}\, \boldsymbol{\mu}^{(a)}_i}\left( v\, f^{(a)}_{d}/e^{(a)}_{i};\, q\right) \label{5dbulkmatterAm}\\
&Z^{(a)}_{bulk,CS} = \prod\limits_{i=1}^{n^{(a)}} \left(T_{\boldsymbol{\mu}^{(a)}_i}\right)^{k^{(a)}_{CS}}\label{5dbulkCSAm}\\
&Z^{(a, b)}_{bulk,bif}= \prod_{i=1}^{n^{(a)}}\prod_{j=1}^{n^{(b)}}\left[N_{\boldsymbol{\mu}^{(a)}_i \boldsymbol{\mu}^{(b)}_j}\left(\mu^{(a)}_{bif}\,e^{(a)}_{i}/e^{(b)}_{j};\, q\right)\right]^{\Delta_{a,b}}\label{5dbifundAm}
\end{align}
Here, $T_{\boldsymbol{\mu}^{(a)}}$ is defined as  $ T_{\boldsymbol{\mu}^{(a)}} =(-1)^{|\boldsymbol{\mu}^{(a)}|} q^{\Arrowvert \boldsymbol{\mu}^{(a)}\Arrowvert^{2}/2}t^{-\Arrowvert \boldsymbol{\mu}^{(a)\, t}\Arrowvert^{2}/2}$.

The Wilson loop factor has the form
\beq\label{WilsonfactorAm}
Y^{(a)}_{5d}(z)\equiv\prod_{i=1}^{n^{(a)}}\prod_{k=1}^{\infty}\frac{1-t\, y^{(a)}_{i,k}/z}{1- y^{(a)}_{i,k}/z}\; .
\eeq

Above, we have defined the variables
\beq\label{ydefAm}
y^{(a)}_{i,k} = e^{(a)}_i\, q^{\boldsymbol{\mu}^{(a)}_{i, k}} \, t^{-k}\;\;\;,\; k=1, \ldots, \infty\;,
\eeq 
where  $\boldsymbol{\mu}^{(a)}_{i, k}$ is the length of the $k$-th row of the partition $\boldsymbol{\mu}^{(a)}_{i}$.\\

Then, having performed all the integrals, the 5d partition function with Wilson loop evaluates to \eqref{Amresult2}, where each vev is now understood to be written in terms of the variables \eqref{ydefAm}.

Two remarks are in order:
First, computing the partition function when the D1 brane wraps an arbitrary non-compact 2-cycle $S^*_a$  is a straightforward exercise, using the JK residue prescription. In the above example, we chose the D1 brane to wrap $S^*_1$, and the resulting partition function was the $qq$-character of the first fundamental representation of  $U_q(\widehat{A_m})$. In general, if the D1 brane wraps the 2-cycle $S^*_a$, the partition function is the $qq$-character of the $a$-th fundamental representation of  $U_q(\widehat{A_m})$.

Second, it is easy to generalize the above discussion to an arbitrary number $L>1$ of D1 branes. This corresponds to considering a Wilson loop valued in the representation ${\textbf{R}}=\left({\textbf{R}}^{(1)}, \ldots , {\textbf{R}}^{(m)}\right)$ of $\prod_{a=1}^m SU(n^{(a)})$. For each node $a$, the representation $\textbf{R}^{(a)}$ is a tensor product of the fundamental representation of $SU(n^{(a)})$, which appears $L^{(a)}$ times. The partition function is again computed in a straightforward way using the JK residue prescription.\\

\vspace{10mm}

------- 3d Gauge Theory and ${\cW}_{q,t}(A_m)$ correlator -------\\

We now make contact with the theory $G^{3d}$ on the vortices of $T^{5d}$. The non-compact D5 branes produce  punctures on the cylinder. Since these  defects factorize, it is sufficient to focus our attention on a single defect, which means we will specialize the number $N_f$ of non-compact D5 branes to a specific number from now on. Imposing the vanishing of D5 brane flux at infinity $[S+S^*]=0$ translates to the condition
\beq\label{conformalAm}
2 \;n^{(a)}-n^{(a-1)}-n^{(a+1)} = N^{(a)}_f\; , \qquad a=1, \ldots, m\; .
\eeq
which further specifies the ranks $n^{(a)}$ of the 5d gauge groups. This constraint also specifies the Chern-Simons term \eqref{5dbulkCSAm} on node $a$ to be $k^{(a)}_{CS}=n^{(a)}-n^{(a+1)}$.  

In the geometry, all the D5 branes previously wrapping compact 2-cycles now bind to D5 branes wrapping non-compact 2-cycles. We can recast these statements in terms of the representation theory of $\fg$; namely, each non-compact D5 brane is labeled by a weight of $A_m$, making up a set $\{\omega_d\}_{d=1}^{N_f}$. As explained in section \ref{ssec:qToda}, such a set needs to satisfy three conditions:
\begin{itemize}
	\item For all $1\leq d \leq N_f$, the weight $\omega_d$ belongs in a fundamental representation of $\fg$.
	\item $\sum_{d=1}^{N_f} \omega_d=0$.
	\item No proper subset of coweights in $\{\omega_d\}$ adds up to $0$. 
\end{itemize}
The classification of such defects was given in \cite{Haouzi:2016ohr}. For the purpose of being explicit and providing closed-form expressions for the 3d partition function at any rank $m$, we will focus our attention on two distinct defects here, so two distinct 3d theories $G^{3d}$. First, we will study a generic defect, sometimes nicknamed the ``full puncture" on $\cC$, and then we will study a highly degenerate defect, sometimes nicknamed the ``simple puncture" on $\cC$.\\

\vspace{10mm}

------ Example 1: The ``Full Puncture" ------\\

\begin{figure}[h!]
	\emph{}
	\centering
	\includegraphics[trim={0 0 0 1cm},clip,width=0.9\textwidth]{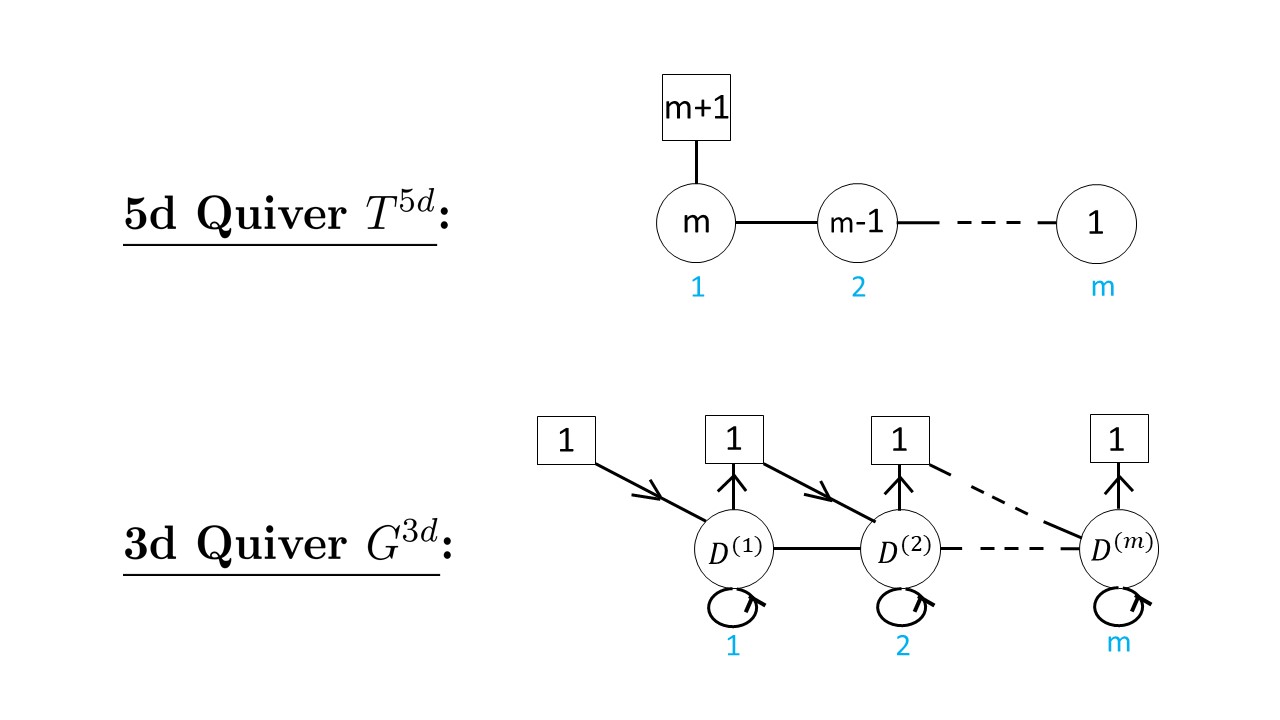}
	\vspace{-10pt}
	\caption{The $A_m$ theory $T^{5d}$ under study, along with the theory $G^{3d}$ on its vortices. This configuration of D5 branes is sometimes referred to as a ``full puncture". The blue numbers label the nodes.} 
	\label{fig:AmFull5d}
\end{figure}

Let us consider the following set $\{\omega_d\}$ of $m+1$ weights:
\begin{align}
	\omega_1 &= [-1,\phantom{-}0, \phantom{-}0, \ldots,   \phantom{-}0,  \phantom{-}0]= -\lambda_1\nonumber\\
	\omega_2 &= [\phantom{-}1, -1, \phantom{-}0, \ldots,   \phantom{-}0,  \phantom{-}0]=-\lambda_1+\alpha_1\nonumber\\
	&  \vdots\nonumber\\ 
	\omega_{m} &= [\phantom{-}0,  \phantom{-}0, \phantom{-}0, \ldots,   \phantom{-}1, -1]= -\lambda_1+\alpha_1+\ldots + \alpha_{m-1}\nonumber\\
	\omega_{m+1} &= [\phantom{-}0,  \phantom{-}0, \phantom{-}0, \ldots,   \phantom{-}0, \phantom{-}1]= -\lambda_1+\alpha_1+\ldots+ \alpha_{m-1}+\alpha_m \label{AmweightFULL}
\end{align}
$\lambda_a$ is the $a$-th fundamental weight, and $\alpha_a$ is the $a$-th positive simple root. Note that the set $\{\omega_d\}$ spans the weight lattice. Each one of the weights $\omega_a$ then represents a distinct D5 brane wrapping a non-compact 2-cycle and some compact 2-cycles.  Correspondingly, we now label the $m+1$ masses by $x_1, x_2, \ldots, x_{m+1}$, respectively. The specific 5d quiver gauge theory $T^{5d}$ engineered by such a choice of weights is shown in figure \ref{fig:AmFull5d}.\\

We further introduce $D^{(a)}$ D3 branes wrapping the compact 2-cycle $S_a$ and one of the two complex lines, say $\mathbb{C}_q$, for a total of $D=\sum_{a=1}^{m}D^{(a)}$ D3 branes. 
In 5d gauge theory terms, we are at the special point on the moduli space where the Coulomb and Higgs branches  of $T^{5d}$ meet, and we further turn on $D$ units of vortex flux. Because of the Omega background on $\mathbb{C}_t$, this effectively shifts the Coulomb parameters $e^{(a)}_{i}$ by $D^{(a)}_i$ units of $t$-flux, since $\mathbb{C}_t$ is the line transverse to the vortex. Effectively, then, we get off the root of the Higgs branch and probe the Coulomb branch of $T^{5d}$ at integer values. For definiteness, let us set the bifundamental masses to be   
\beq\label{bifspecializedAm}
\mu^{(a)}_{bif}= v \; ,\qquad a=1, \ldots, m-1\; ,
\eeq
and we further set the 5d Coulomb moduli to be:
\begin{align}
\label{specializeAm}
e^{(a)}_{i}=x_{m+2-i}\,t^{D^{(a)}_i}v^{-1-a} \; ,\qquad a=1, \ldots, m; \qquad i=1, \ldots, n^{(a)} \; .
\end{align}
Using the definition of the function \eqref{nekrasovN}, it follows that the 5d partition function is now a sum over truncated partitions, of length at most $D^{(a)}_i$; for a detailed derivation, a good reference is \cite{Aganagic:2014oia}. From this point on, we denote the 3d Coulomb parameters as $y^{(a)}_i$, for all $a=1, \ldots, m$ and $i=1, \ldots, D^{(a)}$.\\

The 3d $\cN=2$ theory $G^{3d}$ is defined on $S^1(\widehat{R})\times\mathbb{C}$, and lives on the vortices of $T^{5d}$: it is a quiver gauge theory with gauge group $\prod_{a=1}^m U(D^{(a)})$, and various chiral and antichiral matter multiplets, as pictured in figure \ref{fig:AmFull5d}.\\

The first step is to define a 3d  Wilson loop operator expectation value, as an integral over the Coulomb moduli of $G^{3d}$, where it is understood that none of the contours below enclose poles depending on $z$:
\begin{align}\label{3ddefectexpression2Am}
\left\langle\left[{\widetilde{Y}^{(a)}_{3d}}(z)\right]^{\pm 1} \right\rangle \equiv \left[{\widetilde{Y}^{(a)}_{D1/D5}}(\{x_d\}, z)\right]\; \oint_{\{\overrightarrow{\boldsymbol{\mu}}\}} d{y}\,\left[I^{3d}_{bulk}(y)\, \left[{\widetilde{Y}^{(a)}_{D1/D3}}(y, z)\right]^{\pm 1}\right]  \; .
\end{align}
The bulk contribution
\begin{align}\label{bulk3dAm}
I^{3d}_{bulk}(y)=\prod_{a=1}^{m}\prod_{l=1}^{D^{(a)}}{y^{(a)}_l}^{\left(\tau^{(a)}-1\right)}\;I^{(a)}_{bulk, vec}\cdot I^{(a)}_{bulk, matter}\cdot\prod_{b>a}I^{(a,b)}_{bulk, bif}\; .
\end{align}
is independent of the Wilson of the Wilson loop physics. It contains the 3d F.I. term
\begin{align}\label{FI3dAm}
\prod_{a=1}^{m}\prod_{l=1}^{D^{(a)}}{y^{(a)}_l}^{\left(\tau^{(a)}\right)}
\end{align}
the 3d $\cN=4$ vector multiplet contributions
\begin{align}\label{vec3dAm}
I^{(a)}_{bulk, vec}(y^{(a)})=\prod_{1\leq i\neq j\leq D^{(a)}}\frac{\left(y^{(a)}_{i}/y^{(a)}_{j};q\right)_{\infty}}{\left(t\, y^{(a)}_{i}/y^{(a)}_{j};q\right)_{\infty}}\;\prod_{1\leq i<j\leq D^{(a)}} \frac{\Theta\left(t\,y^{(a)}_{i}/y^{(a)}_{j};q\right)}{\Theta\left(y^{(a)}_{i}/y^{(a)}_{j};q\right)}\; ,
\end{align}
and the bifundamental hypermultiplets
\begin{align}
\label{bif3dAm}
I^{(a,b)}_{bulk, bif}(y^{(a)}, y^{(b)})=\prod_{1\leq i \leq D^{(a)}}\prod_{1\leq j \leq D^{(b)}}\left [ \frac{(v\, t\, y^{(a)}_{i}/y^{(b)}_{j};q)_{\infty}}{(v \, y^{(a)}_{i}/y^{(b)}_{j};q)_{\infty}}\right]^{\Delta_{a, b}}\; .
\end{align}
The $\cN=2$ matter content has a ``handsaw" structure:
\begin{align}\label{matter3dAm}
I^{(a)}_{bulk, matter}(y^{(a)}, \{x_d\})=\prod_{i=1}^{D^{(a)}} \frac{\left(v^{2-a}\, x_{a}/y^{(a)}_{i};q\right)_{\infty}}{\left(v^{-a}\, x_{a+1}/y^{(a)}_{i};q\right)_{\infty}} \; .
\end{align}
The various powers of $v$ in the argument of the $q$-Pochhammer symbols are fixed by the R-symmetry of $G^{3d}$.

The Wilson loop contributions are given 
\beq\label{3dWilsonfactorAm}
{\widetilde{Y}^{(a)}_{D1/D3}}(y^{(a)},z)=\prod_{i=1}^{D^{(a)}}\frac{1-t\, y^{(a)}_{i}/z}{1- y^{(a)}_{i}/z}\; .
\eeq
and by
\beq\label{3dWilsonfactor2Am}
{\widetilde{Y}^{(a)}_{D1/D5}}(\{x_{d}\}, z)=\prod_{j=1}^{n^{(a)}}\left(1- v^{-a}\, x_{m+2-j}/z\right)\; .
\eeq

All in all, the 3d partition function can be written as a twisted $qq$-character of the first fundamental representation of $U_q(\widehat{A_m})$, and has a closed form expression:
\begin{align}
&\left[\widetilde{\chi}^{A_m}\right]_{(1,0,\ldots,0)}^{3d}(z) =\prod_{j=2}^{m+1}\left(1- v^{-1}\, x_{j}/z\right)\left\langle\widetilde{Y}^{(1)}_{D1/D3}(z)  \right\rangle\nonumber\\
&\qquad\qquad + \widetilde{\fq}^{(1)} \, \prod_{j=3}^{m+1}\left(1- v^{-1}\, x_{j}/z\right)\cdot\left(1- v\, x_{1}/z\right) \left\langle\frac{\widetilde{Y}^{(2)}_{D1/D3}(z\, v^{-1})}{\widetilde{Y}^{(1)}_{D1/D3}(z\, v^{-2})}\right\rangle\nonumber\\
&\qquad\qquad + \widetilde{\fq}^{(1)}\widetilde{\fq}^{(2)} \,\prod_{j=4}^{m+1}\left(1- v^{-1}\, x_{j}/z\right)\cdot\left(1- v\, x_{1}/z\right)\cdot\left(1- v\, x_{2}/z\right) \left\langle\frac{\widetilde{Y}^{(3)}_{D1/D3}(z\, v^{-2})}{\widetilde{Y}^{(2)}_{D1/D3}(z\, v^{-3})}\right\rangle\nonumber\\
&\qquad\qquad + \widetilde{\fq}^{(1)}\widetilde{\fq}^{(2)}\widetilde{\fq}^{(3)} \,\prod_{j=5}^{m+1}\left(1- v^{-1}\, x_{j}/z\right)\cdot\left(1- v\, x_{1}/z\right)\cdot\left(1- v\, x_{2}/z\right)\cdot\left(1- v\, x_{3}/z\right)  \left\langle\frac{\widetilde{Y}^{(4)}_{D1/D3}(z\, v^{-3})}{\widetilde{Y}^{(3)}_{D1/D3}(z\, v^{-4})}\right\rangle\nonumber\\
&\qquad\qquad+\ldots\nonumber\\
&\qquad\qquad+\prod_{a=1}^m \widetilde{\fq}^{(a)}\,\prod_{j=1}^{m}\left(1- v\, x_{j}/z\right)\left\langle\frac{1}{\widetilde{Y}^{(m)}_{D1/D3}(z\, v^{-m-1})}\right\rangle \; .\label{3dpartitionfunctionAm}
\end{align}
As described in the main text, this can be proved in a straightforward way by Higgsing the 5d theory. Namely, we use the specialization of the Coulomb parameters \eqref{specializeAm}, and we recall the dictionary from 3d variables to 5d variables (after breaking of the 3d gauge groups due to the choice of contours):
\beq\label{3dto5dAm}
y^{(a)}_{i,k}=  e^{(a)}_{i}\, q^{\boldsymbol{\mu}^{(a)}_{i, k}}\, t^{-k} \; ,\qquad\;\; k=1, \ldots, D^{(a)}_i\; ,
\eeq 
the first step is to show that the 5d $Y$-operator vev is the residue sum of the 3d $Y$-operator vev integral:
\beq\label{YequalityAm}
\left\langle{\widetilde{Y}^{(a)}_{3d}(z)} \right\rangle = c_{3d}\, \left\langle {Y_{5d}(z)} \right\rangle_{e^{(a)}_{i}=\; \eqref{specializeAm}} \; .
\eeq
The proportionality constant $c_{3d}$ is the bulk residue at the empty partition, see \eqref{c3d}.

The second step is to show that each of the $m+1$ terms in the 5d character \eqref{Amresult2} corresponds to a term in the 3d character \eqref{3dpartitionfunctionAm}. 
As an example, let us look at the Wilson loop factors in the first term of the 5d character; there, the Higgsing \eqref{specializeAm} triggers an infinite number of telescopic cancellations:
\begin{align}
Y^{(1)}_{5d}(z) &=\prod_{i=1}^{n^{(1)}}\prod_{k=1}^{\infty}\frac{1-t\, y^{(1)}_{i,k}/z}{1- y^{(1)}_{i,k}/z}\nonumber\\
&= \prod_{i=1}^{n^{(1)}}\prod_{k=1}^{D^{(1)}_i}\frac{1-t\, y^{(1)}_{i,k}/z}{1- y^{(1)}_{i,k}/z}\cdot  \prod_{j=1}^{n^{(1)}}\left(1- v^{-1}\, x_{m+2-j}/z\right)\nonumber\\
&= \left[{\widetilde{Y}^{(1)}_{D1/D3}}(y_{\{\overrightarrow{\boldsymbol{\mu}}\}}, z)\right]\cdot \prod_{j=2}^{m+1}\left(1- v^{-1}\, x_{j}/z\right)\label{fromY5dtoY3dAm}
\end{align}
In the last line, we used the fact that $n^{(1)}=m$ and relabeled the product index. After taking a vacuum expectation value on both sides, it follows that the first term of the 5d partition function is in fact the first term of the 3d partition function.

The remaining $m$ terms work out in the same way, with  cancellations occurring thanks to the 5d hypermultiplet factors (see the case $A_1$ above for an explicit example of this mechanism). Therefore, after one further identifies the 5d gauge coupling with the 3d F.I. term, the equality of the partition functions follows:
\begin{align}\label{CHIequalityAm}
\left[\widetilde{\chi}^{A_m}\right]_{(1,0,\ldots,0)}^{3d}(z)=c_{3d}\;  \left[\chi^{A_m}\right]_{(1,0,\ldots,0)}^{5d}(z)_{e^{(a)}_{i}=\; \eqref{specializeAm}} \; .
\end{align}

We now reinterpret the physics in terms of $A_m$-type $q$-Toda theory on the cylinder $\cC$. The theory enjoys a ${\cW}_{q,t}(A_m)$ algebra symmetry, which is generated by $m$ currents $W^{(s)}(z)$, with $s=2, \ldots, m+1$. To make contact with the 3d gauge theory results above, let us consider the spin 2 operator $W^{(2)}(z)$  operator, whose insertion at a point $z$ on $\cC$ corresponds to the location of a D1 brane. Constructing the operator as the commutant of the screening charges, one finds:
\begin{align}
&W^{(2)}(z) = :\cY^{(1)}(z): + :\cY^{(2)}(v^{-1}z)\left[\cY^{(1)}(v^{-2}z)\right]^{-1}:\nonumber\\
&\qquad\qquad+:\cY^{(3)}(v^{-2}z)\left[\cY^{(2)}(v^{-3}z)\right]^{-1}: +\ldots+:\left[\cY^{(m)}(v^{-m-1}z)\right]^{-1}: ,\label{Amstress}
\end{align}
where $\cY^{(a)}$ is the operator defined in \eqref{YoperatorToda}.\\

We should now  build the vertex operator corresponding to the D5 brane content. Such an operator is fully determined by the set \eqref{AmweightFULL} of weights $\{\omega_d\}$. In particular, the various $v$-shifts inside the argument of the operators are fixed by the R-symmetry of the 3d theory $G^{3d}$. Namely, we first define
\begin{align}
	V_{\omega_1} (x_1)&=\; :\left[\Lambda^{(1)}(x_1)\right]^{-1}: \nonumber\\
	V_{\omega_2} (x_2)&=\; :\Lambda^{(1)}(x_2\, v^{-1})\left[\Lambda^{(2)}(x_2\, v^{-2})\right]^{-1}:\nonumber\\
	V_{\omega_3} (x_3)&=\; :\Lambda^{(2)}(x_3\, v^{-2})\left[\Lambda^{(3)}(x_3\, v^{-3})\right]^{-1}:\nonumber\\
	&\vdots\nonumber\\ 
	V_{\omega_{m+1}} (x_{m+1})&=\; :\Lambda^{(m)}(x_{m+1}\, v^{-m}):\label{Amvertexweight}
\end{align}
where $\Lambda(x)$ is the fundamental weight operator \eqref{coweightvertexdef}. Alternatively, we can define the D5 brane operator using the simple root operators \eqref{rootvertexdef}. The $v$-shifts in the argument of these operators are uniquely determined from the truncation locus \eqref{specializeAm}:
\begin{align}
	V_{\omega_1} (x_1)&=\; :\left[\Lambda^{(1)}(x_1)\right]^{-1}:\nonumber \\
	V_{\omega_2} (x_2)&=\; :\left[\Lambda^{(1)}(x_2)\right]^{-1} E^{(1)}(x_2\,v^{-1}):\nonumber\\
	V_{\omega_3} (x_3)&=\; :\left[\Lambda^{(1)}(x_3)\right]^{-1} E^{(1)}(x_3\,v^{-1})E^{(2)}(x_3\,v^{-2}):\nonumber\\
	&\vdots\nonumber\\ 
	V_{\omega_{m+1}} (x_{m+1})&=\; :\left[\Lambda^{(1)}(x_{m+1})\right]^{-1} E^{(1)}(x_{m+1}\,v^{-1})E^{(2)}(x_{m+1}\,v^{-2})\ldots E^{(m)}(x_{m+1}\,v^{-m}):\label{Amvertexroot}
\end{align}
Either way, the D5 brane operator is defined as
\beq\label{Amgenericvertexexmaple}
V_\eta (u)= \, :\prod_{d=1}^{N_f=m+1} V_{\omega_d}(x_d): \; .
\eeq
The parameters $u$ and $\eta$ are defined through
\begin{align}\label{Amcommomentum}
&\eta = \sum_{d=1}^{m+1}\eta_d\, \omega_d \; ,\\
&x_d = u \, q^{-\eta_d}\; ,\qquad\;\; d=1, \ldots, m+1\; .
\end{align}

We have all the tools to compute the correlator 
\begin{align}
&\left\langle \psi'\left|V_\eta (u)\; \prod_{a=1}^{m} (Q^{(a)})^{D^{(a)}}\; W^{(2)}(z) \right| \psi \right\rangle\nonumber\\ 
&\qquad= \left\langle 0\left|\prod_{d=1}^{m+1} V_{\omega_d}(x_d)\; \prod_{a=1}^{m}\prod_{i=1}^{D^{(a)}} \oint \left(y^{(a)}_i\right)^{\langle\psi, \alpha_a\rangle -1} S(y^{(a)}_i)\, dy^{(a)}_i\; W^{(2)}(z) \right| 0 \right\rangle\label{Amcorrelatordef}
\end{align}
We impose that the contours should not enclose any pole in the $z$ variable.
The state $|\psi\rangle$ is defined such that, for all $a=1, \ldots, m$:
\begin{align}
\alpha_a[0] |\psi\rangle &= \langle\psi, \alpha_a\rangle |\psi\rangle\nonumber\\
\alpha_a[k] |\psi\rangle &= 0\, , \qquad\qquad\;\; \mbox{for} \; k>0.\label{eigenvalueAm}
\end{align}
The $\alpha_a[k]$ generate the deformed Heisenberg algebra \eqref{commutatorgenerators}.

All one needs to do now is compute the various two-points making up the correlator; first, the bulk contributions
\begin{align}
&\prod_{1\leq i< j\leq D^{(a)}}\left\langle S^{(a)}(y^{(a)}_i)\, S^{(a)}(y^{(a)}_j) \right\rangle = \prod_{1\leq i\neq j\leq D^{(a)}}\frac{\left(y^{(a)}_{i}/y^{(a)}_{j};q\right)_{\infty}}{\left(t\, y^{(a)}_{i}/y^{(a)}_{j};q\right)_{\infty}}\;\prod_{1\leq i<j\leq D^{(a)}} \frac{\Theta\left(t\,y^{(a)}_{i}/y^{(a)}_{j};q\right)}{\Theta\left(y^{(a)}_{i}/y^{(a)}_{j};q\right)}\, ,\\
&\prod_{1\leq i \leq D^{(a)}}\prod_{1\leq j \leq D^{(b)}}\left\langle S^{(a)}(y^{(a)}_i)\, S^{(b)}(y^{(b)}_j) \right\rangle = \prod_{1\leq i \leq D^{(a)}}\prod_{1\leq j \leq D^{(b)}}\left [ \frac{(v\, t\, y^{(a)}_{i}/y^{(b)}_{j};q)_{\infty}}{(v \, y^{(a)}_{i}/y^{(b)}_{j};q)_{\infty}}\right]^{\Delta_{a, b}}\, ,\qquad a\neq b\\
&\prod_{i=1}^{D^{(a)}}\left\langle :\prod_{d=1}^{m+1} V_{\omega_d}(x_d):\, S^{(a)}(y^{(a)}_i) \right\rangle = \prod_{i=1}^{D^{(a)}} \frac{\left(v^{2-a}\, x_{a}/y^{(a)}_{i};q\right)_{\infty}}{\left(v^{-a}\, x_{a+1}/y^{(a)}_{i};q\right)_{\infty}} \; .
\end{align}
We write the two-point of the vertex operators with themselves as the factor $A(\{x_d\})$:
\beq\label{vertexvertexAm}
\prod_{i, j}\left\langle V_{\omega_i}(x_i)\,  V_{\omega_j}(x_j) \right\rangle = A(\{x_d\}) \; .
\eeq
As far as the Wilson loop physics is concerned, we first compute the two-point of the stress tensor with the screening currents
\begin{align}
&\prod_{a=1}^{m}\prod_{i=1}^{D^{(a)}}\left\langle S^{(a)}(y_i)\, W^{(2)}(z) \right\rangle =  \widetilde{Y}^{(1)}_{D1/D3}(z)+\frac{\widetilde{Y}^{(2)}_{D1/D3}(z\, v^{-1})}{\widetilde{Y}^{(1)}_{D1/D3}(z\, v^{-2})}\nonumber\\
&\qquad\qquad\qquad\;\;\;+\frac{\widetilde{Y}^{(3)}_{D1/D3}(z\, v^{-2})}{\widetilde{Y}^{(2)}_{D1/D3}(z\, v^{-3})}+\ldots+\frac{1}{\widetilde{Y}^{(m)}_{D1/D3}(z\, v^{-m-1})}\; .
\end{align}
The two-point of the stress tensor with the D5 brane matter vertex operator is more subtle:
\begin{align}
&\left\langle :\prod_{d=1}^{m+1} V_{\omega_d}(x_d):\, W^{(2)}(z) \right\rangle\nonumber\\ 
&\;\;\;= B\left(\{x_d\}, z\right)\cdot\left[\prod_{j=2}^{m+1}\left(1- v^{-1}\, x_{j}/z\right)\right.\nonumber\\
&\;\;\;\;\;\;\;+\prod_{j=3}^{m+1}\left(1- v^{-1}\, x_{j}/z\right)\cdot\left(1- v\, x_{1}/z\right)\nonumber\\
&\;\;\;\;\;\;\;+\prod_{j=4}^{m+1}\left(1- v^{-1}\, x_{j}/z\right)\cdot\left(1- v\, x_{1}/z\right)\cdot\left(1- v\, x_{2}/z\right)\nonumber\\
&\;\;\;\;\;\;\;+\prod_{j=5}^{m+1}\left(1- v^{-1}\, x_{j}/z\right)\cdot\left(1- v\, x_{1}/z\right)\cdot\left(1- v\, x_{2}/z\right)\cdot\left(1- v\, x_{3}/z\right)\nonumber\\
&\;\;\;\;\;\;\;+\ldots\nonumber\\
&\;\;\;\;\;\;\;+\left.\prod_{j=1}^{m}\left(1- v\, x_{j}/z\right)\right]\; ,
\end{align}
where the prefactor is defined as
\beq\label{BdefinedAm}
B\left(\{x_d\}, z\right)\equiv\frac{1}{\prod_{j=2}^{m+1}\left(1- v^{-1}\, x_{j}/z\right)}\exp\left(\sum_{k>0}\, \frac{1}{k} \,\sum_{a=1}^m M_{a1}(q^{k\over 2} , t^{k\over 2})\left[+\left(\frac{v^{1-a}\,x_a}{z}\right)^k-\left(\frac{v^{-1-a}\,x_{a+1}}{z}\right)^k\right]\right) \; .
\eeq

To derive this, one uses the commutator \eqref{commutator3} involving the inverse of the deformed Cartan matrix. We also made use of the identity $\exp(-\sum_{k>0}\frac{x^k}{k})=(1-x)$. 

All in all, we find that the ${\cW}_{q,t}(A_m)$ correlator is the 3d gauge theory partition function
\beq\label{Amcorrelatoris3dindex}
\left\langle \psi'\left|\prod_{d=1}^{m+1} V_{\omega_d}(x_d)\; \prod_{a=1}^{m} (Q^{(a)})^{D^{(a)}}\; W^{(2)}(z) \right| \psi \right\rangle = A(\{x_d\})\, B(\{x_d\}, z)\, \left[\widetilde{\chi}^{A_m}\right]_{(1,0,\ldots,0)}^{3d}(z)\; .
\eeq
The equality of the 5d and 3d partition functions \eqref{CHIequalityAm} further implies:
\begin{align}\label{Amcorrelatoris5d}
\left\langle \psi'\left|\prod_{d=1}^{m+1} V_{\omega_d}(x_d)\; \prod_{a=1}^{m} (Q^{(a)})^{D^{(a)}}\; W^{(2)}(z) \right| \psi \right\rangle = A(\{x_d\})\, B(\{x_d\}, z)\, c_{3d}\left[\chi^{A_m}\right]_{(1,0,\ldots,0)}^{5d}(z)_{e^{(a)}_{i}=\; \eqref{specializeAm}}\; .
\end{align}

As we mentioned in the main text, a natural generalization is to increase the number of D1 branes to $L>1$. For the full puncture we just presented, we explicitly checked the following formula for a variety of D1 branes, finding perfect agreement with our general formula \eqref{correlatoris3dindex}:
\begin{align}
	&\left\langle \psi'\left|\prod_{d=1}^{N_f}V_{\omega_d}(x_d)\; \prod_{a=1}^{m} (Q^{(a)})^{D^{(a)}}\; \prod_{s=2}^{m+1}\prod_{\rho=1}^{L^{(s-1)}}W^{(s)}(z^{(s-1)}_\rho) \right| \psi \right\rangle \nonumber\\
	&\qquad\qquad\qquad=A\left(\{x_d\}\right)\, B\left(\{x_d\},\{z^{(s-1)}_\rho\}\right)\, \left[\widetilde{\chi}^{A_m}\right]_{(L^{(1)},\ldots, L^{(m)})}^{3d}(\{z^{(s-1)}_\rho\})\; .\label{Amcorrelatoris3dindexMORE}
\end{align}

\vspace{10mm}

------- Example 2: The ``Simple Puncture" -------\\

\begin{figure}[h!]
	\emph{}
	\centering
	\includegraphics[trim={0 0 0 1cm},clip,width=0.9\textwidth]{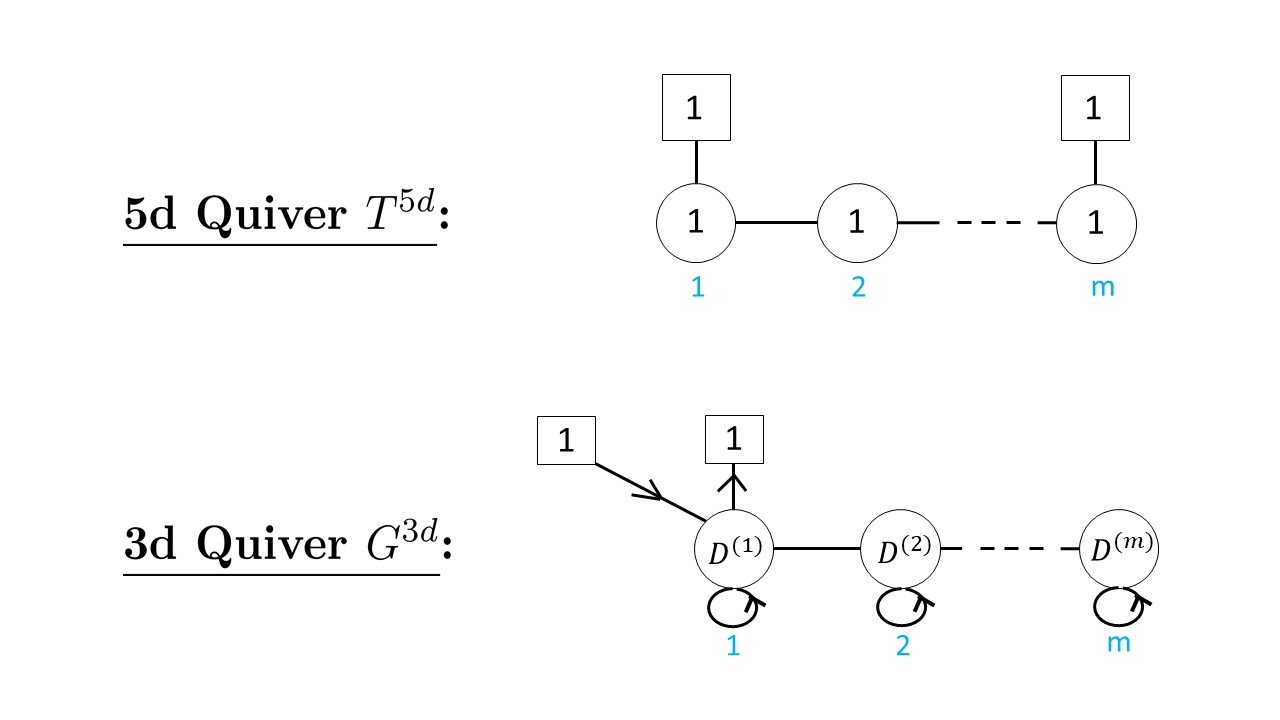}
	\vspace{-10pt}
	\caption{The $A_m$ theory $T^{5d}$ under study, along with the theory $G^{3d}$ on its vortices. This configuration of D5 branes is sometimes referred to as a ``simple puncture". The blue numbers label the nodes.} 
	\label{fig:AmSimple5d}
\end{figure}

Let us now consider a different set $\{\omega_d\}$ of 2 weights only:
\begin{align}
\omega_1 &= [-1,\phantom{-}0, \phantom{-}0, \ldots,   \phantom{-}0,  \phantom{-}0]= -\lambda_1\nonumber\\
\omega_2 &= [\phantom{-}1, \phantom{-}0, \phantom{-}0, \ldots,   \phantom{-}0,  \phantom{-}0]=-\lambda_m+\sum_{a=1}^m\alpha_a\label{AmweightSIMPLE}
\end{align}
This set satisfies the three conditions to represent a D5 brane defect. We label the 2 associated masses by $x_1$ and $x_2$, respectively. The specific 5d quiver gauge theory $T^{5d}$ engineered by such a choice of weights is shown in figure \ref{fig:AmSimple5d}. Because the dimension of the Coulomb branch of $T^{5d}$ is minimal for this choice of set $\{\omega_d\}$  (namely, it has complex dimension $m$), the associated D5 brane defect is sometimes nicknamed a ``simple puncture" on the cylinder $\cC$.\\

Once again, we introduce $D^{(a)}$ D3 branes wrapping the compact 2-cycle $S_a$ and one of the two complex lines, say $\mathbb{C}_q$, for a total of $D=\sum_{a=1}^{m}D^{(a)}$ D3 branes. 
In the 5d theory $T^{5d}$, this effectively shifts the Coulomb parameters $e^{(a)}_{1}$ away from the root of Higgs branch, by $D^{(a)}_1$ units of $t$-flux. For definiteness, let us set the bifundamental masses to be   
\beq\label{bifspecializedAmSIMPLE}
\mu^{(a)}_{bif}= v \; ,\qquad a=1, \ldots, m-1\; ,
\eeq
and we further set the 5d Coulomb moduli to be:
\begin{align}
\label{specializeAmSIMPLE}
e^{(a)}_{1}=x_{2}\,t^{D^{(a)}}v^{-m-1+a} \; ,\qquad a=1, \ldots, m \; .
\end{align}
Using the definition of the function \eqref{nekrasovN}, it follows that the 5d partition function is now a sum over truncated partitions, of length at most $D^{(a)}$. From this point on, we denote the 3d Coulomb parameters as $y^{(a)}_i$, for all $a=1, \ldots, m$ and all $i=1, \ldots, D^{(a)}$.\\

The 3d $\cN=2$ theory $G^{3d}$ is defined on $S^1(\widehat{R})\times\mathbb{C}$, and lives on the vortices of $T^{5d}$: it is a quiver gauge theory with gauge group $\prod_{a=1}^m U(D^{(a)})$, and with 1 chiral and 1 antichiral matter multiplets on the first node, as pictured in figure \ref{fig:AmSimple5d}.\\

The analysis follows closely the one we already carried out in the previous example. The only new physics is in the bulk 3d $\cN=2$ matter content:
\begin{align}\label{matter3dAmSIMPLE}
I^{(1)}_{bulk, matter}(y^{(1)}, \{x_d\})=\prod_{i=1}^{D^{(1)}} \frac{\left(v\, x_{1}/y^{(a)}_{i};q\right)_{\infty}}{\left(v^{-m}\, x_{2}/y^{(1)}_{i};q\right)_{\infty}} \; ,
\end{align}

and in the Wilson loop factor
\beq\label{3dWilsonfactor2AmSIMPLE}
{\widetilde{Y}^{(a)}_{D1/D5}}(\{x_{d}\}, z)=\left(1- v\, x_{2}/z\right)\; .
\eeq

All in all, the 3d partition function with a single D1 brane wrapping $S^*_1$ can be written as a twisted $qq$-character of the first fundamental representation of $U_q(\widehat{A_m})$, and has a closed form expression:
\begin{align}
&\left[\widetilde{\chi}^{A_m}\right]_{(1,0,\ldots,0)}^{3d}(z) =\left(1- v\, x_{2}/z\right)\left\langle\widetilde{Y}^{(1)}_{D1/D3}(z)  \right\rangle\nonumber\\
&\qquad\qquad + \widetilde{\fq}^{(1)} \, \left(1- v\, x_{1}/z\right) \left\langle\frac{\widetilde{Y}^{(2)}_{D1/D3}(z\, v^{-1})}{\widetilde{Y}^{(1)}_{D1/D3}(z\, v^{-2})}\right\rangle\nonumber\\
&\qquad\qquad + \widetilde{\fq}^{(1)}\widetilde{\fq}^{(2)} \,\left(1- v\, x_{1}/z\right)  \left\langle\frac{\widetilde{Y}^{(3)}_{D1/D3}(z\, v^{-2})}{\widetilde{Y}^{(2)}_{D1/D3}(z\, v^{-3})}\right\rangle\nonumber\\
&\qquad\qquad+\ldots\nonumber\\
&\qquad\qquad+\prod_{a=1}^m \widetilde{\fq}^{(a)}\,\left(1- v\, x_{1}/z\right) \left\langle\frac{1}{\widetilde{Y}^{(m)}_{D1/D3}(z\, v^{-m-1})}\right\rangle\; .\label{3dpartitionfunctionAmSIMPLE}
\end{align}
As described in the main text, this can be proved in a straightforward way by Higgsing the 5d theory through \eqref{specializeAmSIMPLE}.
The equality of the partition functions follows:
\begin{align}\label{CHIequalityAmSIMPLE}
\left[\widetilde{\chi}^{A_m}\right]_{(1,0,\ldots,0)}^{3d}(z)=c_{3d}\;  \left[\chi^{A_m}\right]_{(1,0,\ldots,0)}^{5d}(z)_{e^{(a)}_{1}=\; \eqref{specializeAmSIMPLE}} \; .
\end{align}

We now reinterpret the physics in terms of $A_m$-type $q$-Toda theory on the cylinder $\cC$. We only need to write down the new D5 brane vertex operator, since all the other  operators are identical to the ones in the previous example. Recall that the D5 brane vertex operator is fully determined by the set \eqref{AmweightSIMPLE} of weights $\{\omega_d\}$. Namely, we first define
\begin{align}
V_{\omega_1} (x_1)&=\; :\left[\Lambda^{(1)}(x_1)\right]^{-1}:\nonumber \\
V_{\omega_2} (x_2)&=\; :\Lambda^{(1)}(x_2\, v^{-m-1}):\; ,\label{AmvertexweightSIMPLE}
\end{align}
where $\Lambda^{(a)}$ is the $a$-th fundamental weight operator \eqref{coweightvertexdef}. Alternatively, we can define the D5 brane operator using the simple root operators \eqref{rootvertexdef}. The $v$-shifts in the argument of these operators are uniquely determined from the truncation locus \eqref{specializeAmSIMPLE}:
\begin{align}
V_{\omega_1} (x_1)&=\; :\left[\Lambda^{(1)}(x_1)\right]^{-1}: \nonumber\\
V_{\omega_2} (x_2)&=\; :\left[\Lambda^{(m)}(x_{2})\right]^{-1}\prod_{a=1}^m E^{(a)}(x_{2}\,v^{-m-1+a}):\label{AmvertexrootSIMPLE}
\end{align}
Either way, we are now able to construct the D5 brane operator as:
\beq\label{AmgenericvertexexmapleSIMPLE}
V_\eta (u)= \, :\prod_{d=1}^{N_f=2} V_{\omega_d}(x_d): \; .
\eeq
The parameters $u$ and $\eta$ are defined through
\begin{align}\label{AmcommomentumSIMPLE}
&\eta = \sum_{d=1}^{2}\eta_d\, \omega_d \; ,\\
&x_d = u \, q^{-\eta_d}\; ,\qquad\;\; d=1,2\; .
\end{align}

Computing the same two-points as in the previous examples, we find that the following ${\cW}_{q,t}(A_m)$ correlator is a 3d gauge theory partition function:
\beq\label{Amcorrelatoris3dindexSIMPLE}
\left\langle \psi'\left|\prod_{d=1}^{2} V_{\omega_d}(x_d)\; \prod_{a=1}^{m} (Q^{(a)})^{D^{(a)}}\; W^{(2)}(z) \right| \psi \right\rangle = A(\{x_d\})\, B(\{x_d\}, z)\, \left[\widetilde{\chi}^{A_m}\right]_{(1,0,\ldots,0)}^{3d}(z)\; .
\eeq
Furthermore, by our usual argument, the 5d partition function is also equal to this correlator.

\vspace{10mm}

\subsection{Wilson Loop for $G_2$ Theories}
\label{ssec:exampleG2}

\vspace{4mm}

------- 5d Gauge Theory -------\\

Let $X$ be a resolved $D_4$ singularity, and consider type IIB string theory on $X\times\cC\times\mathbb{C}^2$. We consider a nontrivial fibration of $X$ over $\mathbb{C}^2 \times \cC$, where as we go around the origin of one of the lines wrapped by the branes (say,  $\mathbb{C}_q$), the space $X$ goes back to itself, up to $\mathbb{Z}_3$ outer automorphism group action. The $\mathbb{Z}_3$ action will permute the compact 2-cycles representing the ``outer nodes" of the $D_4$ Dynkin diagram, denoted as 1, 3, and 4. For $a=1,\ldots,4$, we introduce $n^{(a)}$ D5 branes wrapping  $S_a^2$ and $\mathbb{C}^2$, such that $n^{(1)}=n^{(3)}=n^{(4)}$. We further introduce ${N^{(a)}_f}$ D5 branes wrapping the dual non-compact 2-cycle $S^*_a$ and $\mathbb{C}^2$, such that $N_f^{(1)}=N_f^{(3)}=N_f^{(4)}$. Finally, we add $L^{(a)}$ D1 branes wrapping $S^*_a$, such that $L^{(1)}=L^{(3)}=L^{(4)}$.

We use the notation $n=n^{(1)}+n^{(2)}$,\;\; $N_f=N_f^{(1)}+N_f^{(2)}$ and $L=L^{(1)}+L^{(2)}$. We set $g_s\rightarrow 0$, which amounts to studying the $(2,0)$ $D_4$ little string on $\cC\times\mathbb{C}^2$ in the presence of $G_2$-type codimension 2 defects (the D5 branes) and point-like defects (the D1 branes). At energies below the string scale, the dynamics in this background are fully captured by the theory on the D5 branes, with D1 brane defects. Because of the nontrivial fibration in the geometry, the theory on the D5 branes is a 5d quiver gauge theory of shape the Dynkin diagram of $G_2$. We call it $T^{5d}$, and it is defined on $S^1(\widehat{R})\times\mathbb{C}^2$, with gauge content  $\prod_{a=1}^{2}U(n^{(a)})$ and flavor content $\prod_{a=1}^{2}U({N^{(a)}_f})$. The D1 branes make up a 1/2-BPS  Wilson loop wrapping the circle and at the origin of $\mathbb{C}^2$.

Then, the second simple root of $D_4$ is in its own orbit under the $\mathbb{Z}_3$-action, so the node 2 of the $G_2$ Dynkin diagram denotes the long simple root. It follows that the node 1 denotes the short simple root.
After putting the theory on $\Omega$-background, the partition function of $T^{5d}$ with Wilson loop is the Witten index \eqref{5dintegral}:
{\allowdisplaybreaks
\begin{align}
\label{5dintegralG2}
&\left[\chi^{G_2}\right]_{(L^{(1)}, L^{(2)})}^{5d}  =\sum_{k^{(1)}, k^{(2)}=0}^{\infty}\;\prod_{a=1}^{2}\frac{\left(\widetilde{\fq}^{(a)}{}\right)^{k^{(a)}}}{k^{(a)}!} \, \oint  \left[\frac{d\phi^{{(a)}}_I}{2\pi i}\right]Z^{(a)}_{vec}\cdot Z^{(a)}_{fund}\cdot Z^{(a)}_{CS} \cdot Z^{(1,2)}_{bif}\cdot \prod_{\rho=1}^{L^{(a)}}  Z^{(a)}_{D1}  \; , \\
&Z^{(1)}_{vec} =\prod_{I, J=1}^{k^{(1)}} \frac{{\widehat{f}}\left(\phi^{(1)}_{I}-\phi^{(1)}_{J}\right)f\left(\phi^{(1)}_{I}-\phi^{(1)}_{J}+ 2\,\epsilon_+ \right)}{f\left(\phi^{(1)}_{I}-\phi^{(1)}_{J}+\epsilon_1 \right)f\left(\phi^{(1)}_{I}-\phi^{(1)}_{J}+\epsilon_2\right)}\nonumber\\
&\qquad\times\prod_{I=1}^{k^{(1)}} \prod_{i=1}^{n^{(1)}} \frac{1}{f\left(\phi^{(1)}_I-a^{(1)}_i+\epsilon_+\right)f\left(\phi^{(1)}_I-a^{(1)}_i-\epsilon_+ \right)}\, ,\\
&Z^{(2)}_{vec} =\prod_{I, J=1}^{k^{(2)}} \frac{{\widehat{f}}\left(\phi^{(2)}_{I}-\phi^{(2)}_{J}\right)f\left(\phi^{(2)}_{I}-\phi^{(2)}_{J}+ 2\,\epsilon_+ + 2\,\epsilon_1\right)}{f\left(\phi^{(2)}_{I}-\phi^{(2)}_{J}+\epsilon_1 + 2\,\epsilon_1\right)f\left(\phi^{(2)}_{I}-\phi^{(2)}_{J}+\epsilon_2\right)}\nonumber\\
&\qquad\times\prod_{I=1}^{k^{(2)}} \prod_{i=1}^{n^{(2)}} \frac{1}{f\left(\phi^{(2)}_I-a^{(2)}_i+\epsilon_+ + \epsilon_1\right)f\left(\phi^{(2)}_I-a^{(2)}_i-\epsilon_+ - \epsilon_1\right)}\, ,\\
&Z^{(a)}_{fund} =\prod_{I=1}^{k^{(a)}} \prod_{d=1}^{N^{(a)}_f} f\left(\phi^{{(a)}}_I-m^{{(a)}}_d\right) \equiv\prod_{I=1}^{k^{(a)}} Q^{(a)}(\phi^{{(a)}}_I)\, ,\\
&Z^{(a)}_{CS}= \prod_{I=1}^{k^{(a)}} e^{k^{(a)}_{CS}\, \phi^{(a)}_I}\\
&Z^{(1, 2)}_{bif} =\prod_{J=1}^{k^{(2)}}\prod_{i=1}^{n^{(1)}} \prod_{p=0}^{2} f\left(\phi^{(2)}_J- m_{bif} - a^{(1)}_i - \epsilon_+ - (1 - p)\,\epsilon_1 \right)\nonumber\\
&\qquad \times\prod_{I=1}^{k^{(1)}}\prod_{j=1}^{n^{(2)}}  f\left(\phi^{(1)}_I+ m_{bif} - a^{(2)}_j + \epsilon_+  \right)\nonumber\\
&\qquad \times \prod_{I=1}^{k^{(1)}}\prod_{J=1}^{k^{(2)}} \frac{f\left(\phi^{(1)}_{I}-\phi^{(2)}_{J}+ m_{bif}+2\epsilon_1 \right)}{f\left(\phi^{(1)}_{I}-\phi^{(2)}_{J} +m_{bif} -\epsilon_1\right)}\nonumber\\
&\qquad \times \prod_{I=1}^{k^{(1)}}\prod_{J=1}^{k^{(2)}} \frac{f\left(\phi^{(1)}_{I}-\phi^{(2)}_{J}+ m_{bif}+\epsilon_2  - \epsilon_1  \right)}{f\left(\phi^{(1)}_{I}-\phi^{(2)}_{J}+ m_{bif}+2\, \epsilon_+ + \epsilon_1  \right)}\, ,\\
&Z^{(1)}_{D1} =
\left[{\prod_{i=1}^{n^{(1)}}} f\left(a^{{(1)}}_i-M^{{(1)}}_\rho\right) {\prod_{I=1}^{k^{(1)}}} \frac{f\left(\phi^{{(1)}}_I-M^{(1)}_\rho+ \epsilon_- \right)f\left(\phi^{{(1)}}_I-M^{(1)}_\rho- \epsilon_- \right)}{f\left(\phi^{{(1)}}_I-M^{(1)}_\rho+ \epsilon_+ \right)f\left(\phi^{{(1)}}_I-M^{(1)}_\rho- \epsilon_+ \right)}\right]\\
&Z^{(2)}_{D1} =\left[{\prod_{i=1}^{n^{(2)}}}  f\left(a^{{(2)}}_i-M^{{(2)}}_\rho\right)  {\prod_{I=1}^{k^{(2)}}} \frac{f\left(\phi^{{(2)}}_I-M^{(2)}_\rho+ \epsilon_- + \epsilon_1\right)f\left(\phi^{{(2)}}_I-M^{(2)}_\rho- \epsilon_- - \epsilon_1\right)}{f\left(\phi^{{(2)}}_I-M^{(2)}_\rho+ \epsilon_+ + \epsilon_1\right)f\left(\phi^{{(2)}}_I-M^{(2)}_\rho- \epsilon_+ - \epsilon_1\right)}\right]\, .
\end{align}}

Note the extra factors in the bifundamental contribution, which manifests themselves as extra Fermi multiplets in the quantum mechanics.

Having written the integrand, one needs to also specify the contours for the integration variables $\phi^{{(a)}}_I$; in the absence of the Wilson loop factor $Z^{(a)}_{D1}$, the poles are classified by $n^{(a)}$-tuples of Young diagrams $\overrightarrow{\boldsymbol{\mu}^{(a)}}=\{\boldsymbol{\mu}^{(a)}_1, \boldsymbol{\mu}^{(a)}_2, \ldots, \boldsymbol{\mu}^{(a)}_{n^{(a)}}\}$. Namely, 
\begin{align}
\label{G2youngtuples}
&\phi^{(a)}_I = a^{(a)}_i +\epsilon_+ -  s_1\, \epsilon_1 - s_2\, \epsilon_2 ,\;\; \text{with}\; (s_1, s_2)\in \boldsymbol{\mu}^{(a)}_i \;\;\; ,\; a=1\; ,\\
&\phi^{(a)}_I = a^{(a)}_i +\epsilon_+ -\epsilon_1 - 3\, s_1\, \epsilon_1 - s_2\, \epsilon_2 , \;\;\text{with}\; (s_1, s_2)\in \boldsymbol{\mu}^{(a)}_i \;\;\; ,\; a=2\; .
\end{align}
After including $Z^{(a)}_{D1}$ in the integrand, there are additional poles to consider, dictated by the JK residue prescription:
\begin{align}
&\phi^{(a)}_I=M^{(a)}_\rho+\epsilon_+ +\epsilon_1 \label{G2newpole1}\\
&\phi^{(a+1)}_J=\phi^{(a)}_I + m_{bif} +2\, \epsilon_+ + \epsilon_1  \label{G2newpole2}\\
&\phi^{(a-1)}_J=\phi^{(a)}_I - m_{bif} + \epsilon_1\label{G2newpole3}
\end{align}
As a consequence, some of the residues will now explicitly depend on the bifundamental mass, and also necessarily on one of the fermion masses $M^{(a)}_\rho$. 
For definiteness, let us  focus on a single D1 brane  wrapping the 2-cycle $S^*_1$, along with a single D1 brane wrapping $S^*_3$ and a single D1 brane wrapping $S^*_4$. Requiring that this brane configuration is left invariant under the $\mathbb{Z}_3$ outer automorphism action, we denote the partition function as $\left[\chi^{G_2}\right]_{(L^{(1)}, L^{(2)})}^{5d}=\left[\chi^{G_2}\right]_{(1,0)}^{5d}$, where the integers $L^{(a)}$ are understood as measuring a D1 brane charge in the coweight lattice of $G_2$. We write the fermion mass $M^{(1)}_1\equiv M$ to simplify the notation.

Then, by the JK residue prescription, there are 6 extra poles to be enclosed by the contours; the rest of the poles should be labeled by Young diagrams \eqref{G2youngtuples}. Explicitly, the six new poles to enclose are at:
\begin{align}
&\phi^{(1)}_1=M+\frac{\epsilon_1+\epsilon_2}{2}\nonumber\\
&\phi^{(2)}_1=M+\frac{5\,\epsilon_1+3\,\epsilon_2}{2}+m\nonumber\\
&\phi^{(1)}_2=M+\frac{7\,\epsilon_1+3\,\epsilon_2}{2}\nonumber\\
&\phi^{(1)}_3=M+\frac{5\,\epsilon_1+3\,\epsilon_2}{2}\nonumber\\
&\phi^{(2)}_2=M+\frac{9\,\epsilon_1+5\,\epsilon_2}{2}+m\nonumber\\
&\phi^{(1)}_4=M+\frac{11\,\epsilon_1+5\,\epsilon_2}{2}
\end{align}
Our labeling specifies the contour prescription, which in turn fully specifies the partition function $\left[\chi^{G_2}\right]_{(1,0)}^{5d}$. We can then evaluate it.\\

We are now able to make contact with the representation theory of quantum affine algebras. First, we switch to K-theoretic fugacities, such as  $z=e^{-\widehat{R}M}$.
As usual, we introduce a defect operator expectation value:
\begin{align}
&\left\langle \left[Y^{(a)}_{5d}(z)\right]^{\pm 1}  \right\rangle  \equiv\sum_{k^{(1)}, k^{(2)}=0}^{\infty}\frac{\left(\widetilde{\fq}^{(1)}{}\right)^{k^{(1)}}\left(\widetilde{\fq}^{(2)}{}\right)^{k^{(2)}}}{k_1!\; k_2!} \nonumber\\
&\qquad\qquad\qquad\;\;\;\times \oint_{\{\overrightarrow{\boldsymbol{\mu}}\}} \,\prod_{b=1}^{2} \left[\frac{d\phi^{{(b)}}_I}{2\pi i}\right]{Z^{(b)}_{vec}}\cdot Z^{(b)}_{fund} \cdot Z^{(b)}_{CS} \cdot \left[Z^{(a)}_{D1}(z)\right]^{\pm 1} \cdot Z^{(1,2)}_{bif} \, .\label{YopG2}
\end{align}
It is understood in the above definition  that the contours enclose poles coming from Young diagrams only, labeled by \eqref{G2youngtuples}. As we have just seen, that is not the correct pole prescription to evaluate the partition function, since there are 6 extra poles to enclose. Nonetheless,  we can write the partition function as a Laurent polynomial in the $Y^{(a)}_{5d}$ operator vevs, as a sum of exactly 7 terms. The partition function then makes up a twisted $qq$-character of the first fundamental representation of $U_q(\widehat{G_2})$, with highest weight $[1,0]$. 
The formulas get involved quickly, so we introduce the following notation: $Y^{(a)}_{5d}(z)\equiv Y^{(a)}_z$, $Q^{(a)}(z)\equiv Q^{(a)}_z$, and $\mu_{bif}\equiv \mu$. The 5d partition function with Wilson loop evaluates to:
\begin{align}
&\left[\chi^{G_2}\right]_{(1,0)}^{5d}(z) =\left\langle Y^{(1)}_z \right\rangle\nonumber\\
&\;\;\; + \widetilde{\fq}^{(1)} \; Q^{(1)}_{z\, \sqrt{t/q}} \left\langle\frac{Y^{(2)}_{z\, \mu\, t/q}}{Y^{(1)}_{z\, t/q}}\right\rangle\nonumber\\
&\;\;\;+ \widetilde{\fq}^{(1)}\widetilde{\fq}^{(2)} \;  Q^{(1)}_{z\, \sqrt{t/q}}\;  Q^{(2)}_{z\, \mu\, \sqrt{t^3/q^5}}\left\langle\frac{Y^{(1)}_{z\, t/q^2}\,Y^{(1)}_{z\, t/q^3}}{Y^{(2)}_{z\,\mu\, t^2/q^4}}\right\rangle\nonumber\\
&\;\;\; + c_{[0,0]}\left[\widetilde{\fq}^{(1)}\right]^2\widetilde{\fq}^{(2)} \;  Q^{(1)}_{z\, \sqrt{t/q}}\;  Q^{(2)}_{z\, \mu\, \sqrt{t^3/q^5}}\; Q^{(1)}_{z\, \sqrt{t^3/q^7}}\left\langle\frac{Y^{(1)}_{z\, t/q^2}}{Y^{(1)}_{z\, t^2/q^4}}\right\rangle\nonumber\\
&\;\;\; + \left[\widetilde{\fq}^{(1)}\right]^3\widetilde{\fq}^{(2)} \;  Q^{(1)}_{z\, \sqrt{t/q}}\;  Q^{(2)}_{z\, \mu\, \sqrt{t^3/q^5}}\; Q^{(1)}_{z\, \sqrt{t^3/q^7}}\; Q^{(1)}_{z\, \sqrt{t^3/q^5}}\left\langle\frac{Y^{(2)}_{z\,\mu\, t^2/q^3}}{Y^{(1)}_{z\, t^2/q^4}\,Y^{(1)}_{z\, t^2/q^3}}\right\rangle\nonumber\\
&\;\;\; + \left[\widetilde{\fq}^{(1)}\right]^3\left[\widetilde{\fq}^{(2)}\right]^2 \;  Q^{(1)}_{z\, \sqrt{t/q}}\;  Q^{(2)}_{z\, \mu\, \sqrt{t^3/q^5}}\; Q^{(1)}_{z\, \sqrt{t^3/q^7}}\; Q^{(1)}_{z\, \sqrt{t^3/q^5}}\; Q^{(2)}_{z\, \mu\, \sqrt{t^5/q^9}}\left\langle\frac{Y^{(1)}_{z\, t^2/q^5}}{Y^{(2)}_{z\,\mu\, t^3/q^6}}\right\rangle\nonumber\\
&\;\;\; + \left[\widetilde{\fq}^{(1)}\right]^4\left[\widetilde{\fq}^{(2)}\right]^2 \;  Q^{(1)}_{z\, \sqrt{t/q}}\;  Q^{(2)}_{z\, \mu\, \sqrt{t^3/q^5}}\; Q^{(1)}_{z\, \sqrt{t^3/q^7}}\; Q^{(1)}_{z\, \sqrt{t^3/q^5}}\; Q^{(2)}_{z\, \mu\, \sqrt{t^5/q^9}}\; Q^{(1)}_{z\, \sqrt{t^5/q^11}}\left\langle\frac{1}{Y^{(1)}_{z\, t^3/q^6}}\right\rangle\label{G2result2}
\end{align} 
Note that the locus of the extra poles is precisely the argument of the fundamental hypermultiplet factors $Q^{(a)}$. We also take note of a prefactor $c_{[0,0]}$ in the fourth term. As advertised in our general discussion \eqref{character5d}, such factors only depend on the $\Omega$-background spacetime equivariant parameters $q$ and $t$. We find in the case at hand that
\beq\label{ccoefG2}
c_{[0,0]}=\frac{\left(q^2-q^{-2}\right)\left(q/t-t/q\right)}{\left(q-q^{-1}\right)\left(q^3/t-t/q^{3}\right)} \; .
\eeq
For completeness, let us also derive from the JK prescription the second fundamental $qq$-character of $U_q(\widehat{G_2})$ (with highest weight $[0, 1]$), corresponding to the D1 brane charge assignment and $L=L^{(2)}=1$. There are now 14 new poles to be enclosed by the contours. Again, these poles can be read off from the argument of the hypermultiplets $Q^{(a)}$:
{\allowdisplaybreaks
\begin{align}
&\left[\chi^{G_2}\right]_{(0,1)}^{5d}(z) =\left\langle Y^{(2)}_z \right\rangle\nonumber\\
&\;\;\; + \widetilde{\fq}^{(2)} \; Q^{(2)}_{z\, \sqrt{t/q^3}} \left\langle\frac{Y^{(1)}_{z\, \mu^{-1}}\, Y^{(1)}_{z\, \mu^{-1}/q}\, Y^{(1)}_{z\, \mu^{-1}/q^2}}{Y^{(2)}_{z\, t/q^3}}\right\rangle\nonumber\\
&\;\;\;+ c_{[1,0]}\, \widetilde{\fq}^{(2)} \widetilde{\fq}^{(1)} \; Q^{(2)}_{z\, \sqrt{t/q^3}}\, Q^{(1)}_{z\, \mu^{-1}\sqrt{t/q^5}} \left\langle\frac{Y^{(1)}_{z\, \mu^{-1}}\, Y^{(1)}_{z\, \mu^{-1}/q}}{Y^{(1)}_{z\, \mu^{-1}t/q^3}}\right\rangle\nonumber\\
&\;\;\;+ c_{[-1,1]}\, \widetilde{\fq}^{(2)} \left[\widetilde{\fq}^{(1)}\right]^2 \; Q^{(2)}_{z\, \sqrt{t/q^3}}\, Q^{(1)}_{z\, \mu^{-1}\sqrt{t/q^5}}\, Q^{(1)}_{z\, \mu^{-1}\sqrt{t/q^3}} \left\langle\frac{Y^{(1)}_{z\, \mu^{-1}}\,Y^{(2)}_{z\, t/q^2}}{Y^{(1)}_{z\, \mu^{-1}t/q^3}\,Y^{(1)}_{z\, \mu^{-1}t/q^2}}\right\rangle\nonumber\\
&\;\;\;+ c_{[2,-1]} \left[\widetilde{\fq}^{(2)}\right]^2 \left[\widetilde{\fq}^{(1)}\right]^2 \; Q^{(2)}_{z\, \sqrt{t/q^3}}\, Q^{(1)}_{z\, \mu^{-1}\sqrt{t/q^5}}\, Q^{(1)}_{z\, \mu^{-1}\sqrt{t/q^3}}\, Q^{(2)}_{z\, \sqrt{t^3/q^7}} \left\langle\frac{Y^{(1)}_{z\, \mu^{-1}}\,Y^{(1)}_{z\,\mu^{-1} t/q^4}}{Y^{(2)}_{z\, t^2/q^5}}\right\rangle\nonumber\\
&\;\;\;+ \widetilde{\fq}^{(2)}\left[\widetilde{\fq}^{(1)}\right]^3 \; Q^{(2)}_{z\, \sqrt{t/q^3}}\, Q^{(1)}_{z\, \mu^{-1}\sqrt{t/q^5}}\, Q^{(1)}_{z\, \mu^{-1}\sqrt{t/q^3}}\, Q^{(1)}_{z\,\mu^{-1} \sqrt{t/q}} \left\langle\frac{Y^{(2)}_{z\, t/q}\,Y^{(2)}_{z\, t/q^2}}{Y^{(1)}_{z\,\mu^{-1} t/q}\, Y^{(1)}_{z\,\mu^{-1} t/q^2}\, Y^{(1)}_{z\,\mu^{-1} t/q^3}}\right\rangle\nonumber\\
&\;\;\;+ c_{[0,0]} \left[\widetilde{\fq}^{(2)}\right]^2 \left[\widetilde{\fq}^{(1)}\right]^3 \; Q^{(2)}_{z\, \sqrt{t/q^3}}\, Q^{(1)}_{z\, \mu^{-1}\sqrt{t/q^5}}\, Q^{(1)}_{z\, \mu^{-1}\sqrt{t/q^3}}\, Q^{(2)}_{z\, \sqrt{t^3/q^7}}\,Q^{(1)}_{z\, \mu^{-1}\sqrt{t^3/q^9}} \left\langle\frac{Y^{(1)}_{z\, \mu^{-1}}}{Y^{(1)}_{z\,\mu^{-1} t^2/q^5}}\right\rangle\nonumber\\
&\;\;\;+ c_{[0,0]'} \left[\widetilde{\fq}^{(2)}\right]^2 \left[\widetilde{\fq}^{(1)}\right]^3 \; Q^{(2)}_{z\, \sqrt{t/q^3}}\, Q^{(1)}_{z\, \mu^{-1}\sqrt{t/q^5}}\, Q^{(1)}_{z\, \mu^{-1}\sqrt{t/q^3}}\, Q^{(2)}_{z\, \sqrt{t^3/q^7}}\,Q^{(1)}_{z\, \mu^{-1}\sqrt{t/q}} \left\langle\frac{Y^{(1)}_{z\, \mu^{-1}t/q^4}\,Y^{(2)}_{z\, t/q}}{Y^{(1)}_{z\,\mu^{-1} t/q}\,Y^{(2)}_{z\, t^2/q^5}}\right\rangle\nonumber\\
&\;\;\;+ c_{[0,0]''} \left[\widetilde{\fq}^{(2)}\right]^2 \left[\widetilde{\fq}^{(1)}\right]^3 \; Q^{(2)}_{z\, \sqrt{t/q^3}}\, Q^{(1)}_{z\, \mu^{-1}\sqrt{t/q^5}}\, Q^{(1)}_{z\, \mu^{-1}\sqrt{t/q^3}}\,Q^{(1)}_{z\, \mu^{-1}\sqrt{t/q}}\, Q^{(2)}_{z\, \sqrt{t^3/q^5}} \left\langle\frac{Y^{(2)}_{z\, t/q^2}}{Y^{(2)}_{z\, t^2/q^4}}\right\rangle\nonumber\\
&\;\;\;+\left[\widetilde{\fq}^{(2)}\right]^3 \left[\widetilde{\fq}^{(1)}\right]^3 \; Q^{(2)}_{z\, \sqrt{t/q^3}}\, Q^{(1)}_{z\, \mu^{-1}\sqrt{t/q^5}}\, Q^{(1)}_{z\, \mu^{-1}\sqrt{t/q^3}}\;\times\nonumber\\
&\qquad\qquad \times Q^{(2)}_{z\, \sqrt{t^3/q^7}}\,Q^{(1)}_{z\, \mu^{-1}\sqrt{t/q}}\, Q^{(2)}_{z\, \sqrt{t^3/q^5}} \left\langle\frac{Y^{(1)}_{z\, \mu^{-1}t/q^4}\,Y^{(1)}_{z\, \mu^{-1}t/q^3}\, Y^{(1)}_{z\, \mu^{-1}t/q^2}}{Y^{(2)}_{z\, t^2/q^5}\, Y^{(2)}_{z\, t^2/q^4}}\right\rangle\nonumber\\
&\;\;\;+ c_{[-2,1]} \left[\widetilde{\fq}^{(2)}\right]^2 \left[\widetilde{\fq}^{(1)}\right]^4 \; Q^{(2)}_{z\, \sqrt{t/q^3}}\, Q^{(1)}_{z\, \mu^{-1}\sqrt{t/q^5}}\, Q^{(1)}_{z\, \mu^{-1}\sqrt{t/q^3}}\;\times\nonumber\\
&\qquad\qquad \times Q^{(2)}_{z\, \sqrt{t^3/q^7}}\,Q^{(1)}_{z\, \mu^{-1}\sqrt{t/q}}\, Q^{(1)}_{z\, \mu^{-1}\sqrt{t^3/q^9}} \left\langle\frac{Y^{(2)}_{z\, t/q}}{Y^{(1)}_{z\,\mu^{-1} t/q}\,Y^{(1)}_{z\,\mu^{-1} t^2/q^5}}\right\rangle\nonumber\\
&\;\;\;+ c_{[1,-1]} \left[\widetilde{\fq}^{(2)}\right]^3 \left[\widetilde{\fq}^{(1)}\right]^4 \; Q^{(2)}_{z\, \sqrt{t/q^3}}\, Q^{(1)}_{z\, \mu^{-1}\sqrt{t/q^5}}\, Q^{(1)}_{z\, \mu^{-1}\sqrt{t/q^3}}\,Q^{(2)}_{z\, \sqrt{t^3/q^7}}\,Q^{(1)}_{z\, \mu^{-1}\sqrt{t/q}}\;\times\nonumber\\ 
&\qquad\qquad\times Q^{(1)}_{z\, \mu^{-1}\sqrt{t^3/q^9}}\, Q^{(2)}_{z\, \sqrt{t^5/q^3}} \left\langle\frac{Y^{(1)}_{z\,\mu^{-1} t/q^3}\, Y^{(1)}_{z\,\mu^{-1} t/q^2}}{Y^{(1)}_{z\,\mu^{-1} t^2/q^5}\,Y^{(2)}_{z\, t^2/q^4}}\right\rangle\nonumber\\
&\;\;\;+ c_{[-1,0]} \left[\widetilde{\fq}^{(2)}\right]^3 \left[\widetilde{\fq}^{(1)}\right]^5 \; Q^{(2)}_{z\, \sqrt{t/q^3}}\, Q^{(1)}_{z\, \mu^{-1}\sqrt{t/q^5}}\, Q^{(1)}_{z\, \mu^{-1}\sqrt{t/q^3}}\,Q^{(2)}_{z\, \sqrt{t^3/q^7}}\,Q^{(1)}_{z\, \mu^{-1}\sqrt{t/q}}\;\times \nonumber\\
&\qquad\qquad\times Q^{(1)}_{z\, \mu^{-1}\sqrt{t^3/q^9}}\, Q^{(2)}_{z\, \sqrt{t^5/q^3}}\, Q^{(1)}_{z\, \mu^{-1}\sqrt{t^3/q^7}} \left\langle\frac{Y^{(1)}_{z\,\mu^{-1} t/q^2}}{Y^{(1)}_{z\,\mu^{-1} t^2/q^5}\,Y^{(1)}_{z\,\mu^{-1} t^2/q^4}}\right\rangle\nonumber\\
&\;\;\;+ \left[\widetilde{\fq}^{(2)}\right]^3 \left[\widetilde{\fq}^{(1)}\right]^6 \; Q^{(2)}_{z\, \sqrt{t/q^3}}\, Q^{(1)}_{z\, \mu^{-1}\sqrt{t/q^5}}\, Q^{(1)}_{z\, \mu^{-1}\sqrt{t/q^3}}\,Q^{(2)}_{z\, \sqrt{t^3/q^7}}\,Q^{(1)}_{z\, \mu^{-1}\sqrt{t/q}}\;\times\nonumber\\ 
&\qquad\qquad\times Q^{(1)}_{z\, \mu^{-1}\sqrt{t^3/q^9}}\, Q^{(2)}_{z\, \sqrt{t^5/q^3}}\, Q^{(1)}_{z\, \mu^{-1}\sqrt{t^3/q^7}}\,Q^{(1)}_{z\, \mu^{-1}\sqrt{t^3/q^5}} \left\langle\frac{Y^{(2)}_{z\, t^3/q^2}}{Y^{(1)}_{z\,\mu^{-1} t^2/q^5}\,Y^{(1)}_{z\,\mu^{-1} t^2/q^4}\, Y^{(1)}_{z\,\mu^{-1} t^2/q^3}}\right\rangle\nonumber\\
&\;\;\;+ \left[\widetilde{\fq}^{(2)}\right]^4 \left[\widetilde{\fq}^{(1)}\right]^6 \; Q^{(2)}_{z\, \sqrt{t/q^3}}\, Q^{(1)}_{z\, \mu^{-1}\sqrt{t/q^5}}\, Q^{(1)}_{z\, \mu^{-1}\sqrt{t/q^3}}Q^{(2)}_{z\, \sqrt{t^3/q^7}}\,Q^{(1)}_{z\, \mu^{-1}\sqrt{t/q}}\;\times\nonumber\\
&\qquad\qquad\times Q^{(1)}_{z\, \mu^{-1}\sqrt{t^3/q^9}}\, Q^{(2)}_{z\, \sqrt{t^5/q^3}}\, Q^{(1)}_{z\, \mu^{-1}\sqrt{t^3/q^7}}\,Q^{(1)}_{z\, \mu^{-1}\sqrt{t^3/q^5}}\, Q^{(2)}_{z\, \sqrt{t^5/q^9}} \left\langle\frac{1}{Y^{(2)}_{z\, t^3/q^6}}\right\rangle\label{G2result22}
\end{align}}
In the above, we encountered the following prefactors:
\begin{align}
&c_{[1,0]}=c_{[-1,1]}=c_{[2,-1]}=c_{[-2,1]}=c_{[1,-1]}=c_{[-1,0]}=\frac{\left(q^3-q^{-3}\right)\left(q/t-t/q\right)}{\left(q-q^{-1}\right)\left(q^3/t-t/q^{3}\right)}\nonumber\\
&c_{[0,0]}=\frac{\left(q^3-q^{-3}\right)\left(q/t-t/q\right)\left(q^5/t-t/q^5\right)\left(q^4/t^2-t^2/q^4\right)}{\left(q-q^{-1}\right)\left(q^3/t-t/q^{3}\right)\left(q^4/t-t/q^4\right)\left(q^5/t^2-t^2/q^5\right)}\nonumber\\
&c_{[0,0]'}=\frac{\left(q^4-q^{-4}\right)\left(q/t-t/q\right)}{\left(q-q^{-1}\right)\left(q^4/t-t/q^{4}\right)}\nonumber\\
&c_{[0,0]''}=-\frac{\left(q^2-q^{-2}\right)\left(q\,t-q^{-1}t^{-1}\right)}{\left(q-q^{-1}\right)\left(q^2/t-t/q^{2}\right)}\label{ccoefG22}
\end{align}
We found a total of 15 terms in the $qq$-character. Indeed, the irreducible representation $V(\lambda_2)$ of $U_q(\widetilde{G_2})$ decomposes into irreducible representations of $U_q(G_2)$ as $V(\lambda_2) = \textbf{14}\oplus \textbf{1}$; this means one necessarily needs to add the trivial representation \textbf{1} (an extra ``null coweight") to the \textbf{14} in order to obtain an irreducible representation of  $U_q(\widetilde{G_2})$.

Performing the integrals, one finds:
\beq\label{5ddefectexpression2G2}
\left\langle \left[Y^{(a)}_{5d}(z)\right]^{\pm 1} \right\rangle = \sum_{\{\overrightarrow{\boldsymbol{\mu}}\}}\left[ Z^{5d}_{bulk} \cdot Y^{(a)}_{5d}(z)^{\pm 1}\right]\, .
\eeq
Each sum is over a collection of 2d partitions, one for each $U(1)$ Coulomb parameter:
\beq
\{\overrightarrow{\boldsymbol{\mu}}\}=\{\boldsymbol{\mu}^{(a)}_i\}_{a=1, 2\, ; \,\; i=1,\ldots,n^{(a)}}\; .
\eeq
The factor  $Z^{5d}_{bulk}$ encodes all the 5d bulk physics. 
It is written in terms of the function \eqref{nekrasovN} as done in the main text.

Meanwhile, the Wilson loop factor after integration has the following form:
\beq\label{WilsonfactorG2}
Y^{(a)}_{5d}(z)\equiv\prod_{i=1}^{n^{(a)}}\prod_{k=1}^{\infty}\frac{1-t\, y^{(a)}_{i,k}/z}{1- y^{(a)}_{i,k}/z}\; .
\eeq

Above, we have defined the variables
\beq\label{ydefG2}
y^{(a)}_{i,k} = e^{(a)}_i\, q^{r^{(a)}{\boldsymbol{\mu}}^{(a)}_{i, k}} \, t^{-k}\;\;\;,\; k=1, \ldots, \infty\;,
\eeq 
where  $\boldsymbol{\mu}^{(a)}_{i, k}$ is the length of the $k$-th row of the partition $\boldsymbol{\mu}^{(a)}_{i}$.\\

Then, having performed all the integrals, the 5d partition function with $L^{(1)}=1$ (respectively $L^{(2)}=1$) evaluates to \eqref{G2result2} (respectively  \eqref{G2result22}), where each vev is now understood to be written in terms of the variables \eqref{ydefG2}.\\

\vspace{10mm}

------- 3d Gauge Theory -------\\

We now make contact with the theory $G^{3d}$ on the vortices of $T^{5d}$. We  fix the total number $N_f$ of D5 branes wrapping the non-compact 2-cycles of the $D_4$ singularity. Further imposing the vanishing of D5 brane flux at infinity translates to the condition $[S+S^*]=0$ (now understood as an equation in the coweight lattice of $G_2$), which is equivalent to
\begin{align}\label{conformalG2}
& 2\;n^{(1)}-n^{(2)} = N^{(1)}_f\\
& 2\;n^{(2)}-3\; n^{(1)} = N^{(2)}_f\; .
\end{align}
These equations further specify the ranks $n^{(a)}$ of the 5d gauge groups. The vanishing D5 flux constraint also specifies the Chern-Simons term on node $a$ to be $k^{(a)}_{CS}=n^{(a)}-n^{(a+1)}$.   In terms of the representation theory of $G_2$, each non-compact D5 brane is labeled by a coweight of $G_2$, making up a set $\{\omega_d\}_{d=1}^{N_f}$. As explained in section \ref{ssec:qToda}, such a set needs to satisfy three conditions:
\begin{itemize}
	\item For all $1\leq d \leq N_f$, the coweight $\omega_d$ belongs in a fundamental representation of $^L\fg$.
	\item $\sum_{d=1}^{N_f} \omega_d=0$.
	\item No proper subset of coweights in $\{\omega_d\}$ adds up to $0$. 
\end{itemize}
The classification of such defects was given in \cite{Haouzi:2017vec}. To mix things up a bit, we will focus our attention on a single defect, defining a particular 3d quiver gauge theory $G^{3d}$, but we will compute both fundamental $qq$-characters for it.\\

\begin{figure}[h!]
	\emph{}
	\centering
	\includegraphics[trim={0 0 0 1cm},clip,width=0.9\textwidth]{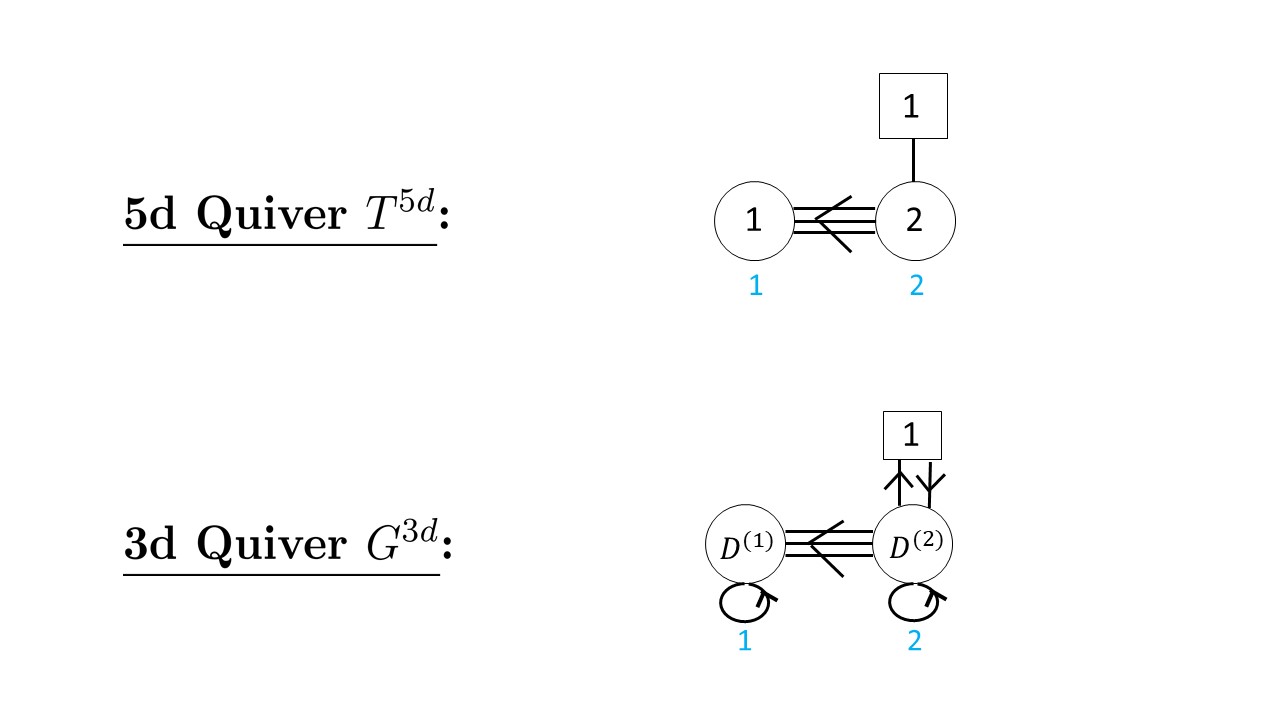}
	\vspace{-10pt}
	\caption{The $G_2$ theory $T^{5d}$ under study, along with the theory $G^{3d}$ on its vortices. The blue numbers label the nodes.} 
	\label{fig:G2Null5d}
\end{figure}

Let us consider the following set $\{\omega_d\}$ of 1 coweight:
\begin{align}\label{G2weightNULL}
\omega &= [ \phantom{-}0, \phantom{-}0]= -\lambda^\vee_2+1\,\alpha^\vee_1+2\, \alpha^\vee_2
\end{align}
$\lambda_a$ is the $a$-th fundamental coweight, and $\alpha_a$ is the $a$-th positive simple coroot. In the geometry, this corresponds to having a single D5 brane wrapping the non-compact 2-cycle $S^*_2$ of the $D_4$ singularity. We label the associated mass by $x$. The specific 5d quiver gauge theory $T^{5d}$ engineered by this coweight is shown in figure \ref{fig:G2Null5d}.\\

We further introduce $D^{(1)}=D^{(3)}=D^{(4)}$ and $D^{(3)}$ D3 branes wrapping the compact 2-cycles and one of the two complex lines, say $\mathbb{C}_q$, for a total of $D=\sum_{a=1}^{4}D^{(a)}$ D3 branes. We require that the D3 branes remain invariant under the $\mathbb{Z}_3$ outer automorphism action, after going around the origin of $\mathbb{C}_q$. 
From now on,  the bifundamental mass is set to    
\beq\label{bifspecializedG2}
\mu_{bif}= v \; ,
\eeq
and we further tune the 5d Coulomb moduli to be
\begin{align}
\label{specializeG2}
&e^{(1)}_{1}=x\,t^{D^{(1)}_1}q^{-1}v^{-2}\\
&e^{(2)}_{1}=x\,t^{D^{(2)}_1}q^{-1}v^{-1}\\
&e^{(2)}_{2}=x\,t^{D^{(2)}_2}q^{-1}v^{-3}
\end{align}
Using the definition of the function \eqref{nekrasovN}, it follows that the 5d partition function is now a sum over truncated partitions, of length at most $D^{(a)}_i$. As usual, we will denote the 3d Coulomb parameters as $y^{(a)}_i$, for all $a=1, 2,$ and $i=1, \ldots, D^{(a)}$.
The 3d $\cN=2$ theory $G^{3d}$ is defined on $S^1(\widehat{R})\times\mathbb{C}$, and lives on the vortices of $T^{5d}$: it is a quiver gauge theory with gauge group $\prod_{a=1}^2 U(D^{(a)})$, with 1 chiral and 1 antichiral matter multiplets on the second node, as pictured in figure \ref{fig:G2Null5d}.

The first step is to define a 3d  Wilson loop operator expectation value, as an integral over the Coulomb moduli of $G^{3d}$, where it is understood that none of the contours below enclose poles depending on $z$:
\begin{align}\label{3ddefectexpression2G2}
\left\langle\left[{\widetilde{Y}^{(a)}_{3d}}\right]^{\pm 1} \right\rangle \equiv \left[{\widetilde{Y}^{(a)}_{D1/D5}}(x, z)\right]\; \oint_{\{\overrightarrow{\boldsymbol{\mu}}\}} d{y}\,\left[I^{3d}_{bulk}(y)\, \left[{\widetilde{Y}^{(a)}_{D1/D3}}(y, z)\right]^{\pm 1}\right]  \; .
\end{align}
The bulk contribution
\begin{align}\label{bulk3dG2}
I^{3d}_{bulk}(y)=\prod_{a=1}^{2}\prod_{l=1}^{D^{(a)}}{y^{(a)}_l}^{\left(\tau^{(a)}-1\right)}\;I^{(a)}_{bulk, vec}\cdot I^{(a)}_{bulk, matter}\cdot I^{(1,2)}_{bulk, bif}\; .
\end{align}
is independent of the Wilson loop physics. It contains the 3d F.I. term
\begin{align}\label{FI3dG2}
\prod_{a=1}^{2}\prod_{l=1}^{D^{(a)}}{y^{(a)}_l}^{\left(\tau^{(a)}\right)}\; ,
\end{align}
the 3d $\cN=4$ vector multiplet contributions
\begin{align}\label{vec3dG2}
&I^{(1)}_{bulk, vec}(y^{(1)})=\prod_{1\leq i\neq j\leq D^{(1)}}\frac{\left(y^{(1)}_{i}/y^{(1)}_{j};q\right)_{\infty}}{\left(t\, y^{(1)}_{i}/y^{(1)}_{j};q\right)_{\infty}}\;\prod_{1\leq i<j\leq D^{(1)}} \frac{\Theta\left(t\,y^{(1)}_{i}/y^{(1)}_{j};q\right)}{\Theta\left(y^{(1)}_{i}/y^{(1)}_{j};q\right)}\\
&I^{(2)}_{bulk, vec}(y^{(2)})=\prod_{1\leq i\neq j\leq D^{(2)}}\frac{\left(y^{(2)}_{i}/y^{(2)}_{j};q^3\right)_{\infty}}{\left(t\, y^{(2)}_{i}/y^{(2)}_{j};q^3\right)_{\infty}}\;\prod_{1\leq i<j\leq D^{(2)}} \frac{\Theta\left(t\,y^{(2)}_{i}/y^{(2)}_{j};q^3\right)}{\Theta\left(y^{(2)}_{i}/y^{(2)}_{j};q^3\right)}\; ,
\end{align}
and the bifundamental hypermultiplets
\begin{align}
\label{bif3dG2}
I^{(1,2)}_{bulk, bif}(y^{(1)}, y^{(2)})=\prod_{1\leq i \leq D^{(1)}}\prod_{1\leq j \leq D^{(2)}}\left [ \frac{(v\, t\, y^{(1)}_{i}/y^{(2)}_{j};q)_{\infty}}{(v \, y^{(1)}_{i}/y^{(2)}_{j};q)_{\infty}}\right]\; .
\end{align}
The $\cN=2$ matter content for the coweight $\omega$ is given by 1 chiral and 1 antichiral multiplets on node 2:
\begin{align}\label{matter3dG2}
I^{(2)}_{bulk, matter}(y^{(2)}, x)=\prod_{i=1}^{D^{(2)}} \frac{\left(\sqrt{t/q}\, x/y^{(2)}_{i};q^3\right)_{\infty}}{\left(\sqrt{t^3/q^5}\, x/y^{(2)}_{i};q^3\right)_{\infty}} \; .
\end{align}
The various powers of $q$ and $t$ in the argument of the $q$-Pochhammer symbols are fixed by the R-symmetry of $G^{3d}$.

The Wilson loop contributions are given by
\beq\label{3dWilsonfactorG2}
{\widetilde{Y}^{(a)}_{D1/D3}}(y^{(a)},z)=\prod_{i=1}^{D^{(a)}}\frac{1-t\, y^{(a)}_{i}/z}{1- y^{(a)}_{i}/z}\; .
\eeq
and by
\begin{align}\label{3dWilsonfactor2G2}
&{\widetilde{Y}^{(1)}_{D1/D5}}(x, z)=\left(1-\sqrt{t^2/q^4}\, x/z\right)\; ,\nonumber\\
&{\widetilde{Y}^{(2)}_{D1/D5}}(x, z)=\left(1-\sqrt{t/q^3}\, x/z\right)\left(1-\sqrt{t^3/q^5}\, x/z\right)\; .
\end{align}

For ease of presentation, we will slightly simplify the notation, as we did in 5d, and write ${\widetilde{Y}^{(a)}_{D1/D3}}(y^{(a)},z)\equiv \widetilde{Y}^{(a)}_z$.
Let us first write down the 3d partition function for the case $L=L^{(1)}=1$. It is a twisted 3d $qq$-character of the first fundamental representation of $U_q(\widehat{G_2})$:
\begin{align}
&\left[\widetilde{\chi}^{G_2}\right]_{(1,0)}^{3d}(z) =\left(1-\sqrt{t^2/q^4}\, x/z\right)\left\langle \widetilde{Y}^{(1)}_z \right\rangle+ \widetilde{\fq}^{(1)} \; \left(1-\sqrt{t^2/q^4}\, x/z\right) \left\langle\frac{\widetilde{Y}^{(2)}_{z\, \sqrt{t/q}}}{\widetilde{Y}^{(1)}_{z\, t/q}}\right\rangle\nonumber\\
&\;\;\;+ \widetilde{\fq}^{(1)}\widetilde{\fq}^{(2)} \;  \left(1-x/z\right)\left\langle\frac{\widetilde{Y}^{(1)}_{z\, t/q^2}\,\widetilde{Y}^{(1)}_{z\, t/q^3}}{\widetilde{Y}^{(2)}_{z\,\sqrt{t^3/q^7}}}\right\rangle
 + c_{[0,0]}\left[\widetilde{\fq}^{(1)}\right]^2\widetilde{\fq}^{(2)} \;  \left(1-x/z\right)\left\langle\frac{\widetilde{Y}^{(1)}_{z\, t/q^2}}{\widetilde{Y}^{(1)}_{z\, t^2/q^4}}\right\rangle\nonumber\\
&\;\;\; + \left[\widetilde{\fq}^{(1)}\right]^3\widetilde{\fq}^{(2)} \;  \left(1-x/z\right)\left\langle\frac{\widetilde{Y}^{(2)}_{z\,\sqrt{t^3/q^5}}}{\widetilde{Y}^{(1)}_{z\, t^2/q^4}\,\widetilde{Y}^{(1)}_{z\, t^2/q^3}}\right\rangle
 + \left[\widetilde{\fq}^{(1)}\right]^3\left[\widetilde{\fq}^{(2)}\right]^2 \;  \left(1-(q^2/t)\, x/z\right)\left\langle\frac{\widetilde{Y}^{(1)}_{z\, t^2/q^5}}{\widetilde{Y}^{(2)}_{z\,\sqrt{t^5/q^{11}}}}\right\rangle\nonumber\\
&\;\;\; + \left[\widetilde{\fq}^{(1)}\right]^4\left[\widetilde{\fq}^{(2)}\right]^2 \;  \left(1-(q^2/t) x/z\right)\left\langle\frac{1}{\widetilde{Y}^{(1)}_{z\, t^3/q^6}}\right\rangle\label{3dpartitionfunctionG2}
\end{align} 

For completeness, we also write down the 3d partition function in the case $L=L^{(2)}=1$. It is a twisted 3d $qq$-character of the second fundamental representation of $U_q(\widehat{G_2})$:

\begin{align}
&\left[\chi^{G_2}\right]_{(0,1)}^{3d}(z) =\left(1-\sqrt{t/q^3}\, x/z\right)\left(1-\sqrt{t^3/q^5}\, x/z\right)\left\langle \widetilde{Y}^{(2)}_z \right\rangle\nonumber\\
&\;\;\; + \widetilde{\fq}^{(2)} \; \left(1-\sqrt{t/q^3}\, x/z\right)\left(1- \sqrt{t/q}\, x/z\right) \left\langle\frac{\widetilde{Y}^{(1)}_{z\, \sqrt{t/q}}\, \widetilde{Y}^{(1)}_{z\, \sqrt{t/q^3}}\, \widetilde{Y}^{(1)}_{z\, \sqrt{t/q^5}}}{\widetilde{Y}^{(2)}_{z\, t/q^3}}\right\rangle\nonumber\\
&\;\;\;+ c_{[1,0]}\, \widetilde{\fq}^{(2)} \widetilde{\fq}^{(1)} \; \left(1-\sqrt{t/q^3}\, x/z\right)\left(1- \sqrt{t/q}\, x/z\right)  \left\langle\frac{\widetilde{Y}^{(1)}_{z\, \sqrt{t/q}}\, \widetilde{Y}^{(1)}_{z\, \sqrt{t/q^3}}}{\widetilde{Y}^{(1)}_{z\, \sqrt{t^3/q^7}}}\right\rangle\nonumber\\
&\;\;\;+ c_{[-1,1]}\, \widetilde{\fq}^{(2)} \left[\widetilde{\fq}^{(1)}\right]^2 \; \left(1-\sqrt{t/q^3}\, x/z\right)\left(1- \sqrt{t/q}\, x/z\right) \left\langle\frac{\widetilde{Y}^{(1)}_{z\, \sqrt{t/q}}\,\widetilde{Y}^{(2)}_{z\, t/q^2}}{\widetilde{Y}^{(1)}_{z\, \sqrt{t^3/q^5}}\,\widetilde{Y}^{(1)}_{z\, \sqrt{t^3/q^7}}}\right\rangle\nonumber\\
&\;\;\;+ c_{[2,-1]} \left[\widetilde{\fq}^{(2)}\right]^2 \left[\widetilde{\fq}^{(1)}\right]^2 \; \left(1-\sqrt{t/q^3}\, x/z\right)\left(1- \sqrt{q^3/t}\, x/z\right)  \left\langle\frac{\widetilde{Y}^{(1)}_{z\, \sqrt{t/q}}\,\widetilde{Y}^{(1)}_{z\,\sqrt{t^3/q^9}}}{\widetilde{Y}^{(2)}_{z\, t^2/q^5}}\right\rangle\nonumber\\
&\;\;\;+ \widetilde{\fq}^{(2)}\left[\widetilde{\fq}^{(1)}\right]^3 \; \left(1-\sqrt{t/q^3}\, x/z\right)\left(1- \sqrt{t/q}\, x/z\right) \left\langle\frac{\widetilde{Y}^{(2)}_{z\, t/q}\,\widetilde{Y}^{(2)}_{z\, t/q^2}}{\widetilde{Y}^{(1)}_{z\,\sqrt{t^3/q^3}}\, \widetilde{Y}^{(1)}_{z\,\sqrt{t^3/q^5}}\, \widetilde{Y}^{(1)}_{z\,\sqrt{t^3/q^7}}}\right\rangle\nonumber\\
&\;\;\;+ c_{[0,0]} \left[\widetilde{\fq}^{(2)}\right]^2 \left[\widetilde{\fq}^{(1)}\right]^3 \; \left(1-\sqrt{t/q^3}\, x/z\right)\left(1- \sqrt{q^3/t}\, x/z\right)  \left\langle\frac{\widetilde{Y}^{(1)}_{z\, \sqrt{t/q}}}{\widetilde{Y}^{(1)}_{z\,\sqrt{t^5/q^{11}}}}\right\rangle\nonumber\\
&\;\;\;+ c_{[0,0]'} \left[\widetilde{\fq}^{(2)}\right]^2 \left[\widetilde{\fq}^{(1)}\right]^3 \; \left(1-\sqrt{t/q^3}\, x/z\right)\left(1- \sqrt{q^3/t}\, x/z\right)  \left\langle\frac{\widetilde{Y}^{(1)}_{z\, \sqrt{t^3/q^9}}\,\widetilde{Y}^{(2)}_{z\, t/q}}{\widetilde{Y}^{(1)}_{z\,\sqrt{t^3/q^3}}\,\widetilde{Y}^{(2)}_{z\, t^2/q^5}}\right\rangle\nonumber\\
&\;\;\;+ c_{[0,0]''} \left[\widetilde{\fq}^{(2)}\right]^2 \left[\widetilde{\fq}^{(1)}\right]^3 \; \left(1-\sqrt{t/q}\, x/z\right)\left(1- \sqrt{q/t}\, x/z\right)  \left\langle\frac{\widetilde{Y}^{(2)}_{z\, t/q^2}}{\widetilde{Y}^{(2)}_{z\, t^2/q^4}}\right\rangle\nonumber\\
&\;\;\;+\left[\widetilde{\fq}^{(2)}\right]^3 \left[\widetilde{\fq}^{(1)}\right]^3 \; \left(1-\sqrt{q/t}\, x/z\right)\left(1- \sqrt{q^3/t}\, x/z\right)  \left\langle\frac{\widetilde{Y}^{(1)}_{z\, \sqrt{t^3/q^5}}\,\widetilde{Y}^{(1)}_{z\, \sqrt{t^3/q^7}}\, \widetilde{Y}^{(1)}_{z\, \sqrt{t^3/q^9}}}{\widetilde{Y}^{(2)}_{z\, t^2/q^5}\, \widetilde{Y}^{(2)}_{z\, t^2/q^4}}\right\rangle\nonumber\\
&\;\;\;+ c_{[-2,1]} \left[\widetilde{\fq}^{(2)}\right]^2 \left[\widetilde{\fq}^{(1)}\right]^4 \; \left(1-\sqrt{t/q^3}\, x/z\right)\left(1- \sqrt{q^3/t}\, x/z\right)  \left\langle\frac{\widetilde{Y}^{(2)}_{z\, t/q}}{\widetilde{Y}^{(1)}_{z\,\sqrt{t^3/q^3}}\,\widetilde{Y}^{(1)}_{z\,\sqrt{t^5/q^{11}}}}\right\rangle\nonumber\\
&\;\;\;+ c_{[1,-1]} \left[\widetilde{\fq}^{(2)}\right]^3 \left[\widetilde{\fq}^{(1)}\right]^4 \; \left(1-\sqrt{q/t}\, x/z\right)\left(1- \sqrt{q^3/t}\, x/z\right)  \left\langle\frac{\widetilde{Y}^{(1)}_{z\,\sqrt{t^3/q^5}}\, \widetilde{Y}^{(1)}_{z\,\sqrt{t^3/q^7}}}{\widetilde{Y}^{(1)}_{z\,\sqrt{t^5/q^{11}}}\,\widetilde{Y}^{(2)}_{z\, t^2/q^4}}\right\rangle\nonumber\\
&\;\;\;+ c_{[-1,0]} \left[\widetilde{\fq}^{(2)}\right]^3 \left[\widetilde{\fq}^{(1)}\right]^5 \; \left(1-\sqrt{q/t}\, x/z\right)\left(1- \sqrt{q^3/t}\, x/z\right)  \left\langle\frac{\widetilde{Y}^{(1)}_{z\,\sqrt{t^3/q^5}}}{\widetilde{Y}^{(1)}_{z\,\sqrt{t^5/q^9}}\,\widetilde{Y}^{(1)}_{z\,\sqrt{t^5/q^{11}}}}\right\rangle\nonumber\\
&\;\;\;+ \left[\widetilde{\fq}^{(2)}\right]^3 \left[\widetilde{\fq}^{(1)}\right]^6 \; \left(1-\sqrt{q/t}\, x/z\right)\left(1- \sqrt{q^3/t}\, x/z\right)  \left\langle\frac{\widetilde{Y}^{(2)}_{z\, t^3/q^2}}{\widetilde{Y}^{(1)}_{z\,\sqrt{t^5/q^7}}\,\widetilde{Y}^{(1)}_{z\,\sqrt{t^5/q^9}}\, \widetilde{Y}^{(1)}_{z\,\sqrt{t^5/q^{11}}}}\right\rangle\nonumber\\
&\;\;\;+ \left[\widetilde{\fq}^{(2)}\right]^4 \left[\widetilde{\fq}^{(1)}\right]^6 \; \left(1-\sqrt{q^3/t}\, x/z\right)\left(1- \sqrt{q^5/t}\, x/z\right)  \left\langle\frac{1}{\widetilde{Y}^{(2)}_{z\, t^3/q^6}}\right\rangle\; .\label{3dpartitionfunctionG22}
\end{align} 

As described in the main text, this can be proved in a straightforward way starting from  5d. Namely, we first use the specialization of the Coulomb moduli \eqref{specializeG2} to prove
\beq\label{YequalityG2}
\left\langle{\widetilde{Y}^{(a)}_{3d}(z)} \right\rangle = c_{3d}\, \left\langle {Y_{5d}(z)} \right\rangle_{e^{(a)}_{i}=\; \eqref{specializeG2}} \; .
\eeq
The proportionality constant $c_{3d}$ is the bulk residue at the empty partition, see \eqref{c3d}.\\

The second step is to show that each term in the 5d $qq$-character gives a term in the 3d $qq$-character, resulting in
\begin{align}\label{CHIequalityG2}
\left[\widetilde{\chi}^{G_2}\right]_{(L^{(1)},L^{(2)})}^{3d}(z)=c_{3d}\;  \left[\chi^{G_2}\right]_{(L^{(1)},L^{(2)})}^{5d}(z)_{e^{(a)}_{i}=\; \eqref{specializeG2}} \; .
\end{align}

\vspace{8mm}

------- ${\cW}_{q,t}(G_2)$ correlator -------\\

We will now recover the 3d physics from $G_2$-type $q$-Toda theory on the cylinder $\cC$. The theory enjoys a ${\cW}_{q,t}(G_2)$ algebra symmetry, which is generated by 2 currents $W^{(s)}(z)$, with $s=2, 3$. To make contact with the 3d gauge theory results above, we first need to construct these two operators, whose insertion at a point $z$ on $\cC$ corresponds to the location of a D1 brane. To proceed, the operators are constructed  as the commutant of the screening charges. The end result is a Laurent polynomial in the $\cY^{(a)}$ operators defined in \eqref{YoperatorToda}. Conveniently, this polynomial can be read off directly from the 3d Wilson loop operators \eqref{3dpartitionfunctionG2} and \eqref{3dpartitionfunctionG22}, respectively: 
\begin{align}
&W^{(2)}(z) = :\cY^{(1)}(z): + :\cY^{(2)}(\sqrt{t/q}\;z)\left[\cY^{(1)}((t/q)z)\right]^{-1}: +\ldots\nonumber\\
&W^{(3)}(z) = :\cY^{(2)}(z): + :\cY^{(1)}( \sqrt{t/q}\;z)\cY^{(1)}( \sqrt{t/q^3}\;z)\cY^{(1)}( \sqrt{t/q^5}\;z)\left[\cY^{(2)}((t/q^3)z)\right]^{-1} : +\ldots\label{G2stress}
\end{align}

We now  build the vertex operator corresponding to the D5 brane content. Such an operator is fully determined by the set \eqref{G2weightNULL} of coweights $\{\omega_d\}$. In our case, we only have one coweight $\omega$, so we only have one operator to define:
\beq\label{G2vertexweight}
V_{\omega} (x) =\; :\left[\Lambda^{(2)}(\sqrt{t/q^3}\;x)\right]^{-1}\, \Lambda^{(2)}((t^2/q^4)\; x): \; ,
\eeq
where $\Lambda^{(a)}$ is the $a$-th fundamental coweight operator \eqref{coweightvertexdef}. Alternatively, we can define the D5 brane operator using the simple root operators \eqref{rootvertexdef}. The $q$ and $t$-shifts in the argument of these operators are uniquely determined from the truncation locus \eqref{specializeG2}:
\beq\label{G2vertexroot}
V_{\omega} (x)=\; :\left[\Lambda^{(2)}(x)\right]^{-1} E^{(2)}(\sqrt{t/q^3}\;x)\;E^{(1)}((t/q^2)\; x)\; E^{(2)}(\sqrt{t^3/q^5}\; x):
\eeq

Then, an explicit computation shows that the following ${\cW}_{q,t}(G_2)$ correlators are equal to a  3d gauge theory partition function:
\begin{align}\label{G2correlatoris3dindex}
&\left\langle \psi'\left|V_{\omega}(x)\, \prod_{a=1}^{m} (Q^{(a)})^{D^{(a)}}\; W^{(2)}(z) \right| \psi \right\rangle =  B_{(1,0)}(x, z)\, \left[\widetilde{\chi}^{G_2}\right]_{(1,0)}^{3d}(z)\\
&\left\langle \psi'\left|V_{\omega}(x)\, \prod_{a=1}^{m} (Q^{(a)})^{D^{(a)}}\; W^{(3)}(z) \right| \psi \right\rangle =  B_{(0,1)}(x, z)\, \left[\widetilde{\chi}^{G_2}\right]_{(0,1)}^{3d}(z)
\end{align}

To establish this, one simply computes the various two-points making up the correlators; note that the usual constant $A(\{x_d\})$ arising from D5 brane vertex operator two-points with themselves is missing here on the right-hand side; this is  because we only inserted a single D5 brane vertex operator $V_{\omega}(x)$ to define $G^{3d}$ in this example. Most of the two-points are straightforward to evaluate, except perhaps for the ones involving $V_{\omega}(x)$. Namely, one computes
\beq
\prod_{a=1}^{2}\prod_{i=1}^{D^{(a)}}\left\langle  V_{\omega}(x)\; S^{(a)}(y^{(a)}_i) \right\rangle = \prod_{i=1}^{D^{(2)}} \frac{\left(\sqrt{t/q}\; x/y^{(2)}_{i};q^3\right)_{\infty}}{\left(\sqrt{t^3/q^5}\; x/y^{(2)}_{i};q^3\right)_{\infty}} \; ,
\eeq
which is exactly the 3d $\cN=2$ matter content \eqref{matter3dG2}.

The two-points of the D5 brane operator  $V_{\omega}(x)$ with the generating currents $W^{(2)}(z)$ and $W^{(3)}(z)$  is needed to identify exactly the prefactors $B_{(1,0)}(x, z)$ and $B_{(0,1)}(x, z)$, respectively. This is done by computing the inverse deformed $G_2$-Cartan matrix and making use of the commutator \eqref{commutator3}:
\begin{align}
&B_{(1,0)}(x, z)\equiv\frac{1}{\left(1-\sqrt{t^2/q^4}\, x/z\right)}\exp\left(\sum_{k>0}\, \frac{1}{k}\left[\frac{t^{3k/2}\left(q^{5k/2}-t^{3k/2}\right)}{q^{2k}\left(q^{3k}-q^{2k}t^k+t^{2k}\right)}\right]\left(\frac{x}{z}\right)^k\right)\nonumber\\
&B_{(0,1)}(x, z)\equiv\frac{1}{\left(1-\sqrt{t/q^3}\, x/z\right)\left(1-\sqrt{t^3/q^5}\, x/z\right)}\exp\left(\sum_{k>0}\, \frac{1}{k}\left[\frac{t^{k}\left(q^{5k/2}-t^{3k/2}\right)\left(q^{k}+t^{k}\right)}{q^{5k/2}\left(q^{3k}-q^{2k}t^k+t^{2k}\right)}\right]\left(\frac{x}{z}\right)^k\right) \; .\label{BdefinedG2}
\end{align}
Quite beautifully, the identity \eqref{G2correlatoris3dindex} follows.\\

Moreover, the equality of the 5d and 3d partition functions \eqref{CHIequalityG2} further implies that the correlators are equal to a 5d partition function at tuned values of the Coulomb moduli:
\begin{align}\label{G2correlatoris5dindex}
&\left\langle \psi'\left|V_{\omega}(x)\, \prod_{a=1}^{m} (Q^{(a)})^{D^{(a)}}\; W^{(2)}(z) \right| \psi \right\rangle =  B_{[1,0]}(x, z)\,c_{3d} \left[{\chi}^{G_2}\right]_{(1,0)}^{5d}(z)_{e^{(a)}_{i}=\; \eqref{specializeG2}}\; ,\\
&\left\langle \psi'\left|V_{\omega}(x)\, \prod_{a=1}^{m} (Q^{(a)})^{D^{(a)}}\; W^{(3)}(z) \right| \psi \right\rangle =  B_{[0,1]}(x, z)\,c_{3d} \left[{\chi}^{G_2}\right]_{(0,1)}^{5d}(z)_{e^{(a)}_{i}=\; \eqref{specializeG2}}\; .
\end{align}

Though we have not explicitly checked the generalization to $L>1$ D1 branes in the $G_2$ case, we also claim  that the following identity holds, based on the arguments in the main text: 
\begin{align}\label{G2correlatoris3dindexMORE}
&\left\langle \psi'\left|V_{\omega}(x)\; \prod_{a=1}^{2} (Q^{(a)})^{D^{(a)}}\; \prod_{s=2}^{3}\prod_{\rho=1}^{L^{(s-1)}}W^{(s)}(z^{(s-1)}_\rho) \right| \psi \right\rangle \nonumber\\
&\qquad\qquad\qquad\qquad\qquad= B\left(x,\{z^{(s-1)}_\rho\}\right)\, \left[\widetilde{\chi}^{G_2}\right]_{(L^{(1)}, L^{(2)})}^{3d}(\{z^{(s-1)}_\rho\})\; .
\end{align}

\vspace{10mm}

\subsection{Wilson Loop for $F_4$ Theories}
\label{ssec:exampleF4}

\vspace{8mm}

------- 5d Gauge Theory -------\\

Let $X$ be a resolved $E_6$ singularity, and consider type IIB string theory on $X\times\cC\times\mathbb{C}^2$. We consider a nontrivial fibration of $X$ over $\mathbb{C}^2 \times \cC$, where as we go around the origin of one of the lines wrapped by the branes (say,  $\mathbb{C}_q$), the space $X$ goes back to itself, up to $\mathbb{Z}_2$ outer automorphism group action. The $\mathbb{Z}_2$ action will permute the compact 2-cycles representing the nodes 3 and 5 (respectively 4 and 6) of the $E_6$ Dynkin diagram\footnote{In our conventions, node 2 is the central trivalent node of the $E_6$ Dynkin diagram, and node 1 is the single node connected to it.}. For $a=1,\ldots,6$, we introduce $n^{(a)}$ D5 branes wrapping  $S_a^2$ and $\mathbb{C}^2$, such that $n^{(3)}=n^{(5)}$ and $n^{(4)}=n^{(6)}$. We further introduce ${N^{(a)}_f}$ D5 branes wrapping the dual non-compact 2-cycle $S^*_a$ and $\mathbb{C}^2$, such that $N_f^{(3)}=N_f^{(5)}$ and $N_f^{(4)}=N_f^{(6)}$. Finally, we add $L^{(a)}$ D1 branes wrapping $S^*_a$, such that $L^{(3)}=L^{(5)}$ and $L^{(4)}=L^{(6)}$.\\

In what follows, we use the notation $n=n^{(1)}+n^{(2)}+n^{(3)}+n^{(4)}$,\;\; $N_f=N_f^{(1)}+N_f^{(2)}+N_f^{(3)}+N_f^{(4)}$ and $L=L^{(1)}+L^{(2)}+L^{(3)}+L^{(4)}$. We set $g_s\rightarrow 0$, which amounts to studying the $(2,0)$ $E_6$ little string on $\cC\times\mathbb{C}^2$ in the presence of $F_4$-type codimension 2 defects (the D5 branes) and point-like defects (the D1 branes). At energies below the string scale, the dynamics in this background are fully captured by the theory on the D5 branes, with D1 brane defects. Because of the nontrivial fibration in the geometry, the theory on the D5 branes is a 5d quiver gauge theory of shape the Dynkin diagram of $F_4$. We call it $T^{5d}$, and it is defined on $S^1(\widehat{R})\times\mathbb{C}^2$, with gauge content  $\prod_{a=1}^{4}U(n^{(a)})$ and flavor content $\prod_{a=1}^{4}U({N^{(a)}_f})$. The D1 branes make up a 1/2-BPS a Wilson loop wrapping the circle and sitting at the origin of $\mathbb{C}^2$.\\

In our notation, the first and second simple roots of $E_6$ are in their own orbit under the $\mathbb{Z}_2$-action, so the nodes 1 and 2 of the $F_4$ Dynkin diagram denote the long simple roots. The third and fifth simple roots get mapped to each other, and so do the fourth and fifth simple roots; as a consequence, we take the nodes 3 and 4 of  the $F_4$ Dynkin diagram to denote the short simple roots.
After putting the theory on $\Omega$-background, the partition function of $T^{5d}$ with Wilson loop is the following Witten index, referring to \eqref{5dintegral} in the main text:
{\allowdisplaybreaks
\begin{align}
&\left[\chi^{F_4}\right]_{(L^{(1)},\ldots, L^{(4)})}^{5d}  =\sum_{k^{(1)},\ldots, k^{(4)}=0}^{\infty}\;\prod_{a=1}^{4}\frac{\left(\widetilde{\fq}^{(a)}{}\right)^{k^{(a)}}}{k^{(a)}!}\;\; \nonumber\\ 
&\qquad\qquad\qquad\;\times\oint  \left[\frac{d\phi^{{(a)}}_I}{2\pi i}\right]Z^{(a)}_{vec}\cdot Z^{(a)}_{fund}\cdot Z^{(a)}_{CS} \cdot Z^{(1,2)}_{bif}\,Z^{(2,3)}_{bif}\,Z^{(3,4)}_{bif}\cdot \prod_{\rho=1}^{L^{(a)}}  Z^{(a)}_{D1} \; ,\nonumber\\
&Z^{(a)}_{vec} =\prod_{I, J=1}^{k^{(a)}} \frac{{\widehat{f}}\left(\phi^{(a)}_{I}-\phi^{(a)}_{J}\right)f\left(\phi^{(a)}_{I}-\phi^{(a)}_{J}+ 2\,\epsilon_++\epsilon_1 \right)}{f\left(\phi^{(a)}_{I}-\phi^{(a)}_{J}+2\,\epsilon_1 \right)f\left(\phi^{(a)}_{I}-\phi^{(a)}_{J}+\epsilon_2\right)}\nonumber\\
&\qquad\times\prod_{I=1}^{k^{(a)}} \prod_{i=1}^{n^{(a)}} \frac{1}{f\left(\phi^{(1)}_I-a^{(1)}_i+\epsilon_++\epsilon_1/2\right)f\left(\phi^{(a)}_I-a^{(a)}_i-\epsilon_+-\epsilon_1/2 \right)}\;\; ,a=1, 2\\
&Z^{(a)}_{vec} =\prod_{I, J=1}^{k^{(a)}} \frac{{\widehat{f}}\left(\phi^{(a)}_{I}-\phi^{(a)}_{J}\right)f\left(\phi^{(a)}_{I}-\phi^{(a)}_{J}+ 2\,\epsilon_+ \right)}{f\left(\phi^{(a)}_{I}-\phi^{(a)}_{J}+\epsilon_1\right)f\left(\phi^{(a)}_{I}-\phi^{(a)}_{J}+\epsilon_2\right)}\nonumber\\
&\qquad\times\prod_{I=1}^{k^{(a)}} \prod_{i=1}^{n^{(a)}} \frac{1}{f\left(\phi^{(a)}_I-a^{(a)}_i+\epsilon_+ \right)f\left(\phi^{(a)}_I-a^{(a)}_i-\epsilon_+ \right)}\;\; ,a=3, 4\\
&Z^{(a)}_{fund} =\prod_{I=1}^{k^{(a)}} \prod_{d=1}^{N^{(a)}_f} f\left(\phi^{{(a)}}_I-m^{{(a)}}_d\right) \equiv\prod_{I=1}^{k^{(a)}} Q^{(a)}(\phi^{{(a)}}_I)\, ,\\
&Z^{(1, 2)}_{bif} =\prod_{J=1}^{k^{(2)}}\prod_{i=1}^{n^{(1)}} f\left(\phi^{(2)}_J- m^{(1)}_{bif} - a^{(1)}_i - \epsilon_+ - \epsilon_1/2 \right)\nonumber\\
&\qquad \times\prod_{I=1}^{k^{(1)}}\prod_{j=1}^{n^{(2)}}  f\left(\phi^{(1)}_I+ m^{(1)}_{bif} - a^{(2)}_j + \epsilon_+ + \epsilon_1/2 \right)\nonumber\\
&\qquad \times \prod_{I=1}^{k^{(1)}}\prod_{J=1}^{k^{(2)}} \frac{f\left(\phi^{(1)}_{I}-\phi^{(2)}_{J}+ m^{(1)}_{bif}+2\epsilon_1 \right)}{f\left(\phi^{(1)}_{I}-\phi^{(2)}_{J} +m^{(1)}_{bif}\right)}\nonumber\\
&\qquad \times \prod_{I=1}^{k^{(1)}}\prod_{J=1}^{k^{(2)}} \frac{f\left(\phi^{(1)}_{I}-\phi^{(2)}_{J}+ m^{(1)}_{bif}+\epsilon_2  \right)}{f\left(\phi^{(1)}_{I}-\phi^{(2)}_{J}+ m^{(1)}_{bif}+2\, \epsilon_+ + \epsilon_1  \right)}\, ,\\
&Z^{(2, 3)}_{bif} =\prod_{J=1}^{k^{(3)}}\prod_{i=1}^{n^{(2)}} f\left(\phi^{(3)}_J- m^{(2)}_{bif} - a^{(2)}_i - \epsilon_+\right)\nonumber\\
&\qquad \times\prod_{I=1}^{k^{(2)}}\prod_{j=1}^{n^{(3)}}\prod_{s=0}^{1}  f\left(\phi^{(2)}_I+ m^{(2)}_{bif} - a^{(3)}_j + \epsilon_+ + (1 -2 \, s)\,\epsilon_1/2 \right)\nonumber\\
&\qquad \times \prod_{I=1}^{k^{(2)}}\prod_{J=1}^{k^{(3)}} \frac{f\left(\phi^{(2)}_{I}-\phi^{(3)}_{J}+ m^{(2)}_{bif}+3\,\epsilon_1/2 \right)}{f\left(\phi^{(2)}_{I}-\phi^{(3)}_{J} +m^{(2)}_{bif}-\epsilon_1/2\right)}\nonumber\\
&\qquad \times \prod_{I=1}^{k^{(2)}}\prod_{J=1}^{k^{(3)}} \frac{f\left(\phi^{(2)}_{I}-\phi^{(3)}_{J}+ m^{(2)}_{bif}+\epsilon_2 -\epsilon_1/2 \right)}{f\left(\phi^{(2)}_{I}-\phi^{(3)}_{J}+ m^{(2)}_{bif}+2\, \epsilon_+ + \epsilon_1/2  \right)}\, ,\\
&Z^{(3, 4)}_{bif} =\prod_{J=1}^{k^{(4)}}\prod_{i=1}^{n^{(3)}} f\left(\phi^{(4)}_J- m^{(3)}_{bif} - a^{(3)}_i - \epsilon_+\right)\nonumber\\
&\qquad \times\prod_{I=1}^{k^{(3)}}\prod_{j=1}^{n^{(4)}} f\left(\phi^{(3)}_I+ m^{(3)}_{bif} - a^{(4)}_j + \epsilon_+  \right)\nonumber\\
&\qquad \times \prod_{I=1}^{k^{(3)}}\prod_{J=1}^{k^{(4)}} \frac{f\left(\phi^{(3)}_{I}-\phi^{(4)}_{J}+ m^{(3)}_{bif}+\epsilon_1 \right)}{f\left(\phi^{(3)}_{I}-\phi^{(4)}_{J} +m^{(3)}_{bif}\right)}\nonumber\\
&\qquad \times \prod_{I=1}^{k^{(3)}}\prod_{J=1}^{k^{(4)}} \frac{f\left(\phi^{(3)}_{I}-\phi^{(4)}_{J}+ m^{(3)}_{bif}+\epsilon_2 \right)}{f\left(\phi^{(3)}_{I}-\phi^{(4)}_{J}+ m^{(3)}_{bif}+2\, \epsilon_+  \right)}\, ,\\
&Z^{(a)}_{CS}= \prod_{I=1}^{k^{(a)}} e^{k^{(a)}_{CS}\, \phi^{(a)}_I}\\
&Z^{(a)}_{D1} =\prod_{i=1}^{n^{(a)}}  f\left(a^{{(a)}}_i-M^{{(a)}}_\rho\right)\nonumber\\ 
&\qquad \times\prod_{I=1}^{k^{(a)}} \frac{f\left(\phi^{{(a)}}_I-M^{(a)}_\rho+ \epsilon_-+\epsilon_1/2 \right)f\left(\phi^{{(a)}}_I-M^{(a)}_\rho- \epsilon_-- \epsilon_1/2 \right)}{f\left(\phi^{{(a)}}_I-M^{(a)}_\rho+ \epsilon_+ + \epsilon_1/2 \right)f\left(\phi^{{(a)}}_I-M^{(a)}_\rho- \epsilon_+ - \epsilon_1/2 \right)}\;\; ,a=1, 2\\
&Z^{(a)}_{D1} =\prod_{i=1}^{n^{(a)}}  f\left(a^{{(a)}}_i-M^{{(a)}}_\rho\right)\nonumber\\ &\qquad \times\prod_{I=1}^{k^{(a)}} \frac{f\left(\phi^{{(a)}}_I-M^{(a)}_\rho+ \epsilon_- \right)f\left(\phi^{{(a)}}_I-M^{(a)}_\rho- \epsilon_- \right)}{f\left(\phi^{{(a)}}_I-M^{(a)}_\rho+ \epsilon_+  \right)f\left(\phi^{{(a)}}_I-M^{(a)}_\rho- \epsilon_+ \right)}\;\; a=3, 4. \label{5dintegralF4}
\end{align}}

Note the extra factors in some of the bifundamental contributions, which manifests themselves as extra Fermi multiplets in the quantum mechanics.\\

We will now exhibit the physics of a Wilson loop on the last node, namely we now focus on $L^{(4)}=1$ and $L^{(a)}=0,\;\;\; a=1, 2, 3$. This corresponds to having a D1 brane wrapping $S^*_4$ and another D1 brane wrapping $S^*_6$ in the $E_6$ geometry. 

Correspondingly, we denote the partition function as $\left[\chi^{F_4}\right]_{(0,0,0,1)}^{5d}$, where the integer $L^{(4)}=1$ is  measuring a D1 brane charge in the coweight lattice of $F_4$. We write the associated fermion mass $M^{(4)}_1\equiv M$ to simplify the notation.

Then, by the JK residue prescription, there are poles labeled by Young diagrams,
\begin{align}
\label{F4youngtuples}
&\phi^{(a)}_I = a^{(a)}_i +\epsilon_+ -\epsilon_1/2 - 2\, s_1\, \epsilon_1 - s_2\, \epsilon_2 ,\;\; \text{with}\; (s_1, s_2)\in \boldsymbol{\mu}^{(a)}_i \;\;\; ,\; a=1, 2\; ,\\
&\phi^{(a)}_I = a^{(a)}_i +\epsilon_+ -  s_1\, \epsilon_1 - s_2\, \epsilon_2 ,\;\; \text{with}\; (s_1, s_2)\in \boldsymbol{\mu}^{(a)}_i \;\;\; ,\; a=3,4\; .
\end{align}
and 25 extra poles depending on the fermion mass $M$.
	
Let us now make contact with the representation theory of quantum affine algebras. First, we switch to K-theoretic fugacities, such as  $z=e^{-\widehat{R}M}$.
As usual, we define a defect operator expectation value:
\begin{align}
&\left\langle \left[Y^{(a)}_{5d}(z)\right]^{\pm 1}  \right\rangle  \equiv\sum_{k^{(1)},\ldots, k^{(4)}=0}^{\infty}\;\prod_{b=1}^{4}\frac{\left(\widetilde{\fq}^{(b)}{}\right)^{k^{(b)}}}{k^{(b)}!} \nonumber\\
&\qquad\qquad\; \times\oint_{\{\overrightarrow{\boldsymbol{\mu}}\}}  \left[\frac{d\phi^{{(b)}}_I}{2\pi i}\right]{Z^{(b)}_{vec}}\cdot Z^{(b)}_{fund}\cdot Z^{(b)}_{CS} \cdot \left[Z^{(a)}_{D1}(z)\right]^{\pm 1} \cdot Z^{(1,2)}_{bif}\,Z^{(2,3)}_{bif}\,Z^{(3,4)}_{bif} \, .\label{YopF4}
\end{align}
It is understood in the above definition  that the contours enclose poles coming from Young diagrams only, labeled by \eqref{F4youngtuples}. As we have just seen, that is not the correct pole prescription to evaluate the partition function, since there are 25 extra poles to enclose. Nonetheless,  we can write the partition function as a Laurent polynomial in the $Y^{(a)}_{5d}$ operator vevs, as a sum of exactly 26 terms. The partition function then makes up a twisted $qq$-character of the fourth fundamental representation of $U_q(\widehat{F_4})$, with highest weight $[0,0,0,1]$. 
The formulas get involved quickly, so we introduce the following notation: $Y^{(a)}_{5d}(z)\equiv Y^{(a)}_z$, $Q^{(a)}(z)\equiv Q^{(a)}_z$, and $\mu^{(a)}_{bif}\equiv \mu_a$. Then, the 5d partition function with Wilson loop evaluates to:
{\allowdisplaybreaks
\begin{align}
\label{F4result2}
&\left[\chi^{F_4}\right]_{(0,0,0,1)}^{5d}(z) =\left\langle Y^{(4)}_z \right\rangle\nonumber\\
&\;\;\; + \widetilde{\fq}^{(4)} \; Q^{(4)}_{z\, \sqrt{t/q}} \left\langle\frac{Y^{(3)}_{z\, \mu_3^{-1}}}{Y^{(4)}_{z\, t/q}}\right\rangle\nonumber\\
&\;\;\;+ \widetilde{\fq}^{(4)}\widetilde{\fq}^{(3)} \; Q^{(4)}_{z\, \sqrt{t/q}}\;Q^{(3)}_{z\,\mu_3^{-1}\sqrt{t/q}} \left\langle\frac{Y^{(2)}_{z\, \mu_3^{-1}\mu_2^{-1}}}{Y^{(3)}_{z\,\mu_3^{-1}\, t/q}}\right\rangle\nonumber\\
&\;\;\;+ \widetilde{\fq}^{(4)}\widetilde{\fq}^{(3)}\widetilde{\fq}^{(2)} \; Q^{(4)}_{z\, \sqrt{t/q}}\;Q^{(3)}_{z\,\mu_3^{-1}\sqrt{t/q}}\;Q^{(2)}_{z\,\mu_3^{-1}\mu_2^{-1}\sqrt{t/q^2}} \left\langle\frac{Y^{(1)}_{z\, \mu_3^{-1}\mu_2^{-1}\mu_1^{-1}}\,Y^{(3)}_{z\,\mu_3^{-1}\, t/q^2}}{Y^{(2)}_{z\,\mu_3^{-1}\mu_2^{-1}\, t/q^2}}\right\rangle\nonumber\\
&\;\;\;+ \widetilde{\fq}^{(4)}\widetilde{\fq}^{(3)}\widetilde{\fq}^{(2)}\widetilde{\fq}^{(1)} \; Q^{(4)}_{z\, \sqrt{t/q}}\;Q^{(3)}_{z\,\mu_3^{-1}\sqrt{t/q}}\;Q^{(2)}_{z\,\mu_3^{-1}\mu_2^{-1}\sqrt{t/q^2}}\;Q^{(1)}_{z\,\mu_3^{-1}\mu_2^{-1}\mu_1^{-1}\sqrt{t/q^2}}           \left\langle\frac{Y^{(3)}_{z\,\mu_3^{-1}\, t/q^2}}{Y^{(1)}_{z\,\mu_3^{-1}\mu_2^{-1}\mu_1^{-1}\, t/q^2}}\right\rangle\nonumber\\
&\;\;\;+\widetilde{\fq}^{(4)}\left[\widetilde{\fq}^{(3)}\right]^2\widetilde{\fq}^{(2)} Q^{(4)}_{z\, \sqrt{t/q}}\;Q^{(3)}_{z\,\mu_3^{-1}\sqrt{t/q}} Q^{(3)}_{z\,\mu_3^{-1} \sqrt{t^3/q^5}}\;Q^{(2)}_{z\,\mu_3^{-1}\mu_2^{-1}\sqrt{t/q^2}}  \left\langle\frac{Y^{(1)}_{z\, \mu_3^{-1}\mu_2^{-1}\mu_1^{-1}}\,Y^{(4)}_{z\, t^2/q^3}}{Y^{(3)}_{z\,\mu_3^{-1} t^2/q^3}}\right\rangle\nonumber\\
&\;\;\;+\widetilde{\fq}^{(4)}\left[\widetilde{\fq}^{(3)}\right]^2\widetilde{\fq}^{(2)}\widetilde{\fq}^{(1)}\; Q^{(4)}_{z\, \sqrt{t/q}}\;Q^{(3)}_{z\,\mu_3^{-1}\sqrt{t/q}}\; Q^{(3)}_{z\,\mu_3^{-1} \sqrt{t^3/q^5}}\;Q^{(2)}_{z\,\mu_3^{-1}\mu_2^{-1}\sqrt{t/q^2}}\;\times\nonumber\\
&\qquad\qquad \times Q^{(1)}_{z\,\mu_3^{-1}\mu_2^{-1}\mu_1^{-1}\sqrt{t/q^2}} \left\langle\frac{Y^{(2)}_{z\, \mu_3^{-1}\mu_2^{-1}\mu_1^{-1}\,t/q^2}\,Y^{(4)}_{z\, t^2/q^3}}{Y^{(1)}_{z\,\mu_3^{-1}\mu_2^{-1}\mu_1^{-1} t/q^2}\, Y^{(3)}_{z\,\mu_3^{-1} t^2/q^3}}\right\rangle\nonumber\\
&\;\;\;+\left[\widetilde{\fq}^{(4)}\right]^2\left[\widetilde{\fq}^{(3)}\right]^2\widetilde{\fq}^{(2)}\; Q^{(4)}_{z\, \sqrt{t/q}} \; Q^{(4)}_{z\, \sqrt{t^5/q^7}}\;Q^{(3)}_{z\,\mu_3^{-1}\sqrt{t/q}}\; Q^{(3)}_{z\,\mu_3^{-1} \sqrt{t^3/q^5}}\;\times\nonumber\\
&\qquad\qquad \times Q^{(2)}_{z\,\mu_3^{-1}\mu_2^{-1}\sqrt{t/q^2}}\; \left\langle\frac{Y^{(1)}_{z\, \mu_3^{-1}\mu_2^{-1}\mu_1^{-1}}}{Y^{(4)}_{z\, t^3/q^4}}\right\rangle\nonumber\\
&\;\;\;+\widetilde{\fq}^{(4)}\left[\widetilde{\fq}^{(3)}\right]^2\left[\widetilde{\fq}^{(2)}\right]^2\widetilde{\fq}^{(1)}\; Q^{(4)}_{z\, \sqrt{t/q}}\;Q^{(3)}_{z\,\mu_3^{-1}\sqrt{t/q}}\; Q^{(3)}_{z\,\mu_3^{-1} \sqrt{t^3/q^5}}\;Q^{(2)}_{z\,\mu_3^{-1}\mu_2^{-1}\sqrt{t/q^2}}\;\times\nonumber\\
&\qquad\qquad \times  Q^{(2)}_{z\,\mu_3^{-1}\mu_2^{-1}\sqrt{t^3/q^6}}\; Q^{(1)}_{z\,\mu_3^{-1}\mu_2^{-1}\mu_1^{-1}\sqrt{t/q^2}} \left\langle\frac{Y^{(3)}_{z\, \mu_3^{-1}t^2/q^3}\,Y^{(4)}_{z\, t^2/q^3}}{Y^{(2)}_{z\,\mu_3^{-1}\mu_2^{-1} t^2/q^4}}\right\rangle\nonumber\\
&\;\;\;+\left[\widetilde{\fq}^{(4)}\right]^2\left[\widetilde{\fq}^{(3)}\right]^2\widetilde{\fq}^{(2)}\widetilde{\fq}^{(1)}\; Q^{(4)}_{z\, \sqrt{t/q}}Q^{(4)}_{z\, \sqrt{t^5/q^7}}\;Q^{(3)}_{z\,\mu_3^{-1}\sqrt{t/q}}\; Q^{(3)}_{z\,\mu_3^{-1} \sqrt{t^3/q^5}}\;\times\nonumber\\
&\qquad\qquad \times Q^{(2)}_{z\,\mu_3^{-1}\mu_2^{-1}\sqrt{t/q^2}}\;Q^{(1)}_{z\,\mu_3^{-1}\mu_2^{-1}\mu_1^{-1}\sqrt{t/q^2}}\; \left\langle\frac{Y^{(2)}_{z\, \mu_3^{-1}\mu_2^{-1}\, t/q^2}}{Y^{(1)}_{z\, \mu_3^{-1}\mu_2^{-1}\mu_1^{-1}\, t/q^2}\,Y^{(4)}_{z\, t^3/q^4}}\right\rangle\nonumber\\
&\;\;\;+\left[\widetilde{\fq}^{(4)}\right]^2\left[\widetilde{\fq}^{(3)}\right]^2\left[\widetilde{\fq}^{(3)}\right]^{2}\widetilde{\fq}^{(1)}\; Q^{(4)}_{z\, \sqrt{t/q}}Q^{(4)}_{z\, \sqrt{t^5/q^7}}\;Q^{(3)}_{z\,\mu_3^{-1}\sqrt{t/q}}\; Q^{(3)}_{z\,\mu_3^{-1} \sqrt{t^3/q^5}}\;\times\nonumber\\
&\qquad\qquad \times Q^{(2)}_{z\,\mu_3^{-1}\mu_2^{-1}\sqrt{t/q^2}}\; Q^{(2)}_{z\,\mu_3^{-1}\mu_2^{-1}\sqrt{t^3/q^6}}\;Q^{(1)}_{z\,\mu_3^{-1}\mu_2^{-1}\mu_1^{-1}\sqrt{t/q^2}}\; \left\langle\frac{Y^{(3)}_{z\, \mu_3^{-1}t^2/q^3}\, Y^{(3)}_{z\, \mu_3^{-1}t^2/q^4}}{Y^{(2)}_{z\, \mu_3^{-1}\mu_2^{-1}\, t^2/q^4}\,Y^{(4)}_{z\, t^3/q^4}}\right\rangle\nonumber\\
&\;\;\;+\widetilde{\fq}^{(4)}\left[\widetilde{\fq}^{(3)}\right]^3\left[\widetilde{\fq}^{(2)}\right]^2\widetilde{\fq}^{(1)}\; Q^{(4)}_{z\, \sqrt{t/q}}\;Q^{(3)}_{z\,\mu_3^{-1}\sqrt{t/q}}\; Q^{(3)}_{z\,\mu_3^{-1} \sqrt{t^3/q^5}}\;  Q^{(3)}_{z\,\mu_3^{-1} \sqrt{t^5/q^9}}\;\times\nonumber\\
&\qquad\qquad \times Q^{(2)}_{z\,\mu_3^{-1}\mu_2^{-1}\sqrt{t/q^2}}\;  Q^{(2)}_{z\,\mu_3^{-1}\mu_2^{-1}\sqrt{t^3/q^6}}\; Q^{(1)}_{z\,\mu_3^{-1}\mu_2^{-1}\mu_1^{-1}\sqrt{t/q^2}} \left\langle\frac{Y^{(4)}_{z\, t^2/q^3}\,Y^{(4)}_{z\, t^3/q^5}}{Y^{(3)}_{z\,\mu_3^{-1} t^3/q^5}}\right\rangle\nonumber\\
&\;\;\;+c_{[0,0]}\left[\widetilde{\fq}^{(4)}\right]^2\left[\widetilde{\fq}^{(3)}\right]^3\left[\widetilde{\fq}^{(2)}\right]^2\widetilde{\fq}^{(1)}\; Q^{(4)}_{z\, \sqrt{t/q}}\; Q^{(4)}_{z\, \sqrt{t^5/q^7}}\;Q^{(3)}_{z\,\mu_3^{-1}\sqrt{t/q}}\; Q^{(3)}_{z\,\mu_3^{-1} \sqrt{t^3/q^5}}\; \times\nonumber\\
&\qquad\qquad \times  Q^{(3)}_{z\,\mu_3^{-1} \sqrt{t^5/q^9}}\; Q^{(2)}_{z\,\mu_3^{-1}\mu_2^{-1}\sqrt{t/q^2}}\;  Q^{(2)}_{z\,\mu_3^{-1}\mu_2^{-1}\sqrt{t^3/q^6}}\; Q^{(1)}_{z\,\mu_3^{-1}\mu_2^{-1}\mu_1^{-1}\sqrt{t/q^2}} \left\langle\frac{Y^{(3)}_{z\,\mu_3^{-1} t^2/q^3}\,Y^{(4)}_{z\, t^3/q^5}}{Y^{(3)}_{z\,\mu_3^{-1} t^3/q^5}\, Y^{(4)}_{z\, t^3/q^4}}\right\rangle\nonumber\\
&\;\;\;+c_{[0,0]'}\left[\widetilde{\fq}^{(4)}\right]^2\left[\widetilde{\fq}^{(3)}\right]^3\left[\widetilde{\fq}^{(2)}\right]^2\widetilde{\fq}^{(1)}\; Q^{(4)}_{z\, \sqrt{t/q}}\; Q^{(4)}_{z\, \sqrt{t^7/q^11}}\;Q^{(3)}_{z\,\mu_3^{-1}\sqrt{t/q}}\; Q^{(3)}_{z\,\mu_3^{-1} \sqrt{t^3/q^5}}\; \times\nonumber\\
&\qquad\qquad \times  Q^{(3)}_{z\,\mu_3^{-1} \sqrt{t^5/q^9}}\; Q^{(2)}_{z\,\mu_3^{-1}\mu_2^{-1}\sqrt{t/q^2}}\;  Q^{(2)}_{z\,\mu_3^{-1}\mu_2^{-1}\sqrt{t^3/q^6}}\; Q^{(1)}_{z\,\mu_3^{-1}\mu_2^{-1}\mu_1^{-1}\sqrt{t/q^2}} \left\langle\frac{Y^{(4)}_{z\, t^2/q^3}}{Y^{(4)}_{z\, t^4/q^6}}\right\rangle\nonumber\\
&\;\;\;+\left[\widetilde{\fq}^{(4)}\right]^2\left[\widetilde{\fq}^{(3)}\right]^4\left[\widetilde{\fq}^{(2)}\right]^2\widetilde{\fq}^{(1)}\; Q^{(4)}_{z\, \sqrt{t/q}}\; Q^{(4)}_{z\, \sqrt{t^5/q^7}}\;Q^{(3)}_{z\,\mu_3^{-1}\sqrt{t/q}}\; Q^{(3)}_{z\,\mu_3^{-1} \sqrt{t^3/q^5}}\; \times\nonumber\\
&\qquad\qquad \times  Q^{(3)}_{z\,\mu_3^{-1} \sqrt{t^5/q^9}}\; Q^{(3)}_{z\,\mu_3^{-1} \sqrt{t^5/q^7}}\; Q^{(2)}_{z\,\mu_3^{-1}\mu_2^{-1}\sqrt{t/q^2}}\;  Q^{(2)}_{z\,\mu_3^{-1}\mu_2^{-1}\sqrt{t^3/q^6}}\; \times\nonumber\\
&\qquad\qquad \times Q^{(1)}_{z\,\mu_3^{-1}\mu_2^{-1}\mu_1^{-1}\sqrt{t/q^2}} \left\langle\frac{Y^{(2)}_{z\,\mu_3^{-1}\mu_2^{-1} t^2/q^3}\,Y^{(4)}_{z\, t^3/q^5}}{Y^{(3)}_{z\,\mu_3^{-1} t^3/q^5}\, Y^{(3)}_{z\,\mu_3^{-1} t^3/q^4}}\right\rangle\nonumber\\
&\;\;\;+\left[\widetilde{\fq}^{(4)}\right]^3\left[\widetilde{\fq}^{(3)}\right]^3\left[\widetilde{\fq}^{(2)}\right]^2\widetilde{\fq}^{(1)}\; Q^{(4)}_{z\, \sqrt{t/q}}\; Q^{(4)}_{z\, \sqrt{t^7/q^{11}}}\; Q^{(4)}_{z\, \sqrt{t^5/q^7}}\;Q^{(3)}_{z\,\mu_3^{-1}\sqrt{t/q}}
\;\; \times\nonumber\\
&\qquad\qquad \times Q^{(3)}_{z\,\mu_3^{-1} \sqrt{t^3/q^5}}\;  Q^{(3)}_{z\,\mu_3^{-1} \sqrt{t^5/q^9}}\; Q^{(2)}_{z\,\mu_3^{-1}\mu_2^{-1}\sqrt{t/q^2}}\;  Q^{(2)}_{z\,\mu_3^{-1}\mu_2^{-1}\sqrt{t^3/q^6}}\; \times\nonumber\\
&\qquad\qquad  \times Q^{(1)}_{z\,\mu_3^{-1}\mu_2^{-1}\mu_1^{-1}\sqrt{t/q^2}} \left\langle\frac{Y^{(3)}_{z\,\mu_3^{-1} t^2/q^3}}{Y^{(4)}_{z\, t^4/q^3}\,Y^{(4)}_{z\, t^4/q^6}}\right\rangle\nonumber\\
&\;\;\;+\left[\widetilde{\fq}^{(4)}\right]^2\left[\widetilde{\fq}^{(3)}\right]^4\left[\widetilde{\fq}^{(2)}\right]^3\widetilde{\fq}^{(1)}\; Q^{(4)}_{z\, \sqrt{t/q}}\; Q^{(4)}_{z\, \sqrt{t^5/q^7}}\;Q^{(3)}_{z\,\mu_3^{-1}\sqrt{t/q}}\; Q^{(3)}_{z\,\mu_3^{-1} \sqrt{t^3/q^5}}\; \times\nonumber\\
&\qquad\qquad \times  Q^{(3)}_{z\,\mu_3^{-1} \sqrt{t^5/q^9}}\; Q^{(3)}_{z\,\mu_3^{-1} \sqrt{t^5/q^7}}\; Q^{(2)}_{z\,\mu_3^{-1}\mu_2^{-1}\sqrt{t/q^2}}\;  Q^{(2)}_{z\,\mu_3^{-1}\mu_2^{-1}\sqrt{t^3/q^6}}\; \times\nonumber\\
&\qquad\qquad \times Q^{(2)}_{z\,\mu_3^{-1}\mu_2^{-1}\sqrt{t^5/q^8}}\; Q^{(1)}_{z\,\mu_3^{-1}\mu_2^{-1}\mu_1^{-1}\sqrt{t/q^2}} \left\langle\frac{Y^{(1)}_{z\,\mu_3^{-1}\mu_2^{-1}\mu_1^{-1} t^2/q^3}\,Y^{(4)}_{z\, t^3/q^5}}{Y^{(2)}_{z\,\mu_3^{-1}\mu_2^{-1} t^3/q^5}}\right\rangle\nonumber\\
&\;\;\;+\left[\widetilde{\fq}^{(4)}\right]^3\left[\widetilde{\fq}^{(3)}\right]^4\left[\widetilde{\fq}^{(2)}\right]^2\widetilde{\fq}^{(1)}\; Q^{(4)}_{z\, \sqrt{t/q}}\; Q^{(4)}_{z\, \sqrt{t^7/q^{11}}}\; Q^{(4)}_{z\, \sqrt{t^5/q^7}}\;Q^{(3)}_{z\,\mu_3^{-1}\sqrt{t/q}}
\;\; \times\nonumber\\
&\qquad\qquad \times Q^{(3)}_{z\,\mu_3^{-1} \sqrt{t^3/q^5}}\;  Q^{(3)}_{z\,\mu_3^{-1} \sqrt{t^5/q^9}}\; Q^{(3)}_{z\,\mu_3^{-1} \sqrt{t^5/q^7}}\; Q^{(2)}_{z\,\mu_3^{-1}\mu_2^{-1}\sqrt{t/q^2}}\;  \times\nonumber\\
&\qquad\qquad  \times Q^{(2)}_{z\,\mu_3^{-1}\mu_2^{-1}\sqrt{t^3/q^6}}\; Q^{(1)}_{z\,\mu_3^{-1}\mu_2^{-1}\mu_1^{-1}\sqrt{t/q^2}} \left\langle\frac{Y^{(2)}_{z\,\mu_3^{-1}\mu_2^{-1} t^2/q^3}}{Y^{(3)}_{z\,\mu_3^{-1} t^3/q^4}\,Y^{(4)}_{z\, t^4/q^6}}\right\rangle\nonumber\\
&\;\;\;+\left[\widetilde{\fq}^{(4)}\right]^2\left[\widetilde{\fq}^{(3)}\right]^4\left[\widetilde{\fq}^{(2)}\right]^3\widetilde{\fq}^{(1)}\; Q^{(4)}_{z\, \sqrt{t/q}}\; Q^{(4)}_{z\, \sqrt{t^5/q^7}}\;Q^{(3)}_{z\,\mu_3^{-1}\sqrt{t/q}}\; Q^{(3)}_{z\,\mu_3^{-1} \sqrt{t^3/q^5}}\; \times\nonumber\\
&\qquad\qquad \times  Q^{(3)}_{z\,\mu_3^{-1} \sqrt{t^5/q^9}}\; Q^{(3)}_{z\,\mu_3^{-1} \sqrt{t^5/q^7}}\; Q^{(2)}_{z\,\mu_3^{-1}\mu_2^{-1}\sqrt{t/q^2}}\;  Q^{(2)}_{z\,\mu_3^{-1}\mu_2^{-1}\sqrt{t^3/q^6}}\; \times\nonumber\\
&\qquad\qquad \times Q^{(2)}_{z\,\mu_3^{-1}\mu_2^{-1}\sqrt{t^5/q^8}}\; Q^{(1)}_{z\,\mu_3^{-1}\mu_2^{-1}\mu_1^{-1}\sqrt{t/q^2}}\;Q^{(1)}_{z\,\mu_3^{-1}\mu_2^{-1}\mu_1^{-1}\sqrt{t^5/q^8}} \left\langle\frac{Y^{(4)}_{z\, t^3/q^5}}{Y^{(1)}_{z\,\mu_3^{-1}\mu_2^{-1}\mu_1^{-1} t^3/q^5}}\right\rangle\nonumber\\
&\;\;\;+\left[\widetilde{\fq}^{(4)}\right]^3\left[\widetilde{\fq}^{(3)}\right]^4\left[\widetilde{\fq}^{(2)}\right]^3\widetilde{\fq}^{(1)}\; Q^{(4)}_{z\, \sqrt{t/q}}\; Q^{(4)}_{z\, \sqrt{t^5/q^7}} \; Q^{(4)}_{z\, \sqrt{t^7/q^{11}}}\;Q^{(3)}_{z\,\mu_3^{-1}\sqrt{t/q}}\;  \times\nonumber\\
&\qquad\qquad \times Q^{(3)}_{z\,\mu_3^{-1} \sqrt{t^3/q^5}}\; Q^{(3)}_{z\,\mu_3^{-1} \sqrt{t^5/q^9}}\; Q^{(3)}_{z\,\mu_3^{-1} \sqrt{t^5/q^7}}\; Q^{(2)}_{z\,\mu_3^{-1}\mu_2^{-1}\sqrt{t/q^2}}\;  \times\nonumber\\
&\qquad\qquad \times Q^{(2)}_{z\,\mu_3^{-1}\mu_2^{-1}\sqrt{t^3/q^6}}\; Q^{(2)}_{z\,\mu_3^{-1}\mu_2^{-1}\sqrt{t^5/q^8}}\; Q^{(1)}_{z\,\mu_3^{-1}\mu_2^{-1}\mu_1^{-1}\sqrt{t/q^2}}\; \left\langle\frac{Y^{(1)}_{z\,\mu_3^{-1}\mu_2^{-1}\mu_1^{-1} t^2/q^3}\, Y^{(3)}_{z\,\mu_3^{-1} t^3/q^5}}{Y^{(2)}_{z\,\mu_3^{-1}\mu_2^{-1} t^3/q^5}\, Y^{(4)}_{z\, t^4/q^6}}\right\rangle\nonumber\\
&\;\;\;+\left[\widetilde{\fq}^{(4)}\right]^2\left[\widetilde{\fq}^{(3)}\right]^4\left[\widetilde{\fq}^{(2)}\right]^3\left[\widetilde{\fq}^{(1)}\right]^2\; Q^{(4)}_{z\, \sqrt{t/q}}\; Q^{(4)}_{z\, \sqrt{t^5/q^7}}\;Q^{(3)}_{z\,\mu_3^{-1}\sqrt{t/q}}\; Q^{(3)}_{z\,\mu_3^{-1} \sqrt{t^3/q^5}}\; \times\nonumber\\
&\qquad\qquad \times  Q^{(3)}_{z\,\mu_3^{-1} \sqrt{t^5/q^9}}\; Q^{(3)}_{z\,\mu_3^{-1} \sqrt{t^5/q^7}}\; Q^{(2)}_{z\,\mu_3^{-1}\mu_2^{-1}\sqrt{t/q^2}}\;  Q^{(2)}_{z\,\mu_3^{-1}\mu_2^{-1}\sqrt{t^3/q^6}}\; \times\nonumber\\
&\qquad\qquad \times Q^{(2)}_{z\,\mu_3^{-1}\mu_2^{-1}\sqrt{t^5/q^8}}\; Q^{(1)}_{z\,\mu_3^{-1}\mu_2^{-1}\mu_1^{-1}\sqrt{t/q^2}}\;Q^{(1)}_{z\,\mu_3^{-1}\mu_2^{-1}\mu_1^{-1}\sqrt{t^5/q^8}}\left\langle\frac{Y^{(4)}_{z\, t^3/q^5}}{Y^{(1)}_{z\,\mu_3^{-1}\mu_2^{-1}\mu_1^{-1} t^3/q^5}}\right\rangle\nonumber\\
&\;\;\;+\left[\widetilde{\fq}^{(4)}\right]^3\left[\widetilde{\fq}^{(3)}\right]^4\left[\widetilde{\fq}^{(2)}\right]^3\left[\widetilde{\fq}^{(1)}\right]^2\; Q^{(4)}_{z\, \sqrt{t/q}}\; Q^{(4)}_{z\, \sqrt{t^5/q^7}} \; Q^{(4)}_{z\, \sqrt{t^7/q^{11}}}\;Q^{(3)}_{z\,\mu_3^{-1}\sqrt{t/q}}\;  \times\nonumber\\
&\qquad\qquad \times Q^{(3)}_{z\,\mu_3^{-1} \sqrt{t^3/q^5}}\; Q^{(3)}_{z\,\mu_3^{-1} \sqrt{t^5/q^9}}\; Q^{(3)}_{z\,\mu_3^{-1} \sqrt{t^5/q^7}}\; Q^{(2)}_{z\,\mu_3^{-1}\mu_2^{-1}\sqrt{t/q^2}} \;  \times\nonumber\\
&\qquad\qquad \times Q^{(2)}_{z\,\mu_3^{-1}\mu_2^{-1}\sqrt{t^3/q^6}}\; Q^{(2)}_{z\,\mu_3^{-1}\mu_2^{-1}\sqrt{t^5/q^8}}\;Q^{(1)}_{z\,\mu_3^{-1}\mu_2^{-1}\mu_1^{-1}\sqrt{t^5/q^8}}\; Q^{(1)}_{z\,\mu_3^{-1}\mu_2^{-1}\mu_1^{-1}\sqrt{t/q^2}}\;  \times\nonumber\\
&\qquad\qquad \times \left\langle\frac{Y^{(3)}_{z\,\mu_3^{-1} t^3/q^5}}{Y^{(1)}_{z\,\mu_3^{-1}\mu_2^{-1}\mu_1^{-1} t^3/q^5}\, Y^{(4)}_{z\, t^4/q^6}}\right\rangle\nonumber\\
&\;\;\;+\left[\widetilde{\fq}^{(4)}\right]^3\left[\widetilde{\fq}^{(3)}\right]^5\left[\widetilde{\fq}^{(2)}\right]^3\left[\widetilde{\fq}^{(1)}\right]^2\; Q^{(4)}_{z\, \sqrt{t/q}}\; Q^{(4)}_{z\, \sqrt{t^5/q^7}} \; Q^{(4)}_{z\, \sqrt{t^7/q^{11}}}\;Q^{(3)}_{z\,\mu_3^{-1}\sqrt{t^7/q^{11}}}\; \times\nonumber\\
&\qquad\qquad \times Q^{(3)}_{z\,\mu_3^{-1}\sqrt{t/q}}\; Q^{(3)}_{z\,\mu_3^{-1} \sqrt{t^3/q^5}}\; Q^{(3)}_{z\,\mu_3^{-1} \sqrt{t^5/q^9}}\; Q^{(3)}_{z\,\mu_3^{-1} \sqrt{t^5/q^7}}\;  \times\nonumber\\
&\qquad\qquad \times Q^{(2)}_{z\,\mu_3^{-1}\mu_2^{-1}\sqrt{t/q^2}} \;  Q^{(2)}_{z\,\mu_3^{-1}\mu_2^{-1}\sqrt{t^3/q^6}}\; Q^{(2)}_{z\,\mu_3^{-1}\mu_2^{-1}\sqrt{t^5/q^8}}\;Q^{(1)}_{z\,\mu_3^{-1}\mu_2^{-1}\mu_1^{-1}\sqrt{t^5/q^8}}\;  \times\nonumber\\
&\qquad\qquad \times  Q^{(1)}_{z\,\mu_3^{-1}\mu_2^{-1}\mu_1^{-1}\sqrt{t/q^2}}\left\langle\frac{Y^{(2)}_{z\,\mu_3^{-1}\mu_2^{-1} t^3/q^5}}{Y^{(1)}_{z\,\mu_3^{-1}\mu_2^{-1}\mu_1^{-1} t^3/q^5}\, Y^{(3)}_{z\,\mu_3^{-1} t^4/q^6}}\right\rangle\nonumber\\ 
&\;\;\;+\left[\widetilde{\fq}^{(4)}\right]^3\left[\widetilde{\fq}^{(3)}\right]^5\left[\widetilde{\fq}^{(2)}\right]^4\left[\widetilde{\fq}^{(1)}\right]^2\; Q^{(4)}_{z\, \sqrt{t/q}}\; Q^{(4)}_{z\, \sqrt{t^5/q^7}} \; Q^{(4)}_{z\, \sqrt{t^7/q^{11}}}\;Q^{(3)}_{z\,\mu_3^{-1}\sqrt{t^7/q^{11}}}\; \times\nonumber\\
&\qquad\qquad \times Q^{(3)}_{z\,\mu_3^{-1}\sqrt{t/q}}\; Q^{(3)}_{z\,\mu_3^{-1} \sqrt{t^3/q^5}}\; Q^{(3)}_{z\,\mu_3^{-1} \sqrt{t^5/q^9}}\; Q^{(3)}_{z\,\mu_3^{-1} \sqrt{t^5/q^7}}\;  \times\nonumber\\
&\qquad\qquad \times Q^{(2)}_{z\,\mu_3^{-1}\mu_2^{-1}\sqrt{t^7/q^{12}}} \; Q^{(2)}_{z\,\mu_3^{-1}\mu_2^{-1}\sqrt{t/q^2}} \;  Q^{(2)}_{z\,\mu_3^{-1}\mu_2^{-1}\sqrt{t^3/q^6}}\; Q^{(2)}_{z\,\mu_3^{-1}\mu_2^{-1}\sqrt{t^5/q^8}}\; \times\nonumber\\
&\qquad\qquad \times Q^{(1)}_{z\,\mu_3^{-1}\mu_2^{-1}\mu_1^{-1}\sqrt{t^5/q^8}}\;  Q^{(1)}_{z\,\mu_3^{-1}\mu_2^{-1}\mu_1^{-1}\sqrt{t/q^2}}\left\langle\frac{Y^{(3)}_{z\,\mu_3^{-1} t^4/q^7}}{Y^{(2)}_{z\,\mu_3^{-1}\mu_2^{-1} t^4/q^7} }\right\rangle\nonumber\\
&\;\;\;+\left[\widetilde{\fq}^{(4)}\right]^3\left[\widetilde{\fq}^{(3)}\right]^6\left[\widetilde{\fq}^{(2)}\right]^4\left[\widetilde{\fq}^{(1)}\right]^2\; Q^{(4)}_{z\, \sqrt{t/q}}\; Q^{(4)}_{z\, \sqrt{t^5/q^7}} \; Q^{(4)}_{z\, \sqrt{t^7/q^{11}}}\;Q^{(3)}_{z\,\mu_3^{-1}\sqrt{t^9/q^{15}}}\; \times\nonumber\\
&\qquad\qquad \times Q^{(3)}_{z\,\mu_3^{-1}\sqrt{t^7/q^{11}}}\; Q^{(3)}_{z\,\mu_3^{-1}\sqrt{t/q}}\; Q^{(3)}_{z\,\mu_3^{-1} \sqrt{t^3/q^5}}\; Q^{(3)}_{z\,\mu_3^{-1} \sqrt{t^5/q^9}}\;  \times\nonumber\\
&\qquad\qquad \times  Q^{(3)}_{z\,\mu_3^{-1} \sqrt{t^5/q^7}}\; Q^{(2)}_{z\,\mu_3^{-1}\mu_2^{-1}\sqrt{t^7/q^{12}}} \; Q^{(2)}_{z\,\mu_3^{-1}\mu_2^{-1}\sqrt{t/q^2}} \;  Q^{(2)}_{z\,\mu_3^{-1}\mu_2^{-1}\sqrt{t^3/q^6}}\;  \times\nonumber\\
&\qquad\qquad \times Q^{(2)}_{z\,\mu_3^{-1}\mu_2^{-1}\sqrt{t^5/q^8}}\; Q^{(1)}_{z\,\mu_3^{-1}\mu_2^{-1}\mu_1^{-1}\sqrt{t^5/q^8}}\;  Q^{(1)}_{z\,\mu_3^{-1}\mu_2^{-1}\mu_1^{-1}\sqrt{t/q^2}}\left\langle\frac{Y^{(4)}_{z t^5/q^8}}{Y^{(3)}_{z\,\mu_3^{-1} t^5/q^8}}\right\rangle\nonumber\\
&\;\;\;+\left[\widetilde{\fq}^{(4)}\right]^4\left[\widetilde{\fq}^{(3)}\right]^6\left[\widetilde{\fq}^{(2)}\right]^4\left[\widetilde{\fq}^{(1)}\right]^2\;Q^{(4)}_{z\, \sqrt{t^{11}/q^{17}}}\; Q^{(4)}_{z\, \sqrt{t/q}}\; Q^{(4)}_{z\, \sqrt{t^5/q^7}} \; Q^{(4)}_{z\, \sqrt{t^7/q^{11}}}\; \times\nonumber\\
&\qquad\qquad \times Q^{(3)}_{z\,\mu_3^{-1}\sqrt{t^9/q^{15}}}\;Q^{(3)}_{z\,\mu_3^{-1}\sqrt{t^7/q^{11}}}\; Q^{(3)}_{z\,\mu_3^{-1}\sqrt{t/q}}\; Q^{(3)}_{z\,\mu_3^{-1} \sqrt{t^3/q^5}}\;  \times\nonumber\\
&\qquad\qquad \times Q^{(3)}_{z\,\mu_3^{-1} \sqrt{t^5/q^9}}\;  Q^{(3)}_{z\,\mu_3^{-1} \sqrt{t^5/q^7}}\; Q^{(2)}_{z\,\mu_3^{-1}\mu_2^{-1}\sqrt{t^7/q^{12}}} \; Q^{(2)}_{z\,\mu_3^{-1}\mu_2^{-1}\sqrt{t/q^2}} \;    \times\nonumber\\
&\;\;\; \times Q^{(2)}_{z\,\mu_3^{-1}\mu_2^{-1}\sqrt{t^3/q^6}}\; Q^{(2)}_{z\,\mu_3^{-1}\mu_2^{-1}\sqrt{t^5/q^8}}\; Q^{(1)}_{z\,\mu_3^{-1}\mu_2^{-1}\mu_1^{-1}\sqrt{t^5/q^8}}\;  Q^{(1)}_{z\,\mu_3^{-1}\mu_2^{-1}\mu_1^{-1}\sqrt{t/q^2}}\left\langle\frac{1}{Y^{(4)}_{z t^6/q^9}}\right\rangle\, .
\end{align}}

Note that the locus of the extra poles is precisely the argument of the fundamental hypermultiplet factors $Q^{(a)}$. We also note the appearance of prefactors $c_{[0,0]}$ and $c_{[0,0]'}$. As advertised in our general discussion \eqref{character5d}, such factors only depend on the $\Omega$-background spacetime equivariant parameters $q$ and $t$. We find in the case at hand:
\begin{align}
&c_{[0,0]}=-\frac{\left(q^2-q^{-2}\right)\left(q\,t-q^{-1}t^{-1}\right)}{\left(q-q^{-1}\right)\left(q^2/t-t/q^{2}\right)}\; ,\nonumber\\
&c_{[0,0]'}=\frac{\left(q^2/t^2-t^2/q^2\right)\left(q^3/t-t/q^3\right)}{\left(q^2/t-t/q^2\right)\left(q^3/t^2-t^2/q^{3}\right)}\; .\label{ccoefF4}
\end{align}

Performing the integrals, one finds
\beq\label{5ddefectexpression2F4}
\left\langle \left[Y^{(a)}_{5d}(z)\right]^{\pm 1} \right\rangle = \sum_{\{\overrightarrow{\boldsymbol{\mu}}\}}\left[ Z^{5d}_{bulk} \cdot Y^{(a)}_{5d}(z)^{\pm 1}\right]\, .
\eeq
Each sum is over a collection of 2d partitions, one for each $U(1)$ Coulomb parameter,
\beq
\{\overrightarrow{\boldsymbol{\mu}}\}=\{\boldsymbol{\mu}^{(a)}_i\}_{a=1, \ldots, 4,\, ; \,\; i=1,\ldots,n^{(a)}}\; .
\eeq
The factor  $Z^{5d}_{bulk}$ encodes all the 5d bulk physics. 
It is written in terms of the function \eqref{nekrasovN} as done in the main text.

Meanwhile, the Wilson loop factor after integration has the form
\beq\label{WilsonfactorF4}
Y^{(a)}_{5d}(z)\equiv\prod_{i=1}^{n^{(a)}}\prod_{k=1}^{\infty}\frac{1-t\, y^{(a)}_{i,k}/z}{1- y^{(a)}_{i,k}/z}\; .
\eeq

Above, we have defined the variables
\beq\label{ydefF4}
y^{(a)}_{i,k} = e^{(a)}_i\, q^{r^{(a)}{\boldsymbol{\mu}}^{(a)}_{i, k}} \, t^{-k}\;\;\;,\; k=1, \ldots, \infty\;,
\eeq 
where  $\boldsymbol{\mu}^{(a)}_{i, k}$ is the length of the $k$-th row of the partition $\boldsymbol{\mu}^{(a)}_{i}$.\\

Then, having performed all the integrals, the 5d partition function with $L=L^{(4)}=1$ evaluates to \eqref{F4result2}, where each vev is now understood to be written in terms of the variables \eqref{ydefF4}.\\

\vspace{10mm}

------- 3d Gauge Theory -------\\

We now make contact with the theory $G^{3d}$ on the vortices of $T^{5d}$. We  fix the total number $N_f$ of D5 branes wrapping the non-compact 2-cycles of the $E_6$ singularity. Further imposing the vanishing of D5 brane flux at infinity translates to the condition $[S+S^*]=0$ (now understood as an equation in the coweight lattice of $F_4$), which is equivalent to:
\begin{align}\label{conformalF4}
& 2\;n^{(1)}- 2 n^{(2)} =  N^{(1)}_f\nonumber\\
& 2\;n^{(2)}-\; n^{(1)}-2\; n^{(3)} = N^{(2)}_f\nonumber\\
& 2\;n^{(3)}- n^{(2)}- n^{(4)} =  N^{(3)}_f\nonumber\\
& 2\;n^{(4)}- n^{(3)} = N^{(4)}_f\; .
\end{align}
These equations further specify the ranks $n^{(a)}$ of the 5d gauge groups. The vanishing D5 flux constraint also specifies the Chern-Simons term on node $a$ to be $k^{(a)}_{CS}=n^{(a)}-n^{(a+1)}$. In terms of the representation theory of $F_4$, each non-compact D5 brane is labeled by a coweight of $F_4$, making up a set $\{\omega_d\}_{d=1}^{N_f}$. As we have seen a few times by now, the set needs to satisfy three conditions:
\begin{itemize}
	\item For all $1\leq d \leq N_f$, the coweight $\omega_d$ belongs in a fundamental representation of $^L\fg$.
	\item $\sum_{d=1}^{N_f} \omega_d=0$.
	\item No proper subset of coweights in $\{\omega_d\}$ adds up to $0$. 
\end{itemize}

\begin{figure}[h!]
	\emph{}
	\centering
	\includegraphics[trim={0 0 0 1cm},clip,width=0.9\textwidth]{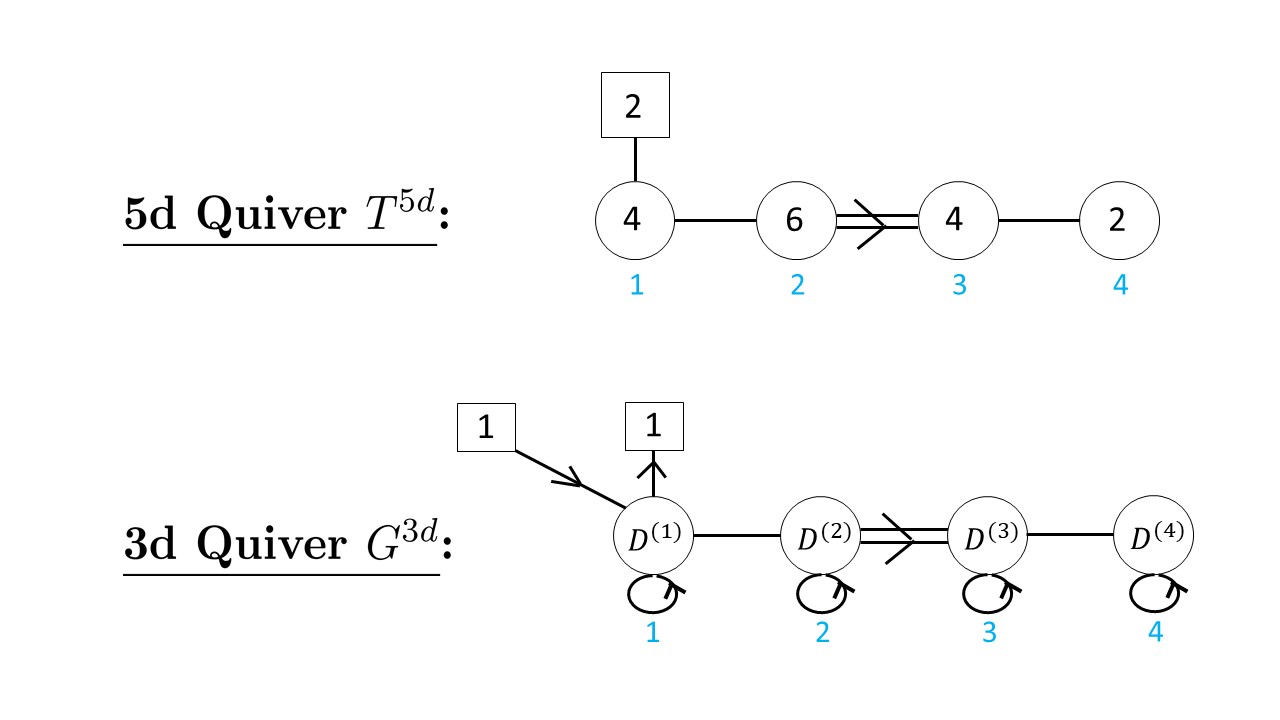}
	\vspace{-10pt}
	\caption{The $F_4$ theory $T^{5d}$ under study, along with the theory $G^{3d}$ on its vortices. The blue numbers label the nodes.} 
	\label{fig:F45d}
\end{figure}

For the purpose of being explicit, we form a defect out of the following set $\{\omega_d\}$ of 2 coweights:
\begin{align}
\omega_1 &= [-1, \phantom{-}0, \phantom{-}0, \phantom{-}0]= -\lambda^\vee_1\nonumber\\
\omega_2 &= [ \phantom{-}1, \phantom{-}0, \phantom{-}0, \phantom{-}0]= -\lambda^\vee_1+4\,\alpha^\vee_1+6\, \alpha^\vee_2+4\,\alpha^\vee_3+2\, \alpha^\vee_4\label{F4weight}
\end{align}
$\lambda_a$ is the $a$-th fundamental coweight, and $\alpha_a$ is the $a$-th positive simple coroot. We label the associated masses by $x_1$ and $x_2$, respectively. The specific 5d quiver gauge theory $T^{5d}$ engineered by these coweights is shown in figure \ref{fig:F45d}.\\

We further introduce $D^{(3)}=D^{(5)}$,\; $D^{(4)}=D^{(6)}$,\; $D^{(1)}$ and $D^{(2)}$ D3 branes wrapping the compact 2-cycles and one of the two complex lines, say $\mathbb{C}_q$, for a total of $D=\sum_{a=1}^{6}D^{(a)}$ D3 branes. We require that the D3 branes remain invariant under the $\mathbb{Z}_2$ outer automorphism action, after going around the origin of $\mathbb{C}_q$. 

From now on, we set the bifundamental masses to be   
\begin{align}\label{bifspecializedF4}
&\mu^{(1)}_{bif}= \sqrt{q^2/t} \; ,\nonumber\\
&\mu^{(2)}_{bif}= \sqrt{q/t} \; ,\nonumber\\
&\mu^{(3)}_{bif}= \sqrt{q/t} \; .
\end{align}
and we further set the 5d Coulomb moduli to be:
\begin{align}\label{specializeF4}
e^{(1)}_{1} &= q^{-1/2} v^{-1} t^{D^{(1)}_1}\, x_2 &  e^{(2)}_1 &=  q^{-1} v^{-2} t^{D^{(2)}_1}\, x_2\nonumber\\
e^{(1)}_2 &=  q^{-3/2} v^{-5} t^{D^{(1)}_2}\, x_2 &  e^{(2)}_2 &= q^{-1} v^{-4} t^{D^{(2)}_2}\, x_2\nonumber\\
e^{(1)}_3 &= q^{-3/2} v^{-7} t^{D^{(1)}_3}\, x_2 &  e^{(2)}_3 &= q^{-2} v^{-6} t^{D^{(2)}_3}\, x_2\nonumber\\
e^{(1)}_4 &= q^{-5/2} v^{-11} t^{D^{(1)}_4}\, x_2 &  e^{(2)}_4 &= q^{-1} v^{-6} t^{D^{(2)}_4}\, x_2\nonumber\\
&\left. \right.  &  e^{(2)}_5 &= q^{-2} v^{-8} t^{D^{(2)}_5}\, x_2\nonumber\\
&\left. \right.  &  e^{(2)}_6 &= q^{-2} v^{-10} t^{D^{(2)}_6}\, x_2\nonumber\\
& \left. \right. &  \left. \right. & \left. \right.\nonumber\\
e^{(3)}_1 &= q^{-1} v^{-3} t^{D^{(3)}_1}\, x_2 &  e^{(4)}_1 &= q^{-1} v^{-4} t^{D^{(4)}_1}\, x_2\nonumber\\
e^{(3)}_2 &=   q^{-1} v^{-5} t^{D^{(3)}_2}\, x_2 &  e^{(4)}_2 &= q^{-2} v^{-8} t^{D^{(4)}_2}\, x_2\\
e^{(3)}_3 &=   q^{-2} v^{-7} t^{D^{(3)}_3}\, x_2 & \left. \right. &\left. \right.\nonumber\\
e^{(3)}_4 &=  q^{-2} v^{-9} t^{D^{(3)}_4}\, x_2 &  \left. \right. &  \left. \right.\nonumber
\end{align}
Using the definition of the function \eqref{nekrasovN}, it follows that the 5d partition function is now a sum over truncated partitions, of length at most $D^{(a)}_i$. From this point on, we denote the 3d Coulomb parameters as $y^{(a)}_i$, for all $a=1, \ldots , 4$, and $i=1, \ldots, D^{(a)}$.\\

The 3d $\cN=2$ theory $G^{3d}$ is defined on $S^1(\widehat{R})\times\mathbb{C}$, and lives on the vortices of $T^{5d}$: it is a quiver gauge theory with gauge group $\prod_{a=1}^4 U(D^{(a)})$, and with 1 chiral and 1 antichiral matter multiplets on the first node, as pictured in figure \ref{fig:F45d}.

The first step is to define a 3d Wilson loop operator expectation value, as an integral over the Coulomb moduli of $G^{3d}$, where it is understood that none of the contours below enclose poles depending on $z$:
\begin{align}\label{3ddefectexpression2F4}
\left\langle\left[{\widetilde{Y}^{(a)}_{3d}}\right]^{\pm 1} \right\rangle \equiv \left[{\widetilde{Y}^{(a)}_{D1/D5}}(x, z)\right]\; \oint_{\{\overrightarrow{\boldsymbol{\mu}}\}} d{y}\,\left[I^{3d}_{bulk}(y)\, \left[{\widetilde{Y}^{(a)}_{D1/D3}}(y, z)\right]^{\pm 1}\right]  \; .
\end{align}
The bulk contribution
\begin{align}\label{bulk3dF4}
I^{3d}_{bulk}(y)=\prod_{a=1}^{4}\prod_{l=1}^{D^{(a)}}{y^{(a)}_l}^{\left(\tau^{(a)}-1\right)}\;I^{(a)}_{bulk, vec}\cdot I^{(a)}_{bulk, matter}\cdot I^{(1,2)}_{bulk, bif}\,I^{(2,3)}_{bulk, bif}\,I^{(3,4)}_{bulk, bif}\; .
\end{align}
is independent of the Wilson loop physics. For the sake of brevity, we will only write down here the $\cN=2$ matter content for the set of coweights $\{\omega_d\}$, given by 1 chiral and 1 antichiral multiplets on the first node:
\begin{align}\label{matter3dF4}
I^{(1)}_{bulk, matter}(y^{(1)}, \{x_d\})=\prod_{i=1}^{D^{(1)}} \frac{\left(\sqrt{t/q}\, x_1/y^{(1)}_{i};q^2\right)_{\infty}}{\left(\sqrt{t^{11}/q^{16}}\, x_2/y^{(1)}_{i};q^2\right)_{\infty}} \; .
\end{align}
The various powers of $q$ and $t$ in the argument of the $q$-Pochhammer symbols are fixed by the R-symmetry of $G^{3d}$.
The Wilson loop contributions are given 
\beq\label{3dWilsonfactorF4}
{\widetilde{Y}^{(a)}_{D1/D3}}(y^{(a)},z)=\prod_{i=1}^{D^{(a)}}\frac{1-t\, y^{(a)}_{i}/z}{1- y^{(a)}_{i}/z}\; .
\eeq
and by
\begin{align}
&{\widetilde{Y}^{(1)}_{D1/D5}}(\{x_d\}, z)=\left(1-\sqrt{t/q^2}\, x_2/z\right)\left(1-\sqrt{t^5/q^8}\, x_2/z\right)\left(1-\sqrt{t^7/q^{10}}\, x_2/z\right)\left(1-\sqrt{t^{11}/q^{16}}\, x_2/z\right)\, ,\nonumber\\
&{\widetilde{Y}^{(2)}_{D1/D5}}(\{x_d\}, z)=\left(1-\sqrt{t^2/q^4}\, x_2/z\right)\left(1-\sqrt{t^4/q^6}\, x_2/z\right)\left(1-\sqrt{t^6/q^{8}}\, x_2/z\right)\nonumber\\
&\qquad\qquad\qquad\qquad\left(1-\sqrt{t^{6}/q^{10}}\, x_2/z\right)\left(1-\sqrt{t^8/q^{12}}\, x_2/z\right)\left(1-\sqrt{t^{10}/q^{14}}\, x_2/z\right)\, ,\nonumber\\
&{\widetilde{Y}^{(3)}_{D1/D5}}(\{x_d\}, z)=\left(1-\sqrt{t^3/q^5}\, x_2/z\right)\left(1-\sqrt{t^5/q^7}\, x_2/z\right)\left(1-\sqrt{t^7/q^{11}}\, x_2/z\right)\left(1-\sqrt{t^{9}/q^{13}}\, x_2/z\right)\, ,\nonumber\\
&{\widetilde{Y}^{(4)}_{D1/D5}}(\{x_d\}, z)=\left(1-\sqrt{t^4/q^6}\, x_2/z\right)\left(1-\sqrt{t^8/q^{12}}\, x_2/z\right)\; .\label{3dWilsonfactor2F4}
\end{align}

For ease of presentation, we will slightly simplify the notation and write ${\widetilde{Y}^{(a)}_{D1/D3}}(y^{(a)},z)\equiv \widetilde{Y}^{(a)}_z$. We now write down the 3d partition function for the case $L=L^{(4)}=1$. It is a twisted 3d $qq$-character for the fourth fundamental representation of $U_q(\widehat{F_4})$:

{\allowdisplaybreaks
\begin{align}
\label{3dpartitionfunctionF4}
&\left[\chi^{F_4}\right]_{(0,0,0,1)}^{5d}(z) =\left(1-\sqrt{t^4/q^6}\, x_2/z\right)\left(1-\sqrt{t^8/q^{12}}\, x_2/z\right)\left\langle \widetilde{Y}^{(4)}_z \right\rangle\nonumber\\
&\;\;\; + \widetilde{\fq}^{(4)} \left(1-\sqrt{t^4/q^6}\, x_2/z\right)\left(1-\sqrt{t^8/q^{12}}\, x_2/z\right) \left\langle\frac{\widetilde{Y}^{(3)}_{z\, \sqrt{t/q^2}}}{\widetilde{Y}^{(4)}_{z\, t/q}}\right\rangle\nonumber\\
&\;\;\;+ \widetilde{\fq}^{(4)}\widetilde{\fq}^{(3)} \left(1-\sqrt{t^4/q^6}\, x_2/z\right)\left(1-\sqrt{t^8/q^{12}}\, x_2/z\right) \left\langle\frac{\widetilde{Y}^{(2)}_{z\, \sqrt{t^2/q^3}}}{\widetilde{Y}^{(3)}_{z\,\sqrt{t^3/q^3}}}\right\rangle\nonumber\\
&\;\;\;+ \widetilde{\fq}^{(4)}\widetilde{\fq}^{(3)}\widetilde{\fq}^{(2)} \left(1-\sqrt{t^4/q^6}\, x_2/z\right)\left(1-\sqrt{t^8/q^{12}}\, x_2/z\right) \left\langle\frac{\widetilde{Y}^{(1)}_{z\, \sqrt{t^3/q^4}}\,\widetilde{Y}^{(3)}_{z\,\sqrt{t^3/q^5}}}{\widetilde{Y}^{(2)}_{z\,\sqrt{t^4/q^6}}}\right\rangle\nonumber\\
&\;\;\;+ \widetilde{\fq}^{(4)}\widetilde{\fq}^{(3)}\widetilde{\fq}^{(2)}\widetilde{\fq}^{(1)} \left(1-\sqrt{t^4/q^6}\, x_2/z\right)\left(1-\sqrt{q^6/t^{4}}\, x_1/z\right)     \left\langle\frac{\widetilde{Y}^{(3)}_{z\,\sqrt{t^3/q^5}}}{\widetilde{Y}^{(1)}_{z\,\sqrt{t^5/q^8}}}\right\rangle\nonumber\\
&\;\;\;+\widetilde{\fq}^{(4)}\left[\widetilde{\fq}^{(3)}\right]^2\widetilde{\fq}^{(2)} \left(1-\sqrt{t^4/q^6}\, x_2/z\right)\left(1-\sqrt{t^8/q^{12}}\, x_2/z\right) \left\langle\frac{\widetilde{Y}^{(1)}_{z\, \sqrt{t^3/q^4}}\,\widetilde{Y}^{(4)}_{z\, t^2/q^3}}{\widetilde{Y}^{(3)}_{z\,\sqrt{t^5/q^7}}}\right\rangle\nonumber\\
&\;\;\;+\widetilde{\fq}^{(4)}\left[\widetilde{\fq}^{(3)}\right]^2\widetilde{\fq}^{(2)}\widetilde{\fq}^{(1)} \left(1-\sqrt{t^4/q^6}\, x_2/z\right)\left(1-\sqrt{q^6/t^{4}}\, x_1/z\right)   \left\langle\frac{\widetilde{Y}^{(2)}_{z\, \sqrt{t^4/q^6}}\,\widetilde{Y}^{(4)}_{z\, t^2/q^3}}{\widetilde{Y}^{(1)}_{z\,\sqrt{t^5/q^8}}\, \widetilde{Y}^{(3)}_{z\,\sqrt{t^5/q^7}}}\right\rangle\nonumber\\    
&\;\;\;+\left[\widetilde{\fq}^{(4)}\right]^2\left[\widetilde{\fq}^{(3)}\right]^2\widetilde{\fq}^{(2)}\left(1-\sqrt{t^4/q^6}\, x_2/z\right)\left(1-\sqrt{t^8/q^{12}}\, x_2/z\right) \left\langle\frac{\widetilde{Y}^{(1)}_{z\, \sqrt{t^3/q^4}}}{\widetilde{Y}^{(4)}_{z\, t^3/q^4}}\right\rangle\nonumber\\
&\;\;\;+\widetilde{\fq}^{(4)}\left[\widetilde{\fq}^{(3)}\right]^2\left[\widetilde{\fq}^{(2)}\right]^2\widetilde{\fq}^{(1)} \left(1-\sqrt{t^4/q^6}\, x_2/z\right)\left(1-\sqrt{q^6/t^{4}}\, x_1/z\right)   \left\langle\frac{\widetilde{Y}^{(3)}_{z\, \sqrt{t^5/q^7}}\,\widetilde{Y}^{(4)}_{z\, t^2/q^3}}{\widetilde{Y}^{(2)}_{z\,\sqrt{t^6/q^{10}}}}\right\rangle\nonumber\\
&\;\;\;+\left[\widetilde{\fq}^{(4)}\right]^2\left[\widetilde{\fq}^{(3)}\right]^2\widetilde{\fq}^{(2)}\widetilde{\fq}^{(1)} \left(1-\sqrt{t^4/q^6}\, x_2/z\right)\left(1-\sqrt{q^6/t^{4}}\, x_1/z\right)   \left\langle\frac{\widetilde{Y}^{(2)}_{z\, \sqrt{t^4/q^6}}}{\widetilde{Y}^{(1)}_{z\, \sqrt{t^5/q^8}}\,\widetilde{Y}^{(4)}_{z\, t^3/q^4}}\right\rangle\nonumber\\
&\;\;\;+\left[\widetilde{\fq}^{(4)}\right]^2\left[\widetilde{\fq}^{(3)}\right]^2\left[\widetilde{\fq}^{(3)}\right]^{2}\widetilde{\fq}^{(1)} \left(1-\sqrt{t^4/q^6}\, x_2/z\right)\left(1-\sqrt{q^6/t^{4}}\, x_1/z\right)  \left\langle\frac{\widetilde{Y}^{(3)}_{z\, \sqrt{t^5/q^7}}\, \widetilde{Y}^{(3)}_{z\, \sqrt{t^5/q^9}}}{\widetilde{Y}^{(2)}_{z\, \sqrt{t^6/q^{10}}}\,\widetilde{Y}^{(4)}_{z\, t^3/q^4}}\right\rangle\nonumber\\
&\;\;\;+\widetilde{\fq}^{(4)}\left[\widetilde{\fq}^{(3)}\right]^3\left[\widetilde{\fq}^{(2)}\right]^2\widetilde{\fq}^{(1)} \left(1-\sqrt{t^4/q^6}\, x_2/z\right)\left(1-\sqrt{q^6/t^{4}}\, x_1/z\right)   \left\langle\frac{\widetilde{Y}^{(4)}_{z\, t^2/q^3}\,\widetilde{Y}^{(4)}_{z\, t^3/q^5}}{\widetilde{Y}^{(3)}_{z\,\sqrt{t^7/q^{11}}}}\right\rangle\nonumber\\
&\;\;\;+c_{[0,0]}\left[\widetilde{\fq}^{(4)}\right]^2\left[\widetilde{\fq}^{(3)}\right]^3\left[\widetilde{\fq}^{(2)}\right]^2\widetilde{\fq}^{(1)} \left(1-\sqrt{t^4/q^6}\, x_2/z\right)\left(1-\sqrt{q^6/t^{4}}\, x_1/z\right)   \left\langle\frac{\widetilde{Y}^{(3)}_{z\,\sqrt{t^5/q^{7}}}\,\widetilde{Y}^{(4)}_{z\, t^3/q^5}}{\widetilde{Y}^{(3)}_{z\,\sqrt{t^7/q^{11}}}\, \widetilde{Y}^{(4)}_{z\, t^3/q^4}}\right\rangle\nonumber\\
&\;\;\;+c_{[0,0]'}\left[\widetilde{\fq}^{(4)}\right]^2\left[\widetilde{\fq}^{(3)}\right]^3\left[\widetilde{\fq}^{(2)}\right]^2\widetilde{\fq}^{(1)} \left(1-\sqrt{t^4/q^6}\, x_2/z\right)\left(1-\sqrt{q^6/t^{4}}\, x_1/z\right)   \left\langle\frac{\widetilde{Y}^{(4)}_{z\, t^2/q^3}}{\widetilde{Y}^{(4)}_{z\, t^4/q^6}}\right\rangle\nonumber\\
&\;\;\;+\left[\widetilde{\fq}^{(4)}\right]^2\left[\widetilde{\fq}^{(3)}\right]^4\left[\widetilde{\fq}^{(2)}\right]^2\widetilde{\fq}^{(1)} \left(1-\sqrt{t^4/q^6}\, x_2/z\right)\left(1-\sqrt{q^6/t^{4}}\, x_1/z\right)   \left\langle\frac{\widetilde{Y}^{(2)}_{z\,\sqrt{t^6/q^{8}}}\,\widetilde{Y}^{(4)}_{z\, t^3/q^5}}{\widetilde{Y}^{(3)}_{z\,\sqrt{t^7/q^{11}}}\, \widetilde{Y}^{(3)}_{z\,\sqrt{t^7/q^{9}}}}\right\rangle\nonumber\\
&\;\;\;+\left[\widetilde{\fq}^{(4)}\right]^3\left[\widetilde{\fq}^{(3)}\right]^3\left[\widetilde{\fq}^{(2)}\right]^2\widetilde{\fq}^{(1)} \left(1-\sqrt{t^4/q^6}\, x_2/z\right)\left(1-\sqrt{q^6/t^{4}}\, x_1/z\right)   \left\langle\frac{\widetilde{Y}^{(3)}_{z\,\sqrt{t^5/q^{7}}}}{\widetilde{Y}^{(4)}_{z\, t^4/q^3}\,\widetilde{Y}^{(4)}_{z\, t^4/q^6}}\right\rangle\nonumber\\
&\;\;\;+\left[\widetilde{\fq}^{(4)}\right]^2\left[\widetilde{\fq}^{(3)}\right]^4\left[\widetilde{\fq}^{(2)}\right]^3\widetilde{\fq}^{(1)} \left(1-\sqrt{t^4/q^6}\, x_2/z\right)\left(1-\sqrt{q^6/t^{4}}\, x_1/z\right)   \left\langle\frac{\widetilde{Y}^{(1)}_{z\,\sqrt{t^7/q^{10}}}\,\widetilde{Y}^{(4)}_{z\, t^3/q^5}}{\widetilde{Y}^{(2)}_{z\,\sqrt{t^8/q^{12}}}}\right\rangle\nonumber\\
&\;\;\;+\left[\widetilde{\fq}^{(4)}\right]^3\left[\widetilde{\fq}^{(3)}\right]^4\left[\widetilde{\fq}^{(2)}\right]^2\widetilde{\fq}^{(1)} \left(1-\sqrt{t^4/q^6}\, x_2/z\right)\left(1-\sqrt{q^6/t^{4}}\, x_1/z\right)   \left\langle\frac{\widetilde{Y}^{(2)}_{z\,\sqrt{t^6/q^{8}}}}{\widetilde{Y}^{(3)}_{z\,\sqrt{t^7/q^{9}}}\,\widetilde{Y}^{(4)}_{z\, t^4/q^6}}\right\rangle\nonumber\\
&\;\;\;+\left[\widetilde{\fq}^{(4)}\right]^2\left[\widetilde{\fq}^{(3)}\right]^4\left[\widetilde{\fq}^{(2)}\right]^3\widetilde{\fq}^{(1)} \left(1-\sqrt{q^{12}/t^8}\, x_1/z\right)\left(1-\sqrt{q^6/t^{4}}\, x_1/z\right)   \left\langle\frac{\widetilde{Y}^{(4)}_{z\, t^3/q^5}}{\widetilde{Y}^{(1)}_{z\,\sqrt{t^9/q^{14}}}}\right\rangle\nonumber\\
&\;\;\;+\left[\widetilde{\fq}^{(4)}\right]^3\left[\widetilde{\fq}^{(3)}\right]^4\left[\widetilde{\fq}^{(2)}\right]^3\widetilde{\fq}^{(1)} \left(1-\sqrt{t^4/q^6}\, x_2/z\right)\left(1-\sqrt{q^6/t^{4}}\, x_1/z\right)   \left\langle\frac{\widetilde{Y}^{(1)}_{z\,\sqrt{t^7/q^{10}}}\, \widetilde{Y}^{(3)}_{z\,\sqrt{t^7/q^{11}}}}{\widetilde{Y}^{(2)}_{z\,\sqrt{t^8/q^{12}}}\, \widetilde{Y}^{(4)}_{z\, t^4/q^6}}\right\rangle\nonumber\\
&\;\;\;+\left[\widetilde{\fq}^{(4)}\right]^2\left[\widetilde{\fq}^{(3)}\right]^4\left[\widetilde{\fq}^{(2)}\right]^3\left[\widetilde{\fq}^{(1)}\right]^2\left(1-\sqrt{q^{12}/t^8}\, x_1/z\right)\left(1-\sqrt{q^6/t^{4}}\, x_1/z\right)\left\langle\frac{\widetilde{Y}^{(4)}_{z\, t^3/q^5}}{\widetilde{Y}^{(1)}_{z\,\sqrt{t^9/q^{14}}}}\right\rangle\nonumber\\
&\;\;\;+\left[\widetilde{\fq}^{(4)}\right]^3\left[\widetilde{\fq}^{(3)}\right]^4\left[\widetilde{\fq}^{(2)}\right]^3\left[\widetilde{\fq}^{(1)}\right]^2\left(1-\sqrt{q^{12}/t^8}\, x_1/z\right)\left(1-\sqrt{q^6/t^{4}}\, x_1/z\right) \left\langle\frac{\widetilde{Y}^{(3)}_{z\,\sqrt{t^7/q^{11}}}}{\widetilde{Y}^{(1)}_{z\,\sqrt{t^9/q^{14}}}\, \widetilde{Y}^{(4)}_{z\, t^4/q^6}}\right\rangle\nonumber\\
&\;\;\;+\left[\widetilde{\fq}^{(4)}\right]^3\left[\widetilde{\fq}^{(3)}\right]^5\left[\widetilde{\fq}^{(2)}\right]^3\left[\widetilde{\fq}^{(1)}\right]^2\left(1-\sqrt{q^{12}/t^8}\, x_1/z\right)\left(1-\sqrt{q^6/t^{4}}\, x_1/z\right)\left\langle\frac{\widetilde{Y}^{(2)}_{z\,\sqrt{t^8/q^{12}}}}{\widetilde{Y}^{(1)}_{z\,\sqrt{t^9/q^{14}}}\, \widetilde{Y}^{(3)}_{z\,\sqrt{t^9/q^{13}}}}\right\rangle\nonumber\\ 
&\;\;\;+\left[\widetilde{\fq}^{(4)}\right]^3\left[\widetilde{\fq}^{(3)}\right]^5\left[\widetilde{\fq}^{(2)}\right]^4\left[\widetilde{\fq}^{(1)}\right]^2\left(1-\sqrt{q^{12}/t^8}\, x_1/z\right)\left(1-\sqrt{q^6/t^{4}}\, x_1/z\right)\left\langle\frac{\widetilde{Y}^{(3)}_{z\,\sqrt{t^9/q^{15}}}}{\widetilde{Y}^{(2)}_{z\,\sqrt{t^{10}/q^{16}}}}\right\rangle\nonumber\\
&\;\;\;+\left[\widetilde{\fq}^{(4)}\right]^3\left[\widetilde{\fq}^{(3)}\right]^6\left[\widetilde{\fq}^{(2)}\right]^4\left[\widetilde{\fq}^{(1)}\right]^2\left(1-\sqrt{q^{12}/t^8}\, x_1/z\right)\left(1-\sqrt{q^6/t^{4}}\, x_1/z\right)\left\langle\frac{\widetilde{Y}^{(4)}_{z t^5/q^8}}{\widetilde{Y}^{(3)}_{z\,\sqrt{t^{11}/q^{17}}}}\right\rangle\nonumber\\
&\;\;\;+\left[\widetilde{\fq}^{(4)}\right]^4\left[\widetilde{\fq}^{(3)}\right]^6\left[\widetilde{\fq}^{(2)}\right]^4\left[\widetilde{\fq}^{(1)}\right]^2\left(1-\sqrt{q^{12}/t^8}\, x_1/z\right)\left(1-\sqrt{q^6/t^{4}}\, x_1/z\right)\left\langle\frac{1}{\widetilde{Y}^{(4)}_{z t^6/q^9}}\right\rangle\; .
\end{align}}

As described in the main text, this can be proved in a straightforward way starting from 5d. Namely, we first use the specialization of the Coulomb moduli \eqref{specializeF4} to prove
\beq\label{YequalityF4}
\left\langle{\widetilde{Y}^{(a)}_{3d}(z)} \right\rangle = c_{3d}\, \left\langle {Y_{5d}(z)} \right\rangle_{e^{(a)}_{i}=\;\eqref{specializeF4}} \; .
\eeq
The proportionality constant $c_{3d}$ is the bulk residue at the empty partition, see \eqref{c3d}.\\

The second step is to show that each term in the 5d $qq$-character gives a term in the 3d $qq$-character, resulting in
\begin{align}\label{CHIequalityF4}
\left[\widetilde{\chi}^{F_4}\right]_{(0,0,0,1)}^{3d}(z)=c_{3d}\;  \left[\chi^{F_4}\right]_{(0,0,0,1)}^{5d}(z)_{e^{(a)}_{i}=\; \eqref{specializeF4}} \; .
\end{align} 

\vspace{10mm}

------- ${\cW}_{q,t}(F_4)$ correlator -------\\

We now reinterpret the physics in terms of $F_4$-type $q$-Toda theory on the cylinder $\cC$. The theory enjoys a ${\cW}_{q,t}(F_4)$ algebra symmetry, which is generated by 4 currents $W^{(s)}(z)$, with $s=2,\ldots, 5$. To make contact with the 3d gauge theory results above, we first need to construct the operator $W^{(5)}(z)$, whose insertion at a point $z$ on $\cC$ corresponds to the location of a D1 brane. To proceed, one constructs the operator as a commutant of the screening charges. The end result is Laurent polynomial in the $\cY^{(a)}$ operators defined in \eqref{YoperatorToda}. Conveniently, this polynomial can be read off directly from the 3d Wilson loop operators \eqref{3dpartitionfunctionF4} : 
\begin{align}\label{F4stress}
&W^{(5)}(z) = :\cY^{(4)}(z): + :\cY^{(3)}(\sqrt{t/q^2}\;z)\left[\cY^{(4)}((t/q)z)\right]^{-1}: +\ldots
\end{align}

We now  build the vertex operator corresponding to the D5 brane content. Such an operator is fully determined by the set \eqref{F4weight} of coweights $\{\omega_d\}$. In the case at hand, we define
\begin{align}\label{F4vertexweight}
V_{\omega_1} (x_1)&=\; :\left[\Lambda^{(1)}(x_1)\right]^{-1}:\nonumber\\
V_{\omega_2} (x_2)&=\; :\Lambda^{(2)}(x_2\, \sqrt{t^{12}/q^{18}}):\; ,
\end{align}
where $\Lambda^{(a)}$ is the $a$-th fundamental coweight operator \eqref{coweightvertexdef}. Then, the D5 brane operator is defined as:
\beq\label{F4genericvertexexmaple}
V_\eta (u)= \, :\prod_{d=1}^{N_f=2} V_{\omega_d}(x_d): \; .
\eeq
The parameters $u$ and $\eta$ are defined through
\begin{align}\label{F4commomentum}
&\eta = \sum_{d=1}^{2}\eta_d\, \omega_d \; ,\\
&x_d = u \, q^{-\eta_d}\; ,\qquad\;\; d=1,2\; .
\end{align}

We find by explicit computation that the following ${\cW}_{q,t}(F_4)$ correlator is a 3d gauge theory partition function:
\beq\label{F4correlatoris3dindex}
\left\langle \psi'\left|\prod_{d=1}^{2} V_{\omega_d}(x_d)\; \prod_{a=1}^{4} (Q^{(a)})^{D^{(a)}}\; W^{(5)}(z) \right| \psi \right\rangle = A(\{x_d\})\, B(\{x_d\}, z)\, \left[\widetilde{\chi}^{F_4}\right]_{(0,0,0,1)}^{3d}(z)\; .
\eeq

To prove the above, one simply computes the various two-points making up the correlator. Most of these two-points are straightforward to evaluate, except perhaps for the ones involving the D5 brane vertex operator. Here, we will only write down explicitly the two-point of this operator with the screenings:
\beq
\prod_{a=1}^{4}\prod_{i=1}^{D^{(a)}}\left\langle  :\prod_{d=1}^{2} V_{\omega_d}(x_d):\; S^{(a)}(y^{(a)}_i) \right\rangle = \prod_{i=1}^{D^{(1)}} \frac{\left(\sqrt{t/q}\, x_1/y^{(1)}_{i};q^2\right)_{\infty}}{\left(\sqrt{t^{11}/q^{16}}\, x_2/y^{(1)}_{i};q^2\right)_{\infty}} \; ,
\eeq
which is exactly the 3d $\cN=2$ matter content \eqref{matter3dF4}.\\

The two-point of the D5 brane operator with the generating current $W^{(5)}(z)$ is needed to identify exactly the prefactor $B(\{x_d\}, z)$. This is done by computing the inverse deformed $F_4$-Cartan matrix and making use of the commutator \eqref{commutator3}. Because the expression for $B(\{x_d\}, z)$ turns out to be rather involved, we will not report it here, though we have calculated it explicitly to establish \eqref{F4correlatoris3dindex}.\\

Moreover, the equality of the 5d and 3d partition functions \eqref{CHIequalityF4} further implies that 
\begin{align}\label{F4correlatoris5dindex}
\left\langle \psi'\left|\prod_{d=1}^{2} V_{\omega_d}(x_d)\; \prod_{a=1}^{4} (Q^{(a)})^{D^{(a)}}\; W^{(5)}(z) \right| \psi \right\rangle = A(\{x_d\})\, B(\{x_d\}, z) \,c_{3d}\, \left[\chi^{F_4}\right]_{(0,0,0,1)}^{5d}(z)_{e^{(a)}_{i}=\; \eqref{specializeF4}}\; .
\end{align}

Though we have not checked explicitly the generalization to $L>1$ D1 branes in the $F_4$ case, we naturally claim that the following identity holds, following the arguments in the main text: 
\begin{align}\label{F4correlatoris3dindexMORE}
&\left\langle \psi'\left|\prod_{d=1}^{N_f} V_{\omega_d}(x_d)\; \prod_{a=1}^{4} (Q^{(a)})^{D^{(a)}}\; \prod_{s=2}^{5}\prod_{\rho=1}^{L^{(s-1)}}W^{(s)}(z^{(s-1)}_\rho) \right| \psi \right\rangle \nonumber\\
&\qquad\qquad\qquad= A(\{x_d\})\; B\left(\{x_d\},\{z^{(s-1)}_\rho\}\right) \left[\widetilde{\chi}^{F_4}\right]_{(L^{(1)},L^{(2)},L^{(3)},L^{(4)})}^{3d}(z^{(s-1)}_\rho)\; .
\end{align}

\pagebreak

\section*{Acknowledgments}
We thank Alexei Morozov and Yegor Zenkevich for their patience in explaining to us their computation of the deformed stress tensor expectation value in $q$-Liouville. We thank Mina Aganagic, Nikita Nekrasov and Luigi Tizzano for discussions and comments at various stages of this project. The research of N. H. is supported  by the Simons Center for Geometry and Physics.

\newpage
\bibliography{summaryqq}
\bibliographystyle{JHEP}

\end{document}